\documentclass[12pt,epsf]{book}
\usepackage{latexsym,amsfonts,amsmath,amscd,amssymb,epsf}
\usepackage[latin1]{inputenc}
\usepackage[american]{babel}
\usepackage[dvips]{graphicx}
\usepackage{bbm}
\usepackage{epsfig}
\usepackage{graphics,color}

\advance \topmargin by -\headheight
\advance \topmargin by -\headsep

\evensidemargin \oddsidemargin
 \def\mod{{\hbox{\rm mod}}} \def\tr{{\hbox{\rm Tr}}}

\def\ie{{\em i.e.}}
\def\ie{\hbox{\it i.e.}}

\def\CC{{\mathchoice
{\rm C\mkern-8mu\vrule height1.45ex depth-.05ex
width.05em\mkern9mu\kern-.05em}
{\rm C\mkern-8mu\vrule height1.45ex depth-.05ex
width.05em\mkern9mu\kern-.05em}
{\rm C\mkern-8mu\vrule height1ex depth-.07ex
width.035em\mkern9mu\kern-.035em}
{\rm C\mkern-8mu\vrule height.65ex depth-.1ex
width.025em\mkern8mu\kern-.025em}}}

\def\RR{{\rm I\kern-1.6pt {\rm R}}}

\def\ZZ{{\rm Z}\kern-3.8pt {\rm Z} \kern2pt}
\def\IB{\relax{\rm I\kern-.18em B}}
\def\ID{\relax{\rm I\kern-.18em D}}
\def\II{\relax{\rm I\kern-.18em I}}
\def\IP{\relax{\rm I\kern-.18em P}}

\def\np{Nucl. Phys.}
\def\pl{Phys. Lett.}
\def\prl{Phys. Rev. Lett.}
\def\pr{Phys. Rev.}

\def\jhep{J. High Energy Phys.}

\newcommand{\beq}{\begin{equation}}
\newcommand{\eeq}{\end{equation}}
\newcommand{\rc}{\nonumber\\}
\newcommand{\bear}{\begin{eqnarray}}
\newcommand{\eear}{\end{eqnarray}}

\newcommand{\sub}[2]{#1_\text{#2}}

\newcommand{\hyph}[1]{$#1$\nobreakdash-\hspace{0pt}}

\newcommand{\cO}{{\cal O}}

\newcommand{\cW}{{\cal W}}

\newcommand{\bbZ}{\mathbb{Z}}
\newcommand{\bbC}{\mathbb{C}}

\def\to{\rightarrow}

\def\tr{{\rm Tr}}
\def\to{\rightarrow}

\newfont{\namefont}{cmr10}
\newfont{\addfont}{cmti7 scaled 1440}
\newfont{\boldmathfont}{cmbx10}
\newfont{\headfontb}{cmbx10 scaled 1728}
\def\ie{{\it i.e.}}

\def\revise#1       {\raisebox{-0em}{\rule{3pt}{1em}}%
                     \marginpar{\raisebox{.5em}{\vrule width3pt\
                     \vrule width0pt height 0pt depth0.5em
                     \hbox to 0cm{\hspace{0cm}{%
                     \parbox[t]{4em}{\raggedright\footnotesize{#1}}}\hss}}}}

\def\calq         {{\cal Q}}

\def\reals        {{\mathbb R}}
\def\zet          {{\mathbb Z}}

\def\ee           {{\rm e}}

\def\tr           {\mathop{\rm Tr}}

\def\sqr#1#2{{\vcenter{\vbox{\hrule height.#2pt
 \hbox{\vrule width.#2pt height#1pt \kern#1pt
 \vrule width.#2pt}\hrule height.#2pt}}}}

\def\a{\alpha}
\def\b{\beta}

\def\be{\begin{equation}}
\def\ee{\end{equation}}

\def\m{\mu}
\def\g{\gamma}

\def \ov{\over}

\def\ep{\epsilon}

\def \p {\phi}
\def \te {\tilde \epsilon}

\def\tr{{\tilde\rho}}
\newcommand{\eel}[1]{\label{#1}\end{equation}}
\newcommand{\bea}{\begin{eqnarray}}
\newcommand{\eea}{\end{eqnarray}}
\newcommand{\eeal}[1]{\label{#1}\end{eqnarray}}

\newcommand{\Nugual}[1]{$\mathcal{N}= #1 $}
\newcommand{\nn}{\nonumber}
\newcommand{\bsp}[1]{\begin{equation} \begin{split} #1 \end{split} \end{equation}}

\def \ov {\over}

\def\ep {\epsilon}

\def \p {\phi}
\def \te {\tilde \epsilon}

\def\m{\mu}

\def\a{\alpha}
\def\b{\beta}
\def\g{\gamma}
\def\G{\Gamma}

\def\te{\theta}

\def\p{\phi}

\def\ep{\epsilon}

\def\be{\begin{equation}}
\def\ee{\end{equation}}

\def \g {\gamma}
\def \G {\Gamma}
\def \k {\kappa}

\def \Tr {{\rm Tr}}

\def \m {\mu}

\def\te{\theta}
\def\g{\gamma}

\def\Ypq{Y^{p,q}}
\def\CYpq{\mathcal{C}Y^{p,q}}
\def\Labc{L^{a,b,c}}

\def\CC{\mathcal{C}}
\def\CP{\mathcal{P}}

\def\tr{{\rm Tr}}

\def\lbldef#1#2{\expandafter\gdef\csname #1\endcsname {#2}}

\def\href#1#2{#2}

\definecolor{M_Beige}         {rgb}{0.96 , 0.96 , 0.86}

\definecolor{M_Brown}         {rgb}{0.65 , 0.16 , 0.16}
\definecolor{M_Gold}          {rgb}{0.12 , 0.84 , 0.30}

\definecolor{M_LemonChiffon}  {rgb}{1.00 , 0.98 , 0.80}

\definecolor{M_Orange}        {rgb}{1.00 , 0.60 , 0.00}

\definecolor{M_Pink}          {rgb}{1.00 , 0.75 , 0.80}

\definecolor{M_Gre}          {rgb}{0.05 ,0.46 , 0.00}

\begin{document}


\begin{titlepage}

\begin{center} 

\vskip.5cm


\vspace{.3cm}

\hspace{0.5cm}
\Large Universidade de Santiago de Compostela

\large Departamento de F\'\i sica de Part\'\i culas 

\end{center}

\vspace{6cm}

\begin{center} 

\LARGE  \bf D-BRANES IN SUPERSYMMETRIC BACKGROUNDS

\end{center}

\vspace{3cm}

\vspace{5cm}

\begin{center} 

{\sf\bf \large Felipe Canoura Fern\'andez}

\sf Santiago de Compostela, xaneiro de 2008.

\end{center} 
\
\newpage
\
\setcounter{page}{0}

\end{titlepage}


\pagestyle{headings}

\setcounter{page}{1}

\tableofcontents


\chapter*{Motivation}
\addcontentsline{toc}{chapter}{Motivation}
\medskip

String Theory was born during the 1960's in the framework of the hadronic physic, as an attempt to explain strong interactions. The idea was to consider the strings as tubes of flux which mediated the hadronic interaction. Very soon that phenomenological idea was obscured by the formulation of the ``Quantum Chromodynamics" or QCD and subsequently by the formulation of the ``Standard Model". This model describes very successfully three of the four fundamental interactions, namely the electromagnetic, weak and strong interactions. Although the consistency of the standard model is based on the existence of a particle (the Higgs boson) which has not been found yet, it seems to be in very good shape experimentally.

Unlike the standard model, the fourth interaction (gravity) has serious inconsistencies. The theory is not renormalisable and (the loop corrections of) any physical quantity that we compute in quantum gravity depends on an infinite number of parameters. However this is a theoretical problem since for any value of the unknown parameters, their effect becomes negligible at observable energies. For this reason the classical limit of the quantum gravity (General Relativity) impressively agrees with experiments.

The real interest in string theory began in the 1980's. By thinking the elementary particles as vibrations of a one-dimensional object (string), the graviton (a massless particle with spin two which mediates the gravity interaction) comes up naturally in the spectrum of the theory. Moreover the theory is consistent once we consider its supersymmetric extension. This is a good point in favour of string theory since it seems to be the most serious candidate for a consistent theory of the quantum gravity. But there is more. String theory does not just contain gravity but it comes inevitably with a large number of particles and interactions which have the same features of the standard model. Standard texts on string theory are \cite{stringbooks}.

Unfortunately, the particles and interactions that string theory predicts are far from unique. There are many possibilities and it is still not clear whether the standard model is among them. At phenomenological level, it is still a challenge the attempt of selecting a vacuum among all the possibilities when one reduces the dimensionality of the spacetime from the critical dimension of the superstrings to four dimensions. Moreover, there are five consistent string theories and they can be thought of as being different perturbative regimes of an still not completely uncovered (beyond the low energy limit) theory, called M-theory, where the fundamental objects turn out to be two-dimensional membranes.  These five string theories are related by a chain of dualities which connect in a non-perturbative way different regimes of M-theory. The low energy effective action of both M-theory and all the consistent string theories is given by the corresponding supergravity. Such supergravities are non-renormalisable but they are relevant for the study of their classical solutions which turn out to be the solitons of the full string theory.

Some of these non-perturbative solitons, usually called D-branes, are extended hyperplanes where the strings can end with Dirichlet boundary conditions (see \cite{johnson} for an introduction on branes). There are two dual descriptions of these objects: they are sources of the closed sector of string theory but their dynamics can be described by the open strings (open string sector of string theory) attached to them. At low energies the dynamics of a D-brane gives rise to a gauge theory living on its worldvolume \cite{givkuta}. These dual description of the branes is a consequence of the open/closed duality present at string level.

Maldacena conjectured in 1997 a specific duality of the kind explained above. The statement \cite {Maldacena:1997re} is that type IIB string theory on $AdS_5 \times S^5$ is dual to four-dimensional ${\cal N}=4$ superconformal Yang-Mills theory with $SU(N_c)$ gauge group. In other words, the closed string sector of string theory quantised on an $AdS_5 \times S^5$ target space is conjectured to be dual to the field theory living on a stack of $N_c$ D3-branes. For a review, see \cite{MAGOO}. This is a holographic duality in the sense that the boundary of the $AdS_5$ space where the gauge theory lives encodes all the bulk information \cite{hologram}. This duality is supposed to hold for generic values of the parameters defining the regime of the two theories. For technical reasons this duality has been more accurately tested so far in the low energy limit. In this limit we can extract information about the strong coupling regime of the field theory by merely performing classical computations in a supergravity background. This is the power of the duality conjectured by Maldacena and that it is known as the AdS/CFT correspondence. An older idea which already signaled the existence of the afterwards conjectured correspondence between a string theory and a gauge theory was suggested by G. 't Hooft \cite{'tHooft:1973jz}. He realised that, by expanding a $U(N_c)$ gauge theory on the dimensionless parameter $1/N_c$ and taking the limit of large $N_c$, we can rearrange the Feynman diagrams as a sum over the genus of the surfaces in which the diagrams can be drawn. This is similar to the computation of string amplitudes where the sum is now over the genus of the possible worldsheets of the string.  By taking the low energy limit in the AdS/CFT correspondence, we are restricting the duality to a subsector of the parameter space of the theories, both in the string theory and in the field theory side, where the result pointed out by 't Hooft can be applied. In this limit the AdS/CFT conjecture states that a solution of supergravity should be dual to a certain supersymmetric gauge theory at strong 't Hooft coupling\footnote{The 't Hooft coupling is the product of the squared gauge coupling by the rank of the gauge group.}. However not all the stringy information of the dual gauge theory is captured by the supergravity solution. One needs to include extra D-branes on the supergravity side in order to extract nontrivial information which does not survive to the low energy limit \cite{ba0}. 

Extensions of the above ideas to more realistic theories (from a phenomenological point of view) have been studied in the last years (see \cite{Aharony:2002up, Bertolini:2003iv} for a review). The reduction of the amount of supersymmetry and the breaking of the conformal invariance would lead to a more (phenomenologically) interesting statement of the duality. The final goal would be to find the stringy dual of QCD, a non-supersymmetric and non-conformal theory. A great effort in searching for ways of breaking softly supersymmetry at a suitable energy scale is being made nowadays. Meanwhile theories with less amount of supersymmetry and without conformal  invariance present some features analogous to QCD, for instance confinement, and the ideas of the duality can be extended here in a proper way. 

In this work we will concentrate basically on the amazing study of the extensions of the AdS/CFT correspondence to more realistic theories. We will focus on supersymmetric solutions of type IIB supergravity which are dual to ${\cal N}=1$ supersymmetric gauge theories in four dimensions. We will search for the possibility of adding supersymmetric D-branes in those backgrounds and we will analyse which nontrivial information of the dual gauge theory we are capturing with these additional degrees of freedom.

\subsubsection{About this thesis}

This Ph.D. thesis is mainly based on papers
\cite{CaEdPaZaRaVa, Canoura:2006es, Benini:2006hh, Benini:2007gx, Canoura:2005fc}. Some of the technical points presented in those papers, which are not relevant for the comprehension of this thesis, have not been included. However we will refer the interested reader to the corresponding paper  whenever a technical point is mentioned. The plan for the rest of the thesis is the following:

In chapter \ref{introduction} we will sketch the bases of the AdS/CFT correspondence. Then we will show with some detail some supersymmetric solutions of type IIB supergravity and their field theory duals, theories with ${\cal N}=1$ supersymmetry in four dimensions. Finally we will explain why we do need to add extra D-branes to the supergravity background and how we can do that preserving (at least) part of the supersymmetry of the background. The main tool will turn out to be a local fermionic symmetry of the worldvolume theory on the branes called kappa symmetry. We will continue with the study of the (probe) limit where the backreaction of the extra D-branes are not taken into account and we will finish by considering the (unquenched) supergravity solutions where the extra D-branes and the supergravity background interact with each other. This chapter provides the basic tools and settles on the notation that we will use in the following.

In chapters \ref{Ypq} and \ref{Labc} we will systematically study supersymmetric embeddings of D-brane probes of different dimensionality in the $AdS_5\times Y^{p,q}$ and $AdS_5 \times \Labc$ backgrounds of type IIB string theory respectively. The main technique employed will be again the kappa symmetry of the probe's worldvolume theory. We will also give insights on the dual interpretation of these extra D-brane probes.

In chapter \ref{KW} we will study the addition of an arbitrary number of backreacting
flavour branes to the Klebanov-Witten theory, making many checks of consistency
between our new type IIB plus branes solution and expectations from field
theory. We will also study generalisations of our method for adding flavours to all \Nugual{1} superconformal field theories that can be realised on D3-branes at the tip of a Calabi-Yau cone. In chapter \ref{KS} we will extend the previous study of adding unquenched flavour branes to the Klebanov-Tseytlin and Klebanov-Strassler backgrounds. We will provide a precise field theory dual and a detailed analysis of the duality cascade which describes its renormalisation group flow. The matching of $\beta$-functions and anomalies between the field theory and the string setup will be presented as well.

In chapter \ref{MN} we will find supersymmetric configurations of a D5-brane probe in the 
Maldacena-N\'u\~nez  background which are extended along one or two of the spatial directions 
of the gauge theory. These embeddings are worldvolume solitons which behave as codimension two
or one defects in the gauge theory dual.

In chapter \ref{conclusions} we will finish with some conclusions.

\chapter{Introduction}
\label{introduction}
\medskip

\setcounter{equation}{0}
\medskip

\section{Introduction to the gauge/gravity correspondence}
\medskip
\setcounter{equation}{0}

The goal of this initial section is to review briefly the AdS/CFT correspondence \cite{Maldacena:1997re, MAGOO, Gubser:1998bc, Witten:1998qj} proposed by Maldacena and its extension to non-conformal and less supersymmetric settings \cite{Itzhaki:1998dd, Girardello:1999bd}. Considered the huge literature on the subject, we will only focus on the supergravity (SUGRA) duals of four dimensional supersymmetric (SUSY) gauge theories. Nice reviews on these topics can be found in \cite{Aharony:2002up, Bertolini:2003iv}. In this chapter we will review the foundations of the gauge/gravity correspondence that we will need to understand the work of this thesis.

The AdS/CFT correspondence is a conjecture which establishes a holographic equivalence between two apparently different theories: type IIB string theory on $AdS_5 \times S^5$ on one side and ${\cal N}=4$ supersymmetric Yang-Mills (SYM) in four dimensions (in the boundary of $AdS_5$) with gauge group $SU(N_c)$ on the other side. This nice correspondence is based on the old open/closed string duality and it can be formulated with the aid of D-branes. D-branes are hypersurfaces where open strings can end. Their dynamics, and hence the dynamics of the corresponding open strings, is described by a (supersymmetric) gauge theory at low energies. However, D-branes are also nonperturbative states of the closed string spectrum (their tensions behave as $1/g_s$, where $g_s$ is the string coupling) and at low energy they are described by solutions of the corresponding supergravity theory.

The strongest version of the correspondence is supposed to hold for generic values of the parameters defining the regime of the two theories. This is the called {\it exact} AdS/CFT correspondence. The parameters which define the regime of type IIB string theory on $AdS_5 \times S^5$ are the string coupling $g_s$ and the (dimensionless) string tension $T=L^2/(2\pi \alpha')$ where $L$ is the common radius of the AdS space and of the $S^5$. Those of the ${\cal N}=4$ SYM in four dimensions with gauge group $SU(N_c)$ are its gauge coupling $g_{YM}$ and the number of colours $N_c$. The dictionary is established in terms of two relations:
\beq  \label{AdS/CFT}
4\pi g_s\,=\, g^2_{YM}\,\, , \qquad \qquad T\,=\,\frac{1}{2\pi} \sqrt{\lambda}\,\, ,
\eeq
where we have defined the 't Hooft coupling $\lambda=g^2_{YM}N_c$.

It is very difficult to test the conjecture at this level since we do not know how to treat string theory for generic values of the string coupling. By setting $g_s \to 0$ (weak coupling limit in string theory) with $T$-fixed and large $N_c$ (planar diagrams limit in the gauge theory) with $g_{YM} \to 0$ ($\lambda=g^2_{YM}N_c$ fixed) \cite{'tHooft:1973jz}, we get the {\it classical} AdS/CFT correspondence. It states that classical (non-interacting strings) type IIB string theory on $AdS_5 \times S^5$ is equivalent to the large $N_c$ limit  with fixed 't Hooft coupling of the field theory.

This is not still enough since we do not even know how to deal with classical string theory in curved backgrounds with RR fluxes. Taking the low energy limit $\alpha' \to 0$ we are going to the weakest version of the correspondence, the {\it low energy} AdS/CFT correspondence. This is the regime where the correspondence has been more accurately tested so far. It states that the dynamics of an ${\cal N}=4$ SYM in four dimensions at strong 't Hooft coupling is captured by the supergravity modes of type IIB supergravity in $AdS_5 \times S^5$, without addition of stringy states.

Explicitly, the gravity dual of ${\cal N}=4$ SYM in four dimensions considered by Maldacena is generated by a stack of $N_c$ D3-branes in flat ten-dimensional space. This configuration preserves sixteen supercharges\footnote{We will comment below on the conformal invariance of this theory. This gives rise to another sixteen conformal supercharges.}. The type IIB supergravity solution of this system reads, in string frame,
\bear  \label{Maldasol}
&&ds^2=h_3(r)^{-1/2} dx^2_{1,3}\,+\,h_3(r)^{1/2}(dr^2\,+\,r^2d\Omega^2_5) \,\, , \rc
&&e^{2\phi}\,=\,e^{2\phi_{\infty}}\,=\, \rm{const.} \,\,  ,  \rc
&&F_5\,=\,(1\,+\, \star)\, dh_3^{-1} \wedge dx^0 \wedge dx^1 \wedge dx^2 \wedge dx^3 \,\, ,
\eear
where $d\Omega^2_5$ is the round metric on $S^5$, $\star$ stands for the Hodge dual in ten dimensions and $h_3$ is an harmonic function of the transverse coordinates
\beq
h_3(r)\,=\,1\,+\,4 \pi g_s N_c \frac{(\alpha')^2}{r^4} \,\, .
\eeq
The previous normalization comes from Dirac quantization of the D3-brane charge. The general quantization condition for a Dp-brane is 
 \beq   \label{quantizationF}
 \int_{S^{8-p}} \star F_{p+2}\,=\,\frac{2\kappa^2N_c}{g_s}T_p\,\, ,
 \eeq
 where the tension of a Dp-brane is given by
 \beq  \label{Dptension}
 T_p\,=\,\frac{\sqrt{\pi}}{\kappa}(4\pi^2\alpha')^{(3-p)/2}
 \eeq
 and $\kappa=8\pi^{7/2}g_s(\alpha')^2$ is the ten-dimensional gravitational constant. 
 
Taking the limit $\alpha' \to 0$ (low energies) we decouple the open and closed string massive modes. Since the Planck length is given by $l^2_P=g_s^{1/2}\alpha'$ and $g_s$ is constant (although in the end we will set $g_s \to 0$, $N_c \to \infty$ with $g_sN_c \sim \lambda$ constant and large enough ), we see that this limit $l_P \to 0$ also decouples the open/closed interactions. The right limiting procedure also involves a {\it near-horizon} limit, $r \to 0$, such that
\beq
U \equiv \frac{r}{\alpha'} \,=\, {\rm{fixed}} \,\, , \qquad r,\alpha' \to 0 \,\, .
\eeq 
Performing this limit in the supergravity solution (\ref{Maldasol}), it can be written as
\beq  \label{Maldasol1}
ds^2\,=\,\frac{U^2}{L^2}dx^2_{1,3}\,+\,\frac{L^2}{U^2} dU^2\,+\,L^2 d\Omega^2_5 \,\, ,
\eeq
 where the scale parameter $L$ introduced before is
 \beq
 L^4\,=\,4\pi g_s N_c (\alpha')^2 \,\, .
 \eeq
 
 The above metric (\ref{Maldasol1}) has constant curvature, ${\cal R} \sim L^{-2}$, in the low energy limit discussed above ($g_s \to 0$, $N_c \to \infty$ with $g_sN_c \sim \lambda$ constant and large enough), in string units. Thus,  the supergravity description is valid for any value of $U$.  Notice that the curvature and the 't Hooft coupling are inversely proportional, $\alpha' {\cal R} \sim \lambda^{-1/2}$. This means that the gauge theory description and the gravity one are complementary and do not overlap. The AdS/CFT correspondence is an example of a strong/weak coupling duality, namely the system is well described by ${\cal N}=4$ $SU(N_c)$ SYM in four dimensions for small values of the 't Hooft coupling while is better described by type IIB string/gravity theory whenever $\lambda$ gets large.

Another interesting point of the correspondence in its weakest version is the perfect matching between the isometries of the supergravity solution and the global symmetries of the field theory. In the case discussed above one can see a particular example. The $AdS_5$ space possesses an $SO(2,4)$ isometry group. The remaining $S^5$ factor of the background provides an extra $SO(6)$ isometry. It is remarkable in the field theory side that $SO(2,4)$ is the conformal group in four dimensions (scale invariance of the theory) while $SO(6) \approx SU(4)$ is exactly the R-symmetry group of ${\cal N}=4$ SYM theory. This shows up a perfect matching of the bosonic symmetries. There are also fermionic symmetries which, together with the bosonic ones, form the supergroup $SU(2,2\vert 4)$. Massless fields in string theory and BPS operators of SYM theory are classified in multiplets of this supergroup \cite{Witten:1998qj}.

It is well-know that type IIB string theory contains a nonperturbative $SL(2,\zet)$ invariance (S-duality) \cite{schwarz95} arising from the compactification of M-theory on a two-torus with modular parameter
\beq
\tau\,=\, \chi \,+\, i e^{-\phi} \,\, ,
\eeq
with $\chi$ the RR-scalar and $\phi$ the dilaton of type IIB. In ${\cal N}=4$ SYM there is a corresponding $SL(2,\zet)$ invariance with modular parameter
\beq
\tau_{YM}\,=\, \frac{\Theta_{YM}}{2\pi}\,+\,\frac{4\pi \, i}{g^2_{YM}} \,\, ,
\eeq
where $\Theta_{YM}$ is a parameter (which corresponds to the Chern-Simons angle) that one can turn on in the lagrangian of the theory. The $SL(2,\zet)$ invariance is realised as a transformation of $\tau_{YM}$ into $-1/\tau_{YM}$ combined with shifts in $\Theta_{YM}$. As we saw in (\ref{Maldasol}), there is a relation between the Yang-Mills coupling, on the gauge theory side, and the string coupling. This relation has to be supplemented by another one that links the $\Theta$-angle with the vacuum expectation value of the RR scalar $\chi$,
\beq \label{AdS/CFT1}
\Theta_{YM}\,=\,2 \pi \chi \,\, ,
\eeq 
such that the $SL(2,\zet)$ symmetry is clearly connected with the usual S-duality in type IIB string theory.

The AdS/CFT correspondence also states that an operator ${\cal O}_i$ in the gauge theory living at the boundary of $AdS_5$ space is associated in a nontrivial way with fluctuations of its dual supergravity field $\Phi_i$ propagating in the bulk of $AdS_5$. The generating functional for correlators in the field theory is related to the type IIB string theory partition function by \cite{Gubser:1998bc, Witten:1998qj}
\beq
{\cal Z}_{\rm{string}} [\Phi_i] \,=\, \left < \exp \left ( \int d^4x \, \varphi_i {\cal O}_i \right )   \right > \,\, ,
\eeq
where the boundary conditions of the supergravity field are given by
\beq  \label{AdS/CFTdimension}
\Phi_i(r,x^{\mu}) \sim \varphi_i(x^{\mu})\, e^{(\Delta_i-4)r} \,\, , 
\eeq
$x^{\mu}$ are gauge theory coordinates living at the boundary and $\Delta_i$ is the conformal dimension of the operator ${\cal O}_i$. This scaling dimension is related to the mass of the corresponding closed string field on $AdS_5 \times S^5$. For a free massive scalar field propagating in $AdS_5$, this relation is 
\beq  \label{AdS/CFTmass}
\Delta\,=\,2\,+\, \sqrt{4\,+\,m^2L^2} \,\, .
\eeq

In the beginning we stressed that we would pay attention to the supergravity duals of four dimensional field theories. For completeness, let us comment some words on the case of considering a stack of $N_c$ Dp-branes (with $p \neq 3$) in flat space. This configuration preserves again sixteen supercharges. Following the previous lines about AdS/CFT correspondence, we would expect that maximally supersymmetric Yang-Mills theory in $p+1$ dimensions with $SU(N_c)$ gauge group be dual to type IIA/IIB string theory in the near horizon limit of the Dp-brane supergravity solution. However, there are some problems to generalise the Maldacena conjecture for $p \neq 3$. First of all, for $p \geq 7$ it is not possible to decouple the open/closed interactions. Moreover, the theory is not scale invariant and the isometry group of the resulting metric has not $AdS$ factor. The near horizon limit of Dp-brane solutions has non-constant curvature for $p \neq 3$ and the dilaton is not constant either. Thus the ranges of validity of the gauge and gravity descriptions become more complicated here and the decoupling limit does not work so cleanly. 

One can try to apply an approach similar to AdS/CFT to study non-conformal and less supersymmetric theories, such as for instance ${\cal N}=1$ supersymmetric gauge theories in four dimensions, starting from D-branes in less supersymmetric backgrounds and breaking eventually conformal invariance. 

The first thing that one can do is to try to reduce the amount of supersymmetry. Given a Sasaki-Einstein five-dimensional manifold $X^5$ one can consider placing a stack of $N_c$ D3-branes at the tip of the (Calabi-Yau) cone over $X^5$. Taking then the Maldacena limit leads to a duality between string
theory on $AdS_5\times X^5$ and a superconformal field theory (SCFT) living in the worldvolume of the D3-branes \cite{gubser}. In subsection \ref{T11introduc} we will review the case in which the Sasaki-Einstein manifold is $X^5=T^{1,1}$ and we will see that it is dual to a four-dimensional ${\cal N}=1$ SCFT coupled to four chiral superfields in the bifundamental representation \cite{Klebanov:1998hh}. In subsection \ref{Ypqintroduc} we will review a new class of Sasaki-Einstein manifolds recently found. They are labeled by two positive integers $X^5=Y^{p,q}$  \cite{GMSW1,GMSW2} and they include the $T^{1,1}$ as a particular case. We will also give some notions of its dual ${\cal N}=1$ superconformal quiver gauge theories \cite{ms,BeFrHaMaSp}. In subsection \ref{Labcintroduc} we will explore the most recent family of Sasaki-Einstein manifolds built, $X^5=L^{a,b,c}$ which contain all the others as a subfamily \cite{CvLuPaPo, MaSp2}. They are also dual to ${\cal N}=1$ superconformal quiver gauge theories \cite{BeKr,FrHaMaSpVeWe,BuFoZa}. These families exhaust all possible toric Calabi-Yau cones on a base with topology $S^2 \times S^3$. Research on AdS/CFT in these SCFT's has led to a better understanding of several important issues such as the appearance of duality cascades, a-maximization, Seiberg duality, etc. In chapters \ref{Ypq}, \ref{Labc} and \ref{KW} we will concentrate on the addition of new degrees of freedom to these supergravity backgrounds and its interpretation in the dual field theory.

The next step is to break conformal invariance. In trying to do this, one finds some problems. The first one is that the dual supergravity solution of a non-conformal gauge theory does not display an AdS-like geometry and, in general, it means that (strictly speaking) holography does not work in these cases. Furthermore, a basic aspect of the AdS/CFT correspondence is the decoupling between open and closed degrees of freedom. In the weakest version of the AdS/CFT correspondence, the gauge theory is supposed to be dual to supergravity, without any addition of string states. In non-conformal theories, the dual gauge theory cannot be decoupled from the bulk if one only deals with supergravity modes. It is believed that a proper duality holds if one lets string states enter into the game but, as it is the case for the original AdS/CFT correspondence, it is much harder to go beyond the supergravity regime and check the duality at string level. This is a crucial point to keep in mind when one studies the gauge/gravity duality in non-conformal theories. However, we do not discuss this further here, and in subsections \ref{fractionalD3}, \ref{deformedconifold} and \ref{MNbackground} of this introductory chapter, as well as in chapters \ref{KW}, \ref{KS} and \ref{MN}, we will try to exploit the power of open/closed string duality and see what we can learn about the dynamics of non-conformal supersymmetric gauge theories from supergravity and vice-versa.

There are several ways to obtain supergravity backgrounds dual to non-conformal four-dimensional ${\cal N}=1$ gauge theories. One way is by introducing fractional D-branes on toric Calabi-Yau three-fold cones. In subsection \ref{fractionalD3} we will study in detail the case of adding fractional D3-branes to the conifold \cite{Klebanov:2000nc} and we will see how the conformal invariance of the $AdS_5 \times T^{1,1}$ background is broken. The same could be done for the $AdS_5 \times Y^{p,q}$ \cite{hek} and $AdS_5 \times L^{a,b,c}$ case. Another way of breaking conformal invariance is by starting with D-branes wrapped on nontrivial supersymmetric cycles of non-compact Calabi-Yau three-fold spaces. This procedure was firstly used by Maldacena and N\'u\~nez to study pure ${\cal N}=1$ SYM in four dimensions \cite{Maldacena:2000yy} and we will review their solution in subsection \ref{MNbackground}. It is worth pointing out here that not all the submanifolds admit wrapped D-branes preserving some amount of supersymmetry. Submanifolds that do preserve it are called {\it supersymmetric} or {calibrated cycles} and are defined by the condition that the worldvolume theory on the D-brane is supersymmetric. These cycles are classified in manifolds with special holonomy.

The breaking of the conformal invariance in a gauge theory leads to a running of the gauge coupling with the energy scale. In gauge/gravity duality the radial coordinate defines the Renormalization Group (RG) scale of the dual gauge theory \cite{Maldacena:1997re, MAGOO, Gubser:1998bc, Witten:1998qj}. In general, for non-conformal theories, the radius-energy relation depends on the phenomenon that one is interested in and accounts for the scheme-dependence in the field theory.

Anomalies are also of great interest in the gauge/gravity duality since the Adler-Bardeen theorem \cite{adler} guarantees that anomaly coefficients computed at one loop are exact, with no radiative corrections. This means that we can compute anomaly coefficients in the field theory at weak coupling and then extrapolate them to strong coupling where we can use dual gravity methods to check the calculation. This allows a nontrivial check of the gauge/gravity duality.

\section{SUSY solutions of type IIB supergravity}  \label{SUSYanalysis}

\medskip
\setcounter{equation}{0}

The aim of this section is to describe some supersymmetric solutions of type IIB supergravity where the geometry is a warped product of the four-dimensional Minkowski space and a six-dimensional Riemannian manifold, ${\cal M}^{1,3} \times {\cal N}^6$. We are interested in classical configurations in which the fermionic fields vanish, and thus the problem of finding supersymmetric solutions reduces to solve the vanishing of the supersymmetric variations of the fermionic fields of type IIB supergravity. We will write down below these transformations and the bosonic action of type IIB supergravity. The fact that a configuration is supersymmetric does not necessarily imply that it is a solution of the supergravity equations of motion. We must check {\it a posteriori} that the equations of motion are solved by the configurations which fulfil the vanishing of the aforementioned supersymmetry transformations of the fermionic fields. 

Type IIB supergravity is a maximal supergravity ({\it i.e.} with 32 supercharges) that can be constructed in ten dimensions \cite{type2b, SUSYIIB, type2b1}. This type IIB theory is chiral and cannot be
obtained by dimensional reduction from eleven dimensions. Nevertheless,
it is related to type IIA sugra by T-duality. The bosonic degrees of
freedom are the metric $G_{MN}$, the dilaton $\phi$, a NSNS two-form
$B_{2}$ whose field strength is $H_3$ ($H_3=dB_2$) and the RR field strengths $F_1$, $F_3$ and $F_5$. The action for these fields reads
(in Einstein frame):
\bear  \label{IIBEF}
S_{IIB}&=&\frac{1}{2\kappa^2_{10}}\int d^{10}x \sqrt{-G}\Big[R-\frac{1}{2}
\partial_{M}\phi\,\partial^M \phi-\frac{1}{2}e^{-\phi}H_{3}^2-
\frac{1}{2} e^{2\phi} F^2_{1}-\frac{1}{2}e^\phi F_{3}^2- \rc
&-&\frac{1}{4} F_{5}^2\,\Big]\,-\,\frac{1}{2 \kappa_{10}^2}\int C_{4}\wedge H_{3}
\wedge F_{3}\,\,,
\eear
where $2\kappa^2_{10}=16\pi G_{N}=(2\pi)^7 g_s^2 (\alpha')^4$ is related to the ten-dimensional gravitational constant and we have chosen the normalization $A_p^2=\frac{1}{p!}A_{M_1\ldots M_p}A^{M_1\ldots M_p}$ for any $p$-form $A_p$. Notice that the last term in $S_{IIB}$ is a Chern-Simons term that involves $C_4$, the RR potential of $F_5$. Apart from the
equations of motion that arise from this action, one has additionally to
impose the self-duality condition $F_{5}=\star F_{5}\ $. For completeness, we write down the equations of motion satisfied by the dilaton and the metric functions:
\bear  \label{IIBeq}
&&\frac{1}{\sqrt{-G}} \partial_M \Big (G^{MN}\, \sqrt{-G}\, \partial_N \phi \Big )\,=\,e^{2\phi}F_1^2\,+\, \frac{1}{2} \Big ( e^{\phi}F_3^2\,-\,e^{-\phi}H_3^2 \Big )\,\, , \rc
&&R_{MN}\,-\, \frac{1}{2}G_{MN}R\,=\,\frac{1}{2} \Big ( \partial_M\phi\partial_N\phi\,-\,\frac{1}{2}G_{MN}\partial_P\phi \partial^P\phi \Big )\,+\,\frac{1}{2}e^{2\phi} \Big ( F^{(1)}_M F^{(1)}_N\,-\,\frac{1}{2}G_{MN}F_1^2 \Big )\,+\,\rc
&&+\frac{1}{2}e^{\phi} \Big ( 3F^{(3)}_{MPQ}F^{(3)PQ}_N\,-\, \frac{1}{2}G_{MN}F_3^2 \Big )\,+\,\frac{5}{4}F^{(5)}_{MPQRS}F^{(5)PQRS}_N\,+\,\rc
&&+\frac{1}{2}e^{-\phi}\Big ( 3H^{(3)}_{MPQ}H^{(3)PQ}_N\,-\, \frac{1}{2}G_{MN}H_3^2 \Big) \,\, . \rc
\eear

The set of Bianchi identities satisfied by the NSNS and RR field strength fluxes of the above supergravity action are the following:
\bear  \label{BI}
dH_3\,&=&\,0 \,\, , \rc
dF_1\,&=&\,0 \,\, , \rc
dF_3\,&=&\,H_3 \wedge F_1 \,\, , \rc
dF_5\,&=&\,H_3 \wedge F_3 \,\, .
\eear
Moreover, the equations of motion for the NSNS and RR forms derived from the action (\ref{IIBEF}) are:
\bear  \label{MAX}
 d \Big ( e^{-\phi} \star H_3 \Big )\,&=&\, e^{\phi}F_1 \wedge \star F_3\,-\,F_5 \wedge F_3 \,\, , \rc
 d \Big ( e^{2\phi} \star F_1 \Big )\,&=&\,-\,e^{\phi}H_3 \wedge \star F_3 \,\, , \rc
 d \Big ( e^{\phi} \star F_3 \Big )\,&=&\, F_5 \wedge H_3 \,\, , \rc
 d \Big ( \star F_5 \Big )\,&=&\,-\, F_3 \wedge H_3 \,\, .
\eear

Let us now consider the supersymmetric variations of the fermionic fields, a
dilatino $\lambda$ and a gravitino $\psi_\mu\ $.
In type IIB string theory, the Killing spinor $\epsilon$ (which parameterises the supersymmetric transformations) is actually composed
by two Majorana-Weyl spinors $\epsilon_1$ and $\epsilon_2$ of well defined
ten-dimensional chirality, which can be arranged as a two-component vector in
the form $\epsilon$=
$\begin{pmatrix}   \label{striibsp}
\epsilon_1\cr\epsilon_2
\end{pmatrix}\,\,$.

Thus, the supersymmetry transformations of the dilatino $\lambda$ and gravitino 
$\psi_\mu$ in type IIB supergravity are (in Einstein frame):
\bear \label{Eframe1}
\delta_{\epsilon}  \lambda&=&{1 \over 2}\Gamma^M \big (  \partial_M\phi-e^{\phi}F_M^{(1)}(i\sigma_2) \big )\epsilon\,-{1 \over 4}{1 \over 3!} \Gamma^{MNP}\big (e^{-{\phi \over 2}} H_{MNP}\sigma_3 +e^{{\phi \over 2}}F_{MNP}^{(3)} \sigma_1 \big )\epsilon   ,  \rc
\delta_{\epsilon}  \psi_M&=&\nabla_M\epsilon+{1 \over 4}e^{\phi}F_M ^{(1)}(i\sigma_2)\epsilon+{1 \over 96}\big (e^{-{\phi \over 2}} H_{NPQ}\sigma_3 -e^{{\phi \over 2}}F_{NPQ}^{(3)} \sigma_1 \big )\big ( \Gamma_M^{\,\, NPQ}-9\delta^N_M\Gamma^{PQ} \big )\epsilon+\,\, \rc
&&+{1 \over 16}{1 \over 5!}F_{NPQRT}^{(5)}\Gamma^{NPQRT}(i\sigma_2)\Gamma_M\epsilon \,\, ,
\eear
where $\Gamma^{M}$ are ten-dimensional Dirac matrices, $\Gamma^{M\dots N}$ stands for their antisymmetric product and $\sigma_i \,\, i=1,2,3$ are Pauli matrices which act on the two-dimensional vector constructed above.

The supersymmetric solutions that we will consider are solutions of the vanishing of the above supersymmetry transformations (and of the supergravity equations of motion as well) which can be interpreted as being dual to a four-dimensional ${\cal N}=1$ gauge theory. As discussed in the previous section, the kind of supergravity solutions dual to conformal field theories can be generated by putting D3-branes at the apex of a Calabi-Yau three-fold and then considering the geometric transition as in \cite{Gopakumar:1998ki,Vafa:2000wi}. The D3-branes deform the geometry and source a RR five-form field strength. In order to break conformal invariance, either fractional branes enter into the game or we have to start from a configuration with branes wrapping a supersymmetric cycle of the geometry. Usually, to solve the vanishing of eqs. (\ref{Eframe1}), one must impose some projections on the Killing spinor. When this happens, not all the supercharges present in the supergravity theory are preserved by the solution. These projections are of the type ${\cal P}\epsilon=\epsilon$. In the cases that we will study they should commute among themselves and each of them halves the number of preserved supercharges. It is known that the number of supersymmetries preserved by a Calabi-Yau three-fold is $1/4$ of the maximally supersymmetric configurations\footnote{In other words, only $1/4$ of the components of the Killing spinor $\epsilon$ are different from zero.}. It leads to $8$ supercharges, hence being dual to a four-dimensional ${\cal N}=1$ SCFT. By breaking conformal invariance we are left with just $4$ supercharges.

In the next subsections we will give some examples of supersymmetric solutions dual to four-dimensional ${\cal N}=1$ gauge theories and in some cases we will also display the Killing spinor which solves eqs. (\ref{Eframe1}) and the projections that it satisfies. We will start from the dual to a SCFT and then we will move on to more realistic solutions, breaking conformal invariance.

We have just summarised very briefly type IIB supergravity. A thorough review on eleven and ten-dimensional supergravities, the
relations among them (Kaluza-Klein reduction, T-duality), solutions of their equations of motion 
from (wrapped-)branes and many other topics on gravity and its relation with
strings can be found in \cite{librortin}.

\subsection{D3-branes on the Conifold: the Klebanov-Witten (KW) model}  \label{T11introduc}

In the same spirit as AdS/CFT \cite{Maldacena:1997re, MAGOO, Gubser:1998bc, Witten:1998qj}, Klebanov and Witten \cite{Klebanov:1998hh} suggested that $N_c$ D3-branes at the singularity of the conifold will result in certain ${\cal N}=1$ superconformal field theory dual to the string theory on $AdS_5 \times X^5$, where $X^5$ is the base of the cone and it was identified as the Sasaki-Einstein manifold $T^{1,1}$ \cite{Klebanov:1998hh,Candelas:1989js}. The conifold (or the cone over $T^{1,1}$) is a non-compact Calabi-Yau three-fold defined by the following equation in $\bbC^4$ (see \cite{Herzog:2002ih} for a review of the conifold):
\beq  \label{zcoor}
z_1\,z_2\,-\,z_3\,z_4\,=\,0\,\, .
\eeq
Since this equation is invariant under a real rescaling of the variables, the conifold is a real cone whose base is the space $T^{1,1}$, an space with topology $S^2 \times S^3$. The metric on the conifold can be written in the form
\beq   \label{ds6}
ds^2_6\,=\,dr^2\,+\,r^2\,ds^2_{T^{1,1}}\,\, ,
\eeq
where
\beq  \label{T11}
ds^2_{T^{1,1}}\,=\,\frac{1}{9} \left ( d\psi^2\,+\,\sum_{i=1}^{2}\cos{\theta_i}d\varphi_i \right )^2\,+\,\frac{1}{6}\sum_{i=1}^{2} (d\theta_i^2\,+\,\sin^2{\theta_i}d\varphi_i^2)
\eeq
is the metric on $T^{1,1}$. Here $\psi$ is an angular coordinate which ranges from $0$ to $4\pi$, while $(\theta_1,\varphi_1)$ and $(\theta_2,\varphi_2)$ parameterise two $S^2$ spheres in a standard way. Therefore, this form of the metric shows that $T^{1,1}$ is a $U(1)$ bundle over the K\"ahler-Einstein space $S^2 \times S^2$ and that its isometry group is $SU(2) \times SU(2) \times U(1)$. Moreover, the coordinates $z_i$ (\ref{zcoor}) can be expressed through the angular variables $\psi, \theta_i, \varphi_i$ and $r$ as follows \footnote{For a thorough study of the complex formulation of the conifold, see \cite{Candelas:1989js}.}:
\bear  \label{complexconifold}
&&z_1\,=\, r^{3/2}e^{\frac{i}{2}(\psi-\varphi_1-\varphi_2)}\sin {\frac{\theta_1}{2}} \sin{\frac{\theta_2}{2}} \,\, , \rc
&&z_2\,=\, r^{3/2}e^{\frac{i}{2}(\psi+\varphi_1+\varphi_2)}\cos {\frac{\theta_1}{2}} \cos{\frac{\theta_2}{2}} \,\, , \rc
&&z_3\,=\, r^{3/2}e^{\frac{i}{2}(\psi+\varphi_1-\varphi_2)}\cos {\frac{\theta_1}{2}} \sin{\frac{\theta_2}{2}} \,\, , \rc
&&z_4\,=\, r^{3/2}e^{\frac{i}{2}(\psi-\varphi_1+\varphi_2)}\sin {\frac{\theta_1}{2}} \cos{\frac{\theta_2}{2}} \,\, .
\eear

Placing $N_c$ D3-branes at the apex of the conical singularity (\ref{zcoor}), they source the RR 5-form flux and warp the geometry. In the near-horizon limit, they give rise to the type IIB supergravity solution:
\bear   \label{solucKW}
&&ds^2\,=\,h(r)^{-1/2}dx^2_{1,3}\,+\,h(r)^{1/2}ds^2_6 \,\, , \rc
&&F_5\,=\,\frac{1}{g_s}(1\,+\, \star) d^4x \wedge dh(r)^{-1} \,\, , \rc
&&h(r)\,=\,\frac{L^4}{r^4}\,\, , \qquad \qquad L^4\,=\,4\pi g_s N_c (\alpha')^2\frac{\pi^3}{\rm{Vol}(T^{1,1})}\,=\,\frac{27\pi g_s N_c (\alpha')^2}{4}\,\, ,
\eear
 with constant dilaton and all the other fields of type IIB supergravity vanishing.  The normalization of the scale factor $L$ is dictated by the quantization of the D3-brane tension $T_p$, 
 \beq  \label{quantizationF5}
  \frac{1}{(4\pi^2\alpha')^2} \int_{T^{1,1}} F_5 \,=\,N_c \,\, .
 \eeq
  
By introducing eq. (\ref{solucKW}) into the supersymmetry transformations (\ref{Eframe1}) one can determine the Killing spinor which lives in this background. A detailed analysis was carried out in \cite{Arean:2006nc}. It is necessary to impose two projections on the Killing spinor and hence this background is $1/4$ supersymmetric, as expected for a Calabi-Yau three-fold. It preserves $8$ supersymmetries so the comparison of the number of preserved supersymmetries allows to conjecture that it is the supergravity dual of an ${\cal N}=1$ SCFT in four dimensions. However there are more evidences of this duality. The field theory was constructed in \cite{Klebanov:1998hh}. It is an $SU(N_c) \times SU(N_c)$ gauge theory coupled to two chiral superfields, $A_i$, in the  $(N_c,\bar{N_c})$ bifundamental representation and two chiral superfields, $B_i$, in the $(\bar{N_c},N_c)$ bifundamental representation of the gauge group. The $A$'s transform as a doublet under one of the global $SU(2)$s while the $B$'s transform as a doublet under the other $SU(2)$. We can motivate the field content of this theory rewriting the complex variables which parameterise the equation of the conifold (\ref{zcoor}) as 
\bear
&&z_1\,=\,A_1\,B_1 \,\, , \qquad \qquad \qquad  z_2\,=\,A_2\,B_2 \,\, , \rc
&&z_3\,=\,A_1\,B_2 \,\, , \qquad \qquad \qquad  z_4\,=\,A_2\,B_1 \,\, .
\eear
The defining equation of the manifold is related to the moduli space of the gauge theory. The anomaly-free $U(1)$ R-symmetry ($U(1)_R$) of the ${\cal N}=1$ superconformal algebra is realised as a common phase rotation of $A_i$, $B_j$ (or of the four coordinates $z_i \to e^{-i\alpha}z_i$). Both $A_i$ and $B_j$ have $1/2$ charge under $U(1)_R$ in order to cancel the anomaly. For consistency of the duality it is necessary to add a marginal superpotential (and so with R-charge $2$). The most general marginal superpotential respecting the global symmetries $SU(2) \times SU(2) \times U(1)_R$ is
 \beq   \label{KWsuper}
 W_{KW}\,=\, \lambda \epsilon^{ij}\epsilon^{kl} \, \tr(A_iB_kA_jB_l) \,\, .
 \eeq
 
 There is another anomaly-free abelian symmetry\footnote{In what follows we will denote it by $U(1)_{B}$.} $U(1)_{\rm{baryon}}$ which shifts $A_i$ and $B_j$ in opposite directions:
 \beq
 A_i \to e^{i\varphi}A_i \,\, , \qquad \qquad B_j \to e^{-i\varphi}B_j \,\, .
 \eeq
 
Therefore, it was proposed in \cite{Klebanov:1998hh} that the $SU(N_c) \times SU(N_c)$ SCFT with this superpotential is dual to type IIB string theory on $AdS_5 \times T^{1,1}$. Although we will not go into details, in \cite{Klebanov:1998hh} the authors gave another argument beyond the simple symmetry analysis which supports the duality. On the gravity side, the geometry of $T^{1,1}$ emerges from $S^5/\bbZ_2$ via blowing-up of the orbifold singularity of $S^5/\bbZ_2$. It can be shown \cite{Klebanov:1998hh} that this mechanism is dual to the RG flow of the gauge theory, which was identified in \cite{Kachru:1998ys}. Perturbing the superpotential of this orbifold configuration in such a way that it breaks conformal symmetry and half of the  supersymmetry, the field theory will flow to an infrared (IR) fixed point where the superpotential is exactly (\ref{KWsuper}).
 
As a final argument in favour of the duality one could discuss the chiral operators of the field theory, namely the gauge invariant operators which have the lowest possible conformal dimension for a given R-charge. In \cite{Klebanov:1998hh} it was argued that the chiral operators of positive R-charge $n$ and dimension $3n/2$ are of the form
 \beq
 C_L^{k_1k_2\ldots k_n}C_R^{l_1l_2\ldots l_n} \tr A_{k_1}B_{l_1}A_{k_2}B_{l_2}\ldots A_{k_n}B_{l_n} \,\, ,
 \eeq
 where $C_L$ and $C_R$ are completely symmetric tensors. These operators are in the $(n+1,n+1)$ representation of $SU(2) \times SU(2)$. In \cite{Ceresole:1999zs} the supergravity modes dual to those chiral operators were studied, showing that they are a mixture of the conformal factors of $AdS_5$ and factors of the $T^{1,1}$ and the RR potential.

\subsection{Adding fractional D3-branes to the Klebanov-Witten model}    \label{fractionalD3}

In this subsection we study the effect of adding $M$ fractional colour D3-branes in the Klebanov-Witten model. These fractional branes are D5-branes located at the tip of the conifold and wrapping the vanishing nontrivial $S^2$ of the $T^{1,1}$. They change the gauge group of the field theory dual to $SU(N_c+M) \times SU(N_c)$. 

First of all we analyse the dual supergravity background. The D5-branes act as sources of the magnetic RR three-form flux through the $S^3$ of the $T^{1,1}$. Therefore, besides the $N_c$ units of RR five-form flux (\ref{quantizationF5}), the supergravity dual involves $M$ units of three-form flux (\ref{quantizationF}):
\beq  \label{KTquantization}
\frac{1}{4\pi^2\alpha'} \int_{S^3} F_3 \,=\,M \,\, .
\eeq

This supergravity solution was constructed in \cite{Klebanov:2000nc} and it is known as the Klebanov-Tseytlin model.

In order to display the supergravity background it is useful to employ the following basis of one-forms on the compact space:
\bear   \label{gbasis}
&&g^1\,=\,{1 \over {\sqrt{2}}} (\omega_2 \,-	\,\sigma_2)\,\, ,
\qquad \qquad g^2\,=\,{1 \over {\sqrt{2}}} (-\omega_1 \,+\,\sigma_1)\,\, \,\, ,
\rc &&g^3\,=\,{-1 \over {\sqrt{2}}} (\omega_2 \,+\,\sigma_2)\,\, , \qquad
\qquad g^4\,=\,{1 \over {\sqrt{2}}} (\omega_1 \,+\,\sigma_1)\,\, , \rc
&&g^5\,=\, \omega_3 \,+\,\sigma_3\,\, ,
\eear
where
\bear  \label{basisforms}
&& \sigma_1\,=\,d\theta_1 \, , \qquad \qquad
\sigma_2\,=\,\sin{\theta_1}\,
d\varphi_1 \, , \qquad \qquad
\sigma_3\,=\,\cos{\theta_1}\,d\varphi_1 \, , \rc
&&\omega_1\,=\,\sin{\psi} \sin{\theta_2}\,
d\varphi_2\,+\,\cos{\psi}\,d\theta_2\,\, ,
\qquad\qquad \,\,\,\,
\omega_2\,=\,-\cos{\psi}
\sin{\theta_2}\, d\varphi_2\,+\,\sin{\psi}\,d\theta_2\,\, , \rc
&&\omega_3\,=\,d\psi\,+\,\cos{\theta_2}\,d\varphi_2 \,\, ,
\eear
are one-forms written in terms of the angular coordinates introduced in the conifold geometry (\ref{T11}). In this basis the metric on $T^{1,1}$ (\ref{T11}) takes the form
\beq
ds^2_{T^{1,1}}\,=\, \frac{1}{9} (g^5)^2\,+\, \frac{1}{6} \sum_{i=1}^{4} (g^i)^2 \,\, .
\eeq

This basis is also useful to write the NSNS two-form flux $B_2$ and the RR three-form flux $F_3$ sourced by the fractional branes. They are magnetic fluxes which must satisfy the Bianchi identities (\ref{BI}). Therefore we need to construct a closed two- and three-form and one may realise that a possibility is
\beq   \label{Upsilon}
\Upsilon_2\,=\, {1 \over 2} (g^1 \wedge g^2 \,+\, g^3 \wedge g^4) \,\, , \qquad
\Upsilon_3\,=\,{1 \over 2} g^5 \wedge (g^1 \wedge g^2 \,+\, g^3 \wedge g^4) \,\, .
\eeq
They are closed by construction and satisfy
\beq  \label{Upsilon1}
\int_{S^2} \Upsilon_2 \,=\, 4\pi \,\, , \qquad \qquad   \int_{S^3} \Upsilon_3 \,=\, 8\pi^2 \,\, ,
\eeq
where the two-cycle $S^2$ is parameterise by $\theta_1 = \theta_2 \equiv \theta$, $\varphi_1 = 2\pi -\varphi_2 \equiv \varphi$, $\psi=\text{const.}$ and the three-cycle $S^3$ by $\theta_2, \varphi_2=\rm{constant}$. Consistency with the Bianchi identities (\ref{BI}) and the quantization condition (\ref{KTquantization}) allows to write the NSNS two-form $B_2$ and the RR three-form $F_3$ as
\bear
&&B_2\,=\,\frac{3g_sM\alpha'}{2}\ln{(r/r_0)} \Upsilon_2 \,\, , \qquad H_3\,=\,dB_2\,=\,\frac{3g_sM\alpha'}{2r}\ln{(r/r_0)}\,dr \wedge \Upsilon_2 \,\, , \rc
&&F_3\,=\, \frac{M\alpha'}{2} \Upsilon_3 \,\, ,
\eear
where $r_0$ is a radial scale (integration constant) introduced in the theory by the fractional branes. Note that
\beq
g_s \star_6 F_3 \,=\,H_3 \,\, , \qquad \qquad g_sF_3\,=\,-\star_6H_3\,\, ,
\eeq
where $\star_6$ is the Hodge dual with respect to the metric $ds^2_6$ (\ref{ds6}). Thus, the complex three-form 
\beq   \label{G3}
G_3\,=\,F_3\,-\, \frac{i}{g_s}H_3
\eeq
satisfies the imaginary self-duality condition $\star_6 G_3=iG_3$. It follows that 
\beq
g_s F_3^{2}\,=\,H_3^2 \,\, ,
\eeq
and analysing the equation of motion for the dilaton (\ref{IIBeq}) we see that we can set consistently to zero the dilaton $\phi$ and the RR scalar $\chi$ ($F_1=d\chi$).

The ten-dimensional metric has the structure of a warped product of $\reals^{1,3}$ and the conifold
\beq  \label{KTmetric}
ds^2\,=\,h(r)^{-1/2}dx^2_{1,3}\,+\,h(r)^{1/2}(dr^2\,+\, r^2\,ds^2_{T^{1,1}}) \,\, ,
\eeq
where the warp factor is obtained by solving the Einstein equations of motion (\ref{IIBeq})
\beq
h(r)\,=\,\frac{27\pi (\alpha')^2}{4r^4} \Big ( g_s N_c\,+\,a(g_sM)^2\ln{r/r_0}\,+\, \frac{a}{4}(g_sM)^2  \Big )\,\, ,
\eeq
with $a=\frac{3}{2\pi}$.

The Klebanov-Tseytlin model is a solution of the type IIB equations of motion and of the Bianchi identities. It preserves $1/8$ of the supersymmetry (see for example \cite{Arean:2006nc}). Again this fact allows us to postulate that the four-dimensional gauge theory dual preserves ${\cal N}=1$ supersymmetry without conformal invariance. We will study below the implications of this statement. 

An important feature of this model is that the RR five-form flux, which can be parameterised (consistently with the quantization condition (\ref{quantizationF5})) as follows
\beq  \label{five-form}
F_5\,=\,\frac{\pi}{4}N_{eff}(r) g^1 \wedge g^2 \wedge g^3 \wedge g^4 \wedge g^5 \,\, ,
\eeq
takes a radial dependence
\beq
N_{eff}\,=\, N_c\,+\, \frac{3}{2\pi}g_sM^2 \ln{(r/r_0)}\,\, .
\eeq
This dependence comes from the fact that the right-hand side of the Bianchi identity (\ref{BI}) involving $F_5$ is non zero in this case. Notice that the five-form flux present at the ultraviolet (UV) may completely disappear by the time we reach a scale where $N_{eff}=0$. A related fact is that the quantity
\beq   \label{SDfield}
b_0\,\equiv \, \frac{1}{4\pi^2\alpha'} \int_{S^2} B_2 \,\, ,
\eeq
which is invariant in string theory as it undergoes a shift of $1$ (due to the quantization condition of $H_3$), is no longer a periodic variable. If we shift $b_0$ by one unit, we see that the shift in the radial variable that realises the same effect is a decreasing of the radius by a factor $\exp{(-2\pi/3g_sM)}$. This implies a decreasing in the five-form flux in $M$ units, $N_{eff} \to N_{eff}-M$. Therefore the integral $\frac{1}{(4\pi^2\alpha')^2}\int_{T^{1,1}}F_5$ is not quantised. On the field theory side this effect is understood as a Seiberg duality \cite{Seiberg:1994pq}. We start from a theory with gauge groups $SU(r_1) \times SU(r_2)$ at some energy scale, with $r_1 > r_2$. The gauge couplings of both gauge groups flow in opposite directions since each gauge group views the fields transforming in the other gauge group as flavour degrees of freedom. The coupling of the gauge group $SU(r_1)$ flows towards strong coupling and before reaching an infinite value, the theory is better described in terms of its Seiberg dual description, which is weakly coupled. The $SU(r_1)$ gauge group has $2r_2$ flavours in the fundamental representation. Under a Seiberg duality this becomes an $SU(2r_2-r_1)$ and the other gauge group remains untouched. Thus, after the Seiberg duality we get $SU(r_2) \times SU(2r_2-r_1)$ which resembles closely the theory we start with. On the field theory side we can read $N_{eff}$ and $M$ from the effective D3-brane and D5-brane charge respectively of the system of fractional D-branes that engineers the field theory: $r_1$ fractional D3-branes of one kind (D5-branes wrapping the $S^2$) and $r_2$ fractional D3-branes of the other kind (D5-branes wrapping the $S^2$ with $-1$ quanta of gauge field flux on the two-cycle). Although we will leave the details of the analysis of the effective charge to chapter \ref{KS}, we advance  here that 
\bear
N_{eff}\,&=&\,b_0\,r_1\,+\,(1-b_0)\,r_2\,\, , \rc
M\,&=&\,r_1\,-\,r_2 \,\, .
\eear
Under the Seiberg duality described above they become 
\bear   \label{KTSD}
N'_{eff}\,&=&\,b_0\,r_2\,+\,(1-b_0)\,(2r_2\,-\,r_1)\,=\,N_{eff}\,-\,M \,\,, \rc
M'\,&=&\,M \,\, ,
\eear
reproducing the SUGRA behaviour. Starting from the UV ($r=r_0$) of the gravity solution and moving to the IR, it is worth pointing out that, after $k$ steps of the logarithmic running (duality cascade) that we have just explained, the radius decreases as $r_k=r_0\exp{(-2\pi k/3g_sM)}$ and the effective number of colours turns out to be $N_{eff}=N_c-kM$. However the rank of the dual gauge groups remains fixed at each step and only changes at the point where we perform the Seiberg duality. In other words, at each step the gauge group can be written as $SU(N_{eff}+M) \times SU(N_{eff})$ only when $b_0=0$. 

The metric (\ref{KTmetric}) has a naked singularity at the value of the radial variable $r=r_s$ where the warp factor becomes zero, $h(r_s)=0$. Then, setting
\beq  \label{KTwarp}
h(r)\,=\,\frac{L^4}{r^4}\ln{(r/r_s)}\,\, , \qquad \qquad L^2\,=\, \frac{9g_sM\alpha'}{2\sqrt{2}} \,\, ,
\eeq
we can write the metric as a purely logarithmic RG cascade:
\beq \label{KTmetric1}
ds^2\,=\,\frac{r^2}{L^2\sqrt{\ln{(r/r_s)}}}ds^2_{1,3}\,+\,\frac{L^2 \sqrt{\ln{(r/r_s)}}}{r^2}dr^2\,+\,L^2\sqrt{\ln{(r/r_s)}}ds^2_{T^{1,1}}\,\, .
\eeq
The nature of this singularity has its origin in the charge quantisation (\ref{KTquantization}). For small values of the radial coordinate $r$ the $S^3$ shrinks to zero. This leads to a divergence of $F_3$ in order to fulfil the quantization condition (\ref{KTquantization}).

The curvature of the metric (\ref{KTmetric1}) decreases for large $r$, so the string corrections to the SUGRA solution become negligible. Even if $g_sM$ is very small, the solution is reliable for sufficiently large radii where $g_sN_{eff} \gg 1$. In this regime the separation between the cascade steps is large and we can compare the $\beta$-functions computed from SUGRA with those of an $SU(N_c+M) \times SU(N_c)$ gauge theory in the UV limit. The integrals over the $S^2$ of the $T^{1,1}$ of the NSNS and RR two-form potentials $B_2$ and $C_2$, the dilaton $\phi$ and the RR scalar $\chi$ are moduli of the type IIB theory on $AdS_5 \times T^{1,1}$. The two gauge couplings and the two $\Theta$-angles of the field theory are related with them in a way which depends on the quantization of string theory in that background. Given the lack of knowledge about the string quantization on $AdS_5 \times T^{1,1}$ together with the fact that the KW model can be obtained as an IR fixed point of the RG flow of the orbifold $AdS_5 \times S^5/\zet_2$, we borrow the holographic relations from those computed in the orbifold theory \cite{Kachru:1998ys, Lawrence:1998ja, Polchinski:2000mx}. For the two gauge couplings they are:
\bear   \label{RGholography}
\frac{4\pi^2}{g_1^2}\,+\, \frac{4\pi^2}{g_2^2}\, &=& \, \frac{\pi e^{-\phi}}{g_s}\,\, , \rc
\frac{4\pi^2}{g_1^2}\,-\, \frac{4\pi^2}{g_2^2}\, &=& \, \frac{e^{-\phi}}{2\pi g_s \alpha'} \left (  \int_{S^2} B_2 \,-\,2\pi^2 \,\, (\rm{mod} \,\, 4\pi^2) \right ) \,\, ,
\eear
where $b_0$ must be defined in the range $[0,1]$ in order to give positive squared couplings. The ambiguity in the last equation is the $2\pi$ periodicity of
$\frac{1}{2\pi\alpha'}\int_{S^2} B_2$ which comes from the quantization
condition on $H_3$. A shift of $2\pi$ amounts to move to a dual description of the gauge theory.
\footnote{In the Klebanov-Witten theory, this is the Seiberg duality.}

In gauge/gravity duality the radial coordinate defines the RG scale of the dual gauge theory \cite{Maldacena:1997re, MAGOO, Gubser:1998bc, Witten:1998qj}. There are several ways of establishing the precise relation. In what follows we adopt the one that typically corresponds to the Wilsonian renormalization group:
\beq
\Lambda \sim r \,\, .
\eeq

Now we are ready to compute the $\beta$-functions of the field theory from the supergravity moduli fields (\ref{RGholography}). The constancy of the dilaton translates into the vanishing of the $\beta$-function for $\frac{8\pi^2}{g_+^2} \equiv \frac{8\pi^2}{g_1^2}+\frac{8\pi^2}{g_2^2}$. The second holographic relation in (\ref{RGholography}) gives rise to a logarithmic running of $\frac{8\pi^2}{g_-^2} \equiv \frac{8\pi^2}{g_1^2}-\frac{8\pi^2}{g_2^2}$ in the $SU(N_c+M) \times SU(N_c)$ gauge theory:
\beq  \label{betafunction}
\frac{8\pi^2}{g_-^2}\,=\,6M \ln{(r/r_s)} \, +\, \rm{const.} \,\, ,
\eeq
since $\ln{(r/r_s)}=\ln{(\Lambda/\mu)}$.

If we compare with the Shifman-Vainshtein $\beta$-function \cite{Shifman} we find that 
\beq
\frac{8\pi^2}{g_-^2}\,=\,M \ln{(\Lambda/\mu)} (3\,+\,2(1-\gamma)) \,\, ,
\eeq
where $\gamma$ is the anomalous dimension of operators $\tr{A_iB_j}$. The conformal invariance of the field theory for $M=0$ and the symmetry under $M \to -M$ require that $\gamma=-\frac{1}{2}+ {\cal O}(M/N_c)^{2n}$ with $n$ a positive integer \cite{Klebanov:2000hb}. Then, 
\beq
\frac{8\pi^2}{g_-^2}\,=\,6M \ln{(\Lambda/\mu)} (1\,+\,{\cal O}(M/N_c)^{2n}) \,\, .
\eeq
Therefore the coefficient $6M$ is in exact agreement with (\ref{betafunction}).

It is worth pointing out again that as the theory flows to the IR, we must perform a Seiberg dualities each time that one of gauge coupling diverges. As we have explained, this duality decreases the rank of the gauge group on which it acts. On the gravity side this effect translates into a decrease of $N_{eff}$ in units of $M$ (\ref{KTSD}). However this cascade must stop before reaching a region where $N_{eff}$ is negative. The fact that the solution  described above is singular in the IR tells us that it has to be modified there. The proper modification goes via the deformation of the conifold as we will explain in the next subsection. 

Finally we want to discuss the chiral anomaly of the Klebanov-Tseytlin model. We show now how the chiral anomaly of an ${\cal N}=1$ cascading gauge theory can be read from the supergravity solution of \cite{Klebanov:2000nc}. In the quantum field theory there are chiral fermions charged under the $U(1)_R$ symmetry and we can understand the R-symmetry breaking as an effect of the chiral anomaly. An standard result of quantum field theory is that in a theory with chiral fermions charged under a global $U(1)$ symmetry of the classical lagrangian, the Noether current $J^{\mu}_R$ associated with an infinitesimal R-symmetry transformation is not generally conserved but instead obeys the equation
\beq  \label{QFTanomaly}
\partial_{\mu}J^{\mu}_R\,=\,\frac{1}{32\pi^2}\sum_f R_f T[\mathcal{R}^{(f)}]\,F^{a}_{\mu \nu} \tilde{F}_{a}^{\mu \nu} \,\, ,
\eeq
where the sum runs over the fermions $f$ circulating in the loop of the relevant triangle anomaly, $R_f$ is the R-charge of the fermion and $T[\mathcal{R}^{(f)}]$ is the Dynkin index of the gauge group representation $ \mathcal{R}^{(f)}$ that the fermion belongs to, normalised as $ T[\mathcal{R}^{(fund.)}]=1$ and $ T[\mathcal{R}^{(adj.)}]=2 N_c$. We follow the convention that fixes the R-charge of the gauginos as $R[\lambda]=1$. In (\ref{QFTanomaly}) $F^{a}_{\mu \nu}$ is the field strength associated to the gauge group, $a$ the gauge index and $\mu, \nu$ are spacetime indices.

In the case of interest we take an $SU(N_c+M) \times SU(N_c)$ gauge theory in the UV. There are two gauge groups, so let us define $F^{a}_{\mu \nu}$ and $G^{a}_{\mu \nu}$ to be the field strengths of $SU(N_c+M)$ and $SU(N_c)$ respectively. In computing the $U(1)_R - SU(N_c+M) - SU(N_c+M)$ and $U(1)_R - SU(N_c) - SU(N_c)$ triangle anomalies we obtain
\beq
\partial_{\mu}J^{\mu}_R\,=\,\frac{M}{16\pi^2}\, \left ( F^{a}_{\mu \nu} \tilde{F}_{a}^{\mu \nu}\,-\,G^{a}_{\mu \nu} \tilde{G}_{a}^{\mu \nu} \right ) \,\, ,
\eeq
or in other words, under an $U(1)_R$ transformation of parameter $\epsilon$, the $\Theta$-angles\footnote{With a conventional normalization, the $\Theta$-angle terms appear in the gauge theory action as $\int d^4x (\frac{\Theta_1}{32\pi^2}F^{a}_{\mu \nu} \tilde{F}_{a}^{\mu \nu}+\frac{\Theta_2}{32\pi^2}G^{a}_{\mu \nu} \tilde{G}_{a}^{\mu \nu})$.} for each gauge group transforms as
\bear
&& \delta_{\epsilon}\Theta_1\,=\,2M\epsilon \,\, , \rc
&& \delta_{\epsilon}\Theta_2\,=\,-2M\epsilon \,\, ,
\eear
or equivalently
\bear  \label{anomalycoeff}
&& \delta_{\epsilon}(\Theta_1+\Theta_2)\,=\,0 \,\, , \rc
&& \delta_{\epsilon}(\Theta_1-\Theta_2)\,=\,4M\epsilon \,\, .
\eear

In order to compare the above analysis with that of the Klebanov-Tseytlin model we will borrow again the holographic relations computed in the orbifold theory \cite{Kachru:1998ys, Lawrence:1998ja, Polchinski:2000mx}\footnote{ Actually, we are not sure about that sign in the first equation below. At any rate, with this minus sign the R-anomaly computation of the supergravity backgrounds in chapters \ref{KW} and \ref{KS} match exactly the field theory computations.}
:
\bear  \label{holographic_theta}
&& \Theta_1\,+\,\Theta_2\,=\,-2\pi \chi \,\, , \rc
&&\Theta_1\,-\,\Theta_2\,=\,\frac{1}{\pi \alpha'} \int_{S^2} C_2 \,\, (\rm{mod}\,\, 4\pi) \,\, .
\eear
The ambiguity of the second equation is subtle: it corresponds to the two kinds of fractional D(-1)-branes appearing in the theory. The angles $\Theta_1$ and $\Theta_2$ come from the imaginary parts of the action of two kinds of fractional Euclidean D(-1) branes. Both of them are then
defined \emph{modulo} $2\pi$ in the quantum field theory. On the string theory side the periodicities exactly match: an Euclidean fractional D(-1)-brane enters the functional
integral with a term $\exp\bigl\{-\frac{8\pi^2}{g_j^2}+i\Theta_j\bigr\}$.
\footnote{We have written the complexified gauge coupling instead of the
supergravity fields for the sake of brevity: the use of the dictionary given in eqs. (\ref{AdS/CFT}) and (\ref{AdS/CFT1}) is
understood.} Hence the imaginary part in the exponent is defined \emph{modulo} $2\pi$
in the quantum string theory.

Although the asymptotic UV metric has a $U(1)_R$ symmetry (dual to the R-symmetry of the gauge theory) associated with rotations of the angular coordinate $\psi/2$ (\ref{T11}, \ref{KTmetric}), the RR two-form $C_2$ does not have this continuous symmetry. Actually, the RR field strength $F_3=dC_2$ does have this symmetry but there is no smooth global expression for $C_2$. Locally we can write
\beq  \label{Z2M}
C_2\,=\,\frac{M\alpha'}{2}\psi \, \Upsilon_2 \,\, ,
\eeq
which is not single-valued as a function of the angular variable $\psi$ and it is not invariant under the $U(1)_R$. Under the transformation $\psi \to \psi+2\epsilon$, the RR potential $C_2$ changes as follows:
\beq
C_2 \to C_2\, + \, M\alpha'\epsilon \, \Upsilon_2 \,\, .
\eeq
Notice that $\int_{S^2}C_2$ is defined modulo $4\pi^2\alpha'$ (\ref{holographic_theta}). Therefore the transformation $\psi \to \psi+2\epsilon$ is a symmetry only when $\epsilon$ is an integer multiple of $\pi/M$. Since $\epsilon$ is defined modulo $2\pi$, we conclude that only a $\zet_{2M}$ subgroup of the $U(1)_R$ is an actual symmetry of the system. 

Let us now compare the anomaly coefficients obtained from field theory computations (\ref{anomalycoeff}) with those given by the holographic relations (\ref{holographic_theta}). The fact that there is no RR scalar field $\chi$ in the Klebanov-Tseytlin model translates into the simple relation of the $\Theta$-angles $\Theta_1=-\Theta_2\equiv \Theta$. Thus, under the $U(1)_R$ rotation $\psi \to \psi+2\epsilon$  the holographic relations yield
\bear
&&\delta_{\epsilon}( \Theta_1\,+\,\Theta_2)\,=\,0 \,\, , \rc
&&\delta_{\epsilon}(\Theta_1\,-\,\Theta_2)\,=\,\frac{1}{\pi \alpha'} \int_{S^2} M\alpha'\epsilon \, \Upsilon_2 \,=\,4M\epsilon\,\, ,
\eear
in perfect agreement with (\ref{anomalycoeff}).

In summary, the chiral anomaly of the $SU(N_c+M) \times SU(N_c)$ gauge theory is encoded in the UV behaviour (large $r$) of the dual {\it classical} supergravity background.

\subsection{Deformation of the conifold: the Klebanov-Strassler (KS) model}  \label{deformedconifold}

It was shown in \cite{Klebanov:2000hb} that the resolution of the naked singularity of the Klebanov-Tseytlin model \cite{Klebanov:2000nc} occurs though the replacement of the conifold (\ref{zcoor}) by the deformed conifold, whose complex structure is described by the following equation in $\bbC^4$:
\beq  \label{zcoorKS}
z_1\,z_2\,-\,z_3\,z_4\,=\,\epsilon^2\,\, .
\eeq
The manifold defined by (\ref{zcoorKS}) has isometry group $SU(2) \times SU(2)$, where the non-abelian factors are understood as left and right multiplication on the matrix $\bigl( \begin{smallmatrix} z_1 & z_4 \\ z_3 & z_2 \end{smallmatrix} \bigr)$. There is also a $U(1)_R$ action given a common phase rotation to all the complex coordinates. However this symmetry is broken to $\zet_2$ by the deformation parameter. The singularity of the conifold is removed through the blowing-up of the $S^3$ of the $T^{1,1}$ at the tip. The equation (\ref{zcoorKS}) was studied in detail in \cite{Candelas:1989js}. Following closely the technique employed there to solve the equation, one can find a parameterisation of the complex variables in terms of the angular variables introduced in (\ref{T11}) and a dimensionless radial coordinate $\tau$ as:
\bear  \label{zcoorKSangular}
&&z_1\,=\,-\epsilon \,e^{-\frac{i}{2}(\varphi_1+\varphi_2)} \left ( e^{(\tau+i\psi)/2}\sin {\frac{\theta_1}{2}} \sin{\frac{\theta_2}{2}} \,-\,e^{-(\tau+i\psi)/2}\cos {\frac{\theta_1}{2}} \cos{\frac{\theta_2}{2}} \right ) \,\, , \rc
&&z_2\,=\,\epsilon \,e^{\frac{i}{2}(\varphi_1+\varphi_2)} \left ( e^{(\tau+i\psi)/2}\cos {\frac{\theta_1}{2}} \cos{\frac{\theta_2}{2}} \,-\,e^{-(\tau+i\psi)/2}\sin {\frac{\theta_1}{2}} \sin{\frac{\theta_2}{2}} \right ) \,\, , \rc
&&z_3\,=\,-\epsilon \,e^{\frac{i}{2}(\varphi_1-\varphi_2)} \left ( e^{(\tau+i\psi)/2}\cos {\frac{\theta_1}{2}} \sin{\frac{\theta_2}{2}} \,+\,e^{-(\tau+i\psi)/2}\sin {\frac{\theta_1}{2}} \cos{\frac{\theta_2}{2}} \right ) \,\, , \rc
&&z_4\,=\,\epsilon \,e^{-\frac{i}{2}(\varphi_1-\varphi_2)} \left ( e^{(\tau+i\psi)/2}\sin {\frac{\theta_1}{2}} \cos{\frac{\theta_2}{2}} \,+\,e^{-(\tau+i\psi)/2}\cos {\frac{\theta_1}{2}} \sin{\frac{\theta_2}{2}} \right ) \,\, .
\eear

The ten-dimensional metric of \cite{Klebanov:2000hb} takes the {\it standard} form
\beq  \label{KSmetric}
ds^2\,=\,h(\tau)^{-1/2}dx^2_{1,3}\,+\,h(\tau)^{1/2}ds^2_6 \,\, ,
\eeq
where $ds^2_6$ is now the metric of the deformed conifold. Using the angular variables of (\ref{T11}) and the basis (\ref{gbasis}), the metric of the deformed conifold is diagonal:
\beq  \label{unflavourdeform}
ds^2_6\,=\,\frac{1}{2}\epsilon^{4/3}K(\tau) \left [  \frac{1}{3K^3(\tau)}(d\tau^2+(g^5)^2)+\cosh^2{\left ( \frac{\tau}{2} \right )} [(g^3)^2+(g^4)^2]+\sinh^2{\left ( \frac{\tau}{2} \right )} [(g^1)^2+(g^2)^2]  \right ] \,\, ,
\eeq
where
\beq  \label{KSfunction}
K(\tau)\,=\, \frac{(\sinh{(2\tau)}-2\tau)^{1/3}}{2^{1/3}\sinh{\tau}}\,\, .
\eeq
In the UV limit (large $\tau$) it is convenient to introduce another radial variable $r$
\beq   \label{newradial}
r^2\,=\,\frac{3}{2^{5/3}}\epsilon^{4/3}e^{2\tau/3} \,\, , 
\eeq
in terms of which the deformed conifold metric for large values of the new radial variable turns into that of the singular conifold (\ref{ds6}), $ds^2_6 \to dr^2\,+\,r^2\,ds^2_{T^{1,1}}$.

At $\tau=0$ the angular metric degenerates into
\beq  \label{KSIRlimit}
d\Omega^2_3\,=\,\frac{1}{2}\epsilon^{4/3} \left ( \frac{2}{3} \right )^{1/3} [\frac{1}{2}(g^5)^2\,+\,(g^3)^2\,+\,(g^4)^2] \,\, ,
\eeq
which is the metric of a round $S^3$ \cite{Candelas:1989js}. The additional two directions corresponding to the $S^2$ fibered over the $S^3$ shrink as \cite{Klebanov:2000hb}
\beq  \label{KSIRlimit1}
\frac{1}{8}\epsilon^{4/3} \left ( \frac{2}{3} \right )^{1/3} \tau^2[(g^1)^2\,+\,(g^2)^2] \,\, .
\eeq

Apart from the metric of the {\it warped} deformed conifold, the model proposed by Klebanov and Strassler in \cite{Klebanov:2000hb} also includes RR field strengths $F_3$, $F_5$ and a Kalb-Ramond potential $B_2$. Let us write them down:
\bear  \label{KTKSforms}
&&F_3\,=\,\frac{M\alpha'}{2} \left [ g^5 \wedge g^3 \wedge g^4 \,(1-F(\tau))\,+\,g^5 \wedge g^1 \wedge g^2\,F(\tau)\,+\,F'(\tau)\,d\tau \wedge (g^1\wedge g^3 \,+\,g^2 \wedge g^4) \right ] \,\, , \rc
&&B_2\,=\,\frac{g_sM\alpha'}{2} \left [ f(\tau)\,g^1 \wedge g^2\,+\,k(\tau)\,g^3 \wedge g^4 \right ] \,\, , \rc
&&F_5\,=\,(1\,+\, \star)\, \frac{g_sM^2(\alpha')^2}{4}l(\tau)g^1 \wedge g^2  \wedge g^3  \wedge g^4  \wedge g^5 \,\, ,
\eear
where
\beq
l\,=\,f\,(1\,-\,F)\,+\,k\,F \,\,.
\eeq
The form of the functions entering into the flux forms was also given in \cite{Klebanov:2000hb}. They were found solving a first-order system of equations derived from a superpotential for the effective radial problem \cite{Pando Zayas:2000sq}:
\bear  \label{firstorder}
&&f'\,=\,(1\,-\,F)\, \tanh^2{(\tau/2)} \,\, , \rc
&&k'\,=\,F\, \coth^2{(\tau/2)} \,\, , \rc
&&F'\,=\,\frac{1}{2}(k\,-\,f) \,\, ,
\eear
and
\beq
h'\,=\,-4(g_sM\alpha')^2\epsilon^{-8/3} \frac{f\,(1-F)+k\,F}{K^2(\tau) \sinh^2{\tau}} \,\,.
\eeq

The Klebanov-Strassler model preserves some amount of supersymmetry and it was shown in \cite{Grana:2000jj} that it is $1/8$ supersymmetric (see \cite{Arean:2006nc} for a explicit calculation). One can also check that the above system of first-order equations is a solution of the second-order equations of motion of type IIB supergravity (\ref{IIBeq} - \ref{MAX}). It is also interesting to point out that (\ref{firstorder}) implies the imaginary self-duality condition of the three-form (\ref{G3}) with respect to the metric of the deformed conifold. For completeness we give the solution of the system (\ref{firstorder}):
\bear
&&F(\tau)\,=\,\frac{\sinh{\tau}-\tau}{2\sinh{\tau}}\,\, , \qquad \qquad \qquad \qquad f(\tau)\,=\,\frac{\tau \coth{\tau}-1}{2\sinh{\tau}}(\cosh{\tau}-1) \,\, , \rc
&&k(\tau)\,=\,\frac{\tau \coth{\tau}-1}{2\sinh{\tau}}(\cosh{\tau}+1)\,\, , \qquad  l(\tau)\,=\,\frac{\tau \coth{\tau}-1}{4\sinh^2{\tau}}(\sinh{2\tau}-2\tau)\,\, , \rc
\eear
where we have tuned the integration constants to avoid singularities in the fluxes.

Once we have solved the system of equations for the three-forms, we can immediately integrate the warp factor. As usual, the boundary condition we have to impose is its vanishing at large $\tau$. Thus the result is \cite{Klebanov:2000hb}:
\beq  \label{KSwarp}
h(\tau)\,=\,(g_sM\alpha')^22^{2/3}\epsilon^{-8/3} I(\tau) \,\, , 
\eeq
where
\beq  \label{KSintegral}
I(\tau)\,=\,\int_{\tau}^{\infty}dx\, \frac{x\coth{x}-1}{\sinh^2{x}}(\sinh{2x}-2x)^{1/3} \,\, .
\eeq
This $I(\tau)$ is non-singular at the tip of the deformed conifold and matches (\ref{KTwarp}) at large $\tau$. Actually this model \cite{Klebanov:2000hb} turns into the Klebanov-Tseytlin one \cite{Klebanov:2000nc} at large $\tau$, once we perform the radial change of variable (\ref{newradial}).

The curvature of the metric (\ref{KSmetric}) is small everywhere for large values of $g_sM$. As we will see below this is the t'~Hooft coupling of the gauge theory far in the IR. As long as this is large, the curvature is small and the supergravity approximation is reliable.  

The theory that we have just described is confining. This means that the quark-antiquark potential is linear with the distance between them. On the gravity side, an external quark is a fundamental string that comes in from infinity and ends on one of the branes which generate the background. When we have a quark-antiquark pair and we separate them by a large distance, we can think about the quark-antiquark potential as the energy of a fundamental string extended on one of the spatial directions where the theory lives. If the tension of that fundamental string is constant, this is a sign of confinement of the dual gauge theory. This is what happens in the metric (\ref{KSmetric}), where for small values of $\tau$ the function multiplying $dx^2_{1,3}$, {\it i.e.} $h^{-1/2}(\tau)$ approaches a constant (\ref{KSwarp}, \ref{KSintegral}). In order to explain more in detail this argument, let us recall that the fundamental string corresponds to the Wilson loop in the fundamental representation of the gauge group. The classic criterion for confinement is that this Wilson loop $W_1(C)$ obeys the area law (in the limit of large area)  \cite{Maldacena:1998im}
\beq  \label{Alaw}
-\ln{<W_1(C)>}\, \approx \,A \,\, , 
\eeq
where $A$ is the area enclosed by the loop $C$ in the gauge theory directions. Considering a Wilson contour at fixed $\tau$ in the warped deformed conifold, the minimal surface bounded by the contour bends towards small $\tau$. If the contour has a very large area, then most of the minimal surface will drift down into the region near $\tau=0$. Being the coefficient of $dx^2_{1,3}$ finite at $\tau=0$ means that the tension of the fundamental string is also finite and so the resulting Wilson loop satisfies the area law (\ref{Alaw}). The tension of the confining string scales as \cite{Klebanov:2000hb}
\beq
T_s\,=\,\frac{\epsilon^{4/3}}{2^{4/3}a_0^{1/2}\pi}\frac{1}{g_sM(\alpha')^2} \,\, ,
\eeq
where $a_0\,=\,I(\tau=0)\approx 0.71805$ (see eq. (\ref{KSintegral})).

A generalisation of the confining string is to consider Wilson loops in antisymmetric tensor representations of the gauge group $SU(M)$ with $q$ indices where $q$ ranges from $1$ to $M-1$. These Wilson loops can be thought of as confining strings which connect $q$ quarks on one end to $q$ antiquarks on the other end. These are the so-called $q$-strings. The case $q=1$ reduces to the fundamental representation and there is a symmetry under $q \to M-q$ which corresponds to replacing quarks by antiquarks. For $q=M$ the quarks combine into a colourless state (a baryon) and the Wilson loop does not satisfy the law area (\ref{Alaw}). The tension of a confining $q$- string in the deformed conifold was computed in \cite{Herzog:2001fq}. There they found that, approximately, this tension goes like
\beq  \label{qstringtension}
T_q\, \sim \,c \sin{\frac{\pi q}{M}} \,\, ,
\eeq
where $c$ is a constant related to an IR scale of the gauge theory. This is in agreement with what is expected for a confining ${\cal N}=1$ supersymmetric $SU(M)$ gauge theory in four dimensions. Notice that as
\beq
T_{q+q'}\, < \, T_q \,+\, T_{q'} \,\, , 
\eeq
the $q$-string will not decay into strings with smaller $q$.

Let us finally discuss briefly the relation between the deformation of the conifold and the pattern of chiral symmetry breaking of the dual field theory. We have already argued that the singularity in the solution of \cite{Klebanov:2000nc} is removed through the blowing-up of the $S^3$ of the $T^{1,1}$ at the tip of the conifold. This blowing-up avoids the divergence of the field strength $F_3$ at $\tau=0$. However, the most powerful argument to see that the conifold is deformed comes from the field theory analysis. The dual field theory in the UV has gauge group $SU(N_c+M) \times SU(N_c)$. We have seen that when the energy scale flows to the IR, it is necessary to perform a Seiberg duality each time that the gauge coupling of one of the gauge groups diverges. We have also seen that in this cascading of Seiberg dualities, the rank of the gauge groups decreases alternatively in $M$ units. This process must stop since negative values of the rank of a gauge group does not make sense. Here we only pay attention to the case in which $N_c$ is multiple of $M$, namely $N_c=pM$. In this particular case the bottom of the cascade is a supersymmetric $SU(2M) \times SU(M)$ gauge theory. The classical moduli space of this theory is modified at the quantum level by nonperturbative effects \cite{Seiberg:1994bz, Dymarsky:2005xt}. The theory acquires a deformed moduli space with $M$ independent branches, each of which has the shape of a deformed conifold (\ref{zcoorKS}). The branches are permuted by the $\zet_{2M}$ R-symmetry (chiral symmetry), which is spontaneously broken down to $\zet_2$. This breaking of the R-symmetry is exactly what one would expect in a pure $SU(M)$ ${\cal N}=1$ Yang-Mills theory in four dimensions. In supersymmetric gluodynamics the breaking of the chiral symmetry is associated with the gluino condensation $<\lambda^2>$ \cite{Klebanov:2002gr, Bertolini:2002xu, anomaly}. A holographic calculation of the gluino condensate was carried out in \cite{Loewy} and the result is
\beq
<\lambda^2> \sim M \frac{\epsilon^2}{(\alpha')^3} \,\, ,
\eeq
which depends on the parameter $\epsilon^2$ of the equation of the deformed conifold (\ref{zcoorKS}).

The chiral symmetry breaking can also be realised in the supergravity solution. Recall that in the UV, only a $\zet_{2M}$ subgroup of the $U(1)_R$ parameterised by $\psi/2$ survives due to the explicit dependence on $\psi$ of the RR two-form $C_2$ (\ref{Z2M}). Recalling also that $\psi$ ranges from $0$ to $4\pi$ we see that in the Klebanov-Strassler model, which depends on $\psi$ through $\cos{\psi}$ and $\sin{\psi}$, the $\zet_{2M}$ symmetry is further broken to $\zet_2$, generated by $\psi \to \psi+2\pi$. Therefore $M$ different vacua come up due to the breaking of the symmetry by IR effects. As a consequence, domain walls appear interpolating among them. They are D5-branes wrapping at $\tau=0$ (domain walls are IR effects) the finite-sized $S^3$ and with the remaining directions along $\reals^{1,3}$. We will study them more in detail in section \ref{WhyDp}.

We have seen that the Klebanov-Strassler model describes some features of an $SU(M)$ ${\cal N}=1$ supersymmetric gauge theory as confinement or the pattern of chiral symmetry breaking. However, it is not its gravitational dual because $SU(M)$ ${\cal N}=1$ supersymmetric gauge theory can be achieved by taking $g_sM \to 0$ and sending the scale of the last step of the cascade $SU(2M) \times SU(M)$ to infinity. Unfortunately, this is the opposite limit of that where the supergravity approximation is reliable ($g_sM \to \infty$), as we discussed previously.

\subsection{D3-branes on the cone over $Y^{p,q}$ manifolds}  \label{Ypqintroduc}

Recently, a new class of Sasaki-Einstein five-dimensional manifolds $Y^{p,q}$, $p$ and $q$
being two coprime positive integers, has been constructed \cite{GMSW1,GMSW2}. They correspond to a new family of solutions of the equations of motion of type IIB supergravity dual to four-dimensional ${\cal N}=1$ SCFT's. They are basically the same as that of the Klebanov-Witten model (\ref{ds6}, \ref{solucKW}) but replacing the Sasaki-Einstein manifolds $T^{1,1}$ by $Y^{p,q}$. From a physical point of view, these solutions are generated by a stack of $N_c$ D3-branes at the apex of the cone over the $Y^{p,q}$ manifold that we will denote as $CY^{p,q}$. Recall that the normalization of the scale factor $L$ is dictated by the quantization of the D3-brane tension (\ref{quantizationF5}). This scale is associated to the radius of the AdS space and changes accordingly to the volume of the internal manifold:
\beq
L^4\,=\,{4\pi^4\over {\rm Vol}(Y^{p,q})}\,g_s\,N_c\,(\alpha')^2\,\,.
\label{L}
\eeq

First of all, we shall briefly review basic features of the $Y^{p,q}$ manifolds. The metric of this Sasaki-Einstein space can be written as \cite{GMSW1, GMSW2}:
\bear
&&ds^2_{Y^{p,q}}\,=\,{1-cy\over
6}\,(d\theta^2\,+\,\sin^2\theta\,d\phi^2)\,+\, {1\over
6\,H^2(y)}\,dy^2\,+\,{H^2(y)\over 6}\, 
(d\beta\,-\,c\cos\theta d\phi)^2\,\rc
&&\qquad\qquad+\,{1\over 9}\,
\big[\,d\psi\,+\,\cos\theta d\phi\,+\,y(d\beta\,-\,c\cos\theta d\phi)\,
\big]^2\,\,,
\label{Ypqmetric}
\eear
$H(y)$ being given by:
\beq
H(y)=\sqrt{{a-3y^2+2cy^3\over 3(1-cy)}}\,\,.
\label{Hfunction}
\eeq
A natural frame for this space reads
\bea
e^1 &=& -\frac{L}{\sqrt{6}}\, \frac{1}{H(y)}\, dy ~, \nonumber \\
e^2 &=& -\frac{L}{\sqrt{6}}\, H(y)\, (d\beta- c\, \cos\theta\,
	d\phi) ~, \nonumber \\ 
e^3 &=& \frac{L}{\sqrt{6}}\, \sqrt{1-c\, y}\, d\theta\,\,, \nonumber \\
e^4 &=& \frac{L}{\sqrt{6}}\, \sqrt{1-c\,y}\sin\theta\, d\phi, \nonumber \\
e^5 &=& {L\over 3}\, \left( d\psi + y\, d\beta +(1-c\, y) \cos\theta\,
d\phi\right) ~.
\label{Ypqvielbein}
\eea

The metrics $ds^2_{Y^{p,q}}$ are Sasaki-Einstein, which means that the
cones $CY^{p,q}$ with metric $dr^2+r^2 ds^2_{Y^{p,q}}$ are Calabi-Yau
manifolds. The metrics in these coordinates neatly display some nice local
features of these spaces. Namely, by writing it as
\beq
ds^2_{Y^{p,q}}\,=\,ds^2_{4} + \left[ \frac{1}{3} d\psi + \sigma
\right]^2 ~,
\label{slice}
\eeq
it turns out that $ds^2_{4}$ is a K\"ahler-Einstein metric with K\"ahler
form $J_4 = \frac{1}{2} d\sigma$. Notice that this is a local splitting
that carries no global information. Indeed, the pair $(ds^2_{4},J_4)$ is
not in general globally defined. The Killing vector $\frac{\partial
~}{\partial\psi}$ has constant norm but its orbits do not close (except
for certain values of $p$ and $q$, see below). It defines a foliation
of $Y^{p,q}$ whose transverse leaves, as we see, locally have a
K\"ahler-Einstein structure. This aspect will be important in chapter \ref{Ypq}.

These $Y^{p,q}$ manifolds are topologically $S^2\times S^3$ and can be
regarded as one-dimensional bundles over manifolds of topology  $S^2\times S^2$.
Their isometry group is $SU(2) \times U(1)^2$. Notice that the metric
(\ref{Ypqmetric}) depends on  two constants $a$ and $c$. The latter,
if different from zero, can be set to one by a suitable rescaling of the
coordinate $y$, although it is sometimes convenient to keep the value of
$c$ arbitrary in order to be able to recover the $T^{1,1}$ geometry, which
corresponds to $c=0$ \footnote{If $c=0$, we can set $a = 3$ by rescaling
$y \to \xi y$, $\beta \to \xi^{-1} \beta$, and $a \to \xi^2 a$. If we
further write $y = \cos \theta_2$ and $\beta = \varphi_2$, identifying $\theta \equiv \theta_1$ and $\phi \equiv \varphi_1$, and
choose the period of $\psi$ to be $4\pi$, the metric goes to that of
$T^{1,1}$ (\ref{T11}).}. If $c\neq 0$, instead, as we have just said we can set 
 $c = 1$ and the parameter $a$ can be written in terms of two
coprime integers $p$ and $q$ (we take $p>q$) as follows:
\beq
a\,=\,{1\over 2}\,-\,{p^2-3q^2\over 4p^3}\,\,\sqrt{4p^2-3q^2}\,\,.
\eeq
Moreover, the coordinate $y$ ranges between the two smaller roots of the
cubic equation
\beq
\calq(y) \equiv a - 3y^2 + 2cy^3\,=\,2c\,\prod_{i=1}^{3}\,
(y-y_i) ~,
\label{calq}
\eeq
\ie\ $\,y_1 \le y \le y_2$ with (for $c=1$):
\bear
&&y_1\,=\,{1\over 4p}\,
\Big(\,2p\,-\,3q\,-\,\sqrt{4p^2-3q^2}\,\Big)\,< 0\,,\rc
&&y_2\,=\,{1\over 4p}\,
\Big(\,2p\,+\,3q\,-\,\sqrt{4p^2-3q^2}\,\Big)\,> 0\,.
\label{y12}
\eear
In order to specify the range of the other variables appearing in the metric,
let us introduce the coordinate $\alpha$ by means of the relation:
\beq
\beta\,=\,-(6\alpha+c\psi)\,\,.
\label{beta-alpha}
\eeq
Then, the coordinates $\theta$, $\phi$, $\psi$ and $\alpha$ span the range:
\beq
0\le \theta\le \pi\,\,,\qquad
0< \phi\le 2\pi\,\,,\qquad
0< \psi\le 2\pi\,\,,\qquad
0< \alpha\le 2\pi \ell\,\,,
\eeq
where $\ell$ is (generically an irrational number) given by:
\beq
\ell\,=\,-{q\over 4p^2\,y_1\,y_2}\,=\,{q\over 3q^2\,-\,2p^2\,+\,p
\sqrt{4p^2\,-\,3q^2}}\,\,,
\label{alphaperiod}
\eeq
the metric (\ref{Ypqmetric}) being periodic in these variables. Notice that,
whenever $c\neq 0$, the coordinate $\beta$ is non-periodic: the periodicities
of $\psi$ and $\alpha$ are not congruent, unless the manifold is quasi-regular,
{\it i.e.}, there exists a positive integer $k$ such that
\beq
k^2 = 4 p^2 - 3 q^2 ~.
\label{qreg}
\eeq

For quasi-regular manifolds, $ds^2_{4}$ in (\ref{slice}) corresponds to a
K\"ahler-Einstein orbifold. Notice that $\ell$ becomes rational and it is
now possible to assign a periodicity to $\psi$ such that $\beta$ ends up
being periodic. If we perform the change of variables (\ref{beta-alpha})
in (\ref{Ypqmetric}), we get
\bear 
&&ds^2_{Y^{p,q}}\,=\,{1-cy\over
6}\,(d\theta^2\,+\,\sin^2\theta\,d\phi^2)\,+\, {1\over
6\,H^2(y)}\,dy^2\,+\,{v(y)\over 9}\, 
(d\psi\,+\,\cos\theta d\phi)^2\,+\,\rc
&&\qquad\qquad+\,w(y)\,
\big[\,d\alpha\,+\,f(y) \left( d\psi
+ \cos\theta d\phi \right) \big]^2\,\,,
\label{alphametric}
\eear
with $v(y)$, $w(y)$ and $f(y)$ given by
\beq
v(y) = {a - 3y^2 + 2cy^3 \over a - y^2} ~, ~~~~~
w(y) = {2 (a - y^2) \over 1 - cy} ~, ~~~~~
f(y) = {ac - 2y + y^2c \over 6 (a - y^2)}~.
\label{qwy}
\eeq

The volume of this manifold can be computed straightforwardly from the metric
(\ref{Ypqmetric}), with the result (for $c=1$):
\beq
{\rm Vol}(Y^{p,q})\,=\,{q^2\over 3p^2}\,\,
{2p+\sqrt{4p^2-3q^2}\over 3q^2-2p^2+p\sqrt{4p^2-3q^2}}\,\,\pi^3\,\,.
\eeq

It will be useful to give a set of complex coordinates describing $CY^{p,q}$. The starting point in
identifying a good set of them is the following set of
closed one-forms \cite{ms} (here we follow the notation of \cite{ben}):
\begin{eqnarray}
{\eta}^1 & = & \frac{1}{\sin\theta}\, d\theta - i d\phi ~, \nonumber\\
\tilde{\eta}^2 & = & -\frac{dy}{H(y)^2} - i (d\beta - c\cos\theta
d\phi) ~, \nonumber\\
\tilde{\eta}^3 & = & 3 \frac{dr}{r} + i \big[ d\psi + \cos\theta d\phi
+ y (d\beta - c\cos\theta d\phi) \big] ~,
\end{eqnarray}
in terms of which, the metric of $CY^{p,q}$ can be rewritten as
\begin{equation}
ds^2 = r^2 \frac{(1-cy)}{6}\, {\sin}^2\theta ~|{\eta}^1|^2 + r^2
\frac{H(y)^2}{6} ~|{\tilde{\eta}^2}|^2 + \frac{r^2}{9}
~|{\tilde{\eta}^3}|^2 ~.
\end{equation}
Unfortunately, ${\tilde{\eta}^2}$ and ${\tilde{\eta}^3}$ are not integrable.
It is however easy to see that integrable one-forms can be obtained by taking
linear combinations of them:
\be
{\eta}^2 = {\tilde{\eta}^2}+c \cos \theta ~{\eta}^1 ~, \qquad
{\eta}^3 = {\tilde{\eta}^3}+\cos\theta ~{\eta}^1 + y ~{\tilde{\eta}^2} ~.
\ee
We can now define ${\eta}^i=dz_i/z_i$ for $i=1,2,3$, where
\beq
z_1 = \tan\frac{\theta}{2}\, e^{-i\phi} ~, \qquad
z_2 = \frac{(\sin\theta)^{c}}{f_1(y)}\, e^{-i\beta} ~, \qquad
z_3 = r^3\, \frac{\sin\theta}{f_2(y)}\, e^{i\psi}\,\,,
\label{complexzs}
\eeq
with $f_1(y)$ and $f_2(y)$ being given by:
\begin{equation}
f_1(y)=\exp\left(\int\frac{1}{H(y)^2}dy\right),\qquad
f_2(y)=\exp\left(\int\frac{y}{H(y)^2}dy\right).
\label{fs}
\end{equation}
By using the form of $H(y)$ written in eq.(\ref{Hfunction}) it is possible
to provide a simpler expression for the functions $f_i(y)$, namely:
\bear
&&{1\over f_1(y)}\,=\,\sqrt{(y-y_1)^{{1\over y_1}}\,
(y_2-y)^{{1\over y_2}}\,(y_3-y)^{{1\over y_3}}}\,\,,\rc
&&{1\over f_2(y)}\,=\,\sqrt{\calq(y)}\,=\,
\sqrt{2c}\,\sqrt{(y-y_1)\,(y_2-y)\,(y_3-y)}\,\,,
\label{yexpression}
\eear
where $\calq(y)$ has been defined in (\ref{calq}), $y_1$ and $y_2$ are
given in eq.(\ref{y12}) and $y_3$ is the third root of the polynomial
$\calq(y)$ which, for $c=1$, is related to $y_{1,2}$ as $y_3={3\over
2} - y_1 - y_2$. The holomorphic three-form of $CY^{p,q}$ simply reads  
\beq
\Omega = -\frac{1}{18} e^{i\psi} r^3 \sqrt{\frac{\calq(y)}{3}}
\sin\theta ~\eta^1 \wedge \eta^2 \wedge \eta^3 = -\frac{1}{18\sqrt{3}}
\frac{dz_1 \wedge dz_2 \wedge dz_3}{z_1 z_2} ~.
\label{threeform}
\eeq
Notice that coordinates $z_1$ and $z_2$ are local complex coordinates on the
transverse leaves of $Y^{p,q}$ (\ref{slice}) with K\"ahler-Einstein metric $ds_4^2$. They are not globally well defined as soon as $z_2$ is periodic in
$\beta$ --which is not a periodic coordinate. Besides, they are meromorphic
functions on $CY^{p,q}$ (the function $z_1$ is singular at $\theta = \pi$
while $z_2$ has a singularity at  $y = y_1$). A set
of holomorphic coordinates on $Y^{p,q}$ was constructed in \cite{BHOP}.

Recall
that the metric of the $Y^{p,q}$ manifold can be written as (\ref{slice}) with $\sigma$ being the one-form given by
\beq
\sigma\,=\,{1\over 3}\,\big[\,\cos\theta d\phi\,+\,y (d\beta\,-\,
c\cos\theta d\phi)\,\big]\,\,.
\eeq
The  K\"ahler form $J_4$ of the four-dimensional K\"ahler-Einstein space is just $J_4=\frac{1}{2}d\sigma$. In the frame (\ref{Ypqvielbein}) it can be written as
\beq
J_4\,=\,{1\over 2}\,d\sigma\,=\,{1\over L^2}\,\big[\,
e^1\wedge e^2\,-\,e^3\wedge e^4\,\big]\,\,.
\eeq
From the Sasaki-Einstein space $Y^{p,q}$  we can construct the Calabi-Yau cone
$CY^{p,q}$, whose metric is just given by: $ds^2_{CY^{p,q}}\,=\,dr^2 +
r^2\,ds^2_{Y^{p,q}}$. The  K\"ahler form $J$ of $CY^{p,q}$ is just:
\beq
J\,=\,r^2\,J_4\,+\,{r\over L}\,dr\wedge e^5\,\,,
\eeq
whose explicit expression in terms of the coordinates is:
\beq  \label{KahlerYpq}
J\,=\,-{r^2\over 6}\,(1-cy)\,\sin\theta d\theta\wedge d\phi\,+\,{1\over
3}\,r dr\wedge (d\psi+\cos\theta d\phi)\,+\,{1\over 6} d(r^2y)\wedge 
(d\beta\,-\,c\cos\theta d\phi)\,\,.
\eeq

We can compute now the Killing spinors for the $AdS_5\times Y^{p,q}$ background by imposing the vanishing of the type IIB supersymmetry transformations (\ref{Eframe1}). We will see that the $AdS_5\times Y^{p,q}$ background preserves eight supersymmetries, in
agreement with the ${\cal N}=1$ superconformal character of the corresponding
dual field theory, which has four ordinary supersymmetries and four
superconformal ones. The result of this calculation
is greatly simplified in some particular basis of frame one-forms, which
we will now specify. In the $AdS_5$ part of the metric 
\beq
ds^2_{\rm{AdS_5}}\,=\, \frac{r^2}{L^2}\,dx^{2}_{1,3}\,+\,\frac{L^2}{r^2}\, dr^2
\eeq
we will choose the
natural basis of vielbein one-forms, namely:
\beq
e^{x^{\alpha}}\,=\,{r\over L}\,\,dx^{\alpha}
\,\,,\,\,\,\,\,\,\,\, (\alpha=0,1,2,3)\,\,,
\,\,\,\,\,\,\,\,\,\,\,\,\,\,\,\,
e^{r}\,=\,{L\over r}\,\,.
\label{AdSvielbein}
\eeq
In the $Y^{p,q}$ directions we will use the frame given in (\ref{Ypqvielbein}). In order to write the expressions of the Killing spinors in a compact form, 
let us define the matrix $\Gamma_{*}$ as:
\beq
\Gamma_{*}\equiv i\Gamma_{x^0x^1x^2x^3}\,\,.
\label{gamma*}
\eeq
Then, the Killing spinors $\epsilon$ of the $AdS_5\times Y^{p,q}$ background 
can be written in terms of a constant spinor $\eta$ as:
\beq
\epsilon\,=\,e^{-{i\over 2}\psi}\, r^{-{\Gamma_{*}\over 2}}\,\,
\Big(\,1\,+\,{\Gamma_r\over 2L^2}\,\,x^{\alpha}\,\Gamma_{x^{\alpha}}\,\,
(1\,+\,\Gamma_{*}\,)\,\Big)\,\,\eta\,\,.
\label{adsspinor}
\eeq
The spinor $\eta$ satisfies the projections :
\beq
\Gamma_{12}\,\eta\,=\,-i\eta\,\,, \qquad\qquad
\Gamma_{34}\,\eta\,=\,i\eta\,\,,
\label{etaspinor}
\eeq
which show that this background preserves eight supersymmetries. Notice
that, since the matrix multiplying $\eta$ in eq.(\ref{adsspinor}) commutes
with $\Gamma_{12}$ and $\Gamma_{34}$, the spinor $\epsilon$ also satisfies
the conditions (\ref{etaspinor}), \ie:
\beq
\Gamma_{12}\,\epsilon\,=\,-i\epsilon\,\,, \qquad\qquad
\Gamma_{34}\,\epsilon\,=\,i\epsilon\,\,.
\label{epsilon-project}
\eeq
In eq. (\ref{adsspinor}) we are parameterising the dependence of $\epsilon$
on the coordinates of  $AdS_5$ as in ref. \cite{LPT}. In order to explore
this dependence in detail, it is interesting to decompose the constant
spinor $\eta$ according to the different eigenvalues of the matrix
$\Gamma_*$:
\beq
\Gamma_{*}\,\eta_{\pm}\,=\,\pm\eta_{\pm}\,\,.
\label{etamasmenos}
\eeq
Using this decomposition we obtain two types of Killing spinors:
\bear
e^{{i\over 2} \psi}\,\epsilon_{-} & = & r^{1/2}\,\eta_-\,\,,\rc
e^{{i\over 2} \psi}\,\epsilon_{+} & = & r^{-1/2}\,\eta_+\,+\,{r^{1/2}
\over L^2}\,\,\Gamma_r\,x^{\alpha}\, \Gamma_{x^{\alpha}}\,\eta_+\,\,.
\label{chiraladsspinor}
\eear
The four spinors $\epsilon_{-}$ are independent of the coordinates
$x^{\alpha}$ and $\Gamma_*\epsilon_{-}=-\epsilon_-$, whereas the
$\epsilon_{+}$'s do depend on the $x^{\alpha}$'s and are not eigenvectors
of $\Gamma_*$. The latter correspond to the four superconformal
supersymmetries, while the $\epsilon_{-}$'s correspond
to the ordinary ones. Notice also that the only dependence of these
spinors on the coordinates of the $Y^{p,q}$ space is through the
exponential of the angle $\psi$ in eq. (\ref{chiraladsspinor}).

In addition to the Poincar\'e coordinates $(x^{\alpha}, r)$ used above to
represent the $AdS_5$ metric, it is also convenient to write it in the
so-called global coordinates, in which $ds^2_{AdS_5}$ takes the form:
\beq
ds^2_{AdS_5}\,=\,L^2\,\Big[ -\cosh^2\varrho \,\,dT^2\,+\,d\varrho^2\,+\,
\sinh^2\varrho\,\,d\Omega_3^2 \Big]\,\,,
\label{globalADS}
\eeq
where $d\Omega_3^2$ is the metric of a unit three-sphere parameterised by
three angles $(\alpha^1, \alpha^2,\alpha^3)$:
\beq
d\Omega_3^2\,=\,(d\alpha^1)^2\,+\,\sin^2\alpha^1\Big(\,
(d\alpha^2)^2\,+\,\sin^2\alpha^2\,(d\alpha^3)^2\,\Big)\,\,,
\eeq
with $0\le\alpha^1,\alpha^2\le \pi$ and $0\le\alpha^3\le 2\pi$. In order
to write down the Killing spinors in these coordinates, we will choose
the same frame as in eq. (\ref{Ypqvielbein}) for the $Y^{p,q}$ part of the metric, while for the $AdS_5$ directions we will use:
\bear
&&e^{T}\,=\,L\cosh\varrho\,dT \,\,,\,\,\,\,\,\,\,\,\,\,\,\,\,\,\,\,
e^{\varrho}\,=\,Ld\varrho\,\,,\rc
&&e^{\alpha^1}\,=\,L\sinh\varrho\,d\alpha^1\,\,,\rc
&&e^{\alpha^2}\,=\,L\sinh\varrho\,\sin\alpha^1\,d\alpha^2\,\,,\rc
&&e^{\alpha^3}\,=\,L\sinh\varrho\,\sin\alpha^1\,\sin\alpha^2\,d\alpha^3\,\,.
\eear
If we now  define the matrix
\beq  \label{matrixgamma}
\gamma_*\,\equiv\,\Gamma_{T}\,\Gamma_{\varrho}\,
\Gamma_{\alpha^1\,\alpha^2\,\alpha^3}\,\,,
\eeq
then, the Killing spinors in these coordinates can be written as
\cite{globalads}:
\beq
\epsilon\,=\,e^{-{i\over 2}\psi}\,
e^{-i\,{\varrho\over 2}\,\Gamma_{\varrho}\gamma_*}\,
e^{-i\,{T\over 2}\,\Gamma_{T}\gamma_*}\,
e^{-{\alpha^1\over 2}\,\Gamma_{\alpha^1 \varrho}}\,
e^{-{\alpha^2\over 2}\,\Gamma_{\alpha^2 \alpha^1}}\,
e^{-{\alpha^3\over 2}\,\Gamma_{\alpha^3 \alpha^2}}\,\eta\,\,,
\label{globalspinor}
\eeq
where $\eta$ is a constant spinor that satisfies the same conditions as
in eq. (\ref{etaspinor}). 

The gauge theory dual to IIB on $AdS_5\times Y^{p,q}$ is by now well
understood. The infinite family of spaces $Y^{p,q}$ was shown to be dual to superconformal
quiver gauge theories \cite{ms,BeFrHaMaSp}. The study of AdS/CFT in these
geometries has shed light in many subtle aspects of SCFT's in four dimensions. Furthermore, the correspondence successfully
passed new tests such as those related to the fact that the central charge
of these theories, as well as the R-charges of the fundamental fields, are
irrational numbers \cite{friends}. Here we quote some of the features that are directly relevant for
the results of chapter \ref{Ypq}. We follow the presentation of ref.~\cite{BeFrHaMaSp}.

\begin{figure}[ht]
\centerline{\epsffile{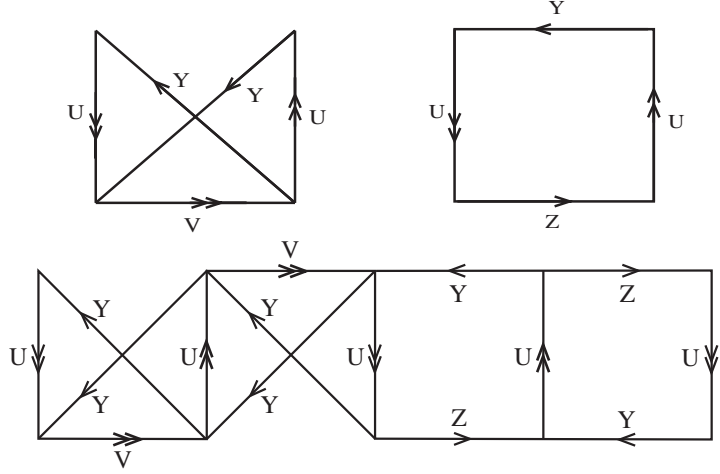}}
\caption{The basic cells $\sigma$ (upper left) and $\tau$ (upper right).
$Y^{p,q}$ quivers are built with $q$ $\sigma$ and $p-q$ $\tau$ unit
cells. The cubic terms in the superpotential (\ref{supYpq}) come from
closed loops of the former and the quartic term arises from the latter.
The quiver for $Y^{4,2}$ is given by $\sigma\tilde\sigma\tau\tilde\tau$
(bottom).}
\label{Y42}
\end{figure}

\begin{table}[ht]
\begin{center}
$$\begin{array}{|c|c|c|c|c|}
\hline
\mathrm{Field}&\mathrm{number}&R -\mathrm{charge}& U(1)_B &  U(1)_F \\
\hline\hline
& & & & \\[-1ex]
Y      & p + q
	& {- 4 p^2 + 3 q^2 + 2 p q + (2 p - q)\sqrt{4 p^2 - 3 q^2} \ov 3 q^2}
	& p - q &- 1 \\[1.5ex]
\hline
& & & & \\[-1ex]
Z      & p - q
	& {- 4 p^2 + 3 q^2 - 2 p q + (2 p + q)\sqrt{4 p^2 - 3 q^2} \ov 3 q^2}
	& p + q &+ 1 \\[1.5ex]
\hline
& & & & \\[-1ex]
U^{\alpha} &  p
	& {2 p (2 p -  \sqrt{4 p^2 - 3 q^2}) \ov 3 q^2} & - p & 0  \\[1.5ex]
\hline
& & & & \\[-1ex]
V^{\beta} &  q
	& {3 q - 2 p + \sqrt{4 p^2 - 3 q^2} \ov 3 q} & q & + 1 \\[1.5ex]
\hline
\end{array}$$
\caption{Charges for bifundamental chiral fields in the quiver dual to
$Y^{p,q}$ \cite{BeFrHaMaSp}.}
\label{charges}
\end{center}
\end{table}

The quivers for $Y^{p,q}$ can be constructed starting with the quiver
of $Y^{p,p}$ which is naturally related to the quiver theory obtained
from $\mathbb C^3/\mathbb Z_{2p}$. The gauge group is $SU(N_c)^{2p}$ and the
superpotential is constructed out of cubic and quartic terms in the
four types of bifundamental chiral fields present: two doublets $U^\a$
and $V^\b$ and two singlets $Y$ and $Z$ of a global $SU(2)$. Namely,
\beq
W=\sum\limits_{i=1}^q \epsilon_{\a\b} (U^\a_i V^\b_i Y_{2i-1}+ V^\a_i U^\b_{i+1}Y_{2i})
+ \sum\limits_{j=q+1}^p \epsilon_{\a\b}Z_j U_{j+1}^\a Y_{2j-1} U_{j}^\b.
\label{supYpq}
\eeq
Greek indices $\a,\b=1,2$ are in $SU(2)$, and Latin subindices $i,j$ refer
to the gauge group where the corresponding arrow originates. Equivalently,
as explained in \cite{hek}, the quiver theory for $Y^{p,q}$ can be
constructed from two basic cells denoted by $\sigma$ and $\tau$, and their
mirror images with respect to a horizontal axis, $\tilde\sigma$ and
$\tilde\tau$ (see Fig. \ref{Y42}). Gluing of cells has to respect the
orientation of double arrow lines corresponding to the $U$ fields. For
example, the quiver $Y^{4,2}$ is given by
$\sigma\tilde\sigma\tau\tilde\tau$. More concrete examples and further
discussion can be found in \cite{BeFrHaMaSp,hek}.

Here we quote a result of \cite{BeFrHaMaSp} which we will largely reproduce
using a study of wrapped branes in chapter \ref{Ypq}. The global $U(1)$ symmetries corresponding
to the factors appearing in the isometry group of the $Y^{p,q}$ manifold are
identified as the R-charge symmetry $U(1)_R$ and a flavour symmetry $U(1)_F$.
There is also a baryonic $U(1)_B$ that becomes a gauge symmetry in the
gravity dual. The charges of all fields in the quiver with respect to
these Abelian symmetries are summarised in Table \ref{charges}.

It is worth noting that the above assignment of charges satisfies a number
of conditions. For example, the linear anomalies vanish
$\mathrm{Tr}\,U(1)_B = \mathrm{Tr}\,U(1)_F = 0$, as well as the cubic
't~Hooft anomaly
$\mathrm{Tr}\,U(1)_B^3$.

\subsection{D3-branes on the cone over $L^{a,b,c}$ manifolds}   \label{Labcintroduc}

A further generalisation of the five-dimensional Sasaki-Einstein manifold led to the construction of the $\Labc$ manifolds \cite{CvLuPaPo, MaSp2}, where $a$, $b$ and $c$ are integers. This family of manifolds contains the $Y^{p,q}$'s as particular cases and exhausts all possible Calabi-Yau cones on a base with topology $S^2 \times S^3$. Considering again a stack of $N_c$ D3-branes at the apex of the cone over the $\Labc$ manifold (we will denote it as $C\Labc$) and taking the geometric transition, we  get a new family of solutions of the equations of motion of type IIB supergravity dual to four-dimensional ${\cal N}=1$ SCFT's. These backgrounds are again of the form $AdS_5 \times \Labc$ with standard RR field strength $F_{5}$. Recall that the normalization of the scale factor $L$, dictated by the quantization of the D3-brane tension (\ref{quantizationF5}), changes accordingly to the volume of the internal manifold:
\beq
L^4\,=\,{4\pi^4\over {\rm Vol}(\Labc)}\,g_s\,N_c\,(\alpha')^2\,\,.
\label{L1}
\eeq

The metric of the Sasaki-Einstein manifold $\Labc$ can be written as \cite{CvLuPaPo, MaSp2}:
\beq
ds^2_{\Labc}\,=\,ds_4^2\,+\,(d\tilde{\tau}\,+\,\tilde{\sigma})^2\,\,,
\eeq
where $ds_4^2$ is a local K\"ahler-Einstein metric, with K\"ahler form
$J_4 = \frac{1}{2} d\tilde{\sigma}$, given by
\bear
&&ds_4^2\,=\,{\rho^2\over 4\Delta_x}\,dx^2\,+\,
{\rho^2\over \Delta_{\theta}}\,d\theta^2\,+\,
{\Delta_x\over \rho^2}\,\Bigg(\,
{\sin^2\theta\over \alpha}\,d\phi\,+\,{\cos^2\theta\over \beta}\,d\psi
\,\Bigg)^2\,\,+\rc
&&\qquad~~~~~~~~+ \, {\Delta_\theta\sin^2\theta\cos^2\theta\over \rho^2}\,\,
\Bigg[\,\bigg(1-{x\over \alpha}\bigg)\,d\phi\,-\,
\bigg(1-{x\over \beta}\bigg)\,d\psi\,\Bigg]^2\,\,,
\eear
and the quantities $\Delta_{x}$, $\Delta_{\theta}$, $\rho^2$ and $\tilde{\sigma}$ are:
\bear
&&\Delta_{x}\,=\,x(\alpha-x)(\beta-x)\,-\,\mu\,\,,\rc
&&\Delta_{\theta}\,=\,\alpha \cos^2\theta \,+\,\beta\sin^2\theta\,\,,
\qquad\rho^2\,=\,\Delta_{\theta}\,-\,x\,\,,\rc
&&\tilde{\sigma}\,=\,\bigg(1-{x\over \alpha}\bigg)\,\sin^2\theta\,d\phi\,+\,
\bigg(1-{x\over \beta}\bigg)\,\cos^2\theta \,d\psi\,\,.
\eear
The ranges of the different coordinates are $0\le \theta\le \pi/2$, $x_1
\le x \le x_2$, $0\le \phi, \psi <2\pi$, where $x_1$ and $x_2$ are the
smallest roots of the cubic equation $\Delta_x=0$. A natural tetrad frame
for this space reads
\bear  \label{Labcframe}
&&e^1\,=\,{\rho\over \sqrt{\Delta_{\theta}}}\,d\theta\,\,,\qquad
e^2\,=\,{\sqrt{\Delta_{\theta}}\,\,\sin\theta\cos\theta\over \rho}
\Bigg(\,\bigg(1-{x\over \alpha}\bigg)\,d\phi\,-\,
\bigg(1-{x\over \beta}\bigg)\,d\psi\,\Bigg)\,\,,\rc
&&e^3\,=\,{\sqrt{\Delta_x}\over \rho}\,\,
\Bigg(\,{\sin^2\theta\over \alpha}\,d\phi\,+\,{\cos^2\theta\over
\beta}\,d\psi\,\Bigg)\,\,,\rc
&&e^4\,=\,{\rho\over 2 \sqrt{\Delta_x}}\,\,dx\,\,, \qquad\qquad
e^5\,=\,\big(\,d\tilde{\tau}+\tilde{\sigma})\,\,.
\eear
Notice that, in this frame, $J_4 = e^1 \wedge e^2 + e^3 \wedge e^4$.
Let us now define $a_i$, $b_i$ and $c_i$ ($i=1,2$) as follows:
\beq
a_i\,=\,{\alpha c_i\over x_i-\alpha}\,\,,\qquad
b_i\,=\,{\beta c_i\over x_i-\beta}\,\,,\qquad
c_i\,=\,{(\alpha-x_i)(\beta-x_i)\over 2(\alpha+\beta)\,x_i\,-\,
\alpha\beta\,-\,3x_i^2}\,\,.
\label{aibici}
\eeq
The coordinate $\tilde{\tau}$ happens to be compact and varies between 0 and $\Delta
\tilde{\tau}$,
\beq  \label{tauperiod}
\Delta \tilde{\tau}\,=\,{2\pi k |c_1|\over b}\,\,,
\qquad k={\rm gcd }\,(a,b)\,\,.
\eeq
The $a_i$, $b_i$ and $c_i$ constants are related to the integers $a,b,c$ of
$\Labc$ by means of the relations:
\beq
a\,a_1\,+\,b\,a_2\,+\,c\,=\,0\,\,,\qquad
a\,b_1\,+\,b\,b_2\,+\,d\,=\,0\,\,,\qquad
a\,c_1\,+\,b\,c_2\,=\,0\,\,,
\label{abc-relations}
\eeq
where $d=a+b-c$. The constants $\alpha$, $\beta$
and $\mu$ appearing in the metric are related to the roots $x_1$, $x_2$  and
$x_3 $ of $\Delta_x$ as
\beq
\mu\,=\,x_1x_2x_3\,\,,\qquad
\alpha+\beta\,=\,x_1+x_2+x_3\,\,,\qquad
\alpha\beta\,=\,x_1x_2+x_1x_3+x_2x_3\,\,.
\label{xs}
\eeq
Moreover, it follows from (\ref{abc-relations}) that all ratios between
the four quantities $a_1c_2-a_2c_1$, $b_1c_2-b_2c_1$, $c_1$, and $c_2$
must be rational. Actually, one can prove that:
\beq
{a_1c_2-a_2c_1\over c_1}\,=\,{c\over b}\,\,,\qquad
{b_1c_2-b_2c_1\over c_1}\,=\,{d\over b}\,\,,\qquad
{c_1\over c_2}\,=-\,{b\over a}\,\,.
\label{ratios}
\eeq
Any other ratio between $(a,b,c,d)$ can be obtained by combining these
equations. In particular, from (\ref{aibici}),  (\ref{xs}) and
(\ref{ratios}), one can rewrite some of these relations as:
\bear
&&{a\over b}\,=\,{x_1\over x_2}\,\,{x_3-x_1\over x_3-x_2}\,\,,\qquad
{a\over c}\,=\,{(\alpha-x_2)(x_3-x_1)\over \alpha(\beta-x_1)}\,\,,\rc
&&{c\over d}\,=\,{\alpha\over \beta}\,\,
{(\beta-x_1)(\beta-x_2)\over (\alpha-x_1)(\alpha-x_2)}\,=\,
{\alpha\over \beta}\,\,{x_3-\alpha\over x_3-\beta}\,\,.
\label{ratios2}
\eear
The manifold has $U(1) \times U(1) \times U(1)$ isometry. It is, thus, toric.
Its volume can be computed from the metric with the result:
\beq
{\rm Vol}(\Labc)\,=\,{(x_2-x_1)(\alpha+\beta-x_1-x_2)\,|c_1|\over
\alpha\beta b}\,\,\pi^3\,\,.
\eeq
Other geometrical aspects of these spaces can be found in
\cite{ms, CvLuPaPo}.

In order to construct a set of local complex coordinates on the Calabi-Yau
cone on $\Labc$, $C\Labc$, let us introduce the following basis of closed
one-forms \footnote{Notice that there are a few sign differences in our conventions as compared to those in \cite{SfZo}.}
\bear
&&\hat\eta_1\,=\,\alpha \,{\cot\theta\over \Delta_{\theta}}\,d\theta\,-\,
{\alpha (\beta-x)\over 2\Delta_x}\,dx\,+\,id\phi\,\,,\rc
&&\hat\eta_2\,=\,-\beta\,{\tan\theta\over \Delta_\theta}\,d\theta\,-\,
{\beta (\alpha-x)\over 2\Delta_x}\,dx\,+\,id\psi\,\,,\rc
&&\hat\eta_3\,=\,{dr\over r}\,+\,id\tilde{\tau}\,+\,
\big(\beta-\alpha)\,{\sin (2\theta)\over 2\Delta_{\theta}}\,d\theta\,+\,
{(\alpha-x)(\beta-x)\over 2\Delta_x}\,dx\,\,.
\eear
From these quantities, it is possible to define a set of $(1,0)$-forms
$\eta_i$ as the following linear combinations:
\beq
\eta_1\,=\,\hat \eta_1-\hat \eta_2\,\,,\qquad
\eta_2\,=\,\hat \eta_1+\hat \eta_2\,\,,\qquad
\eta_3\,=\,3\hat \eta_3\,+\,\hat \eta_1+\hat \eta_2\,\,.
\eeq
One can immediately check that they are integrable, $\eta^i\,=\,{dz^i\over
z^i}$. The explicit form of the complex coordinates $z^i$ is:
\bear
&&z_1\,=\,\tan\theta\,f_1(x)\,e^{i(\phi-\psi)}\,\,,\qquad
z_2\,=\,{\sin(2\theta)\over
f_2(x)\,\Delta_{\theta}}\,e^{i(\phi+\psi)}\,\,,\rc
&&z_3\,=\,r^3\,\sin(2\theta)\,\sqrt{\Delta_{\theta}\Delta_x}\,\,
e^{i(3\tilde{\tau}+\phi+\psi)}\,\,,
\label{zs}
\eear
where
\beq
f_1(x)\,=\,\CP_1(x)^{\alpha\,-\,\beta}\,\,,\qquad
f_2(x)\,=\,\CP_0(x)^{2\,\alpha\,\beta}\,\CP_1(x)^{-(\alpha\,+\,\beta)}\,\,,
\label{f12}
\eeq
and the functions $\CP_q(x)$ are defined as
\beq
\CP_q(x) = \exp\,\left(\,\int\,{x^q\,dx\over 2\,\Delta_x}\,\right)\,=\,
\prod_{i=1}^3\,(x\,-\,x_i)^{{1\over 2} {x_i^q\over \prod_{j\neq i}^3\,(x_i\,-\,x_j)}}
\,\,.
\eeq
In terms of these $(1,0)$-forms, it is now fairly simple to work out the
two-form $\Omega_4$,
\beq
\Omega_4\,=\,3e^{i(\phi+\psi)}\,\sin(2\theta)
\sqrt{\Delta_\theta\Delta_x}\,\,\,\eta_1\wedge\eta_2\,\,,
\eeq
obeying $d\,\Omega_4\,=\,3i\tilde{\sigma}\wedge \Omega_4$. By using these properties
one can verify that the three-form:
\beq
\Omega\,=\,r^2\,e^{3i\tilde{\tau}}\,\Omega_4\wedge \big[\,dr\,+\,ir\,
(d\tilde{\tau}+\tilde{\sigma})\,\big]\,\,,
\eeq
is closed. Moreover, the explicit expression for $\Omega$ in terms of the
above defined closed and integrable $(1,0)$-forms reads
\beq
\Omega\,=\,r^3\,\sin(2\theta)\,
e^{i(3\tilde{\tau}+\phi+\psi)}\,\sqrt{\Delta_\theta\,\Delta_x}\,\,
\eta_1\wedge\eta_2\wedge\eta_3\,\,,
\label{OmegaR}
\eeq
which shows that $\Omega\wedge \eta_i=0$. In terms of the complex coordinates $z_i$, the form $\Omega$ adopts a simple expression from which it is clear
that it is the holomorphic (3,0) form of the Calabi-Yau cone $C\Labc$,
\beq
\Omega\,=\,{dz_1\wedge dz_2\wedge dz_3\over z_1 z_2}\,\,.
\eeq
The expression (\ref{OmegaR}) allows for the right identification of the
angle conjugated to the $R$-symmetry \cite{BeKr}, 
\beq 
\psi' = 3\tilde{\tau} + \phi + \psi ~.
\label{Rangle}
\eeq
Finally, starting from $J_4$, we can write the K\"ahler form $J$ of $C\Labc$,
\beq 
J = r^2\,J_4 + r\,dr \wedge e^5\,\,, \qquad dJ = 0 ~.
\label{Kform}
\eeq
Notice that all the expressions written in this subsection reduce to those of
$\CYpq$ provided
\bear
&&a\,=\,p\,-\,q ~, \qquad b\,=\,p\,+\,q ~, \qquad c\,=\,p ~, \rc
&&3x\,-\,\alpha\,=\,2\,\alpha\,y ~, \qquad \mu\,=\,{4\over 27}\,(1\,-\,a)\,\alpha^3 ~, \\
&&\tilde\theta\,=\,2\theta ~, \qquad \tilde\beta\,=\,-\,(\phi\,+\,\psi) ~,
\qquad \tilde\phi\,=\,\phi\,-\,\psi ~, \nonumber
\eear
while (\ref{Rangle}) provides the right identification with the $U(1)_R$
angle in $\Ypq$.
We shall use this limiting case several times along chapter \ref{Labc} to make
contact\footnote{It is worth pointing out that one should be careful since we are using the same notation for the complex coordinates and for the forms which characterize the complex structure of both manifolds.} with the results found in chapter \ref{Ypq}.

We can perform now an analysis of the Killing spinors of the $AdS_5\times \Labc$ background as we did in the previous subsection (\ref{adsspinor}). They can be written again in terms of a constant spinor $\eta$,
\beq
\epsilon\,=\,e^{{i\over 2}(3\tilde{\tau}+\phi+\psi)}\,r^{-{\Gamma_{*}\over 2}}\,\,
\Big(\,1\,+\,{\Gamma_r\over 2L^2}\,\,x^{\alpha}\,\Gamma_{x^{\alpha}}\,\,
(1\,+\,\Gamma_{*}\,)\,\Big)\,\,\eta\,\,.
\label{adsspinor1}
\eeq
The spinor $\eta$ satisfies the projections
\cite{SfZo}:
\beq
\Gamma_{12}\,\eta\,=\,i\eta\,\,, \qquad\qquad
\Gamma_{34}\,\eta\,=\,i\eta\,\,,
\label{etaspinor1}
\eeq
this implying that $\epsilon$ also satisfies the same projections. Using the decomposition of the constant spinor $\eta$ according to its $\Gamma_{*}$-parity, ~$\Gamma_{*}\,\eta_{\pm}\,=\,\pm\eta_{\pm}$, we obtain again two types of Killing spinors:
\bear
e^{-{i\over 2} (3\tilde{\tau}+\phi+\psi)}\,\epsilon_{-} & = & r^{1/2}\,\eta_-\,\,,\rc
e^{-{i\over 2} (3\tilde{\tau}+\phi+\psi)}\,\epsilon_{+} & = & r^{-1/2}\,\eta_+\,+\,{r^{1/2}
\over L^2}\,\,\Gamma_r\,x^{\alpha}\, \Gamma_{x^{\alpha}}\,\eta_+\,\,.
\label{chiraladsspinor1}
\eear
The spinors $\epsilon_-$ satisfy
$\Gamma_{*}\,\epsilon_-\,=\,-\epsilon_-$, whereas the $\epsilon_+$'s are
not  eigenvectors of $\Gamma_{*}$. The former correspond to  ordinary
supercharges while the latter,  which depend on the $x^\alpha$
coordinates, are related to  the superconformal supersymmetries. The only
dependence on the coordinates of $\Labc$ is through the exponential of
$\psi' = 3\tilde{\tau} +
\phi + \psi$. This angle, as explained above, is identified with the
$U(1)_R$ of the superconformal quiver theory.

It is finally convenient to present the explicit expression for the Killing
spinors when $AdS_5$ is described by its global coordinates (\ref{globalADS}):
\beq
\epsilon\,=\,e^{{i\over 2}(3\tilde{\tau}+\phi+\psi)}\,
e^{-i\,{\varrho\over 2}\,\Gamma_{\varrho}\gamma_*}\,
e^{-i\,{t\over 2}\,\Gamma_{T}\gamma_*}\,
e^{-{\alpha^1\over 2}\,\Gamma_{\alpha^1 \varrho}}\,
e^{-{\alpha^2\over 2}\,\Gamma_{\alpha^2 \alpha^1}}\,
e^{-{\alpha^3\over 2}\,\Gamma_{\alpha^3 \alpha^2}}\,\eta\,\,.
\label{globalspinor1}
\eeq
where $\gamma_*$ is given in (\ref{matrixgamma}) and 
$\eta$ is a constant spinor that satisfies the same conditions as 
in (\ref{etaspinor1}). 

Finally we quote some of the results of the gauge theory dual to IIB on $AdS_5\times \Labc$ that are directly relevant to the understanding of chapter \ref{Labc}. 
The $\Labc$ SCFT's were first constructed in
\cite{BeKr,FrHaMaSpVeWe,BuFoZa}. They are four-dimensional quiver theories
whose main features we would like to briefly remind. 
\begin{table}[ht]
\begin{center}
$$\begin{array}{|c|c|c|c|c|c|}
\hline
\mathrm{Field}&R -\mathrm{charge}&\mathrm{number}&
U(1)_B &  U(1)_{F_1} & U(1)_{F_2} \\
\hline\hline
& & & & & \\[-1ex]
Y  & {2\over 3}\,{x_3\,-\,x_1\over x_3} & b &  a &  1 & 0 \\[1.5ex]
\hline
& & & & & \\[-1ex]
Z  & {2\over 3}\,{x_3\,-\,x_2\over x_3} & a &  b &  0 & k \\[1.5ex]
\hline 
& & & & & \\[-1ex]
U_1 & {2\over 3}\,{\alpha\over x_3}     & d & -c &  0 & l \\[1.5ex]
\hline
& & & & & \\[-1ex]
U_2 & {2\over 3}\,{\beta\over x_3}      & c & -d & -1 & -k-l \\[1.5ex]
\hline
& & & & & \\[-1ex]
V_1 & {2\over 3}\,{2x_3\,+\,x_1\,-\,\beta\over x_3} & c-a & b-c & 0 &
k+l \\[1.5ex]
\hline 
& & & & & \\[-1ex]
V_2 & {2\over 3}\,{2x_3\,+\,x_1\,-\,\alpha\over x_3} & b-c & c-a & -1 &
-l \\[1.5ex]
\hline 
\end{array}$$
\caption{Charges for bifundamental chiral fields in the quiver dual to
$\Labc$ \cite{FrHaMaSpVeWe}.}
\label{charges1}
\end{center}
\end{table}
The gauge theory for
$\Labc$ has $N_g = a + b$ gauge groups and $N_f = a + 3 b$ bifundamental
fields. The latter are summarised in Table~\ref{charges1}. There is a
$U(1)_F^2$ flavour symmetry that corresponds, in the gravity side, to the
subgroup of isometries that leave invariant the Killing spinors. There is
a certain ambiguity in the choice of flavour symmetries in the gauge theory
side, as long as they can mix with the $U(1)_B$ baryonic symmetry group.
This fact is reflected in the appearance of two integers $k$ and $l$ in
the $U(1)_F^2$ charge assignments, whose only restriction is given by the
identity $c\,k\,+\,b\,l\,=\,1$ (here, it is assumed that $b$ and $c$ are
coprime) \cite{FrHaMaSpVeWe}.

The charge assignments in Table~\ref{charges1} fulfil a number of
nontrivial constraints. For example, all linear anomalies vanish,
$\mathrm{Tr}\,U(1)_B =
\mathrm{Tr}\,U(1)_{F_1} = \mathrm{Tr}\,U(1)_{F_2} = 0$. The cubic t'~Hooft anomaly, $\mathrm{Tr}U(1)_B^3$, vanishes as well. The superpotential of the theory has three kind of terms; a quartic one,
\beq
Tr\, Y\,U_1\,Z\,U_2 ~,
\eeq
and two cubic terms,
\beq
Tr\, Y\,U_1\,V_2 ~, \qquad Tr\, Y\,U_2\,V_1 ~.
\eeq
Their R-charge equals two and they are neutral with respect to the baryonic
and flavour symmetries. The number of terms of each sort is uniquely fixed
by the multiplicities of the fields to be, respectively, $2\,a$, $2\,(b-c)$
and $2\,(c-a)$ \cite{FrHaMaSpVeWe}. The total number of terms, then, equals
$N_f - N_g$. In the $\Ypq$ limit, the isometry of the space --thus
the global flavour symmetry-- enhances, $U^1$ and $U^2$ (also $V^1$ and $V^2$)
becoming a doublet under the enhanced $SU(2)$ group. The superpotential
reduces in this limit to the $\Ypq$ expression \cite{BeFrHaMaSp}. More
details about the $\Labc$ superconformal gauge theories can be found in
\cite{BeKr,FrHaMaSpVeWe,BuFoZa}.

\subsection{The Maldacena-N\'u\~nez background}   \label{MNbackground}

The model proposed by Maldacena and N\'u\~nez \cite{Maldacena:2000yy} realises a duality between a supergravity solution of type IIB and a four-dimensional ${\cal N}=1$ super Yang-Mills (SYM) theory with $SU(N_c)$ gauge group. This supergravity solution was found previously by Chamseddine and Volkov \cite{CV} by studying monopole solutions of the ${\cal N}=4$ gauged supergravity in four dimensions with non-Abelian gauge fields.

The setup consists of a stack of $N_c$ D5-branes wrapping a finite, topologically nontrivial and supersymmetric two-cycle of a resolved conifold \cite{Candelas:1989js}. The Calabi-Yau is $1/4$ supersymmetric and the presence of D5-branes further halves the number of supersymmetries (and also spoils conformal symmetry) leaving a total of $4$ supercharges.

On one hand, the backreaction of the branes deforms the geometry and one has a geometric transition as those studied in \cite{Gopakumar:1998ki,Vafa:2000wi}. The final geometry is topologically like a deformed conifold \cite{Candelas:1989js}, namely $\reals \times S^2 \times S^3$. Branes also disappear and are replaced by the fluxes to which the initial branes couple. This generated supergravity background encodes the low energy dynamics of the closed string sector of the theory.

On the other hand, if one looks at the low energy dynamics of open strings on the D5-branes, discarding Kaluza-Klein modes, one finds a Yang-Mills theory living in the $1+3$ unwrapped dimensions. The degrees of freedom of $D=4$, ${\cal N}=1$ SYM can be arranged into a vector multiplet composed by a gauge vector field $A_{\mu}$ (two on-shell bosonic degrees of freedom) and a Majorana spinor $\lambda$ (two on-shell fermionic degrees of freedom), both of them transforming in the adjoint representation of the gauge group. There are no scalar fields, which means that there is no moduli space.

Therefore, in the same spirit as AdS/CFT,  this was the gauge/gravity duality proposed in \cite{Maldacena:2000yy}. The relation is holographic and the non-compact direction of the Calabi-Yau plays the role of the energy scale of the gauge theory. Moreover, it is worth pointing out here that the Maldacena-Nu\~nez (MN) model only describes the IR of ${\cal N}=1$ SYM theory. Its UV completion is instead related to little string theory and the two regimes of the theory are not smoothly connected in terms of a unique solution (they are S-dual to each other). The source of this problem is the bad behaviour (divergence) of the dilaton in the UV, as we will see explicitly below. On the gauge theory side this reflects the difficulties of joining the weak coupling with the strong coupling regime of confining SYM theory in a unifying picture. Such an interpolating picture exists if we compactify one spatial dimension and consider SYM on the cylinder topology ${\mathbb{R}}^{1,2}\times S^1$ \cite{Davies:1999uw}. The supergravity solution dual to this field theory was constructed in \cite{Canoura:2007gw} in terms of $N_c$ M5-branes that wrap a three-cycle with topology $S^2\times S^1$. That solution is valid both to describe the IR of  ${\cal N}=1$ SYM and its UV description (related to NS5-branes in type IIB). Therefore a unique picture connecting the UV and the IR of the gauge theory exists in M-theory. 

Let us summarise now the MN background. The ten-dimensional metric in string frame (and setting $g_s=\alpha'=N_c=1$) is:
\beq
ds^2_{10}\,=\,e^{\phi}\,\,\Big[\,
dx^2_{1,3}\,+\,e^{2h}\,\big(\,d\theta_1^2+\sin^2\theta_1 d\phi_1^2\,\big)\,+\,
dr^2\,+\,{1\over 4}\,(\tilde{w}^i-A^i)^2\,\Big]\,\,,
\label{metricMN}
\eeq
where $\phi$ is the dilaton, $h$ is a function which depends on the radial coordinate $r$, the
one-forms $A^i$ $(i=1,2,3)$ are
\beq
A^1\,=\,-a(r) d\theta_1\,,
\,\,\,\,\,\,\,\,\,
A^2\,=\,a(r) \sin\theta_1 d\phi_1\,,
\,\,\,\,\,\,\,\,\,
A^3\,=\,- \cos\theta_1 d\phi_1\,,
\label{oneform}
\eeq
and the $\tilde{w}^i$'s are  $su(2)$ left-invariant one-forms,
satisfying  $d\tilde{w}^i=-{1\over 2}\,\epsilon_{ijk}\,\tilde{w}^j\wedge \tilde{w}^k$. The $A^i$'s are the components of
the non-abelian gauge vector field of the seven-dimensional gauged supergravity. Moreover, 
the $\tilde{w}^i$'s parameterise the
compactification three-sphere and can be represented in
terms of three angles $\phi_2$, $\theta_2$ and $\psi$:
\bear
\tilde{w}^1&=& \cos\psi d\theta_2\,+\,\sin\psi\sin\theta_2
d\phi_2\,\,,\rc
\tilde{w}^2&=&-\sin\psi d\theta_2\,+\,\cos\psi\sin\theta_2 
d\phi_2\,\,,\rc
\tilde{w}^3&=&d\psi\,+\,\cos\theta_2 d\phi_2\,\,.
\eear
The angles $\theta_i$, $\phi_i$ and $\psi$ take values in the intervals $\theta_i\in [0,\pi]$, 
$\phi_i\in [0,2\pi)$ and $\psi\in [0,4\pi)$. The functions $a(r)$, $h(r)$ and the dilaton $\phi$
are:
\bear   \label{gluinocondensate}
a(r)&=&{2r\over \sinh 2r}\,\,,\rc
e^{2h}&=&r\coth 2r\,-\,{r^2\over \sinh^2 2r}\,-\,
{1\over 4}\,\,,\rc
e^{-2\phi}&=&e^{-2\phi_0}{2e^h\over \sinh 2r}\,\,.
\label{MNsol}
\eear

The solution of the type IIB supergravity also includes a Ramond-Ramond three-form $F_{(3)}$
given by
\beq
F_{(3)}\,=\,-{1\over 4}\,\big(\,\tilde{w}^1-A^1\,\big)\wedge 
\big(\,\tilde{w}^2-A^2\,\big)\wedge \big(\,\tilde{w}^3-A^3\,\big)\,+\,{1\over 4}\,\,
\sum_a\,F^a\wedge \big(\,\tilde{w}^a-A^a\,\big)\,\,,
\label{RRthreeform}
\eeq
where $F^a$ is the field strength of the su(2) gauge field $A^a$, defined as $
F^a\,=\,dA^a\,+\,{1\over 2}\epsilon_{abc}\,A^b\wedge A^c$.

In order to write the Killing spinors of the background in a simple form, let us consider the
frame:
\bear
e^{x^i}&=&e^{{\phi\over 2}}\,d x^i\,\,,
\,\,\,\,\,\,\,(i=0,1,2,3)\,\,,\rc
e^{1}&=&e^{{\phi\over 2}+h}\,d\theta_1\,\,,
\,\,\,\,\,\,\,\,\,\,\,\,\,\,
e^{2}=e^{{\phi\over 2}+h}\,\sin\theta_1 d\phi_1\,\,,\rc
e^{r}&=&e^{{\phi\over 2}}\,dr\,\,,
\,\,\,\,\,\,\,\,\,\,\,\,\,\,
e^{\hat i}={e^{{\phi\over 2}}\over 2}\,\,
(\,\tilde{w}^i\,-\,A^i\,)\,\,,\,\,\,\,\,\,\,(i=1,2,3)\,\,. 
\label{frame}
\eear
Let $\Gamma_{x^i}$ ($i=0,1,2,3$), $\Gamma_{j}$ ($j=1,2$), $\Gamma_{r}$ and
$\hat\Gamma_{k}$ ($k=1,2,3$) be constant Dirac matrices associated to the frame
(\ref{frame}). Then, the Killing spinors of the MN solution satisfy \cite{Nunez:2003cf}:
\bear
&&\Gamma_{x^0\cdots x^3}\,\Gamma_{12}\,\epsilon\,=\,
\Gamma_{r}\hat \Gamma_{123}\,\epsilon\,=\,e^{-\alpha\Gamma_1\hat\Gamma_1}
\epsilon\,=\,\big[\,\cos\alpha\,-\,\sin\alpha\Gamma_1\hat\Gamma_1\,\big]\,
\epsilon\,\,,\rc
&&\Gamma_{12}\,\epsilon\,=\,\hat\Gamma_{12}\,\epsilon\,\,,\rc
&&\epsilon\,=\,i\epsilon^*\,\,,
\label{fullprojection}
\eear
where the angle $\alpha$ is given by
\beq
\sin\alpha\,=\,-{ae^h\over r}\,\,,
\,\,\,\,\,\,\,\,\,\,\,\,
\cos\alpha\,=\,{e^{2h}\,-\,{1\over 4}\,(\,a^2-1\,)\over r}\,\,.
\label{alpha}
\eeq
A simple expression for $\cos\alpha$ as a function of $r$ can be written, namely
\beq
\cos\alpha\,=\,{\rm \coth} 2r\,-\,{2r\over \sinh^22r}\,\,.
\label{alphaexplicit}
\eeq
In the first equation in (\ref{fullprojection}) we have used the fact that $\epsilon$ is a
spinor of definite chirality. Moreover, from the above equations we can obtain the explicit form
of the Killing spinor $\epsilon$. It can be written as:
\beq
\epsilon\,=\,f(r)\,e^{{\alpha\over 2}\,\Gamma_1\hat\Gamma_1}\,\,\,\eta\,\,,
\label{epsiloneta}
\eeq
where $f(r)$ is a commuting function of the radial coordinate, whose explicit expression is
irrelevant in the study that we will perform in chapter \ref{MN}, and $\eta$ is a constant spinor which satisfies:
\beq
\Gamma_{x^0\cdots x^3}\,\Gamma_{12}\,\eta\,=\,\eta\,\,,
\,\,\,\,\,\,\,\,\,\,\,\,
\Gamma_{12}\,\eta\,=\,\hat\Gamma_{12}\,\eta\,\,,
\,\,\,\,\,\,\,\,\,\,\,\,
\eta\,=\,i\eta^*\,\,.
\label{constantfullpro}
\eeq

Apart from the full regular MN solution described above we shall also consider the simpler
background in which the function $a(r)$ vanishes and, thus, the one-form $A$ has only one
non-vanishing component, namely $A^3$. This solution is singular in the IR and coincides with
the regular MN background in the UV region $r\to\infty$. Indeed, by taking $r\to\infty$ in the
expression of $a(r)$ in eq. (\ref{MNsol}) one gets $a(r)\to 0$. Moreover, by neglecting
exponentially suppressed terms one gets:
\beq
e^{2h}\,=\,r\,-\,{1\over 4}\,\,,
\,\,\,\,\,\,\,\,\,\,\,\,\,\,\,\,\,\,(a=0)\,\,,
\label{abelianh}
\eeq
while $\phi(r)$ can be obtained by using the expression of $h$ given in eq. (\ref{abelianh}) on
the last equation in (\ref{MNsol}). The RR three-form $F_{(3)}$ is still given by eq.
(\ref{RRthreeform}), but now $A^1=A^2=0$ and $A^3$ is the same as in eq. (\ref{oneform}). 
We will refer to this solution as the abelian MN background. The metric of this abelian MN
background is singular at $r={1\over 4}$ (by redefining the radial coordinate this singularity
could be moved to $r=0$). Moreover, the Killing spinors in this abelian case can be obtained
from those of the regular background by simply putting $\alpha=0$, which is indeed the value
obtained by taking the $r\to\infty$ limit on the right-hand side of eq. (\ref{alpha}). 

Since $dF_{(3)}=0$, one can find a two-form potential $C_{(2)}$ such that 
$F_{(3)}=dC_{(2)}$. The expression of $C_{(2)}$, which will not be needed in chapter \ref{MN}, can be found
in ref. \cite{Nunez:2003cf}. Moreover, the equation of motion satisfied by $F_{(3)}$ is
$d\star F_{(3)}=0$. Therefore one can write, at least locally, 
$\star F_{(3)}\,=\,d C_{(6)}$, with  $C_{(6)}$ being a six-form potential. The expression of 
$C_{(6)}$ can be taken from the results of ref. \cite{Nunez:2003cf}, namely:
\beq
C_{(6)}\,=\,dx^0\wedge dx^1\wedge dx^2\wedge dx^3\wedge
{\cal C}\,\,,
\label{C6}
\eeq
where ${\cal C}$ is the following two-form:
\bear
{\cal C}&=&-{e^{2\phi}\over 8}\,\,\Big[\,
\Big(\,(\,a^2-1\,)a^2\,e^{-2h}\,-\,16\,e^{2h}\,\Big)\,\cos\theta_1
d\phi_1\wedge dr
\,-\,(\,a^2-1\,)\,e^{-2h}\,\tilde{w}^3\wedge dr\,+\rc
&&+\,a'\,\Big(\,\sin\theta_1 d\phi_1\wedge \tilde{w}^1\,+\,d\theta_1\wedge
\tilde{w}^2\,\Big)\,\Big]\,\,.
\label{calC}
\eear 

It is also interesting to recall the isometries of the abelian and non-abelian metrics. In the
abelian solution $a=0$ the angle $\psi$ does not appear in the expression of the metric 
(\ref{metricMN}) (only $d\psi$ does). Therefore, $\psi$ can be shifted by an arbitrary constant
$\lambda$ as $\psi\to\psi+\lambda$. Actually, this $U(1)$ isometry of the abelian metric is
broken quantum-mechanically to a $\zet_{2N_c}$ subgroup as a consequence of the flux quantization
condition of the RR two-form potential \cite{Maldacena:2000yy, Klebanov:2002gr, Bertolini:2002xu, anomaly, Bertolini:2001qa}.  In the gauge theory side this
isometry can be identified with the $U(1)$ R-symmetry, which is broken in the UV to the same 
$\zet_{2N_c}$ subgroup by a field theory anomaly. On the contrary, the non-abelian metric does
depend on
$\psi$ through $\sin\psi$ and $\cos\psi$ and, therefore, only the discrete $\zet_2$ isometry
$\psi\to\psi+2\pi$ remains when $a\not=0$. This fact has been interpreted \cite{Maldacena:2000yy, Apreda, DiVecchia:2002ks} as
the string theory dual of the spontaneous breaking of the R-symmetry induced by the formation of a gluino condensate $<\lambda^2>$ in the IR. In \cite{Apreda} it was explained that this condensation of gluinos is related to $a(\rho)$ (\ref{gluinocondensate}). Taking also into account its relation to the dynamical scale via $<\tr{\lambda^2}>\approx \Lambda^3_{QCD}$ and introducing the subtraction scale $\mu$ of the gauge theory, it seems natural to identify  \cite{DiVecchia:2002ks}
\beq
\mu^3 \, a(\rho) \,=\, \Lambda^3_{QCD}
\eeq
giving (implicitly) the energy/radius relation between supergravity coordinates and gauge theory scales.

As a consequence of the gluino condensate, the gauge theory has $N_c$ inequivalent vacua. The fact that an $SU(N_c)$ ${\cal N}=1$ SYM theory is characterised by a set of $N_c$ different vacua implies that there exist domain wall configurations that interpolate amongst them. Although we will study them more in detail in section \ref{WhyDp}, let us advance that they are BPS states and preserve half of the supersymmetries of the theory where they live. Their tension is related to the different vacuum expectation values (VEV's) for the gluino condensate at both sides of the domain wall. In subsection \ref{moredefects} we will study a candidate to be domain wall in the MN background. It is a D5-brane wrapping in the far IR the nontrivial three-cycle $S^3$.\\

There exist more checks which show up that some of the properties of an ${\cal N}=1$ SYM theory are encoded by the Maldacena-N\'u\~nez background. Reviews on this topic can be found in \cite{Bertolini:2003iv, Merlatti:2002uz, Imeroni:2003jk, Di Vecchia:2004dg, Paredes:2004xw}. Recall that an ${\cal N}=1$ SYM theory is confining. The computation of the tension of a fundamental string in the MN background was done in \cite{Maldacena:2000yy} finding a finite value, what confirms that the dual gauge theory confines. The tension of a $q$-string in backgrounds dual to four-dimensional $SU(N_c)$ ${\cal N}=1$ gauge theories was computed in \cite{Herzog:2001fq}, with the result displayed in (\ref{qstringtension}). The computation of the $\beta$-function on the gravity side of the duality was performed in \cite{DiVecchia:2002ks}. They found, neglecting subleading exponential corrections, the exact perturbative NSVZ $\beta$-function in the Paulli-Villars renormalization scheme for ${\cal N}=1$ SYM theory. The glueball spectrum of the theory was analysed in \cite{Ametller:2003dj} and a formula for the mass spectrum of the mesons in the quenched approximation (the limit of the gauge theory with $N_f \ll N_c$) was given in \cite{Nunez:2003cf}. In \cite{Casero:2006pt} a supergravity background dual to a four-dimensional ${\cal N}=1$ SQCD with quartic superpotential was proposed and some consistency checks which support the field theory interpretation were studied.

\section{D-branes in supergravity backgrounds}   \label{WhyDp}
\medskip
\setcounter{equation}{0}
According to the gauge/gravity extensions of the AdS/CFT correspondence, the chiral operators (the gauge invariant operators which have the lowest possible conformal dimension for a given R-charge) of an ${\cal N}=1$ SCFT are in one-to-one correspondence with the modes of type IIB supergravity on $AdS_5 \times X^5$, where $X^5$ is a five-dimensional Sasaki-Einstein manifold. However, the massive string modes correspond to (non-chiral) operators in long multiplets whose dimensions diverge for the large 't Hooft coupling limit that is taken in the {\it low energy} version of the AdS/CFT conjecture. Thus, in the limit $g_{YM}^2 N_c \to \infty$ the stringy nature of the dual theory is obscured by the decoupling of the non-chiral operators, which incidentally constitute the majority of possible gauge invariant operators. If we depart from the limit of infinite 't Hooft coupling, then all non-chiral operators do not decouple at all and the spectrum of the gauge theory is presumably related to type IIB string theory on $AdS_5 \times X^5$. Even in the very large 't Hooft coupling limit, it is possible to demonstrate the stringy nature of the dual theory by including extra D-branes wrapped on nontrivial cycles in the $X^5$ manifold. Although in a different context, Witten showed in \cite{ba0} that an ${\cal N}=4$ supersymmetric $SO(2N_c)$ gauge theory (which is dual to type IIB strings on $AdS_5 \times RP^5$) possesses chiral operators of dimension $N_c$, the Pfaffians, whose dual interpretation is provided by a D3-brane wrapping a three cycle of the manifold $RP^5$. This shows that the dual theory cannot be simple supergravity but it must contain D-branes. In principle, one would expect that branes wrapped on nontrivial cycles correspond to states in the conformal field theory that are nonperturbative from the point of view of the $1/N_c$ expansion. In \cite{Gubser:1998fp, ba3} it was shown that wrapped branes could be interpreted indeed as soliton-like states in the large $N_c$ gauge theory for certain ${\cal N}=1$ theories. 

The kind of solitonic-like state in the field theory strongly depends on the dimension of the D-branes  as  an object in the $AdS_5$ space. Consider then a Dp-brane in an $AdS_5 \times X^5$ background of type IIB supergravity. The first thing that one has to do is to study the homology groups of the internal manifold $X^5$. This will give us the different ways in which a Dp-brane can  wrap a nontrivial q-cycle ($q \leq p$). When the cycle is calibrated, the dual state will not spoil supersymmetry while it will do in the opposite case. Let us consider first the case in which the Dp-brane is not extended along the holographic (radial) direction and does not fill either the gauge theory directions completely. From the field theory point of view, the solitonic-like state dual to this brane is a extended $(p-q)$-dimensional object. For instance, baryonic operators are particles, strings are one-dimensional objects whereas domain walls are two-dimensional defects in the gauge theory dual. Let us analyse more in detail each of the three cases and give some evidences of the statement:
\begin{itemize}
 \item A vertex connecting $N_c$ fundamental strings --known as the baryon vertex-- can be identified with a baryon built out of external quarks, since each string ends on a charge in the fundamental
representation of $SU(N_c)$. Such an object can be constructed by wrapping a
D5-brane over the whole five-dimensional compact manifold $X^5$ \cite{ba0}. The argument is that a D5-brane wrapping $X^5$ captures $N_c$ units of the RR five-form flux $F_5$ of the background. There is  a $U(1)$ gauge field on the D5-brane worldvolume (see eq. (\ref{DBIaction})) which couples to $F_5$ and takes $N_c$ units of charge. Since the total charge of a $U(1)$ gauge field must cancel in a closed universe, there must be $-N_c$ unit of charges coming from another source. Such source are $N_c$ elementary strings that end on the D5-brane since each end point is electrically charged with respect to the $U(1)$ field, with charge $+1$ or $-1$ depending on the orientation of each fundamental string. In order to cancel the charge, all must have the same orientation.

Another object of particular interest in quantum field theories that arises when D3-branes are placed at conical singularities (${\cal N}=1$ SCFT's) is given by D3-branes wrapped on
supersymmetric three-cycles; these states are dual to dibaryons built
from chiral fields charged under two different gauge groups of the resulting
quiver theory \cite{ba0, Gubser:1998fp, ba3, ba2}. The argument given in this identification is the matching of the conformal dimension of the dibaryon operators with the mass of wrapped D3-branes using general rules of the AdS/CFT correspondence. We will see explicit examples of this matching as well as of the baryon vertex configuration in chapters \ref{Ypq} and \ref{Labc}.

\item Domain walls and strings (flux tubes) in the field theory side can be introduced holographically as Dp-branes wrapping q-cycles of the internal geometry \cite{ba0, Gubser:1998fp, ba3} with $p-q=2$ or $p-q=1$ respectively. Actually, the superalgebras of ${\cal N}=1$ supersymmetric theories admit central charges associated with objects extended in two or one space directions (codimension one and two, respectively). For instance, in $SU(N_c)$ ${\cal N}=1$ SYM there are 1/2-BPS domain
walls which interpolate between the inequivalent $N_c$ vacua which
come from the spontaneous breaking of the $\zet_{2N_c}$ symmetry (the
non-anomalous subgroup of the $U(1)_R$) to $\zet_2$ by the gaugino condensate.
There can be also BPS codimension two objects, namely strings which have been studied in the context of different ${\cal N}=1$ theories, see \cite{Gorsky:1999hk} and references therein. The physics of such 
objects turns out to be quite rich, including for instance the phenomenon of enhanced (supersized) supersymmetry, also present for domain walls \cite{Shifman:2005st}. We will find potential dual objects to domain-walls and flux-strings mainly in chapters \ref{Ypq}, \ref{Labc} and \ref{MN}.\\
In \cite{Gubser:1998fp} it was also argued that a D3-brane wrapping a nontrivial two-cycle in $X^5$ gives rise to a non BPS object called ``fat" string. We will also find examples of the explicit configuration of that object in chapters $\ref{Ypq}$ and $\ref{Labc}$.
\end{itemize}

Moreover, one can think of modifying the theory by introducing supersymmetric defects of codimension one or two (regions of space-time where some fields are localised), which break the $SO(1,3)$ Lorentz invariance. In particular, on the field theory side one can add spatial defects which
reduce the amount of supersymmetry but nevertheless preserve conformal invariance \cite{dCFT},
giving rise to the so-called ``defect conformal field theories" (dCFT). Since this modifies the lagrangian of the field theory, we expect, on general grounds, that the string theory setup should be modified at infinity.
Therefore, the defects should be dual to D-branes extending infinitely in the holographic direction but without filling completely the gauge theory directions. It is important to point out that the effective gauge coupling (in the Minkowski directions that they fill) of these branes is zero since they are extended along a non-compact direction. Therefore, from the point of view of the dual gauge theory, these objects give rise to the addition of fundamental multiplets to a region (defect) of the spacetime.

A holographic dual of four-dimensional ${\cal N}=4$ super Yang-Mills theory with a
three-dimensional defect was proposed in ref. \cite{KR} by Karch and Randall, who conjectured
that such a dCFT can be realised in string theory by means of a D3-D5 intersection. In the
near-horizon limit the D3-branes give rise to an  $AdS_5\times S^5$ background, in which the
D5-branes wrap an $AdS_4\times S^2$ submanifold. It was argued in ref. \cite{KR} that the AdS/CFT
correspondence acts twice in this system and, apart from the holographic description of the four
dimensional field theory on the boundary of $AdS_5$, the fluctuations of the D5-brane should be
dual to the physics confined to the boundary  of $AdS_4$. The defect conformal field theory associated with the D3-D5 intersection corresponds to  ${\cal N}=4$, $d=4$ super Yang-Mills theory coupled to 
${\cal N}=4$, $d=3$ fundamental hypermultiplets localised at the defect \cite{DeWolfe:2001pq}. These hypermultiplets
arise as a consequence of the strings stretched between the D3- and D5-branes. 

The defect field theories corresponding to other intersections have also been studied in the
literature. For example, from the D1-D3 intersection one gets a four-dimensional CFT with a
hypermultiplet localised on a one-dimensional defect \cite{1defect}. Moreover, the D3-D3
intersection gives rise to a two-dimensional defect in a four-dimensional CFT \cite{CEGK, Arean:2006pk}. 

In chapters \ref{Ypq} and \ref{Labc} we will study embeddings of D-branes which are suitable to introduce defects of codimension one and two in ${\cal N}=1$ SCFT's. Extensions of this analysis to more realistic theories where the conformal symmetry is broken (for instance ${\cal N}=1$ SYM) will be performed in chapter \ref{MN}.

Finally, it deserves special attention the D-branes which fill completely the gauge theory directions and wrap a cycle of the internal manifold. According to the original
proposal of ref.\cite{Karch:2002sh}, these spacetime filling configurations can be used as flavour branes, \ie\ as branes whose fluctuations can be identified with the dynamical mesons of
the gauge theory (see refs.\cite{Nunez:2003cf, Arean:2006pk, mesons, Sakai:2003wu, Kirsch:2005uy, Decays} for the
analysis of the meson spectrum in different theories). These flavour branes must extend along the radial direction from an infinity value since the addition of flavour may modify the lagrangian of the dual field theory. It could be that they reach a minimum $r_0$. It was argued in \cite{Karch:2002sh} that the dual interpretation of this mass scale is given by the mass of the quarks introduced in the gauge theory. It may be possible to set that mass scale to zero and to deal with massless quarks by just taking $r_0 \to 0$, namely extending the flavour brane completely along the radial direction. Since they are extended along a non-compact direction, the gauge theory which support has vanishing four-dimensional effective coupling on the Minkowski directions. Thus, the gauge symmetry on the flavour branes is seen as a flavour symmetry by the four-dimensional gauge theory of interest. In chapters \ref{Ypq} and \ref{Labc} we will also look for embeddings of flavour branes which are suitable to accommodate dynamical quarks in the field theory dual. The construction of supergravity backgrounds dual to ${\cal N}=1$ field theories which include flavour branes will be carry out in chapters \ref{KW} and \ref{KS}.

In the absence of a string theory formulation on backgrounds with Ramond-Ramond forms, the final goal in order to extract valuable information about the stringy spectrum would be to introduce extra D-branes in a supergravity background and to take into account, not only the effects that the D-branes feel coming from the background fields, but also the backreaction undergone by the supergravity background due to the presence of these extra D-branes. The techniques developed so far have only been applied to study backreacted flavour branes, as we will explain in subsection \ref{smearingprocedure}. A generalisation to other kind of D-branes may be possible. However, as a first approach we can neglect the backreaction undergone by the supergravity background. This can be achieved by considering D-brane probes of various dimensions as we will explain in subsection \ref{kappasymmetry}.

\subsection{Effective Dp-brane action}  \label{effectivebraneaction}

In this subsection we want to introduce one of the main tools that we will use along this thesis. It is the effective action that describes the low energy dynamics of small bosonic fluctuations around a classical supergravity Dp-brane solution. This action can be obtained if one requires that the non-linear sigma model describing the propagation of an open string with Dirichlet boundary conditions (and therefore fixed to a Dp-brane) in a general supergravity background is conformally invariant. The constraints in the fields coming from this invariance are the same as the equation of motion derived from the following effective action (in string frame):
\beq   \label{DBIaction}
S_{D_p}\,=\, -T_p \int_{\Sigma_{p+1}} d^{p+1} \xi \, e^{-\phi} \sqrt{-\rm{det}(g+{\cal F})} \,\, .
\eeq
This is the Dirac-Born-Infeld (DBI) action for a Dp-brane. Here $g_{\mu \nu}=\partial_{\mu}X^M \partial_{\nu}X^N\,G_{MN}$ is the pullback of the spacetime metric $G_{MN}$ on the worldvolume $\Sigma_{p+1}$, where greek indices $\mu, \nu \ldots$ are worldvolume indices and $M, N \ldots$ are target spacetime indices. In addition, ${\cal F}=B+2 \pi \alpha' dA$, with $B$ being the pullback of the NSNS two-form, $B_{\mu \nu}=\partial_{\mu}X^M \partial_{\nu}X^N\,B_{MN}$, and $A$ is a worldvolume $U(1)$ gauge field with field strength $F=dA$. $T_p$ is the tension of a Dp-brane (\ref{Dptension}) and $\phi$ is the dilaton field.

The DBI action is an abelian $U(1)$ gauge theory which reduces, to leading order in $\alpha'$, to Yang-Mills (YM) in $p+1$ dimensions with $9-p$ scalar fields when the target space is flat. Let us set to zero the NSNS two-form $B$ and perform the expansion:
\beq
S_{D_p}\, \approx \, - \int_{\Sigma_{p+1}} d^{p+1} \xi \, e^{-\phi} \sqrt{-\rm{det}g} \Big ( \frac{1}{4g^2_{YM}} F_{\mu \nu}F^{\mu \nu} \,+\, \frac{2}{(2 \pi \alpha')^2} \partial_{\mu} X^{i} \partial^{\mu} X_{i} \,+\, {\cal O}(F^4) \Big ) \,\, ,
\eeq
where the index $i$ stands for spacetime indices perpendicular to the Dp-brane. The YM coupling constant $g_{YM}$ is related to the string parameters by
\beq
g^2_{YM}\,=\, (2\pi)^{p-2} \alpha'^{\frac{p-3}{2}} g_s \,\, ,
\eeq
reproducing the relation written in (\ref{AdS/CFT}) in the particular case of D3-branes. By including fermionic superpartners, the low energy action for the D-branes becomes that of SYM theory in $p+1$ dimensions.

Moreover, Dp-branes are objects charged under RR potentials and their action should contain a term of  coupling to these fields. This term must fulfil certain requirements. It must be invariant under worldvolume diffeomorphisms and it should be classically equivalent when the brane propagates in two target space configurations related by a target space gauge symmetry. We will not go into details but one can show that the term
\beq  \label{WZDpbraneaction}
\int_{\Sigma_{p+1}} C \wedge e^{{\cal F}} \,\, ,
\eeq
where $C$ denotes the pullback of the sum of the RR background\footnote{Along this thesis the RR scalar $\chi$ will also be denoted by $C^{0}$.} fields, $C= \sum_{r=0}^{8}C^{(r)}$ and $C^{(r)}$ is a $r$-form, fulfils the demanded requirements. This topological term is called the Wess-Zumino (WZ) action.

If the Dp-brane has RR charge $q_{D_p}$, the total action will be the sum of the DBI and WZ part in the form
\beq
S_{D_p}\,=\, -T_p \int_{\Sigma_{p+1}} d^{p+1} \xi \, e^{-\phi} \sqrt{-\rm{det}(g+{\cal F})} \,+\, q_{D_p} \int_{\Sigma_{p+1}} C \wedge e^{{\cal F}} \,\, .
\eeq
Extreme branes satisfy the BPS bound $T_p= \mid q_{D_p} \mid$ and their action will be given by
\beq   \label{actionintrod}
S_{D_p}\,=\, -T_p \int_{\Sigma_{p+1}} d^{p+1} \xi \, e^{-\phi} \sqrt{-\rm{det}(g+{\cal F})} \,\pm \, T_p \int_{\Sigma_{p+1}} C \wedge e^{{\cal F}} \,\, ,
\eeq
where the sign $+$ stands for branes and the sign $-$ for antibranes. It will also be useful to write down the action in Einstein frame:
\beq   \label{actionintrodEF}
S_{D_p}\,=\, -T_p \int_{\Sigma_{p+1}} d^{p+1} \xi \, e^{\frac{p-3}{4}\phi} \sqrt{-\rm{det}(g^{(E)}+e^{-\phi/2}{\cal F})} \,\pm \, T_p \int_{\Sigma_{p+1}} C \wedge e^{{\cal F}} \,\, ,
\eeq
where $g^{(E)}_{\mu \nu}=e^{-\phi/2}g_{\mu \nu}$ denotes the metric in Einstein frame.

It is possible to generalise the above action and consider a stack of N Dp-branes since they are BPS objects and they can remain statically at any distance of each other. However, this is beyond the scope of the work we want to present in the forthcoming sections.

In short, with the action (\ref{actionintrod}) we can describe, in the low energy limit, the dynamics of the extra D-branes that we are going to place in some supersymmetric supergravity backgrounds dual to four-dimensional ${\cal N}=1$ gauge theories. We will be interested in not to break completely supersymmetry since ultimately we will be concerned about the study of some features of a four-dimensional ${\cal N}=1$ supersymmetric field theory on the supergravity side. Therefore an additional constraint is imposed on the way of placing D-branes in a supergravity background. They must satisfy a local fermionic symmetry called ``kappa symmetry", which we explain in the next subsection.

\subsection{The probe approximation and the kappa symmetry analysis}  \label{kappasymmetry}
As a first approach to understand on the gravity side some features of  the field theory  which need the introduction of additional D-branes, we can neglect the backreaction undergone by the supergravity background. This can be achieved by considering a number of D-branes much smaller than the number of branes which generate the background. This simplification, whose dual counterpart is known as the quenched approximation, is called the probe approximation on the gravity side. Let us explain a bit more in detail the physical meaning of this limit. Let us call $N_{D_p}$ to the number of extra $D_p$ branes we are introducing on the gravity side in order to study some feature of the field theory. We will use $N_c$ to denote the number of branes that engineer the supergravity background. The quenched/probe approximation ($N_{D_p} \ll N_c$) consists of neglecting all the effects of order $N_{D_p}/N_c$ on both sides of the duality. On the gravity side this can be done by considering the $N_{D_p}$ extra branes as brane probes. The meaning of a brane probe is that it does not modify the background configuration but it does couple to the background fields. In consequence we can study its dynamics by means of the effective action displayed in eq. (\ref{actionintrod}). A particular case where the probe limit has a well-known dual interpretation is when we deal with a number $N_f$ of flavour branes. When the ratio $N_f/N_c$ is very small, we are neglecting the effects that include the running of fundamentals in internal loops. Even when these fundamentals are massless, their effects while running in loops are suppressed by a  factor of ${\cal O}(N_f/N_c)$. In the strict 't Hooft limit \cite{'tHooft:1973jz}, if the number of flavours is kept fixed, the corrections due to the quantum dynamics of quarks exactly vanish \cite{Karch:2002sh}.

We are interested in bosonic configurations of the brane probe since we want to keep the classical limit. Hence we set to zero the fermionic fields of the background. However, if we wish that the brane probe does not break supersymmetry completely, we should look at the coupling of the fermionic degrees of freedom of the Dp-branes to the bosonic background fields.  A crucial ingredient (in the Green-Schwarz formulation) is then a local fermionic symmetry of the worldvolume theory called kappa symmetry \cite{bbs, swedes}.  The role played by this symmetry is to eliminate the extra fermionic degrees of freedom which appear when the target space supersymmetry becomes manifest. A nice review of this topic can be found in \cite{Camino Martinez:2002tj}. The idea is to obtain kappa symmetric actions for Dp-branes embedded in a given background. This determines the form of the kappa symmetry matrix $\Gamma_{\kappa}$:
\beq  \label{kappa1}
\Gamma_{\kappa}\,=\,\frac{1}{\sqrt{-\rm{det}(g+{\cal F})}} \sum_{n=0}^{\infty} \frac{1}{2^n n!} \gamma^{\mu_1\nu_1 \ldots \mu_n \nu_n} {\cal F}_{\mu_1 \nu_1} \ldots {\cal F}_{\mu_n \nu_n} \, J^{(n)}_{(p)} \,\, ,
\eeq
where $g$ is the induced metric, $\gamma_{\mu_1 \mu_2\ldots}$ is the antisymmetric product of induced worldvolume Dirac matrices and ${\cal F}= B + 2\pi \alpha' F$ as in eq. (\ref{DBIaction}). In eq. (\ref{kappa1}), $J^{(n)}_{(p)}$ is the following matrix:
\begin{equation}
 J_{(p)}^{(n)}\,= \, \left\{
\begin{aligned} &( \Gamma_{11})^{n+(p-2)/2} \, \Gamma_{(0)} &\qquad &\text{(IIA)} \\
& (-1)^{n} (\sigma_3)^{n+(p-3)/2} \, i \sigma_2 \otimes  \Gamma_{(0)} &\qquad &\text{(IIB)} \,\, , \end{aligned}
\right. \nonumber
\end{equation}
where $\text{IIA}$ and $\text{IIB}$ stand for type IIA and type IIB string theory respectively, $\Gamma_{11}$ is the chiral matrix in ten dimensions and $\Gamma_{(0)}$ is defined by
\beq  \label{kappa2}
\Gamma_{(0)}\,=\, \frac{1}{(p+1)!} \epsilon^{\mu_1 \ldots \mu_{(p+1)}} \gamma_{\mu_1 \ldots \mu_{(p+1)}} \,\, .
\eeq
Recall that in the type IIB string theory the spinor $\epsilon$ is composed by two Majorana-Weyl spinors which can be arranged as a two-component vector. In order to write the expression of $\Gamma_{\kappa}$, it is convenient sometimes to decompose the complex spinor $\epsilon$ in its real and imaginary parts as $\epsilon\,=\,\epsilon_1+\,i\,\epsilon_2$. It is
straightforward to find the following rules to pass from complex to real spinors:
\beq
\epsilon^*\,\leftrightarrow\,\sigma_3\,\epsilon\,\,,
\,\,\,\,\,\,\,\,\,\,\,\,\,\,\,\,\,\,\,
i\epsilon^*\,\leftrightarrow\,\sigma_1\,\epsilon\,\,,
\,\,\,\,\,\,\,\,\,\,\,\,\,\,\,\,\,\,\,
i\epsilon\,\leftrightarrow\,-i\sigma_2\,\epsilon\,\,.
\label{rule}
\eeq

We can write an induced worldvolume Dirac matrix in terms of the ten-dimensional constant gamma matrices $\Gamma_{\underline{M}}$. We need to choose a frame basis $e^{\underline{M}}$ in the background geometry where $G_{MN}=\eta_{MN}\,e^{\underline{M}}\,e^{\underline{N}}$. The $e^{\underline{M}}$ one-forms
are related to the coordinates chosen in the geometry by means of the vielbein
coefficients
$E_{N}^{\underline{M}}$, namely:
\beq
e^{\underline{M}}\,=\,E_{N}^{\underline{M}}\,dX^N\,\,.
\eeq
Then, the induced Dirac matrices on
the worldvolume are defined as
\beq
\gamma_{\mu}\,=\,\partial_{\mu}\,X^{M}\,E_{M}^{\underline{N}}\,\,
\Gamma_{\underline{N}}\,\,.
\label{wvgamma}
\eeq

We mentioned that we must fix the local kappa symmetry in order to remove the extra fermionic degrees of freedom of the worldvolume theory of the Dp-brane. The extra bosonic degrees of freedom are removed by choosing the ``static gauge". In this gauge the worldvolume general coordinate invariance is used to equate $p+1$ of the target space coordinates with the worldvolume coordinates, $X^{\mu}=\xi^{\mu} \, (\mu=0,\cdots,p)$. 

The fact that we are interested in bosonic configurations of D-branes that preserve a fraction of the spacetime supersymmetry imposes projections on the Killing spinor of the background of the kind:
\beq
\G_{\kappa} \epsilon =\pm \epsilon\,,
\label{kappafix4}
\eeq
where $+$ stands for branes and $-$ for anti-branes. In what follows, we will concentrate on the kappa symmetry condition applied to D-branes. The equation (\ref{kappafix4}) can be used to determine the supersymmetric configuration of a brane probe. $\Gamma_{\kappa}$ is a matrix which depends on the embedding of the probe and the condition (\ref{kappafix4}) gives rise to a set of first-order BPS equations which fix, up to integration constants, the embedding of the brane probe. In addition, one has to check that these first-order equations fulfil the second-order equations of motion for the worldvolume bosonic fields and, actually, we will see that they saturate a bound for the energy,
as it usually happens in the case of worldvolume solitons \cite{GGT}.   Moreover, eq. (\ref{kappafix4}) can also be applied  to determine the fraction of spacetime supersymmetry preserved by a bosonic D-brane. For brane probes, eq. (\ref{kappafix4}) is the only supersymmetry condition that arises. However, for supergravity configurations with branes as sources, we have the D-brane 
action coupled to supergravity. Thus, the supersymmetry condition (\ref{kappafix4}) must be complemented with the supersymmetric transformations of the supergravity theory. Therefore, the fraction of supersymmetry preserved by a D-brane is determined by the number of solutions 
of (\ref{kappafix4}), where $\epsilon$ is the Killing spinor of the background. Obviously, to apply the technique sketched above one has to know first the Killing spinor (see the first part of section \ref{SUSYanalysis}).

In chapters \ref{Ypq}, \ref{Labc} and \ref{MN} we will apply the kappa symmetry technique in order to find, in a systematic way, supersymmetric configurations of brane probes which preserve some amount of supersymmetry in different supersymmetric backgrounds. Moreover, in those chapters, we will give a field theory interpretation of these configurations supported with some gauge/gravity checks.

\subsection{Introducing backreacting D-branes: the smearing procedure}  \label{smearingprocedure}

It is interesting to go beyond this `quenched' or `non-backreacting' probe approximation and see what happens when one adds a large number of D-branes, of the same order of magnitude as the number of colour branes which generate the geometry, and the backreaction effects of the D-branes are considered. Indeed, many phenomena that cannot be captured by the quenched approximation might be apparent when a string backreacted background is found. For instance, if we consider a number of flavour branes in a given background of the order of the colour branes, we are working on the so called Veneziano's topological expansion \cite{Veneziano:1976wm}. New physics (beyond the 't Hooft limit \cite {'tHooft:1973jz} where the number of flavours is kept fixed and much smaller than the number of colours) is captured by Veneziano's proposal: we will be able to see this in chapters \ref{KW} and \ref{KS} that consider the backreaction of the flavour branes in the Klebanov-Witten and Klebanov-Strassler models respectively. 

Although all the procedure we are going to carry out below may be generalised for the case of dealing with any kind of D-brane introduced in a given background, we will just pay attention in this introductory chapter to the particular case of flavour D7-branes in the Klebanov-Witten background and its extensions to the $AdS_5 \times Y^{p,q}$ and $AdS_5 \times \Labc$. The explanation of what we will call ``smearing procedure" is particularly simple and well motivated from a geometrical point of view in those cases. Extensions of the smearing procedure to all kind of D-branes, besides the flavour branes, may be a simple generalisation of the ideas that we will discuss in this subsection.

The procedure that we will follow is inspired mostly by 
the papers  \cite{Casero:2006pt, Klebanov:2004ya} and more recently 
\cite{Paredes:2006wb, Murthy:2006xt}. In those papers flavours are added into the dynamics of the dual background via the introduction of $N_f$ spacetime filling flavour branes, whose dynamics is given by a 
Dirac-Born-Infeld action (\ref{actionintrod}). This dynamics is intertwined with the usual Einstein action of type IIB supergravity.

To illustrate the way in which flavour branes will be added, let us  start 
by considering the background of type IIB supergravity studied in subsection \ref{T11introduc}. For the sake of brevity, in the following we will take units is which $g_s=1$, $\alpha'=1$.

We will add $N_f$ spacetime filling D7-branes to this geometry, in a way that preserves some amount of supersymmetry. This problem was studied in \cite{Arean:2004mm,Ouyang:2003df} for the conformal case and in \cite{Sakai:2003wu, Kuperstein:2004hy} for the cascading theory. These authors found calibrated embeddings of D7-branes which preserve (at least some fraction of) the supersymmetry of the background. We will choose to put two sets of D7-branes on the surfaces 
parameterised by
\begin{align} \label{surfaceskappa}
\xi^\alpha_1 &= \{x^0,x^1,x^2,x^3,r,\theta_2,\varphi_2,\psi\} \qquad 
\theta_1=\text{const.} \qquad \varphi_1=\text{const.} \;, \nonumber\\
\xi^\alpha_2 &= \{x^0,x^1,x^2,x^3,r,\theta_1,\varphi_1,\psi\} 
\qquad \theta_2=\text{const.} \qquad \varphi_2=\text{const.} \;.
\end{align}
Note that these two configurations 
are mutually supersymmetric with the background. 
Moreover, since the two embeddings are non-compact, the gauge theory supported on the D7-branes has vanishing four-dimensional effective coupling on the Minkowski directions; therefore the gauge symmetry on them is seen as a flavour symmetry by the four-dimensional gauge theory of interest.
The two sets of flavour branes introduce 
a $U(N_f) \times U(N_f)$ symmetry,
the expected flavour symmetry with 
massless flavours. The configuration with two sets (two branches) 
can be deformed 
to a single set, shifted from the origin, that represents massive flavours, 
and realises the explicit breaking of the flavour 
symmetry to the diagonal vector-like $U(N_f)$. Our configuration 
(eq. (\ref{surfaceskappa})) for probes is nothing else than 
the $z_1=0$ holomorphic embedding of \cite{Ouyang:2003df}.

We will then write an action for a system consisting of type IIB 
supergravity (\ref{IIBEF}) plus D7-branes described by their 
Dirac-Born-Infeld action (in Einstein frame, see eq, (\ref{actionintrodEF})):
\begin{equation} \label{actionnotsmear}
\begin{split} 
S &= \frac{1}{2\kappa_{10}^2} \int d^{10}x \, \sqrt{-G} 
\Big[ R - \frac{1}{2} \partial_M \phi \partial^M \phi -\frac{1}{2} 
e^{2\phi} F_1^2 -\frac{1}{4} F_5^2 \Big] + \\ 
 & \qquad - T_7\sum^{N_f} \int d^{8}x \, e^\phi \Big[ 
\sqrt{-\hat{G}_8^{(1)}} + \sqrt{-\hat{G}_8^{(2)}} \Big] 
 + T_7 \sum^{N_f} \int  C_8 \;.
\end{split}
\end{equation}
Notice that we did not excite the worldvolume gauge fields, but this is a 
freedom of the approach that we adopted. Otherwise one may need to find new suitable kappa symmetric embeddings. 

These two sets of D7-branes are localised in their two 
transverse directions, hence
the equations of motion derived from (\ref{actionnotsmear})
will be quite complicated to solve, due to the presence of source terms 
(Dirac delta functions). 

But we can take some advantage of the fact that we are adding lots of 
flavours. Indeed, since we will 
have many ($N_f\sim N_c$) flavour branes, we might think
about distributing them in a homogeneous way on their respective 
transverse directions.
This `smearing procedure' boils down to approximating
\begin{align}
T_7\sum^{N_f} \int d^{8}x \, e^\phi 
\sqrt{-\hat{G}_8^{(i)}} \quad &\to \quad \frac{T_7 N_f}{4\pi}\int d^{10}x \, \sin\theta_i 
\sqrt{-\hat{G}_8^{(i)}} \,\, , \nonumber\\
T_7 \sum^{N_f} \int  C_8 \quad &\to \quad \frac{T_7 N_f}{4\pi}\int d^{10}x \,
\Big[ Vol(Y_1) + Vol(Y_2) \Big] \wedge C_8 \;,
\label{smearingproc}
\end{align}
with 
$Vol(Y_i)=\sin \theta_i \, d\theta_i\wedge d\varphi_i$ the volume form of the $S^2$'s.

This effectively generates a ten-dimensional action
\begin{equation} \label{action}
\begin{split} 
S &= \frac{1}{2\kappa_{10}^2} \int d^{10}x \, \sqrt{-G} \Big[ R - 
\frac{1}{2} 
\partial_M \phi \partial^M \phi -\frac{1}{2} e^{2\phi} F_1^2 
-\frac{1}{4} F_5^2 \Big] + \\ 
&\qquad -\frac{T_7 N_f}{4\pi} \int d^{10}x \, e^\phi \sum_{i=1,2} \sin\theta_i \,
\sqrt{-\hat{G}_8^{(i)}} 
+ \frac{T_7 N_f}{4\pi} \int \Big[ Vol(Y_1)+Vol(Y_2) \Big] \wedge C_8 \;.
\end{split}
\end{equation}

We can derive in the smeared case the following equations of motion, coming 
from the action (\ref{action}):
\bear  \label{eom+flavour}
&&R_{MN} - \frac{1}{2}G_{MN} R = 
\frac{1}{2} \Big( \partial_M \phi \partial_N \phi -\frac{1}{2} G_{MN} 
\partial_P \phi \partial^P \phi \Big)
+ \frac{1}{2} e^{2\phi} \Big( F_M^{(1)} F_N^{(1)} 
-\frac{1}{2} G_{MN} F_{1}^2 \Big) + \rc
&& \qquad + \frac{1}{96} F_{MPQRS}^{(5)} F_N^{(5)PQRS} + T_{MN}\,\, , \rc
&&\frac{1}{\sqrt{-G}} \partial_M \Big (G^{MN}\, \sqrt{-G}\, \partial_N \phi \Big ) = e^{2\phi} F_1^2 
+ \frac{2\kappa_{10}^2 T_7}{\sqrt{-G}} \frac{N_f}{4\pi} e^\phi 
\sum_{i=1,2} \sin\theta_i \,
\sqrt{-\hat{G}_8^{(i)}} \,\, , \rc
&&d\Big(e^{2\phi} \star F_1 \Big) = 0 \,\, , \rc
&&dF_1 = -2\kappa_{10}^2 T_7 \frac{N_f}{4\pi} 
\Big[ Vol(Y_1)+Vol(Y_2) \Big]  \,\, , \rc
&&dF_5 = 0 \,\, .
\label{eqs2nd}
\eear
The modified Bianchi identity comes from the WZ part of the action 
(\ref{action}).
The contribution to the stress-energy tensor coming from the two 
sets of $N_f$ D7 flavour 
branes is given by
\begin{equation} \label{tmunu}
\begin{split}
T^{MN} &= \frac{2\kappa_{10}^2}{\sqrt{-G}} 
\frac{\delta S^{flavour}}{\delta G_{MN}} = -\frac{N_f}{4\pi} \frac{e^\phi}{\sqrt{-G}} 
\sum_{i=1,2} \sin \theta_i \frac{1}{2} \sqrt{-\hat{G}_8^{(i)}} 
\hat{G}_8^{(i)\alpha\beta} \delta_\alpha^M \delta_\beta^N \,\, ,
\end{split} \end{equation}
where $\alpha,\beta$ are coordinate indices on the D7-brane. 
In chapter \ref{KW} we will solve the equations of motion 
\eqref{eqs2nd}-\eqref{tmunu} and we will 
propose that this type IIB background is dual to the Klebanov-Witten field 
theory when two sets of $N_f$ flavours are added for each gauge group. We 
will actually find BPS equations for the purely bosonic background, by 
imposing that the variations of the dilatino and gravitino (\ref{Eframe1}) vanish. We will verify that these 
BPS first-order equations solve all the equations of motion \eqref{eqs2nd}.

After explaining the strategy that we will adopt to add flavours, let us discuss the significance and effect on the dual gauge theory of the `smearing procedure' introduced above. It is clear that we smear the flavour branes just to be able to write a ten-dimensional action that will produce ordinary differential equations without Dirac delta functions source terms.

The results that we will show in chapters \ref{KW} and \ref{KS} state that many 
properties of the flavoured field theory are still well captured by the 
solutions obtained following this procedure. It is not 
clear what important phenomena on the gauge 
theory we are losing in smearing, but see below for an important subtlety. 

\begin{figure}[ht]
\begin{center}
\includegraphics[width=0.95\textwidth]{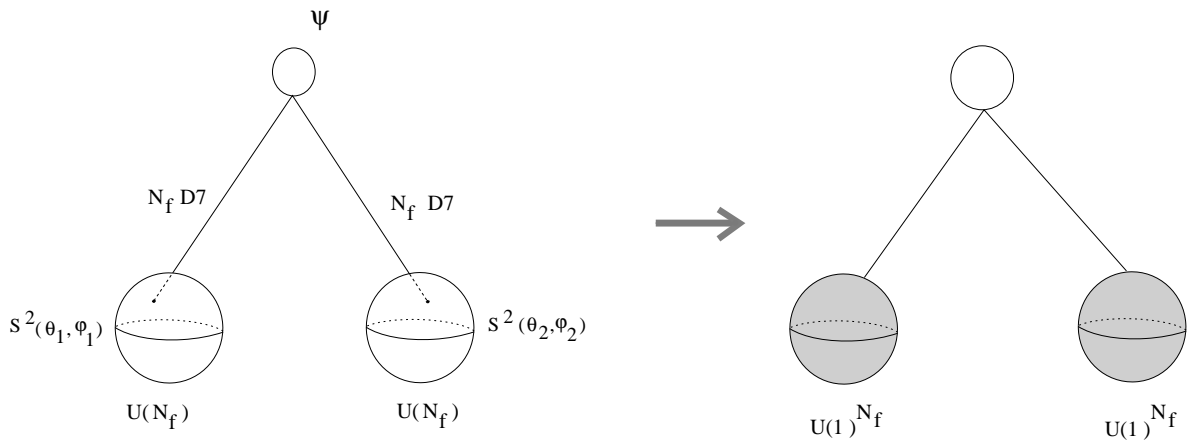}
\end{center}
\caption[smearingfig]{We see on the left side the two stacks of $N_f$ flavour-branes localised on
each of their respective $S^2$'s (they wrap the other $S^2$). The flavour
group is clearly $U(N_f) \times U(N_f)$. After the smearing on the right
side of the figure, this global symmetry is broken to $U(1)^{N_f - 1}\times
U(1)^{N_f -1} \times U(1)_B \times U(1)_A$. \label{smearingfig}} 
\end{figure}

One relevant point to discuss is related to global symmetries. Let us go 
back to the 
weak coupling ($g_s N_c \to 0$) limit, in which we have branes living on a 
spacetime that is the product of four Minkowski directions and the 
conifold. When all the flavour branes of the two separate 
stacks (\ref{surfaceskappa}) are on top 
of each other, the gauge symmetry on the D7's worldvolume 
is given by the product $U(N_f) \times U(N_f)$. When we take the 
decoupling limit for the D3-branes  $\alpha'\to 0$, with fixed $g_s N_c$ and keeping constant the energies of the excitations on the branes, we are left 
with a solution of type IIB supergravity that we propose is dual to 
the Klebanov-Witten field theory with $N_f$ flavours for both gauge groups \cite{Ouyang:2003df}. In this case the flavour symmetry is $U(N_f)\times U(N_f)$, where the axial $U(1)$ is anomalous. When we smear the $N_f$ D7-branes, we are breaking $U(N_f)\to U(1)^{N_f}$ (see Fig. \ref{smearingfig}).

One can also think about the smearing procedure in the following way: usually (unless they are 
D9 branes) the ``localised'' flavour branes break part of the isometries of the 
original background dual to the unflavoured field theory. On the other hand, the ``smeared'' flavour branes reinstate these isometries, which are global symmetries of the field theory dual. In some sense the flavour branes are `deconstructing' these dimensions (or these global groups) for 
the field theory of interest. In the case in which we have a finite number 
of flavours, these manifolds become fuzzy, while for $N_f\to \infty$, we recover 
the full invariance.

Another point that is worth elaborating on is whether there is a limit on the 
number of D7-branes that can be added. Indeed, since a D7-brane is a 
codimension-two object, its gravity 
solution will generate a deficit angle; having many seven branes, will 
basically ``eat-up'' the transverse space. This led to the conclusion that 
solutions that can be globally extended cannot have more than a maximum 
number of twelve D7-branes \cite{Greene:1989ya} (and exactly twenty-four in compact 
spaces). We are adding a number $N_f\to\infty$ of D7-branes, 
certainly larger that the bound mentioned above. However, the smearing procedure distributes the D7-branes all over a two-dimensional compact space, in such a way that the equation for the axion-dilaton is not the one in the vacuum at any point. This avoids the constraint on the number of D7-branes, which came from solving the equation of motion for the axion-dilaton outside sources.

It is possible to extend the smearing procedure of the
D7-brane, which was formulated above for the particular case of
the  $AdS_5\times T^{1,1}$ space, to  the more general case of a
geometry of the type $AdS_5\times M_5$, where $M_5$ is a five-dimensional
compact manifold. Of course, the requirement of supersymmetry restricts
greatly the form of $M_5$. Actually, we will verify that, when $M_5$ is 
Sasaki-Einstein, the formalism can be easily generalised. The five-dimensional manifolds described in subsections \ref{Ypqintroduc} and \ref{Labcintroduc} are particular cases of Sasaki-Einstein space where the generalisation could be applied, as we will see in chapter \ref{KW}. As a result of this generalisation we will get a more intrinsic formulation 
of the smearing, which eventually could be further generalised to other
types of D-branes in different geometries. 

Following the line of thought that led to the action (\ref{action}), let us
assume that, for a general geometry, the effect of the smearing on the WZ
term of the D7-brane action can be modelled by means of the substitution:
\begin{equation} \label{WZaction-general}
S_{WZ}\,=\,T_7\,\,\sum_{N_f}\,\,\int_{{\cal M}_8}\,\,C_8\,\,
\rightarrow\,\,T_7\,\,\int_{{\cal M}_{10}}\,
\Omega\wedge C_8\,\,,
\end{equation}
where $\Omega$ is a two-form which determines the distribution of the RR
charge of the D7-brane in the smearing and ${\cal M}_{10}$ is the full
ten-dimensional manifold. Notice that a well defined $\Omega$ not only must be closed (which is charge conservation) but also exact. Moreover the supersymmetry of this class of solutions forces $\Omega$ to be a real (1,1)-form with respect to the complex structure (as we will see below). For a supersymmetric brane one
expects the charge density  to be equal to  the mass density and, thus, the
smearing of the DBI part of the D7-brane action should be also determined by
the form $\Omega$. Let us explain in detail how this can be done. First of
all, let us suppose that $\Omega$ is {\it decomposable}, \ie\ that it can be
written as the wedge product of two one-forms. In that case, at an arbitrary
point,  $\Omega$ would determine an eight-dimensional orthogonal hyperplane,
which we are going to identify with the tangent space of the D7-brane
worldvolume. A general two-form $\Omega$ will not be decomposable. However,
it can be written as a finite sum of the type:
\beq
\Omega\,=\,\sum\nolimits_i \Omega^{(i)} \;,
\eeq
where each $\Omega^{(i)}$ is decomposable. At an arbitrary point, each of
the  $\Omega^{(i)}$'s is dual to an eight-dimensional hyperplane. Thus,
$\Omega$ will determine locally a collection of eight-dimensional
hyperplanes. In the smearing procedure, to each decomposable component of
$\Omega$ we associate the volume form of its orthogonal complement in ${\cal
M}_{10}$. Thus, the contribution of every $\Omega^{(i)}$ to the DBI action
will be proportional to the ten-dimensional volume element. Since energy is additive, the DBI action is obtained by summing the moduli of each decomposable piece (and not just taking the modulus of $\Omega$). We simply sum the separate contributions because of supersymmetry: the D7-branes do not interact among themselves due to the cancellation of attractive/repulsive forces.
Accordingly, let
us perform the following substitution:
\beq
S_{DBI} = -T_7 \; \sum_{N_f} \int_{{\cal M}_8} d^8\xi \, \sqrt{-\hat G_8} \;\; e^{\phi}
\quad \rightarrow \quad
-T_7 \int_{{\cal M}_{10}} d^{10}x \, \sqrt{-G} \;\; e^{\phi} \; \sum\nolimits_i
\big|\,\Omega^{(i)}\,\big| \;,
\label{DBIaction-general}
\eeq
where $\big|\,\Omega^{(i)}\,\big|$ is the modulus of $\Omega^{(i)}$ and
represents the mass density of the $i^{th}$ piece of $\Omega$ in the
smearing. There is a natural definition of $\big|\,\Omega^{(i)}\,\big|$ 
which is invariant under coordinate transformations. Indeed, let us suppose
that  $\Omega^{(i)}$ is given by:
\beq
\Omega^{(i)}\,=\,{1\over 2!}\,\,\sum_{M,N}\,\,\Omega^{(i)}_{MN}\,
dx^M\wedge dx^N\,\,.
\eeq
Then, $\big|\,\Omega^{(i)}\,\big|$ is defined as follows:
\beq
\big|\,\Omega^{(i)}\,\big|\,\equiv\,
\sqrt{{1\over 2!}\,\Omega^{(i)}_{MN}\,\Omega^{(i)}_{PQ}\,
G^{MP}\,G^{NQ}}\,\,.
\label{modulus}
\eeq

Notice that  $\Omega$ acts as a magnetic source for the
field strength $F_1$. Actually, from the equation of motion of $C_8$ one gets that
$\Omega$ is just the violation of the Bianchi identity for $F_1$, namely:
\beq
dF_1\,=\,-\,\Omega\,\,.
\label{Bianchi-general1}
\eeq
For a supersymmetric configuration the form $\Omega$ is not arbitrary.
Indeed, eq. (\ref{Bianchi-general1}) determines $F_1$ which, in turn, enters
the equation that determines the Killing spinors of the background.
On the other hand, $\Omega$ must come from the superposition (smearing) of kappa symmetric
branes.
When the manifold $M_5$ is Sasaki-Einstein, we will show in subsection \ref{SEBPS} of chapter \ref{KW} that $\Omega$ 
can be determined in terms of the K\"ahler form of the K\"ahler-Einstein base
of $M_5$ and that the resulting DBI+WZ action is a direct generalisation of
the result written in (\ref{action}).  

The last step is to provide a well defined and coordinate invariant way of splitting the charge distribution $\Omega$ in decomposable pieces.
It turns out that {\it the splitting in the minimal number of pieces\footnote{The minimal number of decomposable pieces needed to write a general two-form is half of its rank as a matrix.} compatible with supersymmetry is almost unique}.

In our setup, $\Omega$ lives on the internal six-dimensional manifold, which is complex and $SU(3)$-structure. This means that the internal geometry has an integrable complex structure $\mathcal{I}$ and a non-closed K\"ahler form $\mathcal{J}$ compatible with the metric: $\mathcal{J}_{ab} = g_{ac} \mathcal{I}_b^{\phantom{b}c}$. We can always find a vielbein basis that diagonalizes the metric and block-diagonalizes the K\"ahler form:
\begin{equation} \begin{aligned}
\label{diagonal cplx1}
g &= \sum\nolimits_a e^a \otimes e^a \,\, , \\
\mathcal{J} &= e^1\wedge e^2 + e^3\wedge e^4 + e^5\wedge e^6 \;.
\end{aligned} \end{equation}
This pattern is invariant under the structure group $SU(3)$, as it is also clear by expressing them in local holomorphic basis: $e^{z_i} \equiv e^{2i-1}+i\,e^{2i}$, $\bar e^{\bar z_i} \equiv e^{2i-1}-i\,e^{2i}$, with $i=1,2,3$. One gets the canonical expressions: $g = \sum_i e^{z_i} \otimes_S \bar e^{\bar z_i}$ and $\mathcal{J} = \frac{i}{2} e^{z_i} \wedge \bar e^{\bar z_i}$.

In our class of solutions, the supersymmetry equations force the charge distribution to be a real $(1,1)$-form with respect to the complex structure (see \cite{Grana:2005jc}). Notice that such a property is shared with $\mathcal{J}$. The dilatino equation is $e^\phi \bar F_1 ^{(0,1)} = i \bar\partial \phi$ (which without sources amounts to the holomorphicity of the axion-dilation $\tau = C_0 + i\, e^{-\phi}$). From this one gets
\begin{equation}
\Omega = -dF_1 = 2i\, e^{-\phi} \big( \partial\phi \wedge \bar\partial\phi - \partial\bar\partial \phi \big) \;.
\end{equation} 
It is manifest that $\Omega$ is $(1,1)$ and $\Omega^* = \Omega$. Going to complex components $\Omega = \Omega_{l\bar k} e^{z_l} \wedge \bar e^{\bar z_k}$, the reality condition translates to the matrix $\Omega_{l\bar k}$ being anti-hermitian. Thus it can be diagonalized with an $SU(3)$ rotation of vielbein that leaves \eqref{diagonal cplx1} untouched, and the eigenvalues are imaginary.

Going back to real vielbein and summarizing, there is always a choice of basis which satisfies the diagonalizing condition \eqref{diagonal cplx1} and in which the charge distribution can be written as the sum of three real (1,1) decomposable pieces:
\begin{equation}
\Omega = -\lambda_1 \, e^1\wedge e^2 - \lambda_2 \, e^3\wedge e^4 - \lambda_3 \, e^5\wedge e^6 \;.
\end{equation}
Supersymmetry forces the eigenvalues $\lambda_a$ to be real and, as we will see in chapters \ref{KW} and \ref{KS}, positive. Moreover, as inferred by the previous construction, the splitting is unique as long as the three eigenvalues $\lambda_a$ are different, while there are ambiguities for degenerate values, but different choices give the same DBI action.

We conclude noticing that, in order to extract the eigenvalues $|\lambda_k|=|\Omega^{(k)}|$ it is not necessary to construct the complex basis: one can simply compute the eigenvalues of the matrix $(\Omega)_{MP}g^{PN}$ in any coordinate basis. But in order to compute the stress-energy tensor, the explicit splitting into real (1,1) decomposable pieces is in general required.


\chapter{Supersymmetric Branes on ${\bf AdS_5 \times
Y^{p,q}}$}
\label{Ypq}

\medskip
\setcounter{equation}{0}
\medskip


In this chapter we perform a systematic classification of supersymmetric
branes in  the $AdS_5\times Y^{p,q}$ geometry and we study their field
theoretical interpretation. Geometrical aspects of the $Y^{p,q}$ manifold, as well as the type IIB supergravity background $AdS_5 \times Y^{p,q}$ and its field theory dual were reviewed in subsection \ref{Ypqintroduc}. It is worth mentioning that
the spectrum of type IIB supergravity compactified on $Y^{p,q}$ is not known
due to various technical difficulties including the general form of Heun's
equation \cite{heun}. Therefore, leaving aside the chiral primaries, very little is
known about the gravity modes dual to protected operators in the field
theory. As we explained in section \ref{WhyDp}, our study of supersymmetric objects in the gravity side is a way to obtain information about properties of these operators in the gauge
theory. Important aspects of this duality, relevant in the context of this chapter, have been
further developed in \cite{Gubser:1998fp, ba2, beasley}. They comprise interesting physical objects of these theories such as the baryon vertex, wall defects, the introduction of flavour, ``fat"
strings, etc. It is very remarkable that we are able to provide precise
information about operators with large conformal dimension that grows
like $N_c$. Moreover, we can also extract information about excitations of
these operators.

The main technique that we employ to determine the supersymmetric embeddings
of D-brane probes in the $AdS_5\times Y^{p,q}$ background is
the kappa symmetry of the brane probe studied in subsection \ref{kappasymmetry}. The configurations found by solving the kappa symmetry condition also solve the equations of motion derived from the Dirac-Born-Infeld action of the
probe and, actually, we will verify that they saturate a bound for the energy,
as it usually happens in the case of worldvolume solitons \cite{GGT}. 

\setcounter{equation}{0}
\section{Supersymmetric probes on  $AdS_5\times Y^{p,q}$}
\medskip
\label{ypq}

In the remainder of this chapter we will consider D-brane probes moving in
the $AdS_5\times Y^{p,q}$ background. To write the kappa symmetry matrix (see eq.(\ref{kappa1}))\footnote{We will use a simpler notation for the determinant of the induced metric, namely $\det g \equiv g$.} we
will assume that the worldvolume gauge field $A$ is zero:
\beq
\Gamma_{\kappa}\,=\,{1\over (p+1)!\sqrt{-g}}\,\epsilon^{\mu_1\cdots\mu_{p+1}}\,
(\sigma_3)^{{p-3\over 2}}\,i\sigma_2\,\otimes\,
\gamma_{\mu_1\cdots\mu_{p+1}}\,\,.
\label{gammakappa}
\eeq
This assumption is consistent with the equations of motion of the probe as far as there
are no source terms in the action (\ref{actionintrod}) which could induce a non-vanishing
value of $A$. These source terms must be linear in $A$ and can only appear
in the Wess-Zumino term of the probe action, which is responsible for the
coupling of the probe to the Ramond-Ramond fields of the background. 
In the case under study only $F^{(5)}$ is non-zero and the only linear term in $A$ is of the form $\int A\wedge F^{(5)}$, which is different from zero only for a D5-brane which captures the flux
of $F^{(5)}$. This only happens for the baryon vertex configuration studied
in subsection \ref{baryonvertex}. In all other cases studied in this chapter
one can consistently put the worldvolume gauge field to zero. Nevertheless,
even if one is not forced to do it, in some cases we can switch on the
field $A$ to study how this affects the supersymmetric embeddings. 

As we discuss in subsection \ref{kappasymmetry}, the kappa symmetry condition $\Gamma_{\kappa}\,\epsilon=\epsilon$ imposes a new projection to the Killing spinor $\epsilon$. In general, it will
not be compatible with those already satisfied by $\epsilon$ 
(see eq.(\ref{epsilon-project})). This is so because the new projections
involve matrices which do not commute with those appearing in
(\ref{epsilon-project}). The only way of making these two conditions
consistent with each other is by requiring the vanishing of the
coefficients of those non-commuting matrices, which will give rise to a
set of first-order BPS differential equations. By solving these BPS
equations we will determine the supersymmetric embeddings of the brane
probes that we are looking for. Notice also that the kappa symmetry condition
must be satisfied {\em at any point of the probe worldvolume}. It is a
local condition whose global meaning, as we will see in a moment, has to
be addressed a posteriori. This requirement is not obvious at all since
the spinor $\epsilon$ depends on the coordinates (see eqs.
(\ref{adsspinor}) and (\ref{globalspinor})). However this would be
guaranteed if we could reduce the $\Gamma_{\kappa}\,\epsilon=\epsilon$
projection to some algebraic conditions on the constant spinor $\eta$ of
eqs.(\ref{adsspinor}) and (\ref{globalspinor}). The counting of solutions
of the algebraic equations satisfied by $\eta$ will give us the fraction
of supersymmetry preserved by our brane probe.

\setcounter{equation}{0}
\section{Supersymmetric D3-branes on $AdS_5\times Y^{p,q}$ } 
\medskip
\label{d3}

Let us now apply the methodology just described to find the supersymmetric
configurations of a D3-brane in the $AdS_5\times Y^{p,q}$ background. 
The kappa symmetry matrix in this case can be obtained by putting $p=3$ in
the general expression (\ref{gammakappa}):
\beq
\Gamma_{\kappa}\,=\,-{i\over 4!\sqrt{-g}}\,\epsilon^{\mu_1\cdots\mu_4}\,
\gamma_{\mu_1\cdots\mu_4}\,\,,
\label{Gammad3}
\eeq
where we have used the rule (\ref{rule}) to write the expression of
$\Gamma_{\kappa}$ acting on complex spinors. 
Given that the $Y^{p,q}$ space is topologically $S^2\times S^3$, it is
natural to consider D3-branes wrapping two- and three-cycles in the
Sasaki-Einstein space. A D3-brane wrapping a two-cycle in $Y^{p,q}$
and extended
along one of the spatial directions of $AdS_5$ represents a ``fat" string.
We will study such type of configurations in section \ref{6} where we
conclude that they are not supersymmetric, although we will find stable
non-supersymmetric embeddings of this type. 

In this section we will concentrate on the study of supersymmetric
configurations of D3-branes wrapping a three-cycle of $Y^{p,q}$.
These objects are pointlike from the gauge theory point of view and,
on the field theory side, they correspond to dibaryons constructed from
the different bifundamental fields. In what follows we  will study the
kappa symmetry condition for two different sets of worldvolume coordinates,
which will correspond to two classes of cycles and dibaryons. 

\subsection{Singlet supersymmetric three-cycles }
\label{singlet}

Let us use the global coordinates of eq. (\ref{globalADS}) to parameterise
the $AdS_5$ part of the metric and let us consider the following set of worldvolume coordinates:
\be
\xi^{\mu}=(T,\te,\phi,\beta),
\ee
and the following generic ansatz for the embedding:
\be
y=y(\te,\phi,\beta), \qquad \psi(\te,\phi,\beta).
\ee
The kappa symmetry matrix in this case is:
\beq
\Gamma_{\kappa}\,=\,-iL\,{\cosh\varrho\over 
\sqrt{-g}}\,\Gamma_{T}\,\gamma_{\theta\phi\beta}\,\,.
\eeq
The induced gamma matrices along the $\theta$, $\phi$ and $\beta$ directions
can be straightforwardly obtained from (\ref{wvgamma}), namely:
\bea
{1\over L} \g_\te &=&  \frac{\sqrt{1-c\, y}}{\sqrt{6}}\,\G_3
+\frac13\psi_\te\,\G_5 -{1\over \sqrt{6}\, H}\,y_{\theta}\,\Gamma_1,
\nonumber \\
{1\over L} \g_\p &=& \frac{cH\cos\te}{\sqrt{6}}\,\G_2 + \frac{\sqrt{1-c\,
    y}}{\sqrt{6}}\sin\te\,\G_4 + \frac{1}{3} \left( \psi_\p +(1-c\,
y)\cos\te\right)\,\G_5 -{1\over \sqrt{6}\, H}\,y_{\phi}\,\Gamma_1, 
\nonumber \\
{1\over L} \g_\b &=& -\frac{H}{\sqrt{6}}\,\G_2 + \frac{1}{3}
\left(\psi_\b+y\right)\,\G_5 -{1\over \sqrt{6}\,H}\,y_{\beta}\,\G_1, 
\eea 
where the subscripts in $y$ and $\psi$ denote partial differentiation.
By using this result and the projections (\ref{epsilon-project}) the
action of the antisymmetrised product
$\gamma_{\theta\phi\beta}$ on the Killing spinor $\epsilon$ reads:
\beq
-{i\over L^3}\,\gamma_{\theta\phi\beta}\,\epsilon\,=\,
[\,a_5\,\Gamma_5\,+\,a_1\Gamma_1\,+\,a_3\Gamma_3\,+\,
a_{135}\,\Gamma_{135}\,]\,\epsilon\,\,,
\label{MSantisy}
\eeq
where the coefficients on the right-hand side are given by:
\bear
&&a_5\,=\,{1\over 18}\,\Bigg[\,(y+\psi_{\beta})\,
[\,(1-cy)\,\sin\theta\,+\,c\,y_{\theta}\cos\theta]\,+\rc
&&\qquad\qquad+\,[\,\psi_\phi\,+\,(1-cy)\,\cos\theta\,]\,y_{\theta}\,-\,
\psi_\theta y_\phi\,-\,c\cos\theta 
\psi_\theta y_\beta\,\Bigg]\,\,,\rc
&&a_1\,=\,-{1-cy\over 6\sqrt{6}}\,\,\sin\theta\,\big[\,
{y_\beta\over H}\,-\,iH\,\big]\,\,,\rc
&&a_3\,=\,-{\sqrt{1-cy}\over  6\sqrt{6}}\,
\big[\,y_\phi\,+\,c\cos\theta y_{\beta}\,-\,i\sin\theta y_{\theta}\,
\big]\,\,,\rc
&&a_{135}\,=\,{\sqrt{1-cy}\over 18}\,\,
\Bigg[\,{\sin\theta\over H}\,\big[\psi_{\theta}
y_{\beta}\,-\,(y+\psi_{\beta})\,y_\theta
\,\big]\,+\,H\big[\psi_{\phi}\,+\,(1+c\psi_\beta)\cos\theta\,\big]\,\,+\rc
&&\qquad\qquad+\,\,{i\over H}\,
\Big[(\psi_{\phi}\,+\,(1-cy)\,\cos\theta\,)y_\beta\,-\,
(y+\psi_{\beta})\,y_\phi\,\Big]\,-\,iH\sin\theta\psi_{\theta}\,\,
\Bigg]\,\,.
\label{as}
\eear
As
discussed at the end of section \ref{ypq}, in order to implement the kappa symmetry
projection we must require the vanishing of the terms in (\ref{MSantisy})
which are not compatible with the projection (\ref{epsilon-project}). Since 
the matrices $\Gamma_1$, $\Gamma_3$ and $\Gamma_{135}$ do not commute
with those appearing in the projection (\ref{epsilon-project}), it follows
that we must impose that the corresponding coefficients vanish, \ie:
\beq
a_1 = a_3 = a_{135} = 0\,\,.
\eeq
Let us concentrate first on the condition $a_1 = 0$. By looking at its
imaginary part:
\beq
H(y)=0\,\,,
\eeq
which, in the range of allowed values of $y$, means:
\beq
y = y_1 ~, \qquad {\rm or} \qquad y = y_2 ~.
\eeq
If  $H(y)=0$, it follows by inspection
that $a_1 = a_3 = a_{135} = 0$. Notice that $\psi$ can be an arbitrary
function. Moreover, one can check that:
\beq
\left.\sqrt{-g}\right|_{BPS} \, = \, L^4\,\cosh\varrho \,
\left.{a_5}\right|_{BPS}\,\,.
\eeq
Thus, one has the following equality:
\beq
\left.\Gamma_{\kappa}\,\epsilon\right|_{BPS} \, = \,
\Gamma_{T}\Gamma_5\,\epsilon\,\,,
\eeq
and, therefore, the condition $\Gamma_{\kappa}\epsilon=\epsilon$ becomes
equivalent to
\beq
\Gamma_{T}\Gamma_5\,\epsilon\,=\,\epsilon\,\,.
\label{d3-epsilon-cond}
\eeq
As it happens in the $T^{1,1}$ case \cite{Arean:2004mm}, the compatibility of
(\ref{d3-epsilon-cond}) with the $AdS_5$ structure of the spinor implies
that the D3-brane must be placed at $\varrho=0$, \ie\ at the center of $AdS_5$.
Indeed, as discussed at the end of section \ref{ypq}, we must translate
the condition (\ref{d3-epsilon-cond}) into a condition for the constant
spinor $\eta$ of eq. (\ref{globalspinor}). Notice that $\Gamma_{T}\Gamma_5$
commutes with all the matrices appearing on the right-hand side of
eq. (\ref{globalspinor}) except for $\Gamma_{\varrho}\gamma_*$. Since the
coefficient of $\Gamma_{\varrho}\gamma_*$ in (\ref{globalspinor}) only vanishes
for $\varrho=0$, it follows that only at this point the equation
$\Gamma_{\kappa}\,\epsilon=\epsilon$ can be satisfied for every point
in the worldvolume and reduces to:
\beq
\Gamma_{T}\Gamma_5\,\eta\,=\,\eta\,\,.
\label{d3-eta-cond}
\eeq
Therefore, if we place the D3-brane at the center of the $AdS_5$ space
 and wrap it on the three-cycles at $y=y_{1}$ or
$y=y_{2}$, we
obtain a $\frac{1}{8}$ supersymmetric configuration which preserves the
Killing spinors of the type (\ref{globalspinor}) with $\eta$ satisfying
(\ref{etaspinor}) and the additional condition (\ref{d3-eta-cond}).

The cycles that we have just found have been identified by Martelli and Sparks
as those dual to the dibaryonic operators $\det (Y)$ and $\det (Z)$, made
out of the bifundamental fields that, as the D3-brane wraps the
two-sphere whose isometries are responsible for the global $SU(2)$ group,
are singlets under this  symmetry
\cite{ms}. For this reason we will refer to these cycles as singlet (S)
cycles. Let us recall how this identification is carried out. First of all,
we look at the conformal dimension $\Delta$ of the corresponding dual
operator. Following the general rule of the AdS/CFT correspondence (and
the zero-mode corrections of ref. \cite{ba2}), $\Delta=LM$, where $L$ is
given by (\ref{L}) and $M$ is the mass of the wrapped three-brane. The
latter can be computed as $M=T_3\,V_3$, with $T_3$ being the tension of
the D3-brane ($1/T_3\,=\,8\pi^3(\alpha')^2 g_s$) and $V_3$  the volume
of the three-cycle. If $g_{{\cal C}}$ is the determinant of the spatial
part of the induced metric on the three-cycle ${\cal C}$, one has:
\beq
V_3\,=\,\int_{{\cal C}}\sqrt{g_{{\cal C}}}\,\,\,d^3\xi\,\,.
\eeq
For the singlet cycles ${\rm S}_i$ at $y=y_i$ ($i=1,2$) and $\psi$=constant,
the volume $V_3$ is readily computed, namely:
\beq
V_3^{{\rm S}_i}\,=\,{2L^3\over 3}\,(\,1-cy_i\,)\,|\,y_i\,|\,
(2\pi)^2\,\ell\,\,.
\eeq
Let us define $\lambda_1=+1$, $\lambda_2=-1$. Then, if
$\Delta_i^{{\rm S}}\,\equiv\,\Delta^{{\rm S}_i}$, one has:
\beq
\Delta_i^{{\rm S}}\,=\,{N_c\over 2q^2}\,\Big[\,-4p^2+3q^2+2\lambda_i\,pq\,+\,
(2p-\lambda_i\,q)\,\sqrt{4p^2\,-\,3q^2}\,\,\Big]\,\,.
\eeq
As it should be for a BPS saturated object, the R-charges $R_i$ of the
${\rm S}_i$ cycles are related to $\Delta_i^{{\rm S}}$ as $R_i={2\over 3}\,
\Delta_i^{{\rm S}}$. By comparing the values of $R_i$ with those determined
in \cite{BeFrHaMaSp} from the gauge theory dual (see Table \ref{charges} in chapter \ref{introduction})
one concludes that, indeed, a D3-brane wrapped at $y=y_1$ ($y=y_2$) can be
identified with the operator $\det (Y)$ ($\det (Z)$) as claimed. Another
piece of evidence which supports this claim is the calculation of the
baryon number, that can be identified with the third homology class of
the three-cycle ${\cal C}$ over which the D3-brane is wrapped. This number
(in units of $N_c$) can be obtained by computing the integral over ${\cal C}$
of the pullback of a $(2,1)$ three-form $\Omega_{2,1}$ on $CY^{p,q}$:
\beq
{\cal B}({\cal C})\,=\,\pm i \int_{{\cal C}}\,
P\big[\,\Omega_{2,1}\,\big]_{{\cal C}}\,\,,
\label{baryon-def1}
\eeq
where $P[\cdots]_{{\cal C}}$ denotes the pullback to the cycle ${{\cal C}}$
of the form that is inside the brackets. The sign of the right-hand side
of (\ref{baryon-def1}) depends on the orientation of the cycle. The explicit
expression of $\Omega_{2,1}$ has been determined in ref. \cite{hek}:
\beq
\Omega_{2,1}\,=\,K\,\Big(\,{dr\over r}\,+\,\frac{i}{L}\,
e^5\,\Big)\wedge \omega\,\,,
\eeq
where $e^5$ is the one-form of our vielbein (\ref{Ypqvielbein}) for the
$Y^{p,q}$ space, $K$ is the constant
\beq
K\,=\,{9\over 8\pi^2}\,(p^2-q^2)\,\,,
\eeq
and $\omega$ is the  two-form:
\beq
\omega\,=\,-{1\over (1-cy)^2\, L^2}\,\,\Big[\,e^1\wedge e^2\,+\,e^3\wedge
e^4\,\Big]\,\,.
\eeq
Using $(\theta,\phi,\beta)$ as worldvolume coordinates of the singlet
cycles ${\rm S}_i$,
\beq
P\big[\,\Omega_{2,1}\,\big]_{{\rm S}_i}\,=\,-i\,{K\over 18}\,
{y_i\over 1-cy_i}\,\sin\theta\, d\theta\wedge d\phi\wedge d\beta\,\, .
\eeq
Then,
changing variables from $\beta$ to $\alpha$ by means of
(\ref{beta-alpha}), and taking into account that $\alpha\in [0,2\pi
\ell]$, one gets:
\beq
\int_{{\rm S}_i}\, P\big[\,\Omega_{2,1}\,\big]_{{\rm S}_i}\,=\,-i\,
{8\pi^2\over 3}\, {K \ell y_i\over 1 - c y_i}\,\,.
\eeq
After using the values of $y_1$ and $y_2$ displayed in (\ref{y12}), we
arrive at:
\bear
&&{\cal B}({\rm S}_1)\,=\,- i \int_{{\rm S}_1}\,
P\big[\,\Omega_{2,1}\,\big]_{{\rm S}_1}\,=\,p-q\,\,,\rc\rc
&&{\cal B}({\rm S}_2)\,=\, i \int_{{\rm S}_2}\,
P\big[\,\Omega_{2,1}\,\big]_{{\rm S}_2}\,=\,p+q\,\,.
\eear
Notice the perfect agreement of ${\cal B}({\rm S}_1)$ and ${\cal B}({\rm
S}_2)$ with the baryon numbers of $Y$ and $Z$ displayed in Table
\ref{charges}. 

\subsection{Doublet  supersymmetric three-cycles }
\label{hekconst}

Let us now try to find supersymmetric embeddings of D3-branes on three-cycles
by using a different set of worldvolume coordinates. As in the previous subsection it is convenient to use the global coordinates (\ref{globalADS})
for the $AdS_5$ part of the metric and the following set of worldvolume
coordinates:
\be
\xi^\mu=(T,y,\beta,\psi) ~.
\ee
Moreover, we will adopt the ansatz:
\be
\theta(y,\b, \psi) ~, \qquad \phi(y,\b,\psi) ~.
\label{embdoublet}
\ee
The kappa symmetry matrix $\G_{\k}$ in this case takes the form:
\be
\G_{\k}=-iL\,{\cosh\varrho\over 
\sqrt{-g}}\,\Gamma_{T}\,\gamma_{y\,\beta\,\psi}\,\,,
\ee
and the induced gamma matrices are:
\bea
&&{1\over L}\g_y = -\frac{1}{\sqrt{6}H} \G_1 +
\frac{cH\cos\te}{\sqrt{6}} \p_y \G_2 + \frac{\sqrt{1-c
y}}{\sqrt{6}} \left( \te_y \G_3 +
\p_y\sin\theta\, \G_4 \right) + \frac{1-c\, y}{3}\cos\te \p_y\, \G_5\,\,,
\nonumber \\ \nonumber \\
&&{1\over L}\g_\b= \frac{H}{\sqrt{6}}\left(-1+ c\cos\te\,
\p_\b\right)\, \G_2  +  \frac{\sqrt{1-c\, y}}{\sqrt{6}}\,\te_\b\, \G_3 
+  \frac{\sqrt{1-c\, y}}{\sqrt{6}}\sin\te \,\p_\b\, \G_4 \rc
&&\qquad\qquad+ {1\over 3}\Big( \,y +
(1-c\, y) \cos\te\, \p_\b \Big)\, \G_5\,\,,  \\ \rc
&&{1\over L}\g_\psi= \frac{cH\cos\te}{\sqrt{6}}\,\p_\psi\, \G_2 +
\frac{\sqrt{1-c\, y}}{\sqrt{6}}\, \left( \te_\psi\, \G_3 + \sin\te\,
\p_\psi \,\G_4 \right) + {1\over 3}\left(1 +(1-c)\,
\cos\te\p_\psi\right)\, \G_5\,\,.
\nonumber
\eea
By using again the projections (\ref{epsilon-project}) one easily gets
the action of $\gamma_{y\,\beta\,\psi}$ on the Killing spinor
\beq
-{i\over L^3}\,\gamma_{y\,\beta\,\psi}\,\epsilon\,=\,
\big[\,c_5\,\Gamma_5\,+\,c_1\,\Gamma_1\,+\,c_3\,\Gamma_3\,+\,
c_{135}\,\Gamma_{135}\,\big]\,\epsilon\,\,,
\label{HEKcs}
\eeq
where the different coefficients appearing on the right-hand side of 
(\ref{HEKcs}) are given by:
\bear
&&c_5 = {1\over 18}\,\Bigg[ -1 - \cos\theta (\phi_\psi - c \phi_{\beta}) +
(1-cy) \sin\theta\, \Big[ \theta_y (\phi_\beta - y\phi_\psi) -
\phi_y (\theta_\beta - y\theta_\psi)\,\Big]\,\Bigg]\,,\rc
&&c_1\,=\,-{1-cy\over 6\sqrt{6}}\,\sin\theta\,
\Big[\,{\theta_{\beta}\phi_{\psi}-\theta_{\psi}\phi_{\beta}\over H}\,+\,
iH\,(\theta_y\phi_{\psi}\,-\,\theta_\psi\phi_{y})\,\Big]\,\,,\rc
&&c_3\,=\,-{\sqrt{1-cy}\over 6\sqrt{6}}\,\Big[\,
\theta_{\psi}\,-\,c\cos\theta\,(\theta_{\psi}\phi_\beta\,-\,
\theta_\beta\phi_\psi)\,+\,i\sin\theta\phi_{\psi}\,\Big]\,\,,\rc
&&c_{135} = -{\sqrt{1-cy}\over 18}\,\Bigg[ {\sin\theta\over H}
(\phi_\beta - y\phi_\psi) + H \bigg( \theta_y + \cos\theta \bigg[
\theta_y (\phi_{\psi} - c\phi_\beta) - \phi_y (\theta_{\psi} -
c\theta_\beta) \bigg] \bigg) \rc
&&\qquad\qquad + iH\sin\theta\,\phi_y -
{i\over H}\,\Big[ \theta_{\beta} - y \theta_\psi + (1-cy) \cos\theta
(\theta_\beta \phi_\psi - \theta_{\psi} \phi_\beta) \Big] \Bigg]\,.
\label{expressionHEKcs}
\eear
Again, we notice that the matrices $\Gamma_1$, $\Gamma_3$ and $\Gamma_{135}$
do not commute with the projections (\ref{epsilon-project}). We must impose:
\beq
c_1=c_3=c_{135}=0\,\,.
\eeq
From the vanishing of the imaginary part of $c_3$ we obtain the condition:
\beq
\sin\theta\,\phi_\psi\,=\,0\,\,.
\label{HEKimc3}
\eeq
One can solve the condition (\ref{HEKimc3}) by taking $\sin\theta\,=\,0$,
\ie\ for $\theta=0,\pi$. By inspection one easily realises that $c_1$, $c_3$
and $c_{135}$ also vanish for these values of $\theta$ and for an arbitrary
function $\phi(y,\beta,\psi)$. Therefore, we have the solution
\beq
\theta=0,\pi\,\,,\qquad \phi=\phi(y,\beta,\psi)\,\,.
\eeq
Another possibility is to take $\phi_\psi=0$. In this case one readily
verifies that $c_1$ and $c_3$ vanish if $\theta_\psi=0$. Thus, let us
assume that both $\phi$ and $\theta$ are independent of the angle $\psi$.
From the vanishing of the real and imaginary parts of $c_{135}$ we get two
equations for the functions $\theta=\theta(y,\beta)$ and
$\phi=\phi(y,\beta)$, namely:
\bear
&&\theta_y\,+\,
{\sin\theta\over H^2}\,\phi_{\beta}\,+\,c\cos\theta\,
(\phi_y\,\theta_\beta\,-\,\theta_y\,\phi_\beta)\,=\,0\,\,,\rc
&&\theta_\beta\,-\,H^2\,
\sin\theta\phi_y\,=\,0\,\,.
\label{HEKBPS}
\eear
If the BPS equations (\ref{HEKBPS}) hold, one can verify that the kappa
symmetry condition $\Gamma_{\kappa}\,\epsilon\,=\,\epsilon$ reduces, up
to a sign, to the projection (\ref{d3-epsilon-cond}) for the Killing
spinor. As in the case of the S three-cycles studied in subsection
\ref{singlet}, by using the explicit expression (\ref{globalspinor}) of
$\epsilon$ in terms of the global coordinates of $AdS_5$, one concludes
that the D3-brane must be placed at $\varrho=0$. The corresponding
configuration preserves four supersymmetries. 

In the next subsection we will tackle the problem of finding the general
solution of the system (\ref{HEKBPS}). Here we will analyze the trivial
solution of this system, namely:
\beq
\theta\,=\,{\rm constant}\,\,,\qquad\qquad
\phi\,=\,{\rm constant}\,\,.
\label{HEKconstant}
\eeq
This kind of three-cycle was studied in ref. \cite{hek} by Herzog, Ejaz and Klebanov (see also \cite{BeFrHaMaSp}), who showed that it corresponds to
dibaryons made out of the $SU(2)$
doublet fields $U^{\alpha}$. In what follows we will refer to it as doublet
(D) cycle. Let us review the arguments leading to this identification. First
of all, the volume of the D cycle (\ref{HEKconstant}) can be computed
with the result:
\beq
V_3^{D}\,=\,{L^3\over 3}\,(2\pi)^2\, (y_2-y_1)\,\ell\,\,.
\eeq
By using the values of $y_{1}$ and $y_{2}$ (eq.(\ref{y12})), $L$
(eq.(\ref{L})) and 
$\ell$ (eq.(\ref{alphaperiod})) we find the following value of the conformal
dimension:
\beq
\Delta^{D}\,=\,N_c\,{p\over q^2}\,\Big(\,2p\,-\,\sqrt{4p^2 - 3q^2}\,\,\Big)\,.
\eeq
By comparison with Table \ref{charges} in chapter \ref{introduction}, one can verify that the
corresponding R-charge, namely $2/3\, \Delta^{D}$, is equal to the R-charge of
the field $U^{\alpha}$ multiplied by $N_c$. We can check this identification
by computing the baryon number. Since, in this case, the pullback of
$\Omega_{2,1}$ is:
\beq
P\big[\,\Omega_{2,1}\,\big]_{{\rm D}}\,=\,i\,{K\over 3
(1-cy)^2}\,dy\wedge\,d\alpha\wedge d\psi \,\, ,
\eeq
we get:
\beq
{\cal B}({\rm D})\,=\,- i \int_{{\rm D}}\,
P\big[\,\Omega_{2,1}\,\big]_{{\rm D}}\,=\,-p\,\,,
\eeq
which, indeed, coincides with the baryon number of $U^{\alpha}$ written in
Table \ref{charges}. 

\subsubsection{General integration}

Let us now try to integrate in general the first-order differential system
(\ref{HEKBPS}). With this purpose it is more convenient to describe the
locus of the D3-brane by means of two functions $y = y(\theta,\phi)$,
$\beta =
\beta(\theta,\phi)$. Notice that this is equivalent to the description used
so far (in which the independent variables were $(y, \beta)$), except for
the cases in which $(\theta,\phi)$ or $(y, \beta)$ are constant. The
derivatives in these two descriptions are related by simply inverting the
Jacobian matrix, \ie:
\beq
\begin{pmatrix}
 y_\theta & y_\phi \cr \beta_\theta &
\beta_\phi 
\end{pmatrix}
\,=\,
\begin{pmatrix}
\theta_y &\theta_\beta \cr \phi_y &
\phi_\beta
\end{pmatrix}^{-1} \,\,.
\eeq
By using these equations the first-order system (\ref{HEKBPS}) is equivalent
to:
\beq
\beta_\theta\,=\,{y_\phi\over H^2\sin\theta}\,\,,
\qquad\qquad
\beta_\phi\,=\,c\cos\theta\,-\,{\sin\theta\over H^2}\,y_{\theta}\,\,.
\label{BPSthetaphi}
\eeq
These equations can be obtained directly by using $\theta$ and $\phi$ as
worldvolume coordinates. Interestingly, in this form the BPS equations can
be written as Cauchy-Riemann equations and, thus, they can be integrated in
general. This is in agreement with the naive expectation that, at least
locally, these equations should determine some kind of holomorphic
embeddings. In order to verify this fact, let us introduce new
variables $u_1$ and $u_2$, related to $\theta$ and $y$ as follows:
\begin{equation}
 u_1=\,\log\,
\bigg(\,\tan
\frac{\theta}{2}\,\bigg), \qquad \qquad u_2=\,\log\, \bigg(\,\frac{(\sin
\theta)^c}{f_1(y)}\,\bigg).
\end{equation}
By comparing the above expressions with the definitions of $z_1$ and $z_2$
in eq. (\ref{complexzs}), one gets:
\beq
u_1-i\phi\,=\,\log z_1\,\,, \qquad\qquad
u_2-i\beta\,=\,\log z_2\,\,.
\eeq
The relation between $u_1$ and $\theta$ leads to $du_1 =
d\theta/\sin\theta$, from which it follows that:
\beq
{\partial u_2\over \partial u_1}\,=\,c\cos\theta - {\sin\theta\over
H^2}\,y_{\theta}\,\,, \qquad\qquad\qquad
{\partial \beta\over \partial u_1}\,=\,\sin\theta\,\beta_{\theta}\,\,,
\eeq
and it is easy to demonstrate that the BPS equations (\ref{BPSthetaphi})
can be written as:
\begin{equation}
\frac{\partial u_2}{\partial u_1} = \frac{\partial \beta}{\partial
\phi}\,, \qquad\qquad \frac{\partial u_2}{\partial \phi} = -
\frac{\partial \beta}{\partial u_1}\,,
\end{equation}
these being the Cauchy-Riemann equations for the variables $u_2 - i\beta
= \log z_2$ and $u_1 - i\phi = \log z_1$. Then, the general integral of the
BPS equations is
\begin{equation}
\log z_2\,=\,f(\log z_1)\,\,,
\label{logrelation}
\end{equation}
where $f$ is an arbitrary (holomorphic) function of $\log z_1$. By
exponentiating eq. (\ref{logrelation}) one gets that the general solution of
the BPS equations is a function $z_2=g(z_1)$, in which $z_2$ is an arbitrary
holomorphic function of $z_1$. This result is analogous to what happened for
$T^{1,1}$ \cite{Arean:2004mm}. The appearance of a holomorphic function in the local
complex coordinates $z_1$ and $z_2$ is a consequence of kappa symmetry or,
in other words, supersymmetry. But one still has to check that this equation
makes sense globally. We will come to this point shortly. The simplest case
is that in which $\log z_2$ depends linearly on $\log z_1$, namely
\begin{equation}
\log z_2\,=\,n(\log z_1)\,+\,{\rm const.} ~,
\end{equation}
where $n$ is a constant. By exponentiating this equation we get a relation between 
$z_2$ and $z_1$ of the type:
\beq
z_2\,=\,{\cal C}\,z_1^n ~,
\label{n-winding}
\eeq
where ${\cal C}$ is a complex constant. If we represent this constant as
${\cal C}=Ce^{-i\beta_0}$, the embedding (\ref{n-winding}) reduces to the
following real functions $\beta=\beta(\phi)$ and $y=y(\theta)$:
\bear
\beta&=&n\phi\,+\,\beta_0\,\,,\rc\rc
f_1(y)&=&C\,\,{\big (\,\sin\theta\,\big)^c\over \Big(\tan{\theta\over
2}\Big)^n}\,\,.
\label{embgen}
\eear
This is a nontrivial embedding of a D3-brane probe on $AdS_5 \times Y^{p,q}$.
Notice that in the limit $c \to 0$ one recovers the results of \cite{Arean:2004mm}.
For $c \neq 0$, a key difference arises. As we discussed earlier, $z_2$ is
not globally well defined in $CY^{p,q}$ due to its dependence on $\beta$. 
As a consequence, eqs.(\ref{n-winding})-(\ref{embgen}) describe a
kappa-symmetric embedding for the D3-brane on $Y^{p,q}$ but it does not
correspond to a wrapped brane. The D3-brane spans a submanifold with
boundaries.\footnote{In this respect, notice that it might happen that
global consistency forces, through boundary conditions, the D3-brane
probes to end on other branes.} The only solution corresponding to a probe
D3-brane wrapping a three-cycle is $z_1 = {\rm const.}$ which is the one
obtained in the preceding subsection.

In order to remove $\beta$ while respecting holomorphicity, we
seem to be forced to let $z_3$ enter into the game. The reason is simple,
any dependence in $\beta$ disappears if $z_2$  enters through the
product $z_2 z_3$. This would demand embeddings involving the radius that
we did not consider. In this respect, it is interesting to point out that
this is also the conclusion reached in \cite{BHOP} from a different perspective:
there, the complex coordinates corresponding to the generators of the
chiral ring are deduced and it turns out that all of them depend on $z_1$,
$z_2 z_3$ and $z_3$. It would be clearly desirable to understand these
generalised wrapped D3-branes in terms of algebraic geometry, following
the framework of ref. \cite{beasley} which, in the case of the conifold, 
emphasizes the use of global homogeneous coordinates. Unfortunately, the
relation between such homogeneous coordinates and the chiral fields of
the quiver theory is more complicated in the case of $CY^{p,q}$.

\subsection{The calibrating condition}

Let us now verify that the BPS equations we have obtained ensure that the
three-dimensional submanifolds we have found are calibrated. Given a three-submanifold in $Y^{p,q}$ one can construct its cone
${\cal D}$, which is a four-dimensional submanifold of $CY^{p,q}$. The
calibrating condition for a supersymmetric four-submanifold ${\cal D}$ of
$CY^{p,q}$ is just:
\beq
P\Big[\,{1\over 2}\,J\wedge J\,\Big]_{{\cal D}}\,=\,
{\rm Vol}({\cal D})\,\,,
\label{calibration}
\eeq
where  ${\rm Vol}({\cal D})$ is the volume form of the divisor 
${\cal D}$ and $J$ is the K\"ahler form of $CY^{p,q}$ (\ref{KahlerYpq}). Let us check that the condition (\ref{calibration}) is indeed
satisfied by the cones constructed from our three-submanifolds. In order
to verify this fact it is more convenient to describe the embedding by
means of functions
$y=y(\theta,\phi)$ and $\beta=\beta(\theta,\phi)$. The corresponding BPS
equations are the ones written in (\ref{BPSthetaphi}). By using them one
can verify that the induced volume form for the three-dimensional
submanifold is:
\beq
vol\,=\,{1\over 18}\,\Big|\,(1-cy)\sin\theta\,+\,c\cos\theta y_{\theta}\,+\,
\beta_\theta y_\phi\,-\,y_\theta\beta_\phi\,\Big|_{BPS}\,\,
d\theta\wedge d\phi\wedge d\psi ~.
\eeq
By computing the pullback of $J\wedge J$ one can verify that the calibrating condition
(\ref{calibration}) is indeed satisfied for:
\beq
{\rm Vol}({\cal D})\,=\,-r^3\,dr\wedge vol\,\,,
\eeq
which is just the volume form of ${\cal D}$ with the metric
$ds^2_{CY^{p,q}}$ having a particular orientation. Eq. (\ref{calibration})
is also satisfied for the cones constructed from the singlet and doublet 
three-cycles of subsections \ref{singlet} and \ref{hekconst}. This fact is
nothing but the expression of the local nature of supersymmetry.

\subsection{Energy bound}

The dynamics of the D3-brane probe is governed by the Dirac-Born-Infeld lagrangian that, for the case in which there are not worldvolume gauge
fields, reduces in Einstein frame (see eq. (\ref{actionintrodEF})) to:
\beq
{\cal L}\,=\,-\sqrt{-g}\,\,,
\label{D3lag-den}
\eeq
where we have taken the D3-brane tension equal to one. We have checked that any solution of the first-order
equations (\ref{HEKBPS}) or (\ref{BPSthetaphi}) also satisfies the
Euler-Lagrange equations derived from the lagrangian density
(\ref{D3lag-den}). Moreover, for the static configurations we are
considering here the hamiltonian density ${\cal H}$  is, as expected, 
just ${\cal H}=-{\cal L}$. We are now going to verify that this energy
density satisfies a bound, which is just saturated when the BPS equations
(\ref{HEKBPS}) or (\ref{BPSthetaphi}) hold. In what follows we will take
$\theta$ and $\phi$ as independent variables. For an  arbitrary 
embedding of a D3-brane described by two functions $\beta=\beta(\theta,
\phi)$ and
$y=y(\theta, \phi)$ one can show that ${\cal H}$ can be written as:
\beq
{\cal H}\,=\,\sqrt{{\cal Z}^2\,+\, {\cal Y}^2\,+{\cal W}^2}\,\,,
\eeq
where ${\cal Z}$, ${\cal Y}$ and ${\cal W}$ are given by:
\bear
{\cal Z}&=&{L^4\over 18}\,\Bigg[\,
(1-cy)\sin\theta\,+c\cos\theta y_{\theta}+ 
y_{\phi}\beta_{\theta}-y_{\theta}\beta_{\phi}\,\Bigg]\,\,,\rc
{\cal Y}&=&{L^4\over 18}\,\sqrt{1-cy}\,\,H\,
\Bigg[\,\beta_{\phi}\,-\,c\cos\theta\,+\,{\sin\theta \over H^2}\,y_{\theta}
\,\Bigg]\,\,,\rc
{\cal W}&=&{L^4\over 18}\,\sqrt{1-cy}\,\,H\,
\Bigg[\,\sin\theta\,\beta_{\theta}\,-\,{y_{\phi}\over H^2}\,\Bigg]\,\,.
\eear
Obviously one has:
\beq
{\cal H}\ge \big|\,{\cal Z}\,\big|\,\,.
\label{energybound}
\eeq
Moreover, since 
\beq
{\cal Y}_{\big |\,BPS}\,=\,{\cal W}_{\big |\,BPS}\,=\,0\,\,,
\eeq
the bound saturates when the BPS equations (\ref{BPSthetaphi}) are satisfied.
Thus, the system of
differential equations (\ref{BPSthetaphi}) is equivalent to the condition 
${\cal H}= \big|\,{\cal Z}\,\big|$ (actually ${\cal Z}\ge 0$ if the BPS
equations (\ref{BPSthetaphi}) are satisfied). Moreover, for an arbitrary
embedding
${\cal Z}$ can be written as a total derivative, namely:
\beq
{\cal Z}\,=\,{\partial\over \partial \theta}\,{\cal Z}^{\theta}\,+\,
{\partial\over \partial \phi}\,{\cal Z}^{\phi}\,\,.
\eeq
This result implies that ${\cal H}$ is bounded by the integrand of a topological
charge. The explicit form of ${\cal Z}^{\theta}$ and ${\cal Z}^{\phi}$ is:
\bear
{\cal Z}^{\theta}&=&-{L^4\over 18}\,\Big[\,(1-cy)\cos\theta\,+\,y\,\beta_{\phi}\,
\Big]\,\,,\rc
{\cal Z}^{\phi}&=&{L^4\over 18}\,y\,\beta_{\theta}\,\,.
\eear
In this way, from the point of view of the D3-branes, the configurations
satisfying eq. (\ref{BPSthetaphi})  can be regarded as BPS worldvolume
solitons.

\subsection{BPS fluctuations of dibaryons}  \label{BPSfluctuationdibaryons}

In this subsection we study BPS fluctuations of dibaryon
operators in the $Y^{p,q}$ quiver theory. We start with the
simplest dibaryon which is singlet under $SU(2)$, let us say  $\det Y$. 
To construct excited dibaryons we should replace one of the $Y$ factors 
by any other chiral field  transforming in the same
representation of the gauge groups. For example,  replacing $Y$
by
$YU^{\alpha}V^{\beta}Y$, we get a new operator of the form
\begin{eqnarray}
\epsilon_1\epsilon^2(YU^{\alpha}V^{\beta}Y)Y\cdots Y\,\,,
\end{eqnarray}
where $\epsilon_1$ and $\epsilon^2$ are abbreviations for the
completely anti-symmetric tensors for the respective $SU(N_c)$
factors of the gauge group. Using the identity
\begin{eqnarray}
\epsilon^{a_1\cdots a_{N_c}}\epsilon_{b_1\cdots
b_{N_c}}=\sum_{\sigma}(-1)^{\sigma}\delta^{a_1}_{\sigma(b_1)}\cdots
\delta^{a_{N_c}}_{\sigma(b_{N_c})}\,\,,
\end{eqnarray}
the new operator we get can factorise into the original dibaryon
and a single-trace operator
\begin{eqnarray}
{\rm Tr}(U^{\alpha}V^{\beta}Y)\;{\rm det}\,Y\,\,.
\end{eqnarray}
Indeed for singlet dibaryons, a factorisation of this sort always works.
This fact seems to imply, at least at weak coupling,  that excitation of a
singlet dibaryon can be represented as graviton fluctuations in the
presence of the original dibaryon.

For the case of a dibaryon with $SU(2)$ quantum number the
situation is different. Consider, for simplicity, the state with
maximum $J_3$ of the $SU(2)$
\begin{eqnarray}
\epsilon_1\epsilon^2(U^1\cdots U^1)={\rm det}\,U^1,
\end{eqnarray}
we can replace one of the $U^1$ factors by $U^1\, {\cal O}$, where $\cal O$ is
some operator given by a closed loop in the quiver diagram. As in the case of
a singlet dibaryon, this kind of excitation is factorisable since
all $SU(2)$ indices are symmetric. So this kind of operator should
be identified with a graviton excitation with wrapped D3-brane in the
dual string theory. However, if the $SU(2)$ index of the $U$ field
is changed in the excitation, i.e. $U^1\rightarrow U^2\, {\cal O}$,
then the resulting operator cannot be written as a product of the
original dibaryon and a meson-like operator. Instead it
has to be interpreted as a single particle state in $AdS$. Since
the operator also carries the same baryon number, the natural
conclusion is that the one-particle state is a BPS excitation of
the wrapped D3-brane corresponding to the dibaryon \cite{ba2}.

\begin{figure}[ht]
\centerline{\epsffile{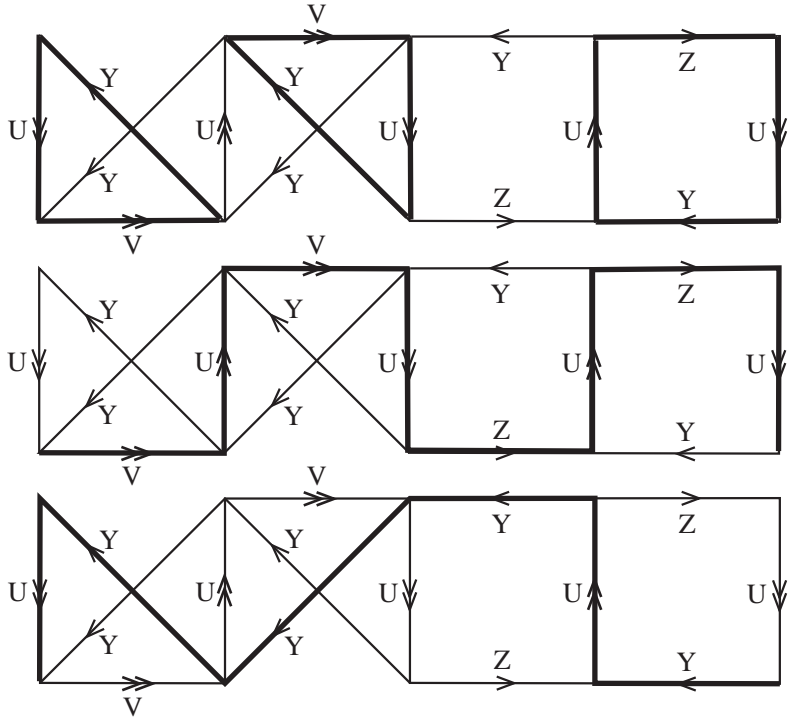}}
\caption{Loops in the $Y^{4,2}$ quiver representing mesonic operators
in the chiral ring. There are short loops such as $UVY$, $VUY$ or $YUZU$
(upper), longest loops as $VUVUZUZU$ (middle) and long loops like $YUYYYU$
(bottom). The representative of each class in the chiral ring is,
respectively, ${\cal O}_1$, ${\cal O}_2$ and ${\cal O}_3$.}
\label{Y42loops}
\end{figure}

In order to classify all these BPS excitations of the dibaryon, we
have to count all possible inequivalent chiral operators $\cal O$
that transform in the bifundamental representation of one of the
gauge group factors of the theory. In $Y^{p,q}$ quiver gauge theory, these
operators correspond to loops in the quiver diagram just like the mesonic
chiral operators discussed in \cite{bk}. The simplest ones are
operators with R-charge 2. They have been thoroughly discussed in
\cite{Benvenuti:2005wi}. They are given by short loops of length 3 or 4 in
the quiver, precisely as those operators entering in the superpotential
(\ref{supYpq}). They are single trace operators of the form
(in what follows we omit the trace and the $SU(2)$ indices) $UVY$,
$VUY$ or $YUZU$ (see the upper quiver in Fig. \ref{Y42loops}).
Since they are equivalent in the chiral ring, we can identify
them as a single operator ${\cal O}_1$. It transforms in the
spin $\frac{1}{2} \otimes \frac{1}{2} = 0 \oplus 1$ representation
of the global $SU(2)$. The scalar component vanishes in the chiral
ring. Thus, we end up with a spin 1 chiral operator with scaling
dimension $\Delta = 3$. Its $U(1)_F$ charge vanishes.

There are also two classes of long loops in the quiver. The first
class, whose representative is named ${\cal O}_2$, has length $2p$,
winds the quiver from the left to the right and is made of $p$ $U$
type operators, $q$ $V$ type operators and $p-q$ $Z$ type operators.
For example, in $Y^{4,2}$, a long loop of this class is $VUVUZUZU$
(middle quiver in Fig. \ref{Y42loops}). It transforms in the spin
$\frac{1}{2} \otimes ... \otimes \frac{1}{2} = \frac{p+q}{2} \oplus
\dots$ representation of $SU(2)$. The dots amount to lower dimensional
representations that vanish in the chiral ring. The resulting operator,
${\cal O}_2$, has spin $\frac{p+q}{2}$. There is another class of
long loops which has length $2p-q$, running along the quiver in the
opposite direction, build with $p$ $Y$ type operators and $p-q$ $U$
type operators. We name its representative as ${\cal O}_3$. 
In the case of $Y^{4,2}$, it is an operator like $YUYYYU$ (bottom
quiver in Fig. \ref{Y42loops}). $SU(2)$ indices, again, have to be
completely symmetrised, the spin being $\frac{p-q}{2}$. Long loops
wind around the quiver and this leads to a nonvanishing value of
$Q_F$ \cite{bk}. The baryonic charge vanishes for any of these loops.
We summarise in Table \ref{chring} the charge assignments for the
three kinds of operators ${\cal O}_i$ \cite{bk}.
\begin{table}
\begin{center}
$$\begin{array}{|c|c|c|c|}
\hline {\rm Operator} & Q_R & Q_F & {\rm Spin} \\
\hline\hline
& & & \\[-1.5ex]
{\cal O}_1 & 2 & 0 & 1 \\[1ex]
\hline
& & & \\[-1.5ex]
{\cal O}_2 & p+q-\frac{1}{3\ell} & p & \frac{p+q}{2}\\[1ex]
\hline
& & & \\[-1.5ex]
{\cal O}_3 & p-q+\frac{1}{3\ell} & -p & \frac{p-q}{2}\\[1ex]
\hline
\end{array}$$
\caption{Charges assignments for the mesonic operators
${\cal O}_i$ that generate the chiral ring.}
\label{chring}
\end{center}
\end{table}
We can see that these operators satisfy the BPS condition $\Delta={3\over
2}\, Q_R$. In fact, they are the building blocks of all other scalar BPS
operators. The general BPS excitation corresponds to operators of the form
\begin{eqnarray}
{\cal O}=\prod_{i=1}^3{\cal O}_{i}^{\,\,n_i}\,\,.
\end{eqnarray}
It is interesting to notice that the spectrum of  fluctuations
of a dibaryon must coincide with the mesonic chiral operators in the 
$Y^{p,q}$ quiver theory. This would provide a nontrivial test of the 
AdS/CFT correspondence. We show this result explicitly via an analysis of
open string fluctuation on wrapped  D3-branes.

Now we are interested in describing the excitations of dibaryon operators from the dual 
string theory. For those excitations that are
factorisable, the dual configurations are just the multi-particle
states of graviton excitations in the presence of a dibaryon. The
correspondence of graviton excitation and mesonic operator were
studied in \cite{bk, ksy}. What we are really interested in
are those non-factorisable operators that can be interpreted as
open string excitations on the D-brane. This can be analyzed by
using the Dirac-Born-Infeld action of the D3-brane. In what follows  we will
focus on the dibaryon made of $U$ fields, which corresponds to the 
three-cycle D studied in subsection \ref{hekconst} which, for convenience,
we will  parameterise with the coordinates $(y,\psi, \alpha)$. The
analysis of the dibaryon made of $V$ field is similar. For our purpose 
we will use, as in eq. (\ref{globalADS}),  the global coordinate system
for the $AdS_5$ part  of the metric and we will take the $Y^{p,q}$ part
as written in eq.  (\ref{alphametric}).  We are interested in the normal
modes of oscillation of the wrapped D3-brane around the solution
corresponding to some fixed worldline in $AdS_5$ and some fixed $\theta$
and $\phi$ on the transverse $S^2$. For such a configuration, the induced
metric on the dibaryon is:
\begin{eqnarray}
L^{-2}ds^2_{ind}=- dT^2+\frac{1}{wv}dy^2+\frac{v}{9}d\psi^2+w(d\alpha+fd\psi)^2\,\,,
\end{eqnarray}
where the functions $v(y)$, $w(y)$ and $f(y)$ have been defined in eq. (\ref{qwy}) (in what
follows of this subsection we will take $c=1$). 

The fluctuations along the transverse $S^2$ are
the most interesting, since they change the $SU(2)$ quantum
numbers and are most readily compared with the chiral primary
states in the field theory.
Without lost of generality, we consider fluctuations around the
north pole of the $S^2$, i.e. $\theta_0=0$. Instead of using
coordinates $\theta$ and $\phi$, it is convenient to go from polar to Cartesian coordinates:
$\zeta^1=\theta\sin\phi$ and $\zeta^2=\theta\cos\phi$. As a further simplification we
perform a shift in the coordinate $\psi$. The action for the D3-brane is:
\beq
S\,=\,-T_3\,\int d^4\xi \sqrt{-\det g}\,+\,T_3\,\int P[C_{(4)}]\,\,.
\eeq
Let us expand the induced metric $g$ around the static configuration as $g=g_{(0)}+\delta
g$, where  $g_{(0)}$ is the zeroth order contribution. The corresponding expansion for the
action takes the form:
\beq
S\,=\,S_0\,-\,{T_3\over 2}\,\int d^4\xi \sqrt{-\det g_{(0)}}\,\,
{\rm Tr}\,\big[g_{(0)}^{-1}\,\delta g\big]\,+\,T_3\,\int P[C_{(4)}]\,\,,
\eeq
where $S_0=-T_3\,\int d^4\xi \sqrt{-\det g_{(0)}}$.  Note that the determinant of the
induced metric at zeroth order is a constant: 
$\sqrt{-{\rm det}(g_{(0)})}=\frac{1}{3}L^4$.
The five-form field strength is 
\begin{eqnarray}
F_5=(1+*)\, 4\sqrt{{\rm det}(G_{Y^{p,q}})}L^4d\theta\wedge d\phi\wedge
dy\wedge d\psi\wedge d\alpha\,\,.
\end{eqnarray}
Moreover, using that ${\rm det}(G_{Y^{p,q}})$ is the determinant of the metric of the $Y^{p,q}$ manifold and $\sqrt{{\rm det}(G_{Y^{p,q}})}=\frac{1-y}{18}\sin\theta$, we
can choose the four-form Ramond-Ramond field to be
\begin{eqnarray}
C_4=\frac{2}{9}(1-y)L^4(\cos\theta-1)\,d\alpha\wedge dy\wedge d\psi
\wedge d\phi\,,
\end{eqnarray}
which is well defined around the north pole of $S^2$. At  quadratic
order, the four form $C_4$ is 
\begin{eqnarray}
C_4=-\sqrt{-{\rm det}\, g_{(0)} }\,\,\frac{1-y}{3}\epsilon_{ij}\,\zeta^i\,d\zeta^j\wedge
d\alpha\wedge dy\wedge d\psi\,\,.
\end{eqnarray}
The contribution from the Born-Infeld part of the effective action is: 
\beq
{\rm Tr}\,\big[g_{(0)}^{-1}\,\delta g\big]=G_{ij}\;g_{(0)}^{\mu\nu}
\,(\partial_{\mu}\zeta^i\partial_{\nu}\zeta^j)+2g^{\mu\nu}_{(0)}\;
G_{\mu i}\,\partial_{\nu}\zeta^i, 
\eeq
where $G$ is the metric of the background, $i,j$ denote the components of
$G$ along the
$\zeta^{1,2}$ directions and the indices $\mu,\nu$ refer to the directions
of the worldvolume of the cycle. The non-vanishing components of $G$ are:
\beq
G_{ij}=\frac{1-y}{6}L^2\delta_{ij}, \quad G_{\psi i}=
-\frac{1}{2}\bigg(wf^2+\frac{v}{9}\bigg)L^2\epsilon_{ij}\,\zeta^j, \quad 
G_{\alpha i}=-\frac{wf}{2}L^2\epsilon_{ij}\,\zeta^j\,\,. 
\eeq
Using these results one can verify that the effective Lagrangian is
proportional to:
\begin{eqnarray}
\sum_{i}L^2\frac{1-y}{6}g_{(0)}^{\mu\nu}
(\partial_{\mu}\zeta^i\partial_{\nu}\zeta^i)+2g^{\mu\nu}_{(0)}\;G_{\mu
i}\,\partial_{\nu}\zeta^i+\frac{2(1-y)}{3}\epsilon_{ij}
\,\zeta^i\,\partial_{T}\zeta^j\,\,.
\end{eqnarray}
The equations of motion for the fluctuation are finally given by
\begin{eqnarray}
{L^2\over 6}\,\partial_{\mu}\bigg((1-y)\,g^{\mu\nu}_{(0)}\,
\partial_{\nu}\zeta^i\,\bigg)+2\partial_{\nu}(g^{\mu\nu}_{(0)}\,G_{\mu
i})-\frac{2(1-y)}{3}\epsilon_{ij}\,\partial_{T}\zeta^j=0\,\,.
\end{eqnarray}
 Introducing  $\zeta^{\pm}=\zeta^1\pm i\zeta^2$, the equations of motion
reduce to
\begin{eqnarray}
\bigg(\nabla^2-\frac{1-y}{6}\partial_{T}^2\,\bigg) \zeta^{\pm}\pm
i\frac{2(1-y)}{3}\partial_{T}\zeta^{\pm}\pm i\partial_{\psi}
\zeta^{\pm}=0\,\,,
\end{eqnarray}
where $\nabla^2$ is the laplacian along the spatial directions of the cycle for the induced
metric $g_{(0)}$. 
The standard strategy to solve this equation is to use separation of
variables as
\begin{eqnarray}
\zeta^{\pm}=\exp(-i\omega T)\exp\bigg(i\frac{m}{\ell}\alpha\bigg)\exp(in\psi)
\,Y^{k\pm}_{mn}(y)\,\,.
\end{eqnarray}
Plugging  this ansatz into the equation of motion, we find
\begin{eqnarray}
&&\frac{1}{1-y}\frac{d}{dy}\bigg[(1-y)w(y)v(y)\frac{d}{dy}
Y^{k\pm}_{mn}(y)\bigg]\\
&=&\bigg[\bigg(\frac{9f^2(y)}{v(y)}+\frac{1}{w(y)}\bigg)\frac{m^2}{\ell^2}-
\frac{18f(y)}{v(y)}\frac{m}{\ell}n+\frac{9}{v(y)}n^2-\omega(\omega\pm
4)\pm\frac{6n}{1-y}\bigg]Y^{k\pm}_{mn}(y)\,\,. \nonumber
\end{eqnarray}
The resulting equation has four
regular singularities at $y=y_1,y_2,y_3$ and $\infty$ and is known as 
Heun's equation (for clarity, in what follows we  omit the
indices in $Y$) \cite{heun}:
\be
\frac{d^2}{dy^2}Y^{\pm}+\bigg(\sum_{i=1}^3
\frac{1}{y-y_i}\bigg)\frac{d}{dy}Y^{\pm}+q(y)Y^{\pm}=0,\label{Fuch}
\ee
where, in our case
\begin{eqnarray}
q(y)&=&\frac{2}{{\cal Q}(y)}\bigg[\mu-\frac{y}{4}\omega(\omega\pm
4)-{1\over 2}\,\sum_{i=1}^3\frac{\alpha_i^2
{\cal Q}'(y_i)}{y-y_i}\bigg]\,\,,\rc
\mu&=&\frac{3}{32}(\frac{m}{\ell}+2n)(\frac{m}{\ell}-6n)+\frac{1}{4}\omega(\omega\pm
4)\mp \frac{3n}{2}\,\,, 
\end{eqnarray}
with ${\cal Q}(y)$ being the function defined in eq. (\ref{calq}). 
Now, given that the R-symmetry is dual to the Reeb Killing vector of
$Y^{p,q}$, namely $2\partial/\partial \psi\,-\,
{1\over 3}\,\partial/\partial \alpha$, we can use the R-charge
$Q_R=2n-m/3\ell$ instead of $n$
as  quantum number. The exponents at the regular singularities $y=y_i$ are
then given by
\begin{eqnarray}
\alpha_i=\pm \frac{1}{2}\frac{(1-y_i)(m/\ell+3Q_R\,y_i)}
{ {\cal Q}'(y_i)}.
\end{eqnarray}
The exponents at $y=\infty$ are $-\frac{\omega}{2}$ and
$\frac{\omega}{2}+2$ for $Y^{+}$, while $-\frac{\omega}{2}+2$ and
$\frac{\omega}{2}$ for $Y^{-}$.
We can transform the singularity from $\{y_1,y_2, y_3,\infty\}$ to
$\{0,1,b=\frac{y_1-y_3}{y_1-y_2},\infty\}$ by introducing a new variable $x$, defined as:
\begin{eqnarray}
x=\frac{y-y_1}{y_2-y_1}.
\end{eqnarray}
It is also convenient to substitute
\begin{eqnarray}
Y=x^{|\alpha_1|}(1-x)^{|\alpha_2|}(b-x)^{|\alpha_3|}\,h(x)\,\,,
\end{eqnarray}
which transforms equation (\ref{Fuch}) into the standard form of the
Heun's equation
\be
\frac{d^2}{dx^2}h(x)+\bigg(\frac{C}{x}+\frac{D}{x-1}+
\frac{E}{x-b}\,\bigg)\frac{d}{dx}h(x)+\frac{AB
x-k}{x(x-1)(x-b)}h(x)=0.\label{Heun}
\ee
Here the Heun's parameters are given by
\begin{eqnarray}
A&=&-\frac{\omega}{2}+\sum_{i=1}^3|\alpha_i|\,, \quad\quad
B=\frac{\omega+ 4}{2}+\sum_{i=1}^3|\alpha_i| \,\,,\rc
C&=&1+2|\alpha_1|\,, \quad\quad D=1+2|\alpha_2|\,, \quad\quad
E=1+2|\alpha_3|\,,
\end{eqnarray}
and
\begin{eqnarray}
k&=&(|\alpha_1|+|\alpha_3|)(|\alpha_1|+|\alpha_3|+1)-|\alpha_2|^2\nonumber\\
&+&b\bigg[(|\alpha_1|+|\alpha_2|)(|\alpha_1|+|\alpha_2|+1)-|\alpha_3|^2\bigg]-{\tilde
\mu}\,\,,\rc
{\tilde \mu}&=&-\frac{1}{y_1-y_2}\bigg(\mu-\frac{y_1}{4}\omega(\omega+
4)\bigg)\nonumber\\
&=&\frac{p}{q}\bigg[\frac{1}{6}(1-y_1)\omega(\omega+
4)-\frac{3}{16}Q_R\,\bigg(Q_R+\frac{4m}{3\ell}\bigg)-
\frac{1}{2}\bigg(Q_R+\frac{m}{3\ell}\bigg)\bigg]
\,\,,\label{mu'}\rc
b&=&\frac{1}{2}\bigg(1+\frac{\sqrt{4p^2-3q^2}}{q}\bigg)\,\,.
\end{eqnarray}
We only presented the equation for $Y^{+}$; the
corresponding equation for $Y^{-}$ can be obtained by replacing
$\omega$ with $\omega-4$ and changing the sign of the last term in
(\ref{mu'}).

Now let us discuss the solutions to this differential equation.
For quantum number $Q_R=2N_c$ (which implies $m=0$), we find all $\alpha_i$
equal to $N_c/2$. If we set $\omega=3N_c$, the Heun's parameters
$A$ and $k$ both vanish. The corresponding solution $h(x)$ is
a constant function. Similarly if $\omega=-3N_c-4$, then $B$ and
$k$ vanish which also implies a constant $h(x)$. The complete
solution of $\zeta^{\pm}$ in these two cases is given by
\begin{eqnarray}
\zeta_1^{\pm}&=&e^{\pm
i(-3N_c T +N_c\psi)}\prod_{i=1}^{3}(y-y_i)^{N_c/2}\,\,,\rc
\zeta_2^{\pm}&=&e^{\pm
i((3N_c+4)T+N\psi)}\prod_{i=1}^{3}(y-y_i)^{N_c/2}\,\,.
\label{zetas}
\end{eqnarray}
These constant solutions represent ground states with fixed quantum
numbers and, since they have the lowest possible dimension for a given
R-charge, they should be identified with the BPS operators. Indeed, in
the solutions (\ref{zetas}) the energy is quantised in units of $3L^{-1}$,
and $3$ is precisely the conformal dimension of ${\cal O}_1$. This
provides a perfect matching of AdS/CFT in this setting.

The situation for quantum numbers $Q_R=N_c(p\pm q\mp1/3\ell)$ and $m=\pm
N_c$ is similar to the case we have just discussed. The solutions for
$h(x)$ are constant with
\beq
\omega=\frac{N_cp}{2}\bigg(3\pm\frac{2p-\sqrt{4p^2-3q^2}}{q}\bigg)\,\,,
\eeq
and
\beq
\omega=-\frac{N_cp}{2}\bigg(3\pm\frac{2p-\sqrt{4p^2-3q^2}}{q}\bigg)-4.
\eeq
We can see that the conformal dimension satisfies
$\Delta=\frac{3}{2}Q_R$. So all these solutions are BPS
fluctuations which should correspond to the operators ${\cal O}_2$
and ${\cal O}_3$. 

An interesting comment is in order at this point. Notice that the 
dibaryon excitations should come out with the multiplicities associated
to the SU(2) spin (see Table \ref{chring}) of the ${\cal O}_i$ operators.
However, in order to tackle this problem, we would need to consider at
the same time the fluctuation of the D3-brane probes and the zero-mode
dynamics corresponding to their collective motion along the sphere with
coordinate $\theta$ and $\phi$ (see ref. \cite{ba2} for a similar
discussion in the conifold case). This is an
interesting problem that we leave open.

\setcounter{equation}{0}
\section {Supersymmetric D5-branes in $AdS_5\times Y^{p,q}$}
\label{d5}
\medskip

In this section we will study the supersymmetric configurations of
D5-branes in the $AdS_5\times Y^{p,q}$ background. First of all, notice
that in this case 
$\Gamma_{\kappa}$ acts on the Killing spinors $\epsilon$ as:

\beq
\Gamma_{\kappa}\,\epsilon\,=\,{i\over 6!\, \sqrt{-g}}\,\,
\epsilon^{\mu_1\cdots\mu_6}\,\gamma_{\mu_1\cdots\mu_6}\,\epsilon^*
\label{Gammakappad5}\,\,,
\eeq
where we have used the relation (\ref{rule}) to translate eq. (\ref{gammakappa}).
The appearance of the complex conjugation on the right-hand side of eq.
(\ref{Gammakappad5}) is crucial in what follows. Indeed, the complex conjugation
does not commute with the projections (\ref{epsilon-project}). Therefore, in order
to construct an additional compatible projection involving the
$\epsilon\to\epsilon^*$ operation we need to include a product of gamma matrices
which anticommutes with both $\Gamma_{12}$ and $\Gamma_{34}$. As in the D3-brane 
case just analyzed, this compatibility requirement between the 
$\Gamma_{\kappa}\,\epsilon = \epsilon$  condition  and (\ref{epsilon-project})
implies a set of differential equations whose solutions, if any, determine the
supersymmetric embeddings we are looking for. 

We will carry out successfully this program only in the case of a D5-brane
extended along a two-dimensional submanifold of $Y^{p,q}$. One expects that these kinds of
configurations represent wall defects in the field theory. When we allow the D5-brane to extend
infinitely in the holographic direction, we get a configuration
dual to a defect conformal field theory. In the remainder of this section we will find the
corresponding configurations of the D5-brane probe. Moreover,  in section
\ref{6} we will find, based on a different set of worldvolume
coordinates, another embedding of this type preserving the same
supersymmetry as the one found in the present section and we will analyze
the effect of adding flux of the worldvolume gauge fields. In section
\ref{6} we will also look at the possibility of having D5-branes wrapped
on a three-dimensional submanifold of 
$Y^{p,q}$. This configuration looks like a domain wall in the
field theory dual and although these configurations are not supersymmetric, we have
been able to  find stable solutions of the equations of motion. The case
in which the D5-brane wraps the entire $Y^{p,q}$ corresponds to the baryon
vertex. In this configuration, studied also in section \ref{6}, the
D5-brane captures the flux of the RR five-form, which acts as a source
for the electric worldvolume gauge field.  We will conclude in section
\ref{6} that this configuration cannot be supersymmetric, in analogy with
what happens in the conifold case \cite{Arean:2004mm}.

\subsection{Wall defect solutions}

We want to find a configuration in which the D5-brane probe wraps a
two-dimensional submanifold of $Y^{p,q}$ and is a codimension one object
in $AdS_5$. Accordingly, let us place the probe at some constant value of
one of the Minkowski coordinates (say $x^3$) and let us extend it along
the radial direction. To describe such an embedding we choose the following
set of worldvolume coordinates for a D5-brane probe
\beq
\xi^{\mu}\,=\,(t,x^1,x^2,r,\te,\p)\,\,,
\eeq
and we adopt the following ansatz:
\beq
y=y(\theta,\phi)\,\,,\qquad
\beta\,=\,\beta(\theta,\phi)\,\,,
\label{D5dwansatz}
\eeq
with $x^3$ and $\psi$ constant. The induced Dirac matrices can be computed
straightforwardly from  eq.  (\ref{wvgamma}). From the general expression (\ref{Gammakappad5}) one readily gets that the kappa symmetry matrix acts on the spinor $\epsilon$ as:
\beq
\Gamma_\kappa\,\epsilon\,=\,{i\over \sqrt{-g}}\,{r^2\over L^2}\,
\Gamma_{x^0x^1x^2r}\,\gamma_{\theta\phi}\,\epsilon^*\,\,.
\eeq
By using the complex conjugate of the projections (\ref{epsilon-project}) one gets:
\beq
{6\over L^2}\,\gamma_{\theta\phi}\,\epsilon^*\,=\,\big[\,
b_I\,+\,b_{15}\,\Gamma_{15}\,+\,b_{35}\,\Gamma_{35}\,+\,b_{13}\,\Gamma_{13}\,\big]\,
\epsilon^*\,\,,
\label{D5gamathetaphi}
\eeq
where the different coefficients are:
\bear
&&b_I\,=\,-i\Big[\,(1-cy)\sin\theta\,+\,c\cos\theta\,y_{\theta}\,+\,
y_\phi\beta_\theta\,-\,y_\theta\beta_\phi\,\Big]\,\,,\rc
&&b_{15}\,=\,-\sqrt{{2\over 3}}\,\,{1\over H}\Big[\,
(1-cy)\cos\theta\,y_\theta\,+\,y\,(\beta_\phi\,y_\theta-\beta_\theta\,y_\phi)\,\Big]\,-\,i
\sqrt{{2\over 3}}\,H\,\cos\theta\,\beta_{\theta}\,\,,\rc
&&b_{35}\,=\,\sqrt{{2\over 3}}\sqrt{1-cy}\,\Big[\,
(1-cy)\cos\theta\,+\,y\beta_\phi\,\Big]\,+\,i
\sqrt{{2\over 3}}\sqrt{1-cy}\,\,y\sin\theta\beta_\theta\,\,,\rc
&&b_{13}\,=\,\sqrt{1-cy}\,\Big[\,{y_\phi\over H}\,-\,H\beta_\theta\,\sin\theta\,\Big]\,+\,
i\sqrt{1-cy}\,\Big[\,{\sin\theta\over H}\,y_{\theta}\,-\,H\,
(c\cos\theta\,-\,\beta_\phi)\,\Big]\,\,.\qquad
\label{D5bs}
\eear
As discussed above,  in this case the action of $\Gamma_{\kappa}$ involves the
complex conjugation, which does not commute with the projections
(\ref{epsilon-project}). Actually, the only term on the right-hand side of
(\ref{D5gamathetaphi}) which is consistent with (\ref{epsilon-project}) is
the one containing $\Gamma_{13}$. Accordingly, we must require:
\beq
b_I\,=\,b_{15}\,=\,b_{35}\,=\,0\,\,.
\eeq
From the vanishing of the imaginary part of $b_{15}$ we get:
\beq
\beta_\theta=0\,\,,
\label{D5beta_theta}
\eeq
while the vanishing of the real part of $b_{15}$ leads to:
\beq
\beta_{\phi}\,=\,-{1-cy\over y}\,\cos\theta\,\,.
\label{betaphi}
\eeq
Notice that $b_{35}$ is zero as a consequence of  equations 
(\ref{D5beta_theta}) and (\ref{betaphi}) which, 
in particular imply that:
\beq
\beta=\beta(\phi)\,\,.
\eeq
Moreover, by using eq. (\ref{D5beta_theta}), the condition $b_I=0$ is equivalent to
\beq
(1-cy)\sin\theta\,+\,(c\cos\theta\,-\,\beta_{\phi})y_\theta\,=\,0\,\,,
\eeq
and plugging the value of $\beta_{\phi}$ from (\ref{betaphi}), one arrives at:
\beq
y_{\theta}\,=\,-(1-cy)\,y\tan\theta\,\,.
\label{D5yeq}
\eeq
In order to implement the kappa symmetry condition at all points of the
worldvolume the phase of $b_{13}$ must be constant. This can be achieved
by requiring that the real part of $b_{13}$ vanishes, which for  
$\beta_\theta=0$   is equivalent to the condition $y_\phi=0$,\,\ie:
\beq
y=y(\theta)\,\,.
\eeq
The equation (\ref{D5yeq}) for $y(\theta)$ is easily integrated, namely:
\beq
{y\over 1-cy}\,=\,k\cos\theta\,\,,
\label{D5ysol}
\eeq
where $k$ is a constant. Moreover, by separating variables  in eq. (\ref{betaphi}), one concludes
that:
\beq
\beta_\phi=m\,\,,
\label{D5phim}
\eeq
where $m$ is a new constant. Plugging (\ref{D5ysol}) into eq.
(\ref{betaphi}) and using the result (\ref{D5phim}) one concludes that
the two constants $m$ and $k$ must be related as:
\beq
km=-1\,\,,
\eeq
which, in particular implies that $k$ and $m$ cannot vanish. 
Thus, the embedding of the D5-brane becomes
\bear
&&\beta=m\phi+\beta_0\,\,,\rc
&&y=-{\cos\theta\over m-c\cos\theta}\,\,.
\label{D5sol}
\eear
Notice that the solution (\ref{D5sol}) is symmetric under the change
$m\to -m$, $\theta\to\pi-\theta$ and $\phi\to 2\pi-\phi$. Thus, from now on we can
assume that $m\ge 0$. 

It is now  straightforward to verify that the BPS equations are equivalent  to
impose the following condition on the spinor $\epsilon$:
\beq
\Gamma_{x^0x^1x^2r13}\epsilon^*\,=\,\sigma \epsilon\,\,,
\label{d5projector}
\eeq
where $\sigma$ is:
\beq
\sigma\,=\,{\rm sign}\,\Big(\,{\cos\theta\over y}\,\Big)\,=\,
-{\rm sign}\Big(\,m\,-\,c\cos\theta\,\Big)\,\,.
\label{D5sigma-sign}
\eeq
Obviously, the only valid solutions are those which correspond to having a constant sign $\sigma$
along the worldvolume. This always happens for $m/c\ge 1$. In this case the minimal
(maximal) value of $\theta$ is $\theta=0$ ($\theta=\pi$) if 
$|m-c||y_1|>1$ ($|m-c||y_2|>1$). Otherwise the angle $\theta$ must be
restricted to lie in the interval $\theta\in [\theta_{1}, \theta_{2}]$, where
$\theta_{1}$ and $\theta_{2}$ are given by:
\beq
\theta_{i}\,=\,\arccos\Big[{my_i\over cy_i-1}\Big]\,\,,
\qquad\qquad (i=1,2)\,\,.
\eeq
Notice that, similarly to what we obtained in the previous section,
eq.(\ref{D5sol}) implies that the configuration we arrived at does not, in
general, correspond to a wrapped brane but to a D5-brane that spans a
two-dimensional submanifold with boundaries. 

Let us now count the number of supersymmetries preserved by our configuration. In
order to do so we must  convert eq. (\ref{d5projector}) into an algebraic
condition on a constant spinor. With this purpose in mind let us write
the general form of
$\epsilon$ as the sum of the two types of spinors written in eq.
(\ref{chiraladsspinor}), namely:

\beq
e^{{i\over 2}\psi}\,\epsilon\,=\,r^{-{1\over 2}}\,\eta_+\,+\,
r^{{1\over 2}}\,\Big(\,{\bar x^3\over L^2}\,
\Gamma_{rx^3}\,\eta_+\,+\,\eta_-\,\Big)\,+
\,{r^{{1\over 2}}\over L^2}\,x^p\,\Gamma_{rx^p}\,\eta_+\,\,,
\label{generalepsilon}
\eeq
where $\bar x^3$ is the constant value of the coordinate $x^3$ in the embedding and the index
$p$ runs over the set  $\{0,1,2\}$. By substituting eq.  (\ref{generalepsilon}) on both sides of
eq. (\ref{d5projector}), one can get the conditions that $\eta_{+}$  and
$\eta_{-}$ must satisfy. Indeed, let us define the operator ${\cal P}$ as follows:
\beq
{\cal P}\,\epsilon\,\equiv\,i\sigma  e^{i\psi_0}\,\Gamma_{rx^3}\,
\Gamma_{1 3}\,\epsilon^*\,\,.
\eeq
Then, one can check that eq. (\ref{d5projector}) is equivalent to:
\bear
&&{\cal P}\,\eta_+\,=\,\eta_+\,\,,\rc\rc
&&(1\,+\,{\cal P}\,)\,\eta_-\,=\,-{2 \bar x^3\over L^2}\,\Gamma_{rx^3}\,\eta_+\,\,.
\label{d5system}
\eear
As ${\cal P}^2=1$, we can classify the four spinors $\eta_-$ according to their 
${\cal P}$-eigenvalue as: ${\cal P}\,\eta_-^{(\pm)}\,=\,\pm\eta_-^{(\pm)}$.
We can now solve the system (\ref{d5system}) by taking $\eta_+=0$ and 
$\eta_-$ equal to one of the two
spinors $\eta_-^{(-)}$ of negative ${\cal P}$-eigenvalue. Moreover, there are other
two solutions which correspond to taking a spinor $\eta_-^{(+)}$ of positive 
${\cal P}$-eigenvalue and a
spinor $\eta_+$  related to the former as:
\beq
\eta_{+}\,=\,{L^2\over \bar x^3}\,\Gamma_{r x^3}\,\,\eta_-^{(+)}\,\,.
\label{secondspinord5}
\eeq
Notice that, according to the first equation in (\ref{d5system}), the spinor
$\eta_+$ must have positive ${\cal P}$-eigenvalue, in agreement with eq.
(\ref{secondspinord5}). All together this configuration preserves four
supersymmetries, \ie\ one half of the supersymmetries of the background, 
as expected for a wall defect.

\subsection{The calibrating condition}
For any two-dimensional submanifold $\tilde{L}$ of $Y^{p,q}$ one can construct its
three-dimensional cone ${\cal L}\subset CY^{p,q}$.  The holomorphic $(3,0)$ form
$\Omega$ of $CY^{p,q}$ can be naturally used to calibrate such submanifolds.
Indeed,  ${\cal L}$ is called a special Lagrangian submanifold of $CY^{p,q}$ if the
pullback of $\Omega$ to ${\cal L}$ is, up to a constant phase, equal to the volume
form of ${\cal L}$, namely:
\beq
P\big[\,\Omega\,\big]_{{\cal L}}\,=\,e^{i\lambda}\,
{\rm Vol}\,({\cal L})\,\,,
\label{D5calibration}
\eeq
where $\lambda$ is constant on ${\cal L}$. If the cone ${\cal L}$ is special
Lagrangian, its base $\tilde{L}$ is said to be special Legendrian. It has been argued in 
ref. \cite{SY} that the supersymmetric configurations of a D5-brane extended along
a two-dimensional submanifold $\tilde{L}$ of a Sasaki-Einstein space are those for which
${\cal L}$ is special Lagrangian. Let us check that this is indeed the case for the
embeddings (\ref{D5sol}). First of all, we notice that the expression of $\Omega$
written in (\ref{threeform}) can be recast as:
\beq
\Omega\,=\,e^{i\psi}\,r^2\,\Omega_{4}\,\wedge
\big[\,dr+\,i\,{r\over L}\, e^5\,\big]\,\,,
\label{newOmega}
\eeq
where $\Omega_{4}$ is the two-form:
\beq
\Omega_{4}\,=\,{1\over L^2}\,\,\big(\, e^1+ie^2\,\big)\wedge
\big(\, e^3-ie^4\,\big)\,\,.
\label{Omega4}
\eeq
In eqs. (\ref{newOmega}) and (\ref{Omega4})  $e^1$, $\cdots$, $e^5$ are the
vielbein one-forms of (\ref{Ypqvielbein}). Moreover, the volume form of
${\cal L}$ can be written as:
\beq
{\rm Vol}\,({\cal L})\,=\,r^2dr\wedge {\rm Vol}\,( \tilde{L})\,\,.
\eeq
For our embeddings (\ref{D5sol}) one can check that:
\beq
{\rm Vol}\,(\tilde{L})\,=\,\,{H\over 6}\,\Big|\,{\cos\theta\over y}\,\Big|\,
\sqrt{1-cy}\,
\Bigg[\,1\,+\,(1-cy)\,{y^2\over H^2}\,\tan^2\theta\,\Bigg]\,d\theta\wedge
d\phi\,\,.
\eeq
It is now straightforward
to verify that our embeddings (\ref{D5sol}) satisfy (\ref{D5calibration}) with
$e^{i\lambda}\,=\,-i\sigma e^{i\psi}$, where $\sigma$ is the constant sign defined
in (\ref{D5sigma-sign}) (recall that in our ansatz (\ref{D5dwansatz}) the
angle $\psi$ is constant). Thus, we conclude that $\tilde{L}$ is special
Legendrian, as claimed. Moreover, one can check that:
\beq
P\big[\,J\,\big]_{{\cal L}}\,=\,0\,\,.
\eeq

\subsection{Energy bound}

Let us consider a generic embedding $y=y(\theta)$, $\beta=\beta(\phi)$ and let
us define the following functions of $\theta$ and $y$
\beq
\Delta_{\theta}\equiv -y(1-cy)\tan\theta\,\,,\qquad
\Delta_{\phi}\equiv-{1-cy\over y}\,\,\cos\theta\,\,.
\eeq
In terms of these functions the BPS equations (\ref{betaphi}) and (\ref{D5yeq})
are simply $y_{\theta}=\Delta_{\theta}$ and
$\beta_{\phi}=\Delta_{\phi}$. We have checked that any solution of this first-order
equations also solves the Euler-Lagrange equations derived from the
Dirac-Born-Infeld lagrangian (\ref{D3lag-den}). Moreover, the 
hamiltonian density 
${\cal H}=\sqrt{-g}$ satisfies a BPS bound as in (\ref{energybound}), where 
${\cal Z}$ is a total derivative. To prove this statement, let us notice that 
${\cal H}$ can be written as:
\bear
&&{\cal H}\,=\,{r^2\over 6}\,{H\over \sqrt{1-cy}}\,
\Big|\,{y\over \cos\theta}\,\Big|\,\sqrt{\Delta_{\phi}^2\,+\,(1-cy)\,
{\cos^2\theta\over y^2 H^2}\,y_{\theta}^2}\,\,\times\rc\rc
&&\qquad\times
\sqrt{(c\cos\theta\,-\,\beta_{\phi})^2\,+\,{\cos^2\theta\over H^2 y^2 (1-cy)}\,
\Delta_{\theta}^2\,+\,{2y^2\over 3H^2}\,(\beta_\phi\,-\,\Delta_\phi)^2}\,\,.
\eear
Let us now rewrite ${\cal H}$ as ${\cal H}=|{\cal Z}|+{\cal S}$, where 
\beq
{\cal Z}\,=\,{r^2\over 6}\,{H\over \sqrt{1-cy}}\,{y\over \cos\theta}\,
\Big[\,{\cos^2 \theta\over y^2 H^2}\,\Delta_{\theta}\,y_{\theta}\,-\,
(c\cos\theta\,-\,\beta_{\phi})\Delta_{\phi}\,\Big]\,\,.
\eeq
One can check that $|{\cal Z}|_{|BPS}\,=\,\sqrt{-g}_{|BPS}$. Moreover, for arbitrary
functions $y=y(\theta)$ and  $\beta=\beta(\phi)$, one can verify that
${\cal Z}$ is a total derivative, namely:
\beq
{\cal Z}\,=\,{\partial \over \partial \theta}\,{\cal Z}^{\theta}\,+\,
{\partial \over \partial \phi}\,{\cal Z}^{\phi}\,\,.
\label{D5calZs}
\eeq
In order to write the explicit expressions of ${\cal Z}^{\theta}$ and
${\cal Z}^{\phi}$, let us define the function $g(y)$ as follows:
\beq
g(y)\,\equiv \, -\int {\sqrt{1-cy}\over H(y)}\,dy\,\,.
\eeq
Then one can verify that eq. (\ref{D5calZs}) is satisfied for ${\cal Z}^{\theta}$ and
${\cal Z}^{\phi}$ given by:
\bear
{\cal Z}^{\theta}&=&{r^2\over 6}\,\sin\theta\,g(y)\,\,,\rc
{\cal Z}^{\phi}&=&{r^2\over 6}\,\Big[\,
-\cos\theta \,g(y)\,\phi\,+\,H(y)\,\sqrt{1-cy}\,\,(c\phi\cos\theta\,-\,\beta)\,\Big]\,\,.
\eear
One can prove that ${\cal H}\ge \big|\,{\cal Z}\,\big|$ is equivalent to:
\bear
&&{\cos^2\theta\over y^2 (1-cy)}\,\Big[\,
\Delta_{\phi}\,\Delta_{\theta}\,+\,(1-cy)\,(c\cos\theta\,-\,\beta_{\phi})\,
y_{\theta}\,\Big]^2\,+\rc
&&\qquad\qquad
{2y^2\over 3}\,\Big[\,\Delta_{\phi}^2\,+\,
{(1-cy)\cos^2\theta\over y^2 H^2}\,y_{\theta}^2\,\Big]\,
[\,\beta_{\phi}\,-\,\Delta_\phi\,]^2\,\ge\, 0\,\,,
\eear
which is always satisfied. 
Moreover, by using that $(c\cos\theta \,-\,\beta_{\phi})_{|BPS}\,=\,\cos\theta/y$, one can
prove that this inequality is saturated precisely when the BPS differential equations are
satisfied.

\setcounter{equation}{0}
\section {Supersymmetric D7-branes in $AdS_5\times Y^{p,q}$}
\label{d7}
\medskip

For a D7-brane the kappa symmetry matrix (\ref{gammakappa}) takes the form:
\be
\G_k = - \frac{i}{8!\sqrt{-g}}\ep^{\m_1\ldots \m_8}\g_{\m_1\ldots
  \m_8},
\label{D7general-gammak}
\ee
where, again, we have used the rules of eq. (\ref{rule}) to write the
expression of $\G_k$ acting on complex spinors. The D7-branes which fill
the four Minkowski spacetime directions and extend along some holographic
non-compact direction can be potentially used as flavour branes,
\ie\ as branes whose fluctuations can be identified with the dynamical mesons of the
gauge theory. In this section we will find a family of these configurations which
preserve four supersymmetries. In section \ref{6} we will determine another family of
supersymmetric spacetime filling configurations of D7-branes and we will also
demonstrate that there are embeddings in which the D7-brane  wraps the entire 
$Y^{p,q}$ space and preserve two supersymmetries.

\subsection{Spacetime filling D7-brane}

Let us choose a system of worldvolume coordinates motivated by the spacetime filling
character of the configuration that we are trying to find, namely:
\be
\xi=(t,x^1,x^2,x^3,y,\b,\te,\p).
\ee
The ansatz we will adopt for the embedding is:
\be
\psi=\psi(\beta,\phi), \qquad r= r(y,\te).
\label{D7ansatz}
\ee
In this case the general expression  of $\Gamma_{\kappa}$
(eq. (\ref{D7general-gammak})) reduces to:
\beq
\Gamma_{\kappa}\,=\,-i\,{r^4\over L^4\sqrt{-g}}\,
\Gamma_{x^0\cdots x^3}\,\gamma_{y\beta\theta\phi}\,.
\eeq
In order to implement the $\Gamma_{\kappa}\,\epsilon\,=\,\epsilon$ condition we
require that the spinor $\epsilon$ is an eigenvector of the matrix $\Gamma_*$
defined in eq. (\ref{gamma*}). Then, according to eq. (\ref{chiraladsspinor}),  
$\Gamma_*\epsilon=-\epsilon$, \ie\ $\epsilon$ is of the form $\epsilon_-$ and,
therefore, it satisfies:
\beq
\Gamma_{x^0\cdots x^3}\,\epsilon_-\,=\,i\epsilon_-\,\,.
\label{D3chirality}
\eeq
Moreover, 
as $\epsilon_-$ has fixed ten-dimensional chirality, the  condition
(\ref{D3chirality}) implies:
\beq
\Gamma_{r5}\epsilon_-\,=\,-i\epsilon_-\,\,.
\label{Gammar5}
\eeq
By using the projection (\ref{D3chirality}), one immediately arrives at:
\beq
\Gamma_{\kappa}\,\epsilon_-\,=\,{r^4\over L^4\sqrt{-g}}\,
\gamma_{y\beta\theta\phi}\,\epsilon_-\,\,.
\label{D7gamma-epsilon-}
\eeq
After using eqs. (\ref{epsilon-project}) and (\ref{Gammar5}),  the action of
$\gamma_{y\beta\theta\phi}$ on
$\epsilon$ can be written as:
\beq
{1\over L^4}\,\gamma_{y\beta\theta\phi}\,\epsilon_-\,=\,
\big[\,d_I\,+\,d_{15}\,\Gamma_{15}\,+\,d_{35}\,\Gamma_{35}\,+\,
d_{13}\Gamma_{13}\,\big]\,\epsilon_-\,\,,
\eeq
where the different coefficients are given by:
\bear
&&d_{I}\,=\,{1-cy\over 36}\,\sin\theta\,+\,{1-cy\over 18}\,\sin\theta\,
(y+\psi_{\beta})\,{r_y\over r}\,-\,{1\over 18}\,
\big[\,(1+c\psi_\beta)\cos\theta+\psi_\phi\,\big]\,{r_\theta\over r}\,\,,
\rc
&&d_{15}\,=\,i\,{1-cy\over 6\sqrt{6}}\,H\,\sin\theta\,
\Big[\,{r_y\over r}\,-\,{y+\psi_\beta\over 3H^2}\,\Big]\,\,,\rc
&&d_{35}\,=\,-i\,{\sqrt{1-cy}\over 6\sqrt{6}}\,\Big[\,\sin\theta\,
{r_\theta\over r}\,+\,{1\over 3}\,
\big(\,(1+c\psi_\beta)\cos\theta\,+\,\psi_{\phi}\,)\,\Big]\,\,,\rc
&&d_{13}\,=\,{{\sqrt{1-cy}\over 18}}\,H\,
\Big[\,\sin\theta\,{y+\psi_\beta\over H^2}\,{r_\theta\over r}\,+\,
\big(\,(1+c\psi_\beta)\,\cos\theta\,+\,\psi_{\phi}\big)\,
{r_y\over r}\,\Big]\,\,.
\eear
As the terms containing the matrices $\Gamma_{15}$, $\Gamma_{35}$ and 
$\Gamma_{13}$ give rise to projections which are not compatible with those 
in eq. (\ref{epsilon-project}), we have to impose that:
\beq
d_{15}\,=\,d_{35}\,=\,d_{13}\,=\,0\,\,.
\eeq
From the vanishing of $d_{15}$ and $d_{35}$ we obtain the following
first-order differential equations
\beq
r_y\,=\,\Lambda_y\,\,, \qquad\qquad
r_\theta\,=\,\Lambda_\theta\,\,,
\label{D7BPSlambda}
\eeq
where we have defined $\Lambda_y$ and $\Lambda_{\theta}$ as:
\bear
&&\Lambda_y\,=\,{r\over 3H^2}\,\big(y+\psi_\beta\big)\,,\rc
&&\Lambda_\theta\,=\,-{r\over 3\sin\theta}\,
\Big[\,(1+c\psi_\beta)\,\cos\theta\,+\,\psi_{\phi}\,\Big]\,\,.
\label{Lambdas}
\eear
Notice that the equations (\ref{D7BPSlambda}) imply that $d_{13}=0$. One can also
check that $r^4\,d_I\,=\,\sqrt{-g}$ if the first-order equations (\ref{D7BPSlambda})
hold and, therefore, one has indeed that $\Gamma_{\kappa}\epsilon_-=\epsilon_-$. 
Thus, any Killing spinor of the type $\epsilon=\epsilon_-$, with 
$\epsilon_-$ as in eq. (\ref{chiraladsspinor}), satisfies the kappa symmetry
condition if the BPS equations  (\ref{D7BPSlambda}) hold. Therefore, these
configurations preserve the four ordinary supersymmetries of the background and, as
a consequence, they are 1/8 supersymmetric.

\subsubsection{Integration of the first-order equations}

Let us now obtain the general solution of the system (\ref{D7BPSlambda}). Our first
observation is that, according to (\ref{D7ansatz}), the only dependence on the
coordinates $\beta$ and $\phi$ appearing in eqs. (\ref{D7BPSlambda}) and
(\ref{Lambdas}) comes from the derivatives of $\psi$. Therefore, for consistency
with the assumed dependence of the functions of the ansatz
(\ref{D7ansatz}),
$\psi_\phi$ and $\psi_\beta$ must be constants. Thus, let us write:
\beq
\psi_\phi\,=\,n_1\,\,,\qquad\qquad
\psi_\beta\,=\,n_2\,\,,
\eeq
which can be  trivially integrated, namely:
\beq
\psi\,=\,n_1\,\phi\,+\,n_2\,\beta\,+\,{\rm constant}\,\,.
\label{D7psi-solution}
\eeq
It is now easy to obtain the function $r(\theta, y)$. 
The equations to integrate are:
\beq
r_y\,=\,{r\over 3H^2}\,(y+n_2)\,\,,\qquad
r_\theta\,=\,-{r\over 3\sin\theta}\,
\Big[\,(1+cn_2)\cos\theta\,\,+\,n_1\,\Big]\,\,.
\label{D7-rtheta-ry}
\eeq
Let us first integrate the equation for $r_\theta$ in 
(\ref{D7-rtheta-ry}). We get:
\beq
r(y,\te)\,=\,{A(y)\over \Big[
\sin {\theta\over 2}\Big]^{{1+n_1+cn_2\over 3}}\,\,
\Big[\cos {\theta\over 2}\Big]^{{1-n_1+cn_2\over 3}}}\,\,,
\eeq
with $A(y)$ a function of $y$ to be determined. Plugging this result in the
equation for $r_y$ in (\ref{D7-rtheta-ry}), we get the following equation for $A$:
\beq
{1\over A}\,{dA\over dy}\,=\,{1\over 3}\,{y+n_2\over H^2}\,\,,
\eeq
which can be integrated immediately, namely:
\beq
A^3(y)\,=\,C\,\Big [f_1(y)\Big]^{n_2}\,f_2(y)\,\,,
\eeq
with $C$ a constant and $f_1(y)$ and $f_2(y)$ being the functions defined in
(\ref{fs}). Then, we can write $r(y,\te)$ as:
\beq
r^3(y,\te)\,=\,C{\Big [f_1(y)\Big]^{n_2}\,f_2(y)
\over \Big[
\sin {\theta\over 2}\Big]^{1+n_1+cn_2}\,\,
\Big[\cos {\theta\over 2}\Big]^{1-n_1+cn_2}}\,\,.
\label{D7r-solution}
\eeq
Several comments concerning the solution displayed in eqs.
(\ref{D7psi-solution}) and (\ref{D7r-solution}) are in order. First of
all, after a suitable change of variables it is easy to verify that for
$c=0$ one recovers from (\ref{D7psi-solution}) and (\ref{D7r-solution})
the family of D7-brane embeddings in
$AdS_5\times T^{1,1}$ found in ref. \cite{Arean:2004mm}. Secondly, the function 
$r(y,\te)$ in (\ref{D7r-solution}) always diverges for some particular
values of $\theta$ and $y$, which means that the probe always extends
infinitely in the holographic direction. Moreover, for some particular
values of $n_1$ and $n_2$ there is a minimal value of the coordinate $r$,
which depends on the integration constant $C$. This fact is important
when one tries to use these D7-brane configurations as flavour branes,
since this minimal value of $r$ provides us with an energy scale, which
is naturally identified with the mass of the dynamical quarks added to the
gauge theory. It is also interesting to obtain the form of the solution
written in eqs. (\ref{D7psi-solution}) and (\ref{D7r-solution}) 
in terms of the complex variables $z_i$ defined in (\ref{complexzs}).
After a simple calculation one can verify that this solution can be
written as a polynomial  equation of the form:
\beq
z_1^{m_1}\,z_2^{m_2}\,z_3^{m_3}\,=\,{\rm constant}\,\,,
\label{D7polynomial}
\eeq
where the $m_i$'s are constants and $m_3\not= 0$.\footnote{
It is natural to expect a condition of the form $f(z_1,z_2,z_3)=0$, where
$f$ is a general holomorphic function of its arguments. However, in order
to be able to solve the  problem analytically we started from a
restrictive ansatz (\ref{D7ansatz}) that, not surprisingly, leads to a
particular case of the expected answer.} The relation
between the $m_i$'s of (\ref{D7polynomial}) and the $n_i$'s of eqs. 
(\ref{D7psi-solution}) and (\ref{D7r-solution}) is:
\beq
n_1\,=\,{m_1\over m_3}\,\,,\qquad
n_2\,=\,{m_2\over m_3}\,\,.
\eeq
Notice that when $n_2=m_2=0$ the dependence on $\beta$ disappears and the
configuration is reminiscent of its analog in the conifold case
\cite{Arean:2004mm}. When $n_2\not= 0$ the D7-brane winds infinitely the
$\psi$-circle.

\subsection{Energy bound}

As it happened in the case of D3- and D5-branes, one can verify that any solution of the first-order equations (\ref{D7BPSlambda}) also solves the equations of motion. We are now going to check that there exists a bound
for the energy which is saturated
by the solutions of the first-order equations  (\ref{D7BPSlambda}). Indeed, let 
$r(y, \theta)$ and $\psi(\beta,\phi)$ be arbitrary functions. The hamiltonian
density ${\cal H}=\sqrt{-g}$ in this case can be written as:
\beq
{\cal H}\,=\,{r^2\over 6}\,\sin\theta\,
\sqrt{\Bigg(r_\theta^2\,+\,(1-cy)\,\Big[H^2\,r_y^2\,+\,{r^2\over 6}\Big]\Bigg)\,
\Bigg(\Lambda_\theta^2\,+\,(1-cy)\,\Big[H^2\,\Lambda_y^2\,+\,{r^2\over 6}\Big]\Bigg)}\,\,,
\eeq
where $\Lambda_y$ and $\Lambda_{\theta}$ are the functions displayed in 
eq. (\ref{Lambdas}). 
Let us rewrite this function ${\cal H}$  
as ${\cal Z}+{\cal S}$, where  ${\cal Z}$ is given
by:
\beq
{\cal Z}\,=\,{r^2\over 6}\,\sin\theta\,\Bigg[\,
r_\theta\Lambda_\theta\,+\,(1-cy)\,\Big(H^2\, r_y\,\Lambda_y\,+\,{r^2\over 6}
\Big)\Bigg] ~.
\label{D7calZ}
\eeq
One can prove that ${\cal Z}$ is a total derivative:
\beq
{\cal Z}\,=\,\partial_{\theta}\,{\cal Z}^\theta\,+\,\partial_y\,{\cal Z}^y\,\,,
\eeq
where ${\cal Z}^\theta$ and ${\cal Z}^y$ are:
\bear
&&{\cal Z}^\theta\,=\,-{r^4\over
72}\,\Big[\,\psi_\phi\,+\,(1+c\psi_\beta)\,\cos\theta\,\Big]\,\,,\rc
&&{\cal Z}^y\,=\,{r^4\over 72}\,(1-cy)\,(y+\psi_\beta)\sin\theta\,\,.
\eear
Moreover, when ${\cal Z}$ is given by (\ref{D7calZ}), one can demonstrate the bound
(\ref{energybound}). Actually, one can show that the condition 
${\cal H}\ge |{\cal Z}|$ is equivalent to the inequality:
\beq
(r_\theta-\Lambda_\theta)^2\,+\,H^2\,(1-cy)\,(r_y-\Lambda_y)^2\,+\,
{H^2\over r^2}\,(r_\theta\,\Lambda_y-r_y\,\Lambda_\theta)^2\,\ge 0\,\,,
\eeq
which is always satisfied and is saturated precisely when the BPS equations 
(\ref{D7BPSlambda}) are satisfied. Notice also that ${\cal Z}_{|BPS}$ is positive.

\setcounter{equation}{0}
\section{Other interesting possibilities}
\label{6}
\medskip

Let us now look at some other configurations of different branes and cycles not 
considered so far. We  first consider D3-branes extended along one of the
Minkowski coordinates and along a two-dimensional submanifold of
$Y^{p,q}$. These configurations represent ``fat" strings  from the point
of view of the gauge theory. We  verify in subsection \ref{fat} that
an embedding of this type breaks completely the supersymmetry, although
there exist stable non-supersymmetric ``fat" strings.  In subsection
\ref{MoreDW} we  find a new configuration of a D5-brane wrapping a
two-dimensional submanifold, whereas in subsection \ref{D5flux} we 
add  worldvolume flux to the wall defect solutions of section \ref{d5}. 
In subsection \ref{d5-3cycle} we  consider the possibility of having
D5-branes wrapping a three-cycle. We  show that such embeddings
cannot be supersymmetric, even though stable solutions of the equations
of motion with these characteristics do exist. In subsection
\ref{baryonvertex} we  analyze the baryon vertex configuration and we  verify that
such embedding breaks supersymmetry completely. In subsection 
\ref{MoreD7} we  explore the existence of spacetime filling
supersymmetric configurations of D7-branes by using a set of worldvolume
coordinates different from those used in section \ref{d7}. Finally, in
subsection \ref{D7Ypq} we  show  that a  D7-brane can wrap the whole
$Y^{p,q}$ space and preserve some fraction of supersymmetry. It can be thought of as a codimension two defect in the gauge theory dual.

\subsection{D3-branes  on a two-submanifold}
\label{fat}

Let us take a D3-brane which is extended along one of the spatial directions of the worldvolume of
the D3-branes of the background (say $x^1$) and wraps a two-dimensional cycle. The worldvolume
coordinates we will take are
\beq
\xi^{\mu}\,=\,(x^0, x^1, \theta, \phi)\,\,,
\eeq
and we will look for embeddings with $x^2$, $x^3$, $r$ and $\psi$ constant and with
\beq
y\,=\,y(\theta,\phi)\,\,,\quad\qquad
\beta\,=\,\beta(\theta,\phi)\,\,.
\eeq
In this case the kappa symmetry matrix acts on $\epsilon$ as:
\beq
\Gamma_{\kappa}\,\epsilon\,=\,-{i\over \sqrt{-g}}\,{r^2\over L^2}\,
\Gamma_{x^0x^1}\,\gamma_{\theta\phi}\,\epsilon\,\,.
\eeq
The expressions of $\gamma_{\theta}$ and $\gamma_{\phi}$ are just those calculated in section \ref{d5} and $\gamma_{\theta\phi}\,\epsilon$ can be obtained by taking the
complex conjugate of  eq. (\ref{D5gamathetaphi}):
\beq
{6\over L^2}\,\gamma_{\theta\phi}\,\epsilon\,=\,\big[\,
b_I^*\,+\,b_{15}^*\,\Gamma_{15}\,+\,b_{35}^*\,\Gamma_{35}\,+\,b_{13}^*\,\Gamma_{13}\,\big]\,
\epsilon\,\,,
\eeq
where the $b$'s are given in eq. (\ref{D5bs}). Since now the complex conjugation does not act on
the spinor $\epsilon$, the only possible projection compatible with those of the background is the
one originated from the term with the unit matrix in the previous expression. Then, we must
require:
\beq
b_{15}=b_{35}=b_{13}=0\,\,.
\eeq
The conditions $b_{15}=0$ and $b_{35}=0$ are equivalent and give rise to eqs. (\ref{D5beta_theta})
and (\ref{betaphi}), which can be integrated as in eq. (\ref{D5sol}). Moreover, the condition
$b_{13}=0$ leads to the equation:
\beq
{y\over H^2}\,y_{\theta}\,=\,\cot\theta\,\,.
\eeq
The integration of this equation can be straightforwardly performed in terms of the function
$f_2(y)$ defined in eq. (\ref{fs}) and can be written as:
\beq
{1\over \sqrt{a-3y^2+2cy^3}}\,=\,k\sin\theta\,\,,
\eeq
with $k$ being a constant of integration, which should be related to the constant $m$ in eq.
(\ref{D5sol}). However, the dependence of $y$ on $\theta$ written in the last equation does not
seem to be compatible with the one of eq. (\ref{D5sol}) (even for $c=0$). Thus, we conclude that
there is no solution for the kappa symmetry condition in this case. 

If we forget about the requirement of supersymmetry it is not difficult to find solutions of the
Euler-Lagrange equations of  motion of the D3-brane probe. Indeed, up to irrelevant global
factors, the lagrangian for the D3-brane considered here is the same as the one corresponding to a
D5-brane extended along a two-dimensional submanifold of $Y^{p,q}$. Thus, 
the embeddings written in eq. (\ref{D5sol}) are stable solutions of the
equations of motion of the D3-brane which represent  a ``fat" string from
the gauge theory point of view.

\subsection{More D5-branes wrapped on a two-cycle}
\label{MoreDW}

Let us consider a D5-brane wrapped on a two-cycle and let us choose the following set of
worldvolume coordinates: $\xi^{\mu}\,=\,(x^0,x^1,x^2,r,\theta,y)$. The embeddings we shall
consider have $x^3$ and $\psi$ constant and $\phi=\phi(\theta, y)$, $\beta=\beta(\theta, y)$. For
this case, one has:
\beq
\Gamma_{\kappa}\,\epsilon\,=\,{i\over \sqrt{-g}}\,{r^2\over L^2}\,\,
\Gamma_{x^0x^1x^2r}\,\,\gamma_{\theta y}\,\epsilon^*\,.
\eeq
where
\beq
{6\over L^2}\,\gamma_{\theta y}\,\,\epsilon^*\,=\,
\Big(\,f_I\,+\,f_{15}\Gamma_{15}\,+\,f_{35}\Gamma_{35}\,+\,
f_{13}\Gamma_{13}\,\Big)\,\epsilon^*\,\,,
\eeq
and the different coefficients are given by:
\bear
&&f_I\,=\,-i\Big(\,(1-cy)\,\sin\theta\,\phi_y\,-\,c\cos\theta\phi_\theta\,
+\,\beta_\theta\,\Big)\,\,,\rc
&&f_{15}\,=\,\sqrt{{2\over 3}}\,{1\over H}\,\Big(\,y\beta_\theta\,+\,
(1-cy)\cos\theta\phi_\theta\,\Big)\,+\,i\sqrt{{2\over 3}}\,H\,\cos\theta\,
\Big(\,\beta_y\,\phi_\theta\,-\,\beta_\theta\,\phi_y\,\Big)\,\,,\\
&&f_{35}\,=\,\sqrt{{2\over 3}}\,\sqrt{1-cy}\,\Big[ \Big(\,y\beta_y\,+\,
(1-cy)\cos\theta\phi_y\,\Big)\,-\,i\,y\,\sin\theta\,
\Big(\,\beta_y\,\phi_\theta\,-\,\beta_\theta\,\phi_y\,\Big) \Big]\,\,,\rc
&&f_{13}\,=\,\sqrt{1-cy}\, \Big[ \Big(\,{1\over H}\,+\,H\sin\theta\,
(\,\beta_y\,\phi_\theta\,-\,\beta_\theta\,\phi_y\,)\Big)
- i\,\Big( {\sin\theta\over H}\,\phi_\theta - H
(\beta_y-c\cos\theta\phi_y)\Big) \Big]\,\,.\nonumber
\eear
The BPS conditions  in this case are the following:
\beq
f_I\,=\,f_{15}\,=\,f_{35}\,=\,0\,\,.
\eeq
From the vanishing of $f_I$ we get the equation:
\beq
\beta_{\theta}\,+\,(1-cy)\sin\theta\,\phi_y\,-
\,c\cos\theta\phi_\theta\,=\,0\,\,.
\label{fIeq}
\eeq
Moreover, the vanishing of $f_{15}$ and $f_{35}$ is equivalent to the equations:
\bear
&&y\beta_{\theta}\,+\,(1-cy)\,\cos\theta\phi_\theta\,=\,0\,\,,\rc
&&y\beta_y\,+\,(1-cy)\cos\theta\,\phi_y\,=\,0\,\,,\rc
&&\beta_y\,\phi_\theta-\beta_\theta\,\phi_y\,=\,0\,\,.
\label{f5eqs}
\eear
Notice that this system of equations is redundant, \ie\ the first two equations are equivalent
if one uses the last one. 
Substituting the value of $\beta_\theta$ as given by  the first equation in 
(\ref{f5eqs}) into  (\ref{fIeq}), one can get a partial differential equation which only
involves derivatives of 
$\phi$, namely:
\beq
\cot\theta\,\phi_{\theta}\,-\,y(1-cy)\,\phi_y\,=\,0\,\,.
\label{D5phi}
\eeq
By using in (\ref{D5phi}) the last equation in (\ref{f5eqs}), one gets:
\beq
\cot\theta\,\beta_{\theta}\,-\,y(1-cy)\,\beta_y\,=\,0\,\,.
\label{D5beta}
\eeq
Eqs. (\ref{D5phi}) and (\ref{D5beta}) can be easily integrated by the method of separation of
variables. One gets
\bear
&&\phi\,=\,A\,\Bigg[{y\over (1-cy)\cos\theta}\,\Bigg]^{\alpha}\,+\,\phi^0\,\,,\rc
&&\beta\,=\,{\alpha \over 1-\alpha}\,A\,\Bigg[{y\over
(1-cy)\cos\theta}\,\Bigg]^{\alpha-1}\,+\,\beta^0\,\,,
\eear
where $A$, $\alpha$, $\phi^0$ and $\beta^0$  are constants of integration and we have used 
eq. (\ref{f5eqs}) to relate the integration constants of $\phi$ and $\beta$. However, in order
to implement the condition $\Gamma_{\kappa}\,\epsilon=\epsilon$, one must require 
the vanishing of the
imaginary part of $f_{13}$. This only happens if $\phi$ and $\beta$ are constant,
\ie\ when $A=0$ in the above solution. One can check that this configuration satisfies the
equations of motion.

\subsection{D5-branes  on a two-submanifold with flux}
\label{D5flux}

We now analyze the effect of adding flux of the worldvolume gauge field $F$
to the configurations of section \ref{d5} \footnote{A nice discussion of supersymmetric
configurations with nonzero gauge field strengths by means of kappa symmetry
can be found in ref. \cite{mmms}.}. Notice that  we now have a non-zero contribution from the Wess-Zumino term of the action, which is of the form\footnote{In this subsection we will rescale the gauge field $F$ given in eqs. (\ref{actionintrod}) and (\ref{kappa1}) by a factor of $2\pi \alpha'$. }:
\beq
{\cal L}_{WZ}\,=\,P[\,C^{(4)}\,]\wedge F\,\,.
\eeq
Let us suppose that we switch on a worldvolume gauge field along the angular directions
$(\theta,\phi)$. We will adopt the ansatz:
\beq
F_{\theta\phi}\,=\,q\,K(\theta,\phi)\,\,,
\label{wvflux}
\eeq
where $q$ is a constant and $K(\theta,\phi)$ a function to be determined. The relevant components
of $P[\,C^{(4)}\,]$ are
\beq
P[\,C^{(4)}\,]_{x^0x^1x^2r}\,=\,h^{-1}\,{\partial x^3\over \partial r}\,\,,
\eeq
where $h=L^4/r^4$.
It is clear from the above expression of ${\cal L}_{WZ}$ that a nonvanishing value of  $q$
induces a dependence of $x^3$ on $r$. In what follows we will assume that $x^3=x^3(r)$, \ie\ that 
$x^3$ only depends on $r$. Let us assume that the angular embedding satisfies the same equations 
as in the case of zero flux. The Lagrangian density in this case is given by:
\beq
{\cal L}\,=\,-h^{-{1\over 2}}\,\sqrt{1+h^{-1}\,(x')^2}\,
\sqrt{g_{\theta\theta}g_{\phi\phi}\,+\,q^2\,K^2}\,+\,q\,h^{-1}x'K\,\,,
\eeq
where $g_{\theta\theta}$ and $g_{\phi\phi}$ are elements of the induced metric, we have denoted
$x^3$ simply by $x$ and the prime denotes derivative with respect to $r$. The equation of motion
of $x$ is:
\beq
-{\sqrt{g_{\theta\theta}g_{\phi\phi}\,+\,q^2\,K^2}\over \sqrt{1+h^{-1}\,(x')^2}}\,
h^{-{3\over 2}}\,x'\,+\,q\,h^{-1}\,K\,=\,{\rm constant}\,\,.
\eeq
Taking the constant on the right-hand side of the above equation equal to zero, we get the
following solution for $x'$:
\beq
x'(r)\,=\,q\,h^{{1\over 2}}\,
{K(\theta,\phi)\over \sqrt{g_{\theta\theta}g_{\phi\phi}}}\,\,.
\eeq
Notice that the left-hand side of the above equation depends only on $r$, whereas the right-hand
side can depend on the angles $(\theta,\phi)$. For consistency the dependence of 
$K(\theta,\phi)$ and $\sqrt{g_{\theta\theta}g_{\phi\phi}}$ on $(\theta,\phi)$ must be the same.
Without lost of generality let us take $K(\theta,\phi)$ to be:
\beq
L^2\,K(\theta,\phi)\,=\,\sqrt{g_{\theta\theta}g_{\phi\phi}}\,\,,
\eeq
where the factor $L^2$ has been introduced for convenience. Using this form of $K$, the
differential equation which determines the dependence of $x^3$ on $r$
becomes:
\beq
x'(r)\,=\,{q\over r^2}\,\,,
\eeq
which can be immediately integrated, namely:
\beq
x(r)\,=\,\bar x^3\,-\,{q\over r}\,\,.
\label{x(r)}
\eeq
Moreover, the expression of $K$ can be obtained by computing the induced metric along the angular
directions.  It takes the form:
\beq
K(\theta)\,=\,\sigma\,{\sqrt{1-cy}\over 6 H(y)}\,\Big[\,H^2(y)\,+\,(1-cy)y^2\,\tan^2\theta\,\Big]\,
{\cos\theta\over y}\,\,,
\eeq
where $y=y(\theta)$ is the function obtained in section \ref{d5} and
$\sigma\,=\,{\rm sign} \Big(\cos\theta/y\Big)$. Actually, notice that $K$ only depends on the angle
$\theta$ and it is independent of $\phi$.

We are now going to verify that the configuration just found is
supersymmetric. The expression of $\Gamma_{\kappa}$ in this case has an additional term due to the
worldvolume gauge field (see eq. (\ref{kappa1})). Actually, it is straightforward to check that in the present case
\beq
\Gamma_{\kappa}\,\epsilon\,=\,{i\over 
\sqrt{-\det (g+F)}}\,\,{r^3\over L^3}\,\Gamma_{x^0x^1x^2}\,
\Bigg[\,\gamma_r\,\gamma_{\theta\phi}\,\epsilon^*\,
-\,\gamma_r\,F_{\theta\phi}\,\epsilon\,\Bigg]\,\,.
\eeq
Notice that $\gamma_r$ is given by:
\beq
\gamma_r\,=\,{L\over r}\,\big(\,\Gamma_r\,+\,{r^2\over L^2}\,x'\,
\Gamma_{x^3}\,\big)\,\,.
\eeq
For the angular embeddings we are considering it is easy to prove from the results of section \ref{d5}
that:
\beq
\gamma_{\theta\phi}\,\epsilon^*\,=\,-i \sigma L^2 K(\theta)\, 
\Gamma_{13}\,\epsilon^*\,\,.
\eeq
By using this result and the value of $F_{\theta\phi}$ (eq. (\ref{wvflux})), one easily verifies
that:
\beq
\Gamma_{\kappa}\,\epsilon\,=\,-{i\over 1+{q^2\over L^4}}\,\,
\Gamma_{x^0x^1x^2 r}\Big[\,
i\sigma\Gamma_{13}\,\epsilon^*\,+\,
{q\over L^2}\, i\sigma\,\Gamma_{r x^3}\,\Gamma_{13}\,\epsilon^* +
{q\over L^2}\,\epsilon\,+\,{q^2\over L^4}\,\Gamma_{rx^3}\,\epsilon\,\Big]\,\,.
\eeq
By using the explicit dependence of $x$ on $r$ (eq. (\ref{x(r)})), one can write the Killing
spinor $\epsilon$ evaluated on the worldvolume as:
\beq
e^{{i\over 2}\psi}\,\epsilon\,=\,r^{-{1\over 2}}\,\Big(\,1\,-\,{ q\over L^2}\,\Gamma_{rx^3}\Big)
\eta_+\,+\,
r^{{1\over 2}}\,\Big(\,{\bar x^3\over L^2}\,
\Gamma_{rx^3}\,\eta_+\,+\,\eta_-\,\Big)\,+
\,{r^{{1\over 2}}\over L^2}\,x^p\,\Gamma_{rx^p}\,\eta_+\,\,,
\label{epsilonflux}
\eeq
where the constant spinors $\eta_{\pm}$ are the ones defined in eq. (\ref{etamasmenos}).
Remarkably, one finds that the condition $\Gamma_{\kappa}\epsilon=\epsilon$ is verified if
$\eta_{+}$ and $\eta_{-}$ satisfy the same system (\ref{d5system}) as is the case of zero
flux. 

\subsection{D5-branes wrapped on a three-cycle}
\label{d5-3cycle}

We will now try to find supersymmetric configurations of D5-branes wrapping a three cycle of the 
$Y^{p,q}$ space. Let us choose the following set of worldvolume coordinates 
$\xi^{\mu}=(x^0, x^1, x^2, y, \beta, \psi)$  and consider an embedding with $x^3$ and $r$ constant, 
 $\theta=\theta(y,\beta)$ and $\phi=\phi(y,\beta)$. In this case:
\beq
\Gamma_{\kappa}\,\epsilon\,=\,{i\over \sqrt{-g}}\,\,
{r^3\over L^3}\,\,\Gamma_{x^0x^1x^2}\,\gamma_{y\beta\psi}\,
\epsilon^*\,\,.
\eeq
The value of $\gamma_{y\beta\psi}\,\epsilon^*$ can be obtained by taking the complex conjugate 
of eq. (\ref{HEKcs}). As $c_1=c_3=0$ when $\theta_{\psi}=\phi_{\psi}=0$, we can write:
\beq
{i\over L^3}\,\gamma_{y\beta\psi}\,\epsilon^*\,=\,
\big[\,c_5^*\,\Gamma_5\,+\,
c_{135}^*\,\Gamma_{135}\,\big]\,\epsilon^*\,\,.
\eeq
The only possible BPS condition compatible with the projections satisfied by $\epsilon$ is
$c_5=0$, which leads to a projection of the type 
\beq
\Gamma_{x^0x^1x^2}\Gamma_{135}\,\epsilon^*\,=\,\lambda \epsilon\,\,,
\eeq
where $\lambda$ is a phase.
Notice that, however, as the spinor $\epsilon$ contains a factor $e^{-{i\over 2}\psi}$, the
two sides of the above  equation depend differently on $\psi$ due to the complex
conjugation appearing on the left-hand side ($\lambda$ does not depend on $\psi$). Thus,
these configurations cannot be supersymmetric. We could try to use another set of
worldvolume coordinates, in particular one which does not include $\psi$. After some
calculation one can check that there is no consistent solution. 

For the ansatz considered above the lagrangian density of the D5-brane is, up to irrelevant
factors, the same as the one obtained in subsection \ref{hekconst} 
 for a D3-brane wrapping a  three-dimensional submanifold of $Y^{p,q}$. Therefore
any solution of the  first-order equations (\ref{HEKBPS}) gives rise to an embedding
of a D5-brane which  solves the equations of motion and saturates an energy bound. This
last fact implies that the D5-brane configuration is stable, in spite of
the fact that it is not supersymmetric.  

\subsection{The baryon vertex}
\label{baryonvertex}

If a D5-brane wraps the whole $Y^{p,q}$ space, the flux of the Ramond-Ramond five form 
$F^{(5)}$ that it captures acts as a source for the electric worldvolume gauge field which, in
turn, gives rise  to a bundle of fundamental strings emanating from the D5-brane. This is
the basic argument of  Witten's construction of the baryon vertex
\cite{ba0}, which we will explore in detail now. In this case the probe
action  must include the worldvolume gauge field $F$ in both the
Born-Infeld and Wess-Zumino terms. It takes the form\footnote{In this subsection we will rescale again the gauge field $F$ given in eqs. (\ref{actionintrod}) and (\ref{kappa1}) by a factor of $2\pi \alpha'$. }:
\beq
S\,=\,-T_5\,\int d^6\xi\,\sqrt{-\det (g+F)}\,+\,
T_5\int d^6\xi \,\,\,A\wedge F^{(5)}\,\,,
\label{baryonaction}
\eeq
where $T_5$ is the tension of the D5-brane and $A$ is the one-form potential for $F$ ($F=dA$). In order to analyze the contribution of the Wess-Zumino term in
(\ref{baryonaction})  let us rewrite the standard expression of $F^{(5)}$ as:
\beq
F^{(5)}\,=\,{L^4\over 27}\,\,(1-cy)\,\sin\theta\,
dy\wedge d\beta\wedge d\theta\wedge d\phi \wedge d\psi\,+\,
{\rm Hodge}\,\,\, {\rm dual}\,\,,
\label{baryonF5}
\eeq
where, for simplicity we are taking the string coupling constant $g_s$ equal to one. Let us
also choose the following set of worldvolume coordinates:
\beq
\xi^{\mu}\,=\,(x^0,y,\beta,\theta,\phi,\psi)\,\,.
\label{baryon-coordinates}
\eeq
It is clear from the expressions of $F^{(5)}$ in (\ref{baryonF5}) and  of the
Wess-Zumino term in (\ref{baryonaction}) that, for consistency, we must turn on the time
component of the field $A$. Actually, we will adopt the following ansatz:
\beq
r\,=\,r(y)\,\,,\qquad\qquad
A_0\,=\,A_0(y)\,\,.
\label{baryon-ansatz}
\eeq
The action (\ref{baryonaction}) for such a configuration can be written as:
\beq
S\,=\,{T_5 L^4\over 108}\,\,V_4\,\,\int dx^0 dy\,\,
{\cal L}_{eff}\,\,,
\label{Leff}
\eeq
where the volume $V_4$ is :
\beq
V_4\,=\,6\int d\alpha\, d\psi\, d\phi\,
d\theta\sin\theta\,=\,96\pi^3\,\ell\,,
\eeq
and the effective lagrangian density ${\cal L}_{eff}$  is given by:
\beq
{\cal L}_{eff}\,=\,(1-cy)\,\Bigg[\,-H\,
\sqrt{{r^2\over H^2}\,+\,6\,(r')^2\,-\,6\,(F_{x^0 y})^2}\,+\,4A_0\,\Bigg]\,\,.
\eeq
Notice that, for our ansatz (\ref{baryon-ansatz}), the electric field 
is $F_{x^0 y}=-\partial_y A_0$. 
Let us now introduce the displacement field, defined as:
\beq
D(y) \equiv {\partial {\cal L}_{eff}\over \partial F_{x^0y}}\,=\,
{6(1-cy)HF_{x^0 y}\over 
\sqrt{{r^2\over H^2}\,+\,6\,(r')^2\,-\,6\,(F_{x^0 y})^2}}\,\,.
\label{D}
\eeq
From the equations of motion of the system it is straightforward to determine $D(y)$.
Indeed, the variation of $S$ with respect to $A_0$ gives rise to the Gauss' law:
\beq
{dD(y)\over dy}\,=\,-4(1-cy)\,\,,
\eeq
which can be immediately integrated, namely:
\beq
D(y)\,=\,-4\bigg(\,y-{cy^2\over 2}\,\bigg)\,+\,{\rm constant}\,\,.
\label{D(y)}
\eeq
By performing a Legendre transform in (\ref{Leff}) we can obtain
the energy of the configuration:
\beq
E\,=\,{T_5 L^4\over 108}\,\,V_4\,\,\int  dy\,\,
{\cal H}\,\,,
\eeq
where ${\cal H}$ is given by:
\beq
{\cal H}\,=\,
(1-cy)\,H\,\sqrt{{r^2\over H^2}\,+\,6\,(r')^2\,-\,6\,(F_{x^0 y})^2}\,+\,
D(y)\,F_{x^0 y}\,\,.
\eeq
Moreover, the relation (\ref{D}) between $D(y)$ and $F_{x^0 y}$ can be inverted, with the
result:
\beq
F_{x^0 y}\,=\,{1\over 6}\,\,
{\sqrt{{r^2\over H^2}\,+\,6\,(r')^2}\over
\sqrt{{D^2\over 6}\,+\,(1-cy)^2\,H^2}}
\,\,D\,\,.
\label{invertedD}
\eeq
Using the relation (\ref{invertedD}) we can rewrite ${\cal H}$ as:
\beq
{\cal H}\,=\,\sqrt{{D^2\over 6}\,+\,(1-cy)^2\,H^2}\,\,\,
\sqrt{{r^2\over H^2}\,+\,6\,(r')^2}\,\,,
\eeq
where $D(y)$ is the function of the $y$ coordinate displayed in (\ref{D(y)}). 
The Euler-Lagrange equation derived from ${\cal H}$ is a second-order differential equation
for the function $r(y)$. This equation is rather involved and we will not
attempt to solve it
here. In a supersymmetric configuration one expects that there exists  a first-order
differential equation for $r(y)$ whose solution also solves the equations
of motion. This first-order equation has been found in
ref. \cite{severalbaryon} for the $AdS_5\times S^5$ background. We have
not been able to find such first-order equation in this 
$AdS_5\times Y^{p,q}$ case. A similar negative result was obtained in \cite{Arean:2004mm} for the 
$AdS_5\times T^{1,1}$ background. This result is an indication that this baryon vertex
configuration is not supersymmetric. Let us check explicitly  this fact by analyzing the
kappa symmetry condition. In our case
$\Gamma_{\kappa}\,\epsilon$ reduces to:
\beq
\Gamma_{\kappa}\,\epsilon\,=\,-{i\over \sqrt{-\det (\,g+F\,)}}\,\,\Bigg[\,
{r\over L}\,\Gamma_{x^0}\,\gamma_{y\beta\theta\phi\psi}\,\,\epsilon^*\,-\,
F_{x^0 y}\,\gamma_{\beta\theta\phi\psi}\,\,\epsilon\,\Bigg]\,\,.
\label{baryonkappa}
\eeq
The two terms on the right-hand side of (\ref{baryonkappa}) containing the antisymmetrised
products of gamma matrices can be written as:
\bear
&&\gamma_{y\beta\theta\phi\psi}\,\,\epsilon^*\,=\,
{L^5\over 108}\,(1-cy)\sin\theta\,\Big(\,\Gamma_{5}\,-\,
\sqrt{6}\,H\,{r'\over r}\,\Gamma_{r15}\,\Big)\,\epsilon^*\,\,,\rc
&&\gamma_{\beta\theta\phi\psi}\,\,\epsilon\,=\,
-{L^4\over 18\sqrt{6}}\,(1-cy)\,H\,\sin\theta\,\Gamma_{15}\,\epsilon\,\,.
\eear
By using this result, we can write $\Gamma_{\kappa}\,\epsilon$ as:
\beq
\Gamma_{\kappa}\,\epsilon\,=\,
-{i\,L^4\,(1-cy) \over \sqrt{-\det (\,g+F\,)}}\sin\theta\Bigg[
{r\over 108}\,\Gamma_{x^0}\Gamma_5\,\epsilon^*+
{H\over 18\sqrt{6}}\,\Big(\,F_{x^0y}\,\Gamma_{15}\,\epsilon\,-\,
r'\,\Gamma_{x^0r15}\,\epsilon^*\,\Big)\Bigg]\,\,.
\label{baryongammakappa}
\eeq
In order to solve the $\Gamma_{\kappa}\,\epsilon\,=\,\epsilon$ equation we
shall impose, as in ref. \cite{Susybaryon}, an extra projection such that
the contributions of the worldvolume gauge field  $F_{x^0y}$ and of $r'$
in (\ref{baryongammakappa}) cancel  each other.  This can be achieved
by imposing that $\Gamma_{x^0r}\,\epsilon^*\,=\,\epsilon$  and that
$F_{x^0y}=r'$. Notice that the condition $\Gamma_{x^0r}\,\epsilon^*\,=\,\epsilon$
corresponds to having fundamental strings in the radial direction, as expected for a baryon
vertex configuration. Moreover, as the spinor $\epsilon$ has fixed ten-dimensional
chirality, this extra projection implies that 
$i\Gamma_{x^0}\Gamma_5\,\epsilon^*\,=\,-\epsilon$ which, in turn, is needed to satisfy the 
$\Gamma_{\kappa}\,\epsilon\,=\,\epsilon$ equation. However, the condition 
$\Gamma_{x^0r}\,\epsilon^*\,=\,\epsilon$  is incompatible with the conditions
(\ref{epsilon-project}) and, then, it cannot be imposed on the Killing
spinors. Thus, as in the analysis of \cite{Arean:2004mm}, we conclude from this
incompatibility argument (which is more general than the particular
ansatz we are adopting here) that the baryon vertex configuration breaks
completely the supersymmetry of the
$AdS_5\times Y^{p,q}$ background. 

\subsection{More spacetime filling D7-branes}
\label{MoreD7}

Let us adopt $\xi^{\mu}=(x^0, x^1, x^2,x^3,  y, \beta, \psi, r)$ as our set of worldvolume
coordinates for a D7-brane probe and let us consider a configuration with
$\theta=\theta(y,\beta)$ and $\phi=\phi(y,\beta)$.  In this case:
\beq
\Gamma_{\kappa}\,=\,-{i\over \sqrt{-g}}\,\,
{r^4\over L^4}\,\,\Gamma_{x^0x^1x^2x^3}\,\gamma_{y\beta\psi r}\,\,.
\eeq
Let us take $\epsilon=\epsilon_-$, where $\Gamma_*\epsilon_-=-\epsilon_-$(see eq.
(\ref{chiraladsspinor})). 
As $\gamma_r={L\over r}\,\Gamma_r$, we can write:
\beq
{r\over L^4}\,\,
\gamma_{y\beta\psi r}\,\epsilon_-\,=\,-\big[c_5\,+\,c_{135}\,\Gamma_{13}\,\big]\,
\epsilon_-\,\,,
\eeq
where the coefficients $c_5$ and $c_{135}$ are exactly those written in eq.
(\ref{expressionHEKcs}) for the D (doublet) three-cycles. The BPS
condition is just $c_{135}=0$, which leads to  the system of differential
equations (\ref{HEKBPS}). Thus, in this case the D7-brane extends 
infinitely in the radial direction and wraps a three-dimensional
submanifold of the 
$Y^{p,q}$ space  of the
type studied in subsection \ref{hekconst}. These embeddings preserve four supersymmetries.

\subsection{D7-branes wrapped on $Y^{p,q}$}
\label{D7Ypq}

Let us take a D7-brane which wraps the entire $Y^{p,q}$ space and is extended along two 
spatial directions. The set of worldvolume coordinates we will use in this case are
$\xi^{\mu}\,=\,(x^0,x^1,r,\theta,\phi,y,\beta,\psi)$ and we will assume that $x^2$ and $x^3$ are
constant. The matrix $\Gamma_{\kappa}$ in this case is:
\beq
\Gamma_{\kappa}\,=\,-{i\over \sqrt{-g}}\,\,
\gamma_{x^0x^1r\theta\phi y\beta\psi}\,\,.
\eeq
Acting on a spinor $\epsilon$ of the background one can prove that
\beq
\Gamma_{\kappa}\,\epsilon\,=\,i\Gamma_{x^0x^1r 5}\,\epsilon\,\,,
\eeq
which can be solved by a spinor $\epsilon_-=r^{{1\over 2}}\,\,e^{-{i\over 2}\psi}\,\eta_-$, with
 $\eta_-$ satisfying the additional projection $\Gamma_{x^0x^1r
5}\,\eta_-\,=\, -i\eta_-$. Thus this configuration preserves two
supersymmetries.

\setcounter{equation}{0}
\section{Summary and Discussion}
\label{conclusions}
\medskip

Let us briefly summarise the results of this chapter. Using
kappa symmetry as the central tool, we have  systematically studied
supersymmetric embeddings of branes in the $AdS_5\times Y^{p,q}$
geometry.  Our  study focused on three kinds of branes D3, D5 and D7.

{\it D3-branes: }This is the case that we studied most exhaustively. For 
D3-branes wrapping three-cycles in $Y^{p,q}$ we first reproduced 
all the results present in the literature. In particular, using
kappa symmetry, we obtained two kinds of  supersymmetric cycles:
localised at $y_1$ and $y_2$  \cite{ms} and  localised in the  round
$S^2$ \cite{BeFrHaMaSp,hek}. For these branes we found perfect agreement
with the field theory results.  Moreover, we also found a new class of
supersymmetric embeddings of D3-branes in this background. They do not
correspond to dibaryonic operators since the D3-brane does not wrap a
three-cycle.  The field theory interpretation of these new embeddings is
not completely clear to us due to various issues with global properties.
We believe that they might be a good starting point  to find candidates
for representatives of the integer part of the third homology group of
$Y^{p,q}$, just like the analogous family of cycles found in
\cite{ba2,Arean:2004mm} were representative of the integer  part
$H_3(T^{1,1}, \zet)$. It would be important to understand these
wrapped D3-branes in terms of algebraic geometry as well as in terms of  
operators in the field theory dual, following the framework of
ref. \cite{beasley} which, in the case of the conifold, emphasizes the use
of global homogeneous coordinates. It is worth stressing that such global 
homogeneous coordinates exist in any toric variety \cite{Cox} but the
relation to the field theory operators is less clear in $CY^{p,q}$. 
We analyzed the spectrum of excitations of a wrapped  D3-brane
describing an $SU(2)$-charged dibaryon and found perfect agreement with
the field theory expectations.  We considered other embeddings and found
that  a D3-brane wrapping a two-cycle in $Y^{p,q}$ is not a supersymmetric
state but, nevertheless, it is stable. In the field theory this 
configuration describes a ``fat" string.

{\it D5-branes:}  The embedding that we paid the most attention to is a
D5-brane extended along a two-dimensional submanifold in $Y^{p,q}$ and 
having codimension one in $AdS_5$. In the field theory this is the kind
of brane that represents a wall defect. When we allow the D5-brane to extend
infinitely in the holographic direction, we get a configuration
dual to a defect conformal field theory.  We showed explicitly that such
configuration  preserves four supersymmetries and saturates the expected
energy bound. For this configuration we also considered  turning on a
worldvolume flux and found that it can be done in a supersymmetric way.
The flux in the  worldvolume of the brane provides a bending  of the
profile of the wall, analogously to what happens in $AdS_5\times T^{1,1}$ \cite{Arean:2004mm}. We  showed the consistency of similar embeddings in which the D5-brane wraps a different two-dimensional
submanifold in $Y^{p,q}$. We also considered D5-branes wrapping
three-cycles. This configuration looks like a domain wall in the
field theory dual and, although it  cannot be supersymmetric,  it is stable. Finally, we
considered a D5-brane wrapping the whole $Y^{p,q}$, which corresponds  to  the baryon vertex. We verified that, as in the case of $T^{1,1}$, it is not a supersymmetric configuration.

{\it D7-branes:} With the aim of  introducing mesons in the
corresponding field theory,  we  considered spacetime filling D7-branes.
We explicitly showed that such configurations preserve  four
supersymmetries and found  the precise embedding in terms of the radial
coordinate. We  found an interpretation of the embedding equation in
terms  of complex coordinates.  We also analyzed other spacetime filling
D7-brane embeddings. Finally, we considered a D7-brane that  wraps
$Y^{p,q}$ and is codimension two in $AdS_5$.  This configuration looks,
from the field theory point of view, as a string (one-dimensional defect) and preserves two
supersymmetries. 

Part of our analysis of some branes could be made more precise. In particular, it would be interesting 
to understand the new family of supersymmetric embeddings of D3-branes
in terms of algebraic geometry as well  as in terms of
operators in the field theory. We did not present an analysis of the
spectrum of  excitations for all of the branes. In particular, we would
like to understand the excitations of the spacetime filling  D7-branes
and  the baryon vertex better. However, the study of the excitations of the flavour branes will turn out to be more relevant in the background that we will display in chapter \ref{KW}, where their backreaction is taken into account.

In the next chapter we will present a similar systematic study of supersymmetric embeddings of branes in the $AdS_5\times \Labc$ geometry, using again kappa symmetry as the central tool.


\chapter{Supersymmetric Branes on ${\bf AdS_5 \times
\Labc}$}
\label{Labc}

\medskip

\setcounter{equation}{0}
\medskip


In this chapter, we aim at exhausting the study of D-brane probes at the tip of
toric Calabi-Yau cones on a base with topology $S^2 \times S^3$ initiated in
\cite{Arean:2004mm}. Thus, we perform the same analysis as in the previous chapter but now in $\Labc$ theories. We will skip over many details\footnote{One should be careful with not getting confused with the notation, which is similar to that used in chapter \ref{Ypq}.} since the analysis is pretty closed to the one in chapter \ref{Ypq}. Furthermore, geometrical aspects of the $\Labc$ manifold as well as the type IIB supergravity background $AdS_5 \times \Labc$ and its field theory interpretation were reviewed in subsection \ref{Labcintroduc}.

In order to determine the supersymmetric embeddings of D-brane probes we
employ again kappa symmetry, as explained in subsection \ref{kappasymmetry}. This condition gives rise to a set of first-order BPS differential equations whose solutions determine the details of the embedding. As well, they solve the equations of motion derived from the DBI action of the probe while saturating a bound for the energy \cite{GGT}.

\setcounter{equation}{0}
\section{D3-branes on three-cycles}
\label{D3s}
\medskip

In this section we consider D3-brane probes wrapping three-cycles of $\Labc$.
These are pointlike objects from the gauge theory point of view, corresponding
to dibaryons constructed from the different bifundamental fields of the quiver
theory. There are other configurations of physical interest that we will not
discuss in this chapter. Though, we will briefly discuss their most salient
features in section \ref{FinalRemarks}.

Given a D3-brane probe wrapping a supersymmetric three-cycle $\CC$, the conformal dimension $\Delta$ of the corresponding dual operator is
proportional to the volume of the wrapped three-cycle, as we already used in the previous chapter:
\beq
\Delta\,=\,{\pi\over 2}\,{N_c\over L^3}\,\,
{{\rm Vol}({\cal C})\over {\rm Vol}({ \Labc})}\,\,.
\eeq
Since the $R$-charge of a protected operator is related to its dimension
by $R={2\over 3}\Delta$, we can readily compute the $R$-charge of the dibaryon
operators. We also used in the previous chapter that the baryon number associated to the D3-brane probe wrapping $\CC$ (in units of $N_c$) can be obtained as the integral over
the cycle of the pullback of a $(2,1)$-form $\Omega_{2,1}$:
\beq
{\cal B}({\cal C})\,=\,\pm i \int_{{\cal C}}\,
P\big[\,\Omega_{2,1}\,\big]_{{\cal C}}\,\,.
\label{baryon-def}
\eeq
The explicit form of $\Omega_{2,1}$ is:
\beq
\Omega_{2,1}\,=\,{K\over  \tilde{\rho}^4}\,\Big(\,{dr\over r}\,+\,i\,
e^5\,\Big) \wedge \left(\,e^1\wedge e^2\,-\,e^3\,\wedge e^4\,\right)\,\,,
\label{Odosuno}
\eeq
where $K$ is a constant that will be determine below and we are using the frame displayed in (\ref{Labcframe}). Armed with these
expressions, we can extract the relevant gauge theory information of the
configurations under study.

\subsection{$U_1$ dibaryons}
\label{U1}

Let us take the worldvolume coordinates (defined in subsection \ref{Labcintroduc}) for the D3-brane probe to be
$\xi^{\mu}\,=\,(T,x,\psi,\tilde{\tau})$, with $\theta\,=\,\theta_0$ and
$\phi\,=\,\phi_0$ constant, and let us assume that the brane is located
at a fixed point in $AdS_5$ (\ref{globalADS}). The action of the kappa symmetry matrix (\ref{Gammad3}) on
the Killing spinor reads
\beq
\Gamma_{\kappa}\,\epsilon\,=\,-{i \over 4!\sqrt{ -\det g}}\,
\epsilon^{\mu_1\cdots\mu_4}\,\gamma_{\mu_1\cdots\mu_4}\,\epsilon\,=\,
-{iL^4 \over \sqrt{ -\det g}}\,
\big[\,a_5\Gamma_{T5}\,+\,a_{135}\,\Gamma_{T135}\,\big]\,\epsilon\,\,,
\label{U1-antisymmetrised}
\eeq
where
\beq
a_5\,=\,-i{\cosh\varrho\over 2\beta}\,\cos^2\theta\,\,,\qquad
a_{135}\,=\,-{\cosh\varrho\over 4 \sqrt{\Delta_x}}\,\,
\bigg(1-{x\over \beta}\bigg)\,\,\sqrt{\Delta_\theta}\,\sin (2\theta)\,\,.
\eeq
Compatibility of (\ref{U1-antisymmetrised}) with the projections
(\ref{etaspinor1}) demands $a_{135}\,=\,0$.
Since $\Delta_\theta$ cannot vanish for positive $\alpha$ and $\beta$, this
condition implies $\sin(2\theta)\,=\,0$, \ie\, $\theta=0$ or $\pi/2$. Due to the
fact that, for these configurations, the determinant of the induced metric is:
\beq
-\det g\,=\,{L^8\over 4}\,\Bigg[\,
{\Delta_\theta\sin^2(2\theta)\over 4\Delta_x}\,
\bigg(1-{x\over \beta}\bigg)^2\,+\,{\cos^4\theta\over \beta^2}\,\Bigg]\,
\cosh^2\varrho\,\,,
\eeq
we must discard the $\theta=\pi/2$ solution since the volume of the cycle
would vanish in that case. Thus, the D3-brane probe is placed at
$\theta=0$ and the kappa symmetry condition
$\Gamma_{\kappa}\,\epsilon\,=\,\epsilon$ reduces to the new projection:
\beq
\Gamma_{T 5}\,\epsilon\,=-\,\epsilon\,\,,
\eeq
which can only be imposed at the center of $AdS_5$. The corresponding
configuration preserves four supersymmetries.

Given that the volume of ${\cal U}_1$ can be
easily computed with the result
\beq
{\rm Vol}(\,{\cal U}_1\,)\,=\,{\pi L^3\over \beta}\,
(x_2-x_1)\,{\Delta \tilde{\tau}\over k}\,\,,
\eeq
the corresponding value for the $R$-charge is:
\beq
R_{{\cal U}_1}\,=\,{2\over 3}\,{\alpha\over \alpha+\beta-x_1-x_2}\,N_c\,=\,
{2\alpha\over 3x_3}\,N_c\,\,,
\eeq
where we have used the second relation in (\ref{xs}). This result agrees with
the value expected for the operator $\det (U_1)$ \cite{FrHaMaSpVeWe}.
Let us now compute the baryon number associated to the D3-brane probe
wrapping ${\cal U}_1$. For the ${\cal U}_1$ cycle, we get
\beq
{\cal B}({\cal U}_1)\,=\,i\int_{{\cal U}_1}\,
P\big[\,\Omega_{2,1}\,\big]_{{\cal U}_1}\,=\,
-\,{2\pi^2\over \alpha\beta}\,{c\over a\,b}\,K\,\,,
\eeq
where we have used the second identity in (\ref{ratios2}). 
From the field theory analysis \cite{FrHaMaSpVeWe} it is known that the baryon
number of the $U_1$ field should be $-c$ (see Table \ref{charges1} in chapter \ref{introduction}). We
can use this result to fix the constant $K$ to:
\beq
K\,=\,-\,{\alpha\beta\over 2\pi^2}\,a\,b\,\,.
\label{Kconstant}
\eeq
Once it is fixed, formulas (\ref{baryon-def}) and (\ref{Odosuno}) allow us to
compute the baryon number of any D3-brane probe wrapping a three-cycle.

\subsection{$U_2$ dibaryons}


Let us again locate the D3-brane probe at a fixed point in $AdS_5$ and take the
following set of worldvolume coordinates $\xi^{\mu}\,=\,(T,x,\phi,\tilde{\tau})$,
with constant $\theta\,=\,\theta_0$ and $\psi\,=\,\psi_0$. The kappa symmetry matrix now acts on the Killing spinor as
\beq
\Gamma_{\kappa} \epsilon\,=\,-{iL^4 \over \sqrt{ -\det g}}\,
\big[\,b_5\Gamma_{T5}\,+\,b_{135}\,\Gamma_{T135}\,\big]\,\epsilon\,\,,
\eeq
where
\beq
b_5\,=\,-i{\cosh\varrho\over 2\alpha}\,\sin^2\theta\,\,,\qquad
b_{135}\,=\,{\cosh\varrho\over 4 \sqrt{\Delta_x}}\,\,
\bigg(1-{x\over \alpha}\bigg)\,\,\sqrt{\Delta_\theta}\,\sin (2\theta)\,\,.
\eeq
The BPS condition is $b_{135}\,=\,0$, which can only be satisfied if $\sin (2\theta)=0$. We have to select now the solution $\theta\,=\,{\pi\over 2}$
if we want to have a non-zero volume for the cycle. The above condition
defines the ${\cal U}_2$ cycle. The associated R-charge can be computed as
above and reads:
\beq
R_{{\cal U}_2}\,=\,{2\beta\over 3x_3}\,N_c\,\,,
\eeq
in precise agreement with the gauge theory result \cite{FrHaMaSpVeWe}.
The baryon number reads
\beq
{\cal B}({\cal U}_2)\,=\,
i\int_{{\cal U}_2}\,P\big[\,\Omega_{2,1}\,\big]_{{\cal U}_2}\,=\,
-c\,{\beta\over \alpha}\,\,{(\alpha-x_1)(\alpha-x_2)\over
(\beta-x_1)(\beta-x_2)}\,\,,
\eeq
where we have used (\ref{Kconstant}) and, after using the third identity in
(\ref{ratios2}), we get:
\beq
{\cal B}({\cal U}_2)\,=\,-d\,=\,-(a+b-c)\,\,,
\eeq
in agreement with the field theory result \cite{MaSp2} (see Table\,\ref{charges1} in chapter \ref{introduction}). If we consider the
case $a = p-q$, $b = p+q$ and $c = p$, which amounts to $\Ypq$, 
a $U(1)$ factor
of the isometry group enhances to $SU(2)$ and these dibaryons are
constructed out of a doublet of bifundamental fields.

\subsection{$Y, Z$ dibaryons}


We now take the following set of worldvolume coordinates $\xi^{\mu}\,=\,(T,\theta,\psi,\tilde{\tau})$ and 
the embedding $x = x_0$ and $\psi'\,=\,\psi'_0$, where $\psi'_0$ is 
a constant and $\psi' = 3\tilde{\tau} + \phi + \psi$ is the angle conjugated to
the $U(1)_R$ charge (see eq. (\ref{Rangle})). We implement this embedding in our coordinates by
setting
$\phi(\psi,\tilde{\tau}) = \psi'_0 - 3\tilde{\tau} - \psi$. In this case
\beq
\Gamma_{\kappa} \epsilon\,=\,-{iL^4 \over \sqrt{ -\det g}}\,
\big[\,c_3\Gamma_{T3}\,+\,c_5\Gamma_{T5}\,+\,c_{135}\,\Gamma_{T135}\,\big]\,
\epsilon\,\,,
\eeq
where
\bear
&&c_3\,=\,3i\,{\tilde{\rho} \cosh\varrho\over 2\alpha\beta}\,\,\sin(2\theta)\,
\sqrt{\Delta_x}\,\,,\rc
&&c_5\,=\,i\,{\cosh\varrho\over 2\alpha\beta}\,\sin(2\theta)\,
\left( 3 x^2 - 2 (\alpha + \beta) x + \alpha \beta \right)\,,\rc
&&c_{135}\,=\,{\cosh\varrho\over \alpha\beta}\,{\alpha\,\cos^2\theta\,
(1-3\sin^2\theta)\,-\beta\,\sin^2\theta\,(1-3\cos^2\theta)\over
\sqrt{\Delta_\theta}}\,\sqrt{\Delta_x}\,\,.
\eear
The BPS conditions are, as before, $c_3\,=\,c_{135}\,=\,0$. Clearly these conditions are satisfied only if $\Delta_x\,=\,0$, or, in other words, when
\beq
x\,=\,x_1\,,\,x_2\,\,.
\eeq
Notice that the value of $\psi'_0$ is undetermined. The induced volume
takes the form:
\beq
\sqrt{-\det g}\,\big|_{x=x_i}\,=\,{L^4\over 2\alpha\beta}\,
\left| 3 x_i^2 - 2 (\alpha + \beta) x_i + \alpha \beta \right|\,\sin(2\theta)\,
\cosh\varrho\,\,.
\eeq
As before, the compatibility with the $AdS_5$ SUSY requires that $\varrho=0$. 
Let us denote by ${\cal X}_i$ the cycle with $x=x_i$. We get that the volumes are given by:
\beq
{\rm Vol}\big({\cal X}_i\big)\,=\,{\pi\over k\,\alpha\,\beta}\,
\left| 3 x_i^2 - 2 (\alpha + \beta) x_i + \alpha \beta \right|\,
\Delta\tilde{\tau}\, L^3\,\,.
\eeq
From this result we get the corresponding values of the $R$-charges,
namely:
\beq
R_{{\cal Y}}\,=\,{2N_c\over 3}\,{x_3-x_1\over x_3}\,\,,\qquad\qquad
R_{{\cal Z}}\,=\,{2N_c\over 3}\,{x_3-x_2\over x_3}\,\,,
\eeq
where ${\cal Y} = {\cal X}_1$ and ${\cal Z} = {\cal X}_2$.
Let us now compute the baryon number of these cycles. 
The pullback of the three-form $\Omega_{2,1}$ to the cycles with $x=x_i$ and
$\psi'=\psi'_0$ is:
\beq
P\big[\,\Omega_{2,1}\,\big]_{x=x_i}\,=\,
iK\,{\left( 3 x_i^2 - 2 (\alpha + \beta) x_i + \alpha \beta \right)\over
2\alpha\beta}\,\,
{\sin(2\theta)\over \tilde{\rho}^4}\,\,
d\theta\wedge d\psi\wedge d\tilde{\tau}\,\,,
\eeq
where $K$ is the constant written in (\ref{Kconstant}). We
obtain: 
\beq
{\cal B}({\cal X}_i)\,=\,-i\,\int_{{\cal X}_i}\,
P\big[\,\Omega_{2,1}\,\big]_{{\cal X}_i}\,=\,{\pi\over k\,\alpha\beta}\,K\,
{3 x_i^2 - 2 (\alpha + \beta) x_i + \alpha \beta\over
(\alpha\,-\,x_i)\,(\beta\,-\,x_i)}\,\Delta\tilde{\tau}\,\,.
\eeq
Taking into account the third identity in (\ref{ratios2}), we get:
\beq
{\cal B}({\cal Y})\,=\,a\,\,,\qquad\qquad
{\cal B}({\cal Z})\,=\,b\,\,,
\eeq
as it should \cite{FrHaMaSpVeWe} (see Table \ref{charges1} in chapter \ref{introduction}). 

\subsection{Generalised embeddings}  \label{Generalizedembeddings}


In this subsection we show that there are generalised embeddings of D3-brane
probes that can be written in terms of the local complex coordinates (\ref{zs})
as  holomorphic embeddings or divisors of $C\Labc$. Let us consider, for
example, $(T,x,\psi,\tilde{\tau})$ as worldvolume coordinates and the ansatz
\beq
\theta=\theta(x,\psi)\,\,,\qquad\qquad
\phi=\phi(x,\psi)\,\,.
\label{D3-general-ansatz}
\eeq
This ansatz is a natural generalisation of the one used in subsection \ref{U1}.
The case where the worldvolume coordinate $\psi$ is changed by $\phi$, can
be easily addressed by changing $\alpha \to \beta$ and $\theta \to \pi/2
- \theta$.
Putting the D3-brane at the center of $AdS_5$, we get that the kappa symmetry
condition is given by an expression as in (\ref{U1-antisymmetrised})
\beq
\Gamma_{\kappa} \epsilon\,=\,-{iL^4 \over \sqrt{ -\det g}}\,
\big[\,a_5\Gamma_{T5}\,+\,a_{135}\,\Gamma_{T135}\,\big]\,\epsilon\,\,,
\label{U1g-antisymmetrised}
\eeq
where $a_5$ and $a_{135}$ are now given by:
\bear
&&a_5\,=\,-{i\over 2}\,\bigg[\,{\cos^2\theta\over \beta}\,+\,
{\sin^2\theta\over \alpha}\,\phi_\psi\,+\,\sin(2\theta)\,\bigg\{
\bigg(\,1-\,{x\over \beta}\,\bigg)\,\theta_x\,-\,\bigg(\,1-\,{x\over
\alpha}\,\bigg)\, \big(\theta_x\phi_\psi-\theta_\psi\phi_x\big)\,\bigg\}
\,\bigg]\,\,,\rc
&&a_{135}\,=\,-\sqrt{{\Delta_\theta\over \Delta_x}}\,
{\sin(2\theta)\over 4}\,\bigg[\,
1-\,{x\over \beta}-\bigg(\,1-\,{x\over \alpha}\,\bigg)\,\phi_\psi\,\bigg]+
\sqrt{{\Delta_x\over \Delta_\theta}}\,\bigg[\,
{\cos^2\theta\over \beta}\theta_x\,\,+\rc
&&\qquad\qquad+\,{\sin^2\theta\over \alpha}\,
\big(\theta_x\phi_\psi-\theta_\psi\phi_x\big)\,\bigg]\,+\,
{i\over 2}\,\bigg[\,\sqrt{\Delta_x\Delta_\theta}\,\,{\sin(2\theta)\over
\alpha\beta}\,\phi_x\,-\,{\tilde{\rho}^2\over \sqrt{\Delta_x\Delta_\theta}}\,
\theta_{\psi}\,\bigg]\,\,.\qquad\qquad
\eear
The BPS condition $a_{135}=0$ reduces to the following pair of differential
equations:
\bear
&&{\cos^2\theta\over \beta}\theta_x+ {\sin^2\theta\over \alpha}\,
\big(\theta_x\phi_\psi-\theta_\psi\phi_x\big)\,=\,
{\Delta_\theta\over \Delta_x}\,
\bigg[\,
1-\,{x\over \beta}-\bigg(\,1-\,{x\over \alpha}\,\bigg)\,\phi_\psi\,\bigg]
\,\,{\sin(2\theta)\over 4}\,\,,\rc
&&\tilde{\rho}^2\theta_{\psi}\,=\,
{\Delta_x\Delta_\theta\over \alpha\beta}\,\,\sin(2\theta)\, \phi_x\,\,.
\label{D3-BPS-general}
\eear
The integral of the above equations can be simply written as:
\beq
z_2\,=\,f(z_1)\,\,,
\eeq
where $z_1$ and $z_2$ are the local complex coordinates of $C\Labc$ written in eq. (\ref{zs}) and 
$f(z_1)$ is an arbitrary holomorphic function. Actually, if $\xi^{\mu}$ is an
arbitrary worldvolume coordinate, one has:
\beq
\partial_{\xi^{\mu}}\,z_2\,=\,f'(z_1)\,\partial_{\xi^{\mu}}\,z_1\,\,.
\eeq
One can eliminate the function $f$ in the above equation by considering the
derivatives with respect to two worldvolume coordinates $\xi^{\mu}$ and
$\xi^{\nu}$. One gets:
\beq
\partial_{\xi^{\mu}}\,\log z_2\,\,
\partial_{\xi^{\nu}}\,\log z_1\,=\,
\partial_{\xi^{\nu}}\,\log z_2\,\,
\partial_{\xi^{\mu}}\,\log z_1\,\,.
\label{infalible}
\eeq
Taking $\xi^{\mu}=x$ and $\xi^{\nu}=\psi$ in the previous equation and
considering that the other coordinates $\theta$ and $\phi$ entering $z_1$
and $z_2$ depend on $(x,\psi)$ (as in the ansatz (\ref{D3-general-ansatz})),
one can prove that (\ref{infalible}) is equivalent to the system of BPS equations (\ref{D3-BPS-general}).

We have checked that the Hamiltonian density of a static D3-brane probe of
the kind discussed in this section satisfies a bound that is saturated when
the BPS equations (\ref{D3-BPS-general}) hold, as it happened in section \ref{d3}. This comes from the fact that,
from the point of view of the probes, these configurations can be regarded
as BPS worldvolume solitons. We have also checked that these generalised
embeddings are calibrated
\beq
P\Big[\,{1\over 2}\,J\wedge J\,\Big]_{{\cal D}}\,=\,
{\rm Vol}({\cal D})\,\,,
\label{calibration1}
\eeq
where  ${\rm Vol}({\cal D})$ is the volume form of the divisor 
${\cal D}$, namely 
${\rm Vol}({\cal D})\,=\,r^3\,dr\wedge {\rm Vol}({\cal C})$ and the K\"ahler form $J$ is displayed in (\ref{Kform}) . It is
important to remind at this point that supersymmetry holds locally but it
is not always true that a general embedding makes sense globally. We have
seen examples of this feature in chapter \ref{Ypq}.

\setcounter{equation}{0}
\section{D5-branes}
\label{D5s}
\medskip

Let us consider a D5-brane probe that creates a codimension one defect on the
field theory. It represents a wall defect in the gauge theory side. 

We choose the following set of worldvolume coordinates: $\xi^{\mu}\,=\,(t,x^1,x^2,r, \theta,\phi)$, and we will adopt the ansatz $x=x(\theta,\phi)$,
$\psi\,=\,\psi(\theta,\phi)$, $\tilde{\tau}\,=\,\tilde{\tau}(\theta,\phi)$
with $x^3$ constant. The kappa
symmetry matrix (\ref{Gammakappad5}) acts on the spinor $\epsilon$ as:
\beq
\Gamma_\kappa\,\epsilon\,=\,{i\over \sqrt{-\det g}}\,{r^2\over L^2}\,
\Gamma_{x^0x^1x^2r}\,\gamma_{\theta\phi}\,\epsilon^*\,=\,
{i\over \sqrt{-\det g}}\,r^2\,\Gamma_{x^0x^1x^2r}\,\big[\,
b_I\,+\,b_{15}\,\Gamma_{15}\,+\,b_{35}\,\Gamma_{35}\,+\,b_{13}
\,\Gamma_{13}\,\big]\,
\epsilon^*\,\,,
\label{D5gamathetaphi1}
\eeq
where 
\bear
&&b_I\,=\,{i\over 2}\,\bigg[\,\sin(2\theta)\,\bigg(\,1-\,{x\over \alpha}\,-
\bigg(\,1-\,{x\over \beta}\,\bigg)\,\psi_{\phi}\,\bigg)\,-\,
{\sin^2\theta\over \alpha}\,x_{\theta}\,+\,{\cos^2\theta\over \beta}\,
\big(\,\psi_\theta\, x_\phi\,-\,\psi_\phi\,
x_\theta\,\big)\,\bigg]\,\,,\rc
&&b_{15}\,=\,{\tilde{\rho}\over \sqrt{\Delta_{\theta}}}\,
\bigg[\,\bigg(\,1-\,{x\over \alpha}\,\bigg)\,\sin^2\theta+
\bigg(\,1-\,{x\over \beta}\,\bigg)\cos^2\theta\,\psi_{\phi}\,
+\,\tilde{\tau}_{\phi}\,\bigg]
\,\,-\rc
&&\qquad
-{i\over 2}\,\sin(2\theta)\,{\sqrt{\Delta_{\theta}}\over \tilde{\rho}}
\,\,\Bigg[\,\,
\bigg(\,1-\,{x\over \alpha}\,\bigg)
\bigg[\,\tilde{\tau}_{\theta}\,+\,\bigg(\,1-\,{x\over
\beta}\,\bigg)\,\psi_\theta\,\bigg]\,+\,
\bigg(\,1-\,{x\over \beta}\,\bigg)\bigg(\,
\tilde{\tau}_{\phi}\,\psi_{\theta}\,-\,\tilde{\tau}_{\theta}\,\psi_{\phi}\,\bigg)
\,\,\Bigg]
\,,\rc
&&b_{35}\,=\,{\sqrt{\Delta_x}\over \tilde{\rho}}\,\Bigg[\,\,
{\alpha-\beta\over 4\alpha\beta}\,
\sin^2(2\theta)\,\psi_{\theta}\,-\,{\sin^2\theta\over
\alpha}\,\tilde{\tau}_{\theta}\,+\,{\cos^2\theta\over \beta}\,
\bigg(\,\tilde{\tau}_{\phi}\,\psi_{\theta}\,-\,\tilde{\tau}_{\theta}\,\psi_{\phi}\,\bigg)
\,\,\Bigg]
\,\,+\rc
&&\qquad
+\,{i\over 2}\,{\tilde{\rho}\over \sqrt{\Delta_x}}\,
\bigg[\bigg(\,1-\,{x\over \alpha}\,\bigg)\,\sin^2\theta\,x_\theta\,-\,
\bigg(\,1-\,{x\over \beta}\,\bigg)\cos^2\theta
\big(\,\psi_\theta\, x_\phi\,-\,\psi_\phi\, x_\theta\,\big)\,
+\,\tilde{\tau}_{\phi}\,x_{\theta}\,-\,\tilde{\tau}_{\theta}\,x_{\phi}
\,\bigg]\,\,,\rc
&&b_{13}\,=\,{1\over 4}\,\sqrt{{\Delta_\theta\over \Delta_x}}\,
\sin(2\theta)\,\bigg[\,
\bigg(\,1-\,{x\over \alpha}\,\bigg)\,x_\theta\,+\,
\bigg(\,1-\,{x\over \beta}\,\bigg)
\big(\,\psi_\theta\, x_\phi\,-\,\psi_\phi\, x_\theta\,\big)\,\bigg]\,\,+
\rc
&&\qquad
+\,\sqrt{{\Delta_x\over \Delta_\theta}}\,\bigg[\,
{\sin^2\theta\over \alpha}\,+\,{\cos^2\theta\over \beta}\,\psi_{\phi}\,
\bigg]\,+\,{i\over 2}\,\bigg[\,{\tilde{\rho}^2\over  \sqrt{\Delta_x\Delta_\theta}}\,
x_{\phi}\,-\,\sqrt{\Delta_x\Delta_\theta}\,\,
{\sin(2\theta)\over \alpha\beta}\,\psi_{\theta}\,\bigg]\,\,.
\label{D5-coefficients}
\eear
The BPS conditions are $b_I\,=\,b_{15}\,=\,b_{35}\,=\,0$. The imaginary part
of $b_{15}$ is zero if $\psi_{\theta}=\tilde{\tau}_{\theta}=0$, \ie,
$\psi\,=\,\psi(\phi)$, $\tilde{\tau}\,=\,\tilde{\tau}(\phi)$. Let us assume that this is
the case and define the quantities $n$ and $m$ as:
\beq
\psi_{\phi}\,=\,n\,\,,
\qquad\qquad
\tilde{\tau}_{\phi}\,=\,m\,\,.
\label{D5-psi-tau}
\eeq
Clearly $n$ and $m$ are independent of the angle $\theta$. Moreover, from
the vanishing of the real part of $b_{15}$ and of the imaginary part of 
$b_{35}$ we get an algebraic equation for $x$, which can be solved as:
\beq
x\,=\,\alpha\beta\,\,{\sin^2\theta\,+\,n\cos^2\theta\,+\,m
\over\beta\sin^2\theta \,+\,n\alpha\cos^2\theta}\,\,.
\label{D5-xtheta}
\eeq
On the other hand, when $\psi_\theta=0$  and $\psi_{\phi}\,=\,n$ the
vanishing of $b_I$ is equivalent to the equation:
\beq
\bigg[\,{\sin^2\theta\over \alpha}\,+\,{\cos^2\theta\over \beta}\,
n\,\bigg]\,x_{\theta}\,=\,
\sin(2\theta)\,\bigg(\,1-\,{x\over \alpha}\,-
\bigg(\,1-\,{x\over \beta}\,\bigg)\,n\,\bigg)\,\,,
\eeq
which is certainly satisfied by our function 
(\ref{D5-xtheta}). For an embedding satisfying (\ref{D5-psi-tau}) and
(\ref{D5-xtheta}) one can check that 
$\sqrt{-\det g}\,=\,r^2\,|\,b_{13}\,|$. Therefore, for these embeddings, 
$\Gamma_{\kappa}$ acts on the Killing spinors as:
\beq
\Gamma_{\kappa}\,\epsilon\,=\,i e^{i\delta_{13}}\,\Gamma_{x^0x^1x^2r}\,
\Gamma_{13}\,\epsilon^*\,\,,
\eeq
where $\delta_{13}$ is the phase of $b_{13}$. In order to implement
correctly  the kappa symmetry condition
$\Gamma_{\kappa}\,\epsilon\,=\,\epsilon$, the phase $\delta_{13}$ must be
constant along the worldvolume of the probe. By inspecting the form of
the coefficient $b_{13}$  in (\ref{D5-coefficients}), one readily
concludes that $b_{13}$ must be real, which happens only when
$x_{\phi}=0$. Moreover, it follows from (\ref{D5-xtheta}) that $x$ is
independent of $\phi$ only when $n$ and $m$ are constant. Thus, $\psi$
and $\tilde{\tau}$ are linear functions of the angle $\phi$, namely:
\beq
\psi\,=\,n\phi\,+\,\psi_0\,\,,
\qquad\qquad
\tilde{\tau}\,=\,m\phi\,+\,\tilde{\tau}_0\,\,,
\eeq
where $\psi_0$ and $\tilde{\tau}_0$ are constant. Notice that in these
conditions the equation 
$\Gamma_{\kappa}\,\epsilon\,=\,\epsilon$ reduces to 
\beq
i \,\Gamma_{x^0x^1x^2r}\,
\Gamma_{13}\,\epsilon^*\,=\,\epsilon\,\,.
\label{D5-condition}
\eeq
Due to the presence of the complex conjugation, (\ref{D5-condition})
is only consistent if the R-charge angle $\psi'=3\tilde{\tau}+\phi+\psi$ is
constant along the worldvolume (see the expression of $\epsilon$ in
(\ref{adsspinor1})). This in turn gives rise to an additional
restriction to the possible supersymmetric embeddings. Indeed, the
condition 
$3\tilde{\tau}+\phi+\psi\,=\,\psi'_0\,=\,{\rm constant}$ implies that the
constants $n$ and $m$ satisfy
\beq
3m+n+1\,=\,0\,\,.
\eeq
Thus, the possible supersymmetric  embeddings of the D5-brane are
labeled by a constant $n$ and are given by:
\bear
&&\psi\,=\,n\phi\,+\,\psi_0\,\,,
\qquad\qquad
\tilde{\tau}\,=\,-{n+1\over 3}\,\,\phi\,+\,\tilde{\tau}_0\,\,,\rc
&&x\,=\,{\alpha\beta\over 3}\,\,
{2-n-3(1-n)\,\cos^2\theta\over \beta\,+\,
(n\alpha\,-\,\beta)\,\cos^2\theta}\,\,.
\eear
It can be now checked as in section \ref{d5} that the
projection (\ref{D5-condition}) can be converted into a set of  algebraic
conditions on the constant spinors $\eta_{\pm}$ of
(\ref{chiraladsspinor1}). These conditions involve a projector which
depends on the constant R-charge angle $\psi_0'=3\tilde{\tau}_0+\psi_0$ and has
four possible solutions. Therefore these embeddings are 1/8
supersymmetric.

The configuration obtained in this section can be also shown to saturate
a Bogomol'nyi bound in the worldvolume theory of the D5-brane probes, as it happened in section \ref{d5}. This amounts to a point of view in which the solution is seen as a worldvolume
soliton. 

Other configurations of physical interest can be considered at this point.
Most notably, we expect to find stable non-supersymmetric configurations of
D5-branes wrapping three cycles of $\Labc$. A similar solution where the
D5-brane probe wraps the entire $\Labc$ manifold, thus corresponding to the
baryon vertex of the gauge theory, should also be found. We will not include here
the detailed analysis of these aspects.

\setcounter{equation}{0}
\section{Spacetime filling  D7-branes}
\label{D7s}
\medskip

Let us consider a D7-brane probe that fills the four Minkowski gauge theory
directions while possibly extending along the holographic direction. These
configurations are relevant to add flavour to the gauge theory. In particular,
the study of fluctuations around them provides the meson spectrum. We start
from the following set of worldvolume coordinates $\xi^{\mu}\,=\, (x^0,x^1,x^2,x^3,x,\psi,\theta,\phi)$ and the ansatz $r\,=\,r(x,\theta)$,
$\tilde{\tau}=\tilde{\tau}(\psi,\phi)$. The kappa symmetry matrix (\ref{D7general-gammak}) in this case reduces to:
\beq
\Gamma_{\kappa}\,\epsilon\,=\,-i\,{r^4\over L^4\sqrt{-\det g}}\,
\Gamma_{x^0\cdots x^3}\,\gamma_{x\psi\theta\phi}\,\epsilon\,\,.
\eeq
Let us assume that the Killing spinor $\epsilon$ satisfies the condition 
$\Gamma_*\epsilon=-\epsilon$, \ie\ $\epsilon$ is of the form $\epsilon_-$ (see eq. (\ref{chiraladsspinor1})) and,
therefore, one has:
\beq
\Gamma_{x^0\cdots x^3}\,\epsilon_-\,=\,i\epsilon_-\,\,,
\label{D3chirality1}
\eeq
which implies $\Gamma_{r5}\epsilon_-\,=\,i\epsilon_-$. Then:
\beq
\Gamma_{\kappa}\,\epsilon_-\,=\,{r^4\over \sqrt{-\det g}}\,
\big[\,d_I\,+\,d_{15}\,\Gamma_{15}\,+\,d_{35}\,\Gamma_{35}\,+\,
d_{13}\Gamma_{13}\,\big]\,\epsilon_-\,\,.
\label{D7gamma-epsilon-}
\eeq
In order to express these coefficients in a compact form, let us define
$\Lambda_x$ and $\Lambda_{\theta}$ as:
\bear
&&\Lambda_x\,=\,-{1\over 2\Delta_x}\bigg[\,\big(\,\alpha-x
\,\big)\,
\big(\,\beta-x\,\big)\,+\,\alpha\big(\,\beta-x\,\big)
\tilde{\tau}_{\phi}\,+\,\beta\big(\,\alpha-x\,\big)\tilde{\tau}_{\psi}\,\bigg]\,
\,,\rc
&&\Lambda_\theta\,=\,{1\over \Delta_\theta}\bigg[\,
\big(\alpha-\beta\big)\,\,
\sin\theta\cos\theta\,+\,\alpha\,\cot\theta\,\,\tilde{\tau}_{\phi}\,-\,
\beta\tan\theta\,\,\tilde{\tau}_{\psi}\,\bigg]\,\,.
\label{d7Lambda}
\eear
Then:
\bear
&&d_I\,=\,{\sin\theta\cos\theta\over 2\alpha\beta}\,\,
\bigg[\,\tilde{\rho}^2\,+\,\Delta_\theta\,\,\Lambda_\theta\,{r_\theta\over r}\,+\,
4\,\Delta_x\,\,\Lambda_x\,{r_x\over r}\,\bigg]\,\,,\rc
&&d_{15}\,=\,i\tilde{\rho}\,\,
{\sin\theta\cos\theta\over 2\alpha\beta}\,\,\sqrt{\Delta_{\theta}}\,\,\bigg[\,
{r_\theta\over r}\,-\,\Lambda_{\theta}\,\bigg]\,\,,\rc
&&d_{35}\,=\,-\tilde{\rho} \,\,{\sin\theta\cos\theta\over \alpha\beta}\,\,
\sqrt{\Delta_{x}}\,\,\,\,\bigg[\,
{r_x\over r}\,-\,\Lambda_{x}\,\bigg]\,\,,\rc
&&d_{13}\,=\,i\,{\sin\theta\cos\theta\over \alpha\beta}\,\,
\sqrt{\Delta_{\theta}\Delta_{x}}\,\,\,\bigg[\,
\Lambda_{x}\,{r_\theta\over r}\,-\,\Lambda_{\theta}\,
{r_x\over r}\,\bigg]\,\,. 
\eear
The BPS conditions are clearly $d_{15}\,=\,d_{35}\,=\,d_{13}\,=\,0$.
From the vanishing of $d_{15}$ and $d_{35}$ we get the following
first-order equations:
\beq
{r_\theta\over r}\,=\,\Lambda_\theta\,\,,\qquad\qquad
{r_x\over r}\,=\,\Lambda_x\,\,.
\label{d7BPS}
\eeq
Notice that $d_{13}=0$ as a consequence of these equations. 
By looking at the explicit form of our ansatz and at the
expression of $\Lambda_\theta$ and $\Lambda_x$ in (\ref{d7Lambda}), one realises
that the only dependence on the angles $\phi$ and $\psi$ in the first-order
equations (\ref{d7BPS}) comes from the partial derivatives of $\tilde{\tau}(\psi,\phi)$.
For consistency these derivatives must be constant, \ie\
$\tilde{\tau}_{\psi}\,=\,n_{\psi}$, $\tilde{\tau}_{\phi}\,=\,n_{\phi}$, where $n_{\psi}$
and $n_{\phi}$ are constants. These equations can be trivially integrated:
\beq
\tilde{\tau}(\psi,\phi)\,=\,n_{\psi}\,\psi\,+\,n_{\phi}\,\phi\,+\,\tilde{\tau}_0\,\,.
\label{d7tau}
\eeq
Notice that $\tilde{\tau}(\psi,\phi)$ relates angles whose periods are not congruent (see eq. (\ref{tauperiod})).
Thus, the D7-brane spans a submanifold that is not, in general, a cycle.
It is worth reminding that this is not a problem for flavour branes. 
If the BPS conditions (\ref{d7BPS}) hold one can check that 
$r^4d_I\,=\,\sqrt{-\det g}$ and, therefore, one has indeed that 
$\Gamma_{\kappa}\epsilon=\epsilon$ for any Killing spinor
$\epsilon=\epsilon_-$, with $\epsilon_-$ as in (\ref{chiraladsspinor1}).
Thus these configurations preserve the four ordinary supersymmetries
of the background.

In order to get the dependence of $r$ on $\theta$ and $x$ it is
interesting to notice that, if $\tilde{\tau}(\psi, \phi)$ is given by
(\ref{d7tau}), the integrals of 
$\Lambda_{\theta}$ and $\Lambda_{x}$ turn out to be:
\bear
&&\int\,\Lambda_{\theta}\,d\theta\,=\,\log
\Bigg[\,{(\sin\theta)^{n_\phi}\,(\cos\theta)^{n_\psi}\over
\Delta_{\theta}^{{n_\phi+n_\psi+1\over 2}}}\,\Bigg]\,\,,\rc\rc
&&\int\,\Lambda_{x}\,dx\,=\,\log
\Bigg[\,{[f_1(x)]^{{n_\phi-n_\psi\over 2}}\over
\Delta_x^{{1\over 6}}\,\,
[f_2(x)]^{{n_\phi+n_\psi\over 2}+{1\over 3}}}\,\Bigg]\,\,,
\eear
where $f_1(x)$ and $f_2(x)$ are the functions defined in (\ref{f12}). 
From this result it straightforward to obtain the general solution of 
$r(\theta,x)$:
\beq
r(\theta,x)\,=\,C\,
{(\sin\theta)^{n_\phi}\,(\cos\theta)^{n_\psi}\over
\Delta_{\theta}^{{n_\phi+n_\psi+1\over 2}}}\,\,\,
{[f_1(x)]^{{n_\phi-n_\psi\over 2}}\over
\Delta_x^{{1\over 6}}\,\,
[f_2(x)]^{{n_\phi+n_\psi\over 2}+{1\over 3}}}\,\,,
\label{r-theta-x}
\eeq
where $C$ is a constant. Notice that the function $r(x,\theta)$ diverges for some particular values of $\theta$ and $x$. This means that the probe always extends infinitely in the holographic direction. For particular values of
$n_\phi$ and $n_\psi$ there is a minimal value of the coordinate $r$,
$r_\star$, which depends on the integration constant $C$. If one uses
this probe as a flavour brane, $r_\star$ provides an energy scale that
is naturally identified with the mass of the dynamical quarks added to the
gauge theory.

It is finally interesting to write the embedding characterized by eqs. 
(\ref{d7tau}) and (\ref{r-theta-x}) in terms of the complex coordinates $z_1$,
$z_2$ and  $z_3$ defined in eq. (\ref{zs}). Indeed, one can check that this
embedding can be simply written as:
\beq
z_1^{m_1}\,z_2^{m_2}\,z_3^{m_3}\,\,=\,{\rm constant}\,\,,
\eeq
where $m_3\not=0$. The relation between the exponents $m_i$ and the 
constants $n_\psi$ and $n_\phi$ is the following:
\beq
{m_1\over m_3}\,=\, {3\over 2}\,\,(\,n_\psi\,-\,n_\phi)\,\,,
\qquad\qquad
{m_2\over m_3}\,=\, -{3\over 2}\,\,(\,n_\psi\,+\,n_\phi)\,-\,1\,\,.
\eeq
By using  the Dirac-Born-Infeld
action of the D7-brane, it is again possible to show that  there
exists a bound for the energy which is saturated for BPS configurations.

\setcounter{equation}{0}
\section{Final Remarks}
\label{FinalRemarks}
\medskip

In this chapter we have worked out supersymmetric configurations involving D-brane probes in $AdS_5 \times \Labc$. Our study focused on three kinds of branes, namely D3, D5 and D7. We have dealt with embeddings corresponding to dibaryons, defects and flavour branes in the gauge theory. For
D3-branes wrapping three-cycles in $\Labc$ we first reproduced all quantum
numbers of the bifundamental chiral fields in the dual quiver theory. We
also found a new class of supersymmetric embeddings of D3-branes in this background that we identified with a generic holomorphic embedding. The
three-cycles wrapped by these D3-branes are calibrated. In the case of
D5-branes, we found an embedding that corresponds to a codimension one
defect in $AdS_5$. From the point of view of the D5-branes, it can be seen as a BPS saturated
worldvolume soliton. We finally found a spacetime filling D7-brane probe
configuration that can be seen to be holomorphically embedded in the
Calabi-Yau, and is a suitable candidate to introduce flavour in the
quiver theory.

Other interesting configurations have been considered. We would only list their main features:

{\it ``Fat" strings} ~If we take a D3-brane with worldvolume coordinates
$(x^0,x^1,\theta,\phi)$ and consider an embedding of the form $x =
x(\theta,\phi)$ and $\psi = \psi(\theta,\phi)$, with the remaining scalars
constant, we see that there is no solution preserving kappa symmetry.
However, we have obtained a ``fat" string solution by wrapping a probe
D3-brane on a two-cycle, which is the same considered in section \ref{D5s}
for a D5-brane probe. This configuration is not supersymmetric but it
is stable.

{\it D5 on a three-cycle} ~We have found an embedding corresponding to
D5-branes that wrap a three-cycle in $\Labc$. They are codimension one
in the gauge theory coordinates. These configurations happen to be non
supersymmetric yet stable. These could be thought of as being a domain wall of the dual field theory.

{\it D5 on a two-cycle} ~We studied another embedding where a D5-brane
probe wraps a two-cycle in $\Labc$ while it extends along the radial
coordinate. For this embedding, $\phi$, $\psi$, $x^3$ and $\tilde{\tau}$ are
held constant. This is a supersymmetric configuration. We also considered  
turning on a worldvolume flux  in the case studied in section \ref{D5s},
and found that it can be done in a supersymmetric way. The flux in the 
worldvolume of the brane provides a bending  of the profile $x^3$ of the
wall, analogously to what happened in subsection \ref{D5flux} for this kind of configurations.

{\it Another spacetime filling D7} ~We considered a different spacetime
filling D7-brane that extends infinitely in the radial direction and wraps
a three-cycle holomorphically embedded in $\Labc$ of the type studied
in subsection \ref{Generalizedembeddings}. It preserves four supersymmetries.

{\it D7 on $\Labc$} ~We finally studied a D7-brane probe wrapping the
entire $\Labc$ space and extended along the radial coordinate. From the
point of view of the gauge theory, this is a string-like configuration
that preserves two supersymmetries.

We would like to understand the excitations of the spacetime filling  D7-branes
and  the baryon vertex better. However, the study of the excitations of the flavour branes will turn out to be more relevant in the background that we will display in chapter \ref{KW}, where their backreaction is taken into account.


\chapter{Unquenched Flavours in the KW Model}
\label{KW}

\medskip
\setcounter{equation}{0}
\medskip

In this chapter we will propose a type IIB string dual to the field theory of 
Klebanov-Witten described in subsection \ref{T11introduc}, in the case in which a large number 
of flavours (comparable to the number of colours) is added to each gauge group. Therefore, we are going beyond the probe approximation and we will be working on the so called Veneziano's topological expansion \cite{Veneziano:1976wm} (see subsection \ref{smearingprocedure}), unlike in chapters \ref{Ypq} and \ref{Labc}. 

We will study in detail the dual field theory to the supergravity
solutions mentioned above, making a considerable number of matchings. 
The field theories turn out to have a positive $\beta$-function along the
RG flow, exhibiting a Landau pole in the UV. In the IR we still have a strongly coupled
field theory, which is ``almost conformal''.

We will also generalise all these results to the interesting case
of a large class of different \Nugual{1} superconformal field theories, deformed by the addition of flavours. In particular we will be able to add flavours to every gauge theory whose dual is $AdS_5 \times M_5$, where $M_5$ is a five-dimensional Sasaki-Einstein manifold. The backgrounds introduced in subsections \ref{Ypqintroduc} and \ref{Labcintroduc} are included in that family of dual solutions.

Finally, a possible way of handling the massive flavour case is undertaken.

\section{Adding Flavors to the Klebanov-Witten Field Model}
\label{sect2}
\medskip
\setcounter{equation}{0}

\subsection{What to Expect from Field Theory Considerations}

In this first part we will address in detail the problem of adding a large number of backreacting non-compact D7-branes to the Klebanov-Witten type IIB supergravity solution introduced in subsection \ref{T11introduc}. Before presenting the solution and describing how it is obtained, we would like to have a look at the dual field theory and sketch which are the features that we expect.

The addition of flavours, transforming in the fundamental and antifundamental representations of the gauge groups, can be addressed by including D7-brane probes into the geometry, following the procedure proposed in \cite{Karch:2002sh}.
This was done in \cite{Ouyang:2003df}, where the embedding of the flavour branes and the corresponding superpotential for the fundamental and antifundamental superfields were found. The D7-branes have four Minkowski directions parallel to the stack of D3-branes transverse to the conifold, whereas the other four directions are embedded holomorphically in the conifold. In particular, D7-branes describing massless flavours can be introduced by considering the holomorphic non-compact embedding $z_1=0$ (see eq. (\ref{complexconifold})). The flavours, which correspond to 3-7 and 7-3 strings, are massless because the D7-branes intersect the D3-branes. Note that the D7-branes have two branches, described by $z_1=z_3=0$ and $z_1=z_4=0$, each one corresponding to a stack. The presence of two branches is required by RR tadpole cancellation: in the field theory this amounts to adding flavours in vector-like representations to each gauge group, hence preventing gauge anomalies.
The fundamental and antifundamental chiral superfields of the two gauge groups will be denoted as $q$, $\tilde{q}$ and $Q$, $\tilde{Q}$ respectively, and the gauge invariant and flavour invariant superpotential proposed in \cite{Ouyang:2003df} is 
\begin{equation}
W=W_{KW}+W_f\;,
\end{equation}
where $W_{KW}$ is the $SU(2)\times SU(2)$ invariant Klebanov-Witten superpotential for the bifundamental fields given in eq. (\ref{KWsuper}).
For a stack of flavour branes, it is conventional to take the coupling between bifundamentals and quarks at a given point of $S^2$ as
\begin{equation}
W_f = h_1 \:\tilde{q}^a A_1 Q_a + h_2 \:\tilde{Q}^a B_1 q_a \;.
\end{equation}
This coupling between bifundamental fields and the fundamental and antifundamental flavours arises from the D7-brane embedding $z_1=0$. The explicit indices are flavour indices. This superpotential, as well as the holomorphic embedding $z_1=0$, explicitly breaks the $SU(2)\times SU(2)$ global symmetry (this global symmetry will be recovered after the smearing).

\begin{figure}[ht]
\begin{center}
\includegraphics[width=0.4\textwidth]{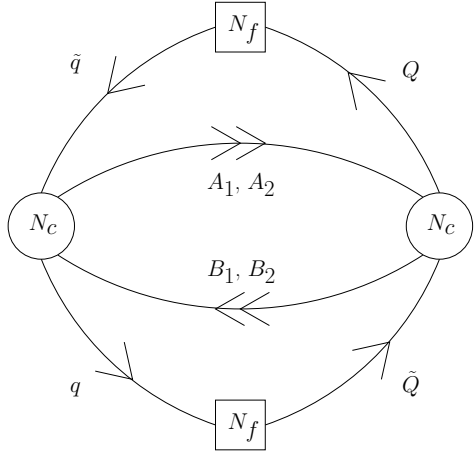}
\end{center}
\caption[Quiver diagram]{Quiver diagram of the Klebanov-Witten gauge theory with flavours. Circles are gauge groups while squares are non-dynamical flavour groups. \label{quiverdiagram}} 
\end{figure}

The field content and the relevant gauge and flavour symmetries of the theory are summarised in Table \ref{fieldcontent} and depicted in the quiver diagram in Fig. \ref{quiverdiagram}.

\begin{table}[ht]
\begin{center}
\begin{tabular}{c|c|c|c|c|c|c|}
 & $SU(N_c)^2$  & $SU(N_f)^2$ &  $SU(2)^2$ & $U(1)_R$ & $U(1)_B$ & $U(1)_{B'}$ \\
\hline &&&&&&\\
$A$ & $(N_c,\overline{N_c})$  & $(1,1)$ & $(2,1)$ & $1/2$ & $0$ & $1$ \\
$B$ & $(\overline{N_c},N_c)$  & $(1,1)$ & $(1,2)$ & $1/2$ & $0$ & $-1$ \\
$q$ & $(N_c,1)$  & $(\overline{N_f},1)$ & $(1,1)$ & $3/4$ & $1$ & $1$ \\
$\tilde{q}$ & $(\overline{N_c},1)$  & $(1,N_f)$ & $(1,1)$ & $3/4$ & $-1$ & $-1$ \\
$Q$ & $(1,N_c)$  & $(1,\overline{N_f})$ & $(1,1)$ & $3/4$ & $1$ & $0$ \\
$\tilde{Q}$ & $(1,\overline{N_c})$  & $(N_f,1)$ & $(1,1)$ & $3/4$ & $-1$ & $0$ \\
\end{tabular}
\end{center}
\caption[Field content]{Field content and symmetries of the KW field theory with massless flavours. \label{fieldcontent}} 
\end{table}

The $U(1)_R$ R-symmetry is preserved at the classical level by the inclusion of  D7-branes embedded in such a way to describe massless flavours, as it can be seen from the fact that the equation $z_1=0$ is invariant under the rotation $z_i\to e^{-i\alpha}z_i$ and the D7-brane wrap the R-symmetry circle. Nevertheless the $U(1)_R$ turns out to be anomalous after the addition of flavours, due to the nontrivial $C_0$ gauge potential sourced by the D7-brane. The baryonic symmetry $U(1)_B$ inside the flavour group is anomaly free, being vector-like.

As it was noted in \cite{Ouyang:2003df}, the theory including D7-brane probes is also invariant under a rescaling $z_i\to \beta z_i$, therefore the field theory is scale invariant in the probe approximation. In this limit the scaling dimension of the bifundamental fields is $3/4$ and the one of the flavour fields is $9/8$, as required by power counting in the superpotential.
Then the $\beta$-function for the holomorphic gauge couplings in the Wilsonian scheme is 
\begin{equation} \label{betas}
\beta_{\frac{8\pi^2}{g_i^2}} = -\frac{16\pi^2}{g_i^3} \beta_{g_i}=-\frac{3}{4} N_f \,\, , \qquad \qquad \beta_{\lambda_i} = \frac{1}{(4\pi)^2} \frac{3N_f}{2N_c} \lambda_i^2 \;,
\end{equation}
with $\lambda_i= g_i^2 N_c$ the 't Hooft couplings.
In the strict planar 't Hooft limit (zero order in $N_f/N_c$), the field theory has a fixed point specified by the aforementioned choice of scaling dimensions, since the beta functions of the superpotential couplings and the 't Hooft couplings are zero. As soon as $N_f/N_c$ corrections are taken into account, the field theory has no fixed points for nontrivial values of all couplings. Rather, it displays a ``near conformal point'' with vanishing $\beta$-functions for the superpotential couplings, but non-vanishing $\beta$-functions for the 't Hooft couplings.
In a $N_f/N_c$ expansion, formula \eqref{betas} holds at order $N_f/N_c$ \textit{if} the anomalous dimensions of the bifundamental fields $A_j$ and $B_j$ do not get corrections at this order. \emph{A priori} it is difficult to expect such a behaviour from string theory, since the stress-energy tensor of the flavour branes will induce backreaction effects on the geometry at linear order in $N_f/N_c$, differently from the fluxes, which will backreact at order $(N_f/N_c)^2$.

Moreover, since we are adding flavours to a conformal theory, we can naively expect a Landau pole to appear in the UV. Conversely, we expect the theory to be slightly away from conformality in the far IR.

\subsection{The Setup and the BPS Equations}

The starting point for adding backreacting branes to a given background is
the identification of the supersymmetric embeddings in that background,
that is the analysis of brane probes.  In \cite{Arean:2004mm}, by imposing
kappa symmetry on the brane worldvolume, the supersymmetric embeddings displayed in eq. (\ref{surfaceskappa}) for D7-branes on the Klebanov-Witten background
were found:
\bsp{
\xi^\alpha_1 &= \{x^0,x^1,x^2,x^3,r,\theta_2,\varphi_2,\psi\} \qquad
\theta_1=\text{const.} \qquad \varphi_1=\text{const.} \,\, , \\
\xi^\alpha_2 &= \{x^0,x^1,x^2,x^3,r,\theta_1,\varphi_1,\psi\} \qquad
\theta_2=\text{const.} \qquad \varphi_2=\text{const.}
}
They are precisely the two branches of the supersymmetric embedding
$z_1=0$ first proposed in \cite{Ouyang:2003df}. Each branch realises a
$U(N_f)$ symmetry group, giving the total flavour symmetry group
$U(N_f)\times U(N_f)$ of massless flavours (a diagonal axial $U(1)_A$ is
anomalous in field theory).
We choose these embeddings because of the following properties: they reach the
tip of the cone and intersect the colour D3-branes; wrap the $U(1)_R$
circle corresponding to rotations $\psi\to\psi+\alpha$; are invariant
under radial rescalings. So they realise in field theory massless flavours,
without breaking explicitly the $U(1)_R$ and the conformal symmetry.
Actually, they are both broken by quantum effects. Moreover the
configuration does not break the $\mathbb{Z}_2$ symmetry of the conifold
solution which corresponds to exchanging the two gauge groups.

The fact that we must include both branches is due to D7-charge
tadpole cancellation, which is dual to the absence of gauge anomalies in
field theory. An example of a (non-singular) two-submanifold in the conifold
geometry is $S^2=\{\theta_1=\theta_2, \, \varphi_1=2\pi-\varphi_2, \, \psi=\text{const},
\, r=\text{const}\}$. The charge distributions of the two branches are
\begin{equation}
\omega^{(1)} = \sum\nolimits_{N_f} \delta^{(2)}(\theta_1,\varphi_1) \, d\theta_1\wedge d\varphi_1 \,\, , \qquad \qquad \omega^{(2)} = \sum\nolimits_{N_f} \delta^{(2)}(\theta_2,\varphi_2) \, d\theta_2\wedge d\varphi_2 \;,
\end{equation}
where the sum is over the various D7-branes, possibly localised at different points, and a correctly normalised scalar delta function (localised on an eight-submanifold) is $\delta^{(2)}(x) \sqrt{-\hat{G}_8}/\sqrt{-G}$.
Integrating the two D7-charges on the two-submanifold we get:
\begin{equation}
\int_{S^2} \omega^{(1)} = - N_f \,\, , \qquad \qquad \int_{S^2} \omega^{(2)} = N_f \;.
\end{equation} 
Thus, whilst the two branches have separately non-vanishing tadpole, putting an equal number of them on the two sides, the total D7-charge cancels. This remains valid for all (non-singular) two-submanifolds.

The embedding can be deformed into a single D7-brane that only reaches a minimum
radius, and realises a merging of the two branches. This corresponds to
giving mass to flavours and explicitly breaking the flavour symmetry to
$SU(N_f)$ and the R-symmetry completely. These embeddings were also found
in \cite{Arean:2004mm}.

Each embedding preserves the same four supercharges, irrespectively of
where the branes are located on the two two-spheres parameterised by
$(\theta_1,\varphi_1)$ and $(\theta_2,\varphi_2)$. Thus we can smear the
distribution and still preserve the same amount of supersymmetry. The
two-form charge distribution is readily obtained to be the same as the
volume forms on the two two-spheres in the geometry, and through the
modified Bianchi identity it sources the flux $F_1$.%
\footnote{The modified Bianchi identity of $F_1$ is obtained from the
Wess-Zumino action term (\ref{WZaction-general}) with $F_1 = -e^{-2\phi}\ast F_9$.}
We expect to obtain a solution where all the functions have only radial
dependence, thanks to the smearing procedure. Moreover we were careful in never breaking the $\mathbb{Z}_2$
symmetry that exchanges the two spheres. The natural ansatz is\footnote{See subsection \ref{T11introduc} to get used to the notation.}:
\begin{align}  \label{ansatzKW}
\begin{split}
ds^2 &= h(r)^{-1/2} dx_{1,3}^2 + h(r)^{1/2} \Bigg ( dr^2 + \\
&\qquad+\,\frac{e^{2g(r)}}{6} \sum_{i=1,2} ( d\theta_i^2 + \sin^2 \theta_i \,
d\varphi_i^2)  + \frac{e^{2f(r)}}{9} (d\psi + \sum_{i=1,2} \cos\theta_i \,
d\varphi_i)^2 \Bigg ) \,\, , 
\end{split} \rc
\phi &= \phi(r) \,\, , \rc
F_5 &= K(r) \, h(r)^{3/4} \Big( e^{x^0x^1x^2x^3r} -
e^{\theta_1\varphi_1\theta_2\varphi_2\psi}
\Big) \,\, , \rc
F_1 &= \frac{N_f}{4\pi} \bigl( d\psi + \cos\theta_1\,
d\varphi_1 + \cos\theta_2\, d\varphi_2 \bigr) = \frac{3N_f}{4\pi} \, h(r)^{-1/4} e^{-f(r)} \, e^{\psi} \,\, , \rc
dF_1 &= -\frac{N_f}{4\pi} \bigl(\sin\theta_1 \, d\theta_1\wedge d\varphi_1 +
\sin\theta_2 \, d\theta_2\wedge d\varphi_2 \bigr) \;,
\end{align}
where the unknown functions are $h(r)$, $g(r)$, $f(r)$, $\phi(r)$ and
$K(r)$ and the vielbein that we have chosen is:
\begin{align} \label{T11frame}
\begin{split}
e^{x^i} &= h^{-1/4} \, dx^i \,\, ,\\
e^{\theta_i} &= \frac{1}{\sqrt{6}} h^{1/4} e^g \, d\theta_i \,\, ,\\
e^{\psi} &= \frac{1}{3} h^{1/4} e^f \,
( d\psi + \cos\theta_1\, d\varphi_1 + \cos\theta_2\, d\varphi_2) \;.
\end{split} \begin{split}
e^r &= h^{1/4} \, dr \,\, ,\\
e^{\varphi_i} &= \frac{1}{\sqrt{6}} h^{1/4} e^g \, \sin\theta_i d\varphi_i \,\, ,\\
\phantom{X}&
\end{split} \end{align}
With this ansatz the field equation $d\big(e^{2\phi}\ast F_1)=0$ is
automatically satisfied, as well as the self-duality condition $F_5=\ast
F_5$. The Bianchi identity $dF_5=0$ together with the Dirac quantisation condition (\ref{quantizationF5}) and the fact that $Vol(T^{1,1})= \frac{16}{27}\pi^3$ give:
\begin{equation}
K \, h^2 \, e^{4g+f} = 27\pi N_c\;,
\end{equation}
and $K(r)$ can be solved. 

We impose that the ansatz preserves the same four supersymmetries as the D7-brane probes on the Klebanov-Witten solution. In the next section we will perform a careful analysis of the supersymmetry variations of the dilatino and gravitino (\ref{Eframe1}) for the ansatz sketched above. Actually, the first-order BPS differential equations which arise from the vanishing of the SUSY variations turn out to be the same for the introduction of backreacted flavour branes in all manifolds of the sort $AdS_5 \times M_5$, with $M_5$ a Sasaki-Einstein space. Let us just show here the solution and put the analysis off for the next section:
\bear \label{BPSsystem}
g' &=& e^{f-2g} \,\, ,\rc
f' &= & e^{-f} (3\,-\,2\,e^{2f-2g})\, -\, \frac{3N_f}{8\pi} e^{\phi-f} \,\, , \rc
\phi' &=& \frac{3N_f}{4\pi} e^{\phi-f} \,\, , \rc
h' &=& -27\pi N_c \, e^{-f-4g} \,\, .
\eear

Notice that taking $N_f=0$ in the BPS system \eqref{BPSsystem} we simply
get equations for a deformation of the Klebanov-Witten solution without
any addition of flavour branes. Solving the system we find both the
original KW background and the solution for D3-branes at a conifold
singularity, as well as other solutions which correspond on the gauge
theory side to giving VEV to dimension six operators. These solutions 
were considered in \cite{PandoZayas:2001iw,Benvenuti:2005qb} and were shown to follow from our system in \cite{Benini:2006hh}. 

In order to be sure that the BPS equations  (\ref{BPSsystem}) capture the correct dynamics, we have to check that the Einstein, Maxwell 
and dilaton equations are solved. This can be done even before finding
actual solutions of the BPS system. We checked that the first-order system
\eqref{BPSsystem} (and the Bianchi identity) in fact \textit{implies} the
second-order Einstein, Maxwell  and dilaton differential equations. An analytic
general proof will be given in subsection \ref{BPS-Einstein}. We did not explicitly check the Dirac-Born-Infeld equations for the D7-brane distribution. We expect them to be solved
because of kappa symmetry (supersymmetry) on their worldvolume.

\subsubsection*{Solution with General Couplings}

We can generalise our set of solutions by switching on non-vanishing VEV's for the bulk gauge potentials $C_2$ and $B_2$. We show that this can be done without modifying the previous set of equations and the two parameters are present for every solution of them. The condition is that the gauge potentials are flat, that is with vanishing field strength. 

Let us switch on the following fields:
\begin{equation}
C_2 = c \, W_2 \,\, , \qquad \qquad B_2 = b \, W_2 \;,
\end{equation}
where the two-form $W_2$ is Poincar\'e dual\footnote{Actually, it is a rescaling of the two-form $\Upsilon_2$ introduced in eq. (\ref{Upsilon}).} to the two-cycle $S^2$. We see that $F_{3}=0$ and $H_{3}=0$. So the supersymmetry variations are not modified, neither are the gauge invariant field strength definitions. In particular the BPS system \eqref{BPSsystem} does not change.

Consider the effects on the action (the argument is valid both for localised and smeared branes). It can be written as a bulk term plus the D7-brane terms:
\begin{equation}
S = S_{bulk} - T_7 \int d^8\xi \, e^\phi \, \sqrt{-\det(\hat G_8 + \mathcal{F})} + T_7 \int \Big[ \sum\nolimits_q \hat C_q \wedge e^\mathcal{F} \Big]_8 \;,
\end{equation} 
with $\mathcal{F}=\hat B_2 + 2\pi\alpha'\, F$ is the D7 gauge invariant field strength and the hat means that the pullback is taken. To get solutions of the kappa symmetry conditions and of the equations of motion, we must take $F$ such that
\begin{equation}
\mathcal{F}=\hat B_2 + 2\pi\alpha'\, F =0\;.
\end{equation}
Notice that there is a solution for $F$ because $B_2$ is flat: $d\hat B_2=\widehat{dB_2}=0$. With this choice kappa symmetry is preserved as before, since it depends on the combination $\mathcal{F}$. The dilaton equation is fulfiled. The Bianchi identities and the bulk field strength equations of motion are not modified, since the WZ term only sources $C_8$. The stress-energy tensor is not modified, so the Einstein equations are fulfiled. The last steps are the equations of $B_2$ and $A_1$ (the gauge potential on the D7-brane). For this, notice that they can be written as:
\begin{align}
d \frac{\delta S}{\delta F} &= 2\pi\alpha' \: d \frac{\delta S_{brane}}{\delta \mathcal{F}} =0\,\, , \rc
\frac{\delta S}{\delta B_2} &= \frac{\delta S_{bulk}}{\delta B_2} + \frac{\delta S_{brane}}{\delta \mathcal{F}} =0\:.
\end{align}
The first is solved by $\mathcal{F}=0$ since in the equation all the terms are linear or higher order in $\mathcal{F}$. This is because the brane action does not contain terms linear in $\mathcal{F}$, and this is true provided $C_6=0$ (which in turn is possible only if $C_2$ is flat). The second equation then reduces to $\frac{\delta S_{bulk}}{\delta B_2}=0$, which amounts to $d(e^{-\phi}\ast H_3)=0$ and is solved.

As we will see in subsection \ref{field theory analysis}, being able to switch on arbitrary constant values $c$ and $b$ for the (flat) gauge potentials, we can freely tune the two gauge couplings (actually the two renormalization invariant scales $\Lambda$'s) and the two theta angles \cite{Klebanov:1998hh,Morrison:1998cs}. This turns out to break the $\mathbb{Z}_2$ symmetry that exchanges the two gauge groups, even if the breaking is mild and only affects $C_2$ and $B_2$, while the metric and all the field strengths continue to have that symmetry. However this does not modify the behaviour of the gauge theory.

\subsection{The Solution in Type IIB Supergravity}
\label{solKW}

The BPS system \eqref{BPSsystem} can be solved through the change of the
radial variable
\begin{equation}
e^f \frac{d}{dr} \equiv \frac{d}{d\rho} \qquad \Rightarrow \qquad e^{-f}dr
= d\rho \;.
\end{equation}
We get the new system:
\bear
\dot{g} &=& e^{2f-2g}\,\, , \label{eqng} \\
\dot{f} &=& 3\, - \,2 e^{2f-2g} \,-\, \frac{3N_f}{8\pi} \, e^\phi \,\, , \label{eqnf} \\
\dot{\phi} &=& \frac{3N_f}{4\pi} e^\phi \,\, , \label{eqnphi} \\
\dot{h} &=& -27\pi N_c \, e^{-4g} \,\, , \label{eqnh} 
\eear
where derivatives are taken with respect to $\rho$.

Equation \eqref{eqnphi} can be solved first. By absorbing an integration
constant in a shift of the radial coordinate $\rho$, we get
\begin{equation}
e^{\phi} = - \frac{4\pi}{3 N_f} \frac{1}{\rho} \qquad \qquad \Rightarrow
\qquad \qquad \rho<0 \;.
\label{dilatonKW}
\end{equation}
The solution is thus defined only up to a maximal radius
$\rho_\text{MAX}=0$ where the dilaton diverges. As we will see, it
corresponds to a Landau pole in the UV of the gauge theory.
On the contrary for $\rho\to -\infty$, which corresponds in the gauge
theory to the IR, the string coupling goes to zero. Note
however that the solution could stop at a finite negative
$\rho_\text{MIN}$ due to integration constants or, for example, more dynamically,  due to the presence of massive flavours. Then define
\begin{equation}
u = 2f-2g \qquad \qquad \Rightarrow \qquad \qquad \dot{u} = 6(1-e^u) +
\frac{1}{\rho}\;,
\end{equation}
whose solution is
\begin{equation}
e^u = \frac{-6\rho \, e^{6\rho}}{(1-6\rho) e^{6\rho} + c_1} \;.
\end{equation}

The constant of integration $c_1$ cannot be reabsorbed, and according to
its value the solution dramatically changes in the IR. A systematic
analysis of the various behaviours was presented in \cite{Benini:2006hh}. The
value of $c_1$ determines whether there is a (negative) minimum value for
the radial coordinate $\rho$. The requirement that the function $e^u$ be
positive defines three cases:
\begin{equation} \nn \begin{split}
-1<c_1<0 \qquad &\rightarrow \qquad \sub{\rho}{MIN} \leq \rho \leq 0 \,\, , \\
c_1 =0 \qquad &\rightarrow \qquad -\infty < \rho \leq 0 \,\, ,\\
c_1 > 0 \qquad &\rightarrow \qquad -\infty < \rho \leq 0 \;.
\end{split} \end{equation}
In the case $-1<c_1<0$, the minimum value $\rho_\text{MIN}$ is given by an
implicit equation. It can be useful to plot this value as a function of
$c_1$:
\\[0.3cm]
\parbox[c]{0.55\textwidth}{\quad\includegraphics[width=0.45\textwidth]{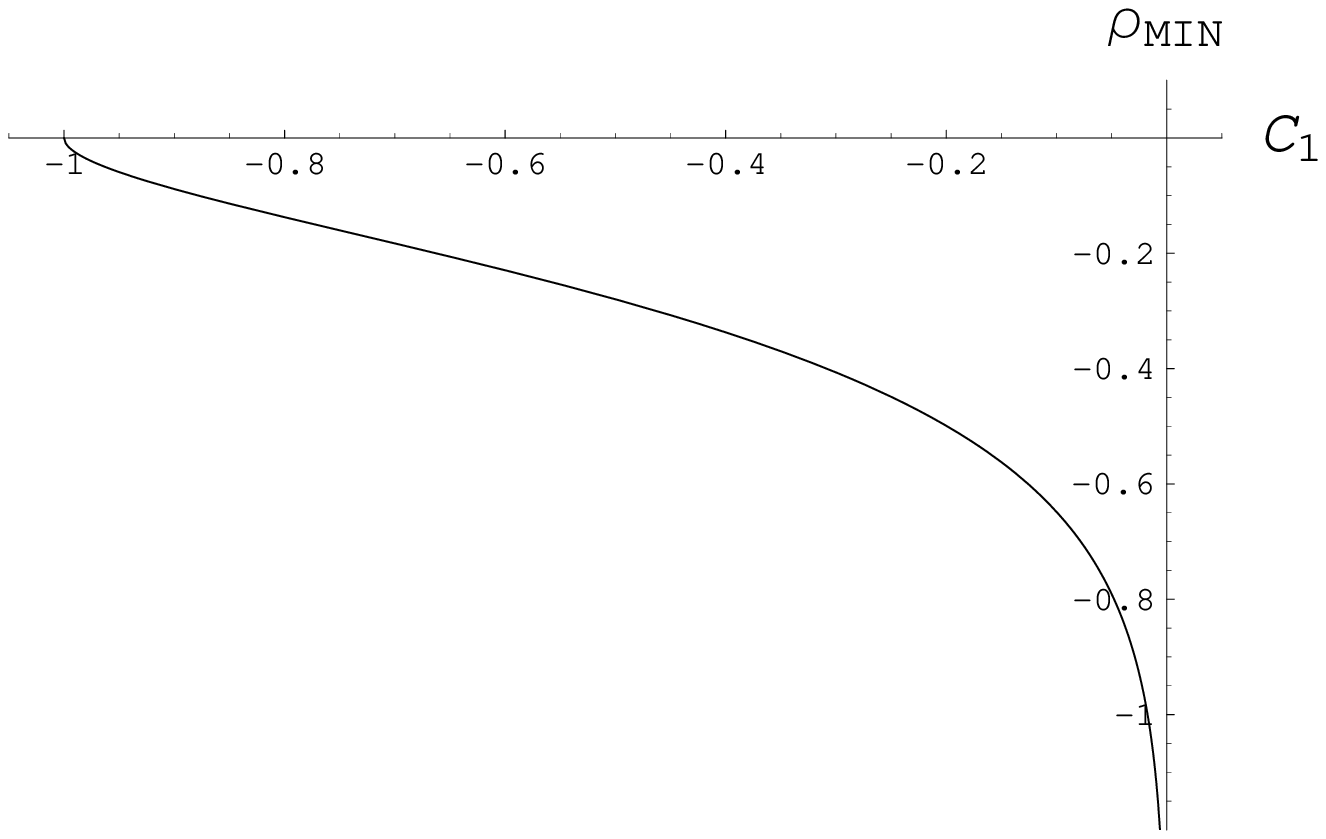}}
\parbox[c]{0.3\textwidth}{\begin{equation}
0 = (1-6\rho_\text{MIN}) \, e^{6\rho_\text{MIN}} + c_1\,\, . \nonumber
\end{equation}
}
\\[0.3cm]
As it is clear from the graph, as $c_1\to-1^+$ the range of the solution
in $\rho$ between the IR and the UV Landau pole shrinks to zero size,
while in the limit $c_1\to 0^-$ we no longer have a minimum radius.

The functions $g(\rho)$ and $f(\rho)$ can be analytically integrated,
while the warp factor $h(\rho)$ and the original radial coordinate
$r(\rho)$ cannot (in the particular case $c_1=0$ we found an explicit
expression for the warp factor). By absorbing an irrelevant integration
constant into a rescaling of $r$ and $x^{0,1,2,3}$, we get:
\bear
e^g &=& \Big[ (1\,-\,6\rho)\,e^{6\rho} \,+\, c_1 \Big ]^{1/6} \,\, , \label{egKW}\\
e^f &=& \sqrt{-6\rho} \, e^{3\rho} \Big [ (1\,-\,6\rho)e^{6\rho}\, +\, c_1
\Big ]^{-1/3} \,\, , \label{efKW} \\
h(\rho) &=& -27\pi N_c \int_0^\rho e^{-4g} \,+\, c_2 \,\, , \label{warp}\\
r(\rho) &=& \int^\rho e^f \, d\tilde{\rho}\;.
\eear
This solution is a very important result of this chapter. We accomplished in finding a
supergravity solution describing a (large) $N_f$ number of backreacting
D7-branes, smeared on the background produced by D3-branes at the tip of a
conifold geometry.

The constant $c_1$ and $c_2$ correspond in field theory to switching on VEV's for relevant operators, as we will see in subsection \ref{irtheory}. Moreover, in the new radial coordinate $\rho$, the metric reads
\begin{equation} 
ds^2 \,=\, h^{-\frac{1}{2}} \, dx_{1,3}^2 \,+\, h^{\frac{1}{2}} \, e^{2f} \, \left ( d\rho^2 \, +\,
\frac{e^{2g-2f}}{6} 
\sum_{i=1,2} ( d\theta_i^2 + \sin^2 \theta_i \, d\varphi_i^2)
\,+\, \frac{1}{9} (d\psi + \sum_{i=1,2} \cos\theta_i \, d\varphi_i)^2 \right ) \;.
\end{equation}

\subsection{Analysis of the Solution: Asymptotics and Singularities} \label{asymp analysis}

We perform here an analysis of the solutions of the BPS system, focusing mainly on the case with $c_1=0$ and we study the asymptotics in the IR and in the UV. In this subsection we will make use of the following formula for the Ricci scalar curvature (in string frame), which can be obtained for solutions of the BPS system:
\begin{equation} \label{ricci scalar}
R=-2 \, \frac{3N_f}{4\pi} \, h^{-1/2} e^{-2g+\frac{1}{2}\phi} \, \bigg[7 + 4 \frac{3N_f}{4\pi} e^{2g-2f+\phi}  \bigg]\;.
\end{equation}

Although the warp factor $h(\rho)$ cannot be analytically integrated in general, it can be if the integration constant $c_1$ is equal to $0$.
Indeed, introducing the \emph{incomplete gamma function}, defined as follows:
\begin{equation}
\Gamma[a,x] \equiv \int_x^\infty t^{a-1} e^{-t} dt \xrightarrow[x\to-\infty]{} e^{i2\pi a} e^{-x} \left ( \frac{1}{x}\right)^{1-a} \Big\{ 1 + \mathcal{O}\Big(\frac{1}{x}\Big) \Big\}\;,
\end{equation} 
we can integrate 
\begin{equation} \begin{split}
h(\rho) &= - 27\pi N_c \int d\rho \frac{e^{-4\rho}}{(1-6\rho)^{2/3}} + c_2 =\\
&= \frac{9}{2}\pi N_c (\frac{3}{2e^2})^{1/3} \Gamma[\frac{1}{3},-\frac{2}{3}+4\rho] + c_2 \simeq \\
&\simeq \frac{27}{4} \pi N_c (-6\rho)^{-2/3} e^{-4\rho}  \:\;\text{for} \:\;\rho \to -\infty\;.
\end{split} \end{equation} 
The warp factor diverges for $\rho\to -\infty$, and the integration constant $c_2$ disappears in the IR. Moreover, if we integrate the proper line element $ds$ from a finite point to $\rho=-\infty$, we see that the throat has an \emph{infinite invariant length}.

The function $r(\rho)$ cannot be given as an analytic integral but, using the asymptotic behaviour of $e^f$ for $\rho\to -\infty$ and setting $c_1=0$, we can approximately integrate it: 
\begin{equation}
r(\rho) \simeq  6^{1/6} \Big[ (-\rho)^{1/6} e^\rho + \frac{1}{6} \Gamma[\frac{1}{6},-\rho] \Big] + c_3 \,\, ,
\end{equation} 
in the IR. Fixing $r\to 0$ when $\rho\to -\infty$ we set $c_3=0$. We approximate further on
\begin{equation}\label{r(rho)}
r(\rho) \simeq 
 (-6\rho)^{1/6} e^\rho\;. 
\end{equation}  
Substituting $r$ in the asymptotic behaviour of the functions appearing in the metric, we find that up to logarithmic corrections of relative order $1/|\log(r)|$:
\begin{equation} 
\begin{split}  \label{asympKW}
e^{g(r)} &\simeq e^{f(r)} \simeq r \,\, , \\
h(r) &\simeq \frac{27\pi N_c}{4} \frac{1}{r^4}\;.
\end{split} 
\end{equation}
Therefore the geometry in the case $c_1=0$ approaches $AdS_5 \times T^{1,1}$ with logarithmic corrections in the IR limit $\rho\to -\infty$.
 
In the UV limit and coming back to the general case with $c_1 \neq 0$, the solutions with backreacting flavours have a Landau pole ($\rho\to 0^-$) since the dilaton diverges (see (\ref{dilatonKW})). The asymptotic behaviours of the functions appearing in the metric are: 
\begin{align}
e^{2g} &\simeq (1+c_1)^{1/3}  \Big[ 1 - \frac{6\rho^2}{1+c_1} + \mathcal{O}(\rho^3) \Big] \,\, , \\
e^{2f} &\simeq - 6\rho \, (1+c_1)^{-2/3} \Big[ 1 + 6\rho + \mathcal{O}(\rho^2) \Big] \,\, , \\
h &\simeq c_2 + 27\pi N_c (1+c_1)^{-2/3} \Big[ -\rho -\frac{4}{1+c_1} \rho^3 + \mathcal{O}(\rho^4) \Big]\,\,.
\end{align}
Note that we have used (\ref{warp}) for  the warp factor.
One concludes  that $h(\rho)$ is monotonically decreasing with $\rho$; if it is positive at some radius, then it is positive  down to the IR. If the integration constant $c_2$ is larger than zero, $h$ is always positive and approaches $c_2$ at the Landau pole (UV). If $c_2=0$, then $h$ goes to zero at the pole. If $c_2$ is negative, then the warp factor vanishes at $\sub{\rho}{MAX}<0$ before reaching the pole (and the curvature diverges there). The physically relevant solutions seem to have $c_2>0$.

The curvature invariants, evaluated in string frame, diverge when $\rho\to 0^-$, indicating that the supergravity description cannot be trusted in the UV. For instance the Ricci scalar $R\sim (-\rho)^{-5/2}$ if $c_2 \neq 0$, whereas $R\sim (-\rho)^{-3}$ if $c_2 = 0$.
If $c_2<0$, then the Ricci scalar $R\sim (\sub{\rho}{MAX}-\rho)^{-1/2}$ when $\rho\to \sub{\rho}{MAX}^-$.

It is worth mentioning that the IR ($\rho\to -\infty$) limit of the geometry of the flavoured solutions is independent of the number of flavours, if we neglect logarithmic corrections to the leading term. Indeed, at the leading order, flavours decouple from the theory in the IR (see the discussion below eq. \eqref{betas}). The counterpart in our supergravity plus branes solution is evident when we look at the BPS system \eqref{BPSsystem}: when $\rho\to -\infty$ the $e^{\phi}$ term disappears from the system together with all the backreaction effects of the D7-branes. Therefore the system reduces to the unflavoured one. The asymptotics of the functions appearing in the metric in the IR limit $\rho\to -\infty$ imply that some of the curvature invariants that one can construct diverge, irrespective of the value of the integration constant $c_1$. Thus, the supergravity description presents a singularity and some care is needed when computing observables from it.  

Using the criterion in \cite{Maldacena:2000mw}, that proposes the IR singularity to be physically acceptable if $g_{tt}$ is bounded near the IR problematic point, we observe that these
singular geometries are all acceptable. Gauge theory physics can be read from these supergravity backgrounds. We call them ``good singularities''.

\subsection{Detailed Study of the Dual Field Theory} \label{field theory analysis}

In this subsection we are going to undertake a detailed analysis of the dual gauge theory features reproduced by the supergravity solution. The first  issue we want to address is what is the effect of the smearing on the gauge theory dual.

As we wrote above, the addition to the supergravity solution of one stack of localised non-compact D7-branes at $z_1=0$ introduces in the field theory flavours coupled through a superpotential term
\begin{equation}
W =\lambda \, \Tr(A_i B_k A_j B_l) \, \epsilon^{ij} \epsilon^{kl} + h_1 \, \tilde{q}^a A_1 Q_a + h_2 \,\tilde{Q}^a B_1 q_a \;,
\end{equation} 
where we explicitly write the flavour indices $a$. For this particular embedding the two branches are localised, say, at $\theta_1=0$ and $\theta_2=0$ respectively on the two spheres. One can exhibit a lot of features in common with the supergravity plus D7-branes solution:
\begin{itemize}
\item the theory has $U(N_f)\times U(N_f)$ flavour symmetry (the diagonal axial $U(1)_A$ is anomalous), each group corresponding to one branch of D7-branes;
\item putting only one branch there are gauge anomalies in the quantum field theory and a tadpole in supergravity, while for two branches they cancel;
\item adding a mass term for the fundamentals the flavour symmetry is broken to the diagonal $U(N_f)$, while in supergravity there are embeddings moved away from the origin for which the two branches merge.
\end{itemize}

The $SU(2) \times SU(2)$ part of the isometry group of the background without
D7-branes is broken by the presence of localised branes. It amounts to separate rotations of the two $S^2$ in the geometry and shifts the
location of the branches. Its action is realised through the
superpotential, and exploiting its action we can obtain the
superpotential for D7-branes localised in other places.
The two bifundamental doublets $A_j$ and $B_j$ transform as spinors of the respective $SU(2)$. So the flavour superpotential term for a configuration in which the two branches are located at $x$ and $y$ on the two spheres can be obtained by identifying two rotations that bring the north pole to $x$ and $y$. There is of course a $U(1)\times U(1)$ ambiguity in this. Then we have to act with the corresponding $SU(2)$ matrices $U_x$ and $U_y$ on the vectors $(A_1,A_2)$ and $(B_1,B_2)$ (which transform in the $(\mathbf{2},1)$ and $(1,\mathbf{2})$  representations) respectively, and select the first vector component. In summary we can write
\footnote{In the case in which the two gauge couplings and theta angles are equal, we could appeal to the $\mathbb{Z}_2$ symmetry that exchanges them to argue $|h_1|=|h_2|$, but no more because of the ambiguities.}
\begin{equation}
W_f = h_1 \: \tilde q^x \Bigl[ U_x \binom{A_1}{A_2} \Bigr]_1 Q_x + h_2 \: \tilde Q^y \Bigl[ U_y \binom{B_1}{B_2} \Bigr]_1 q_y \;,
\end{equation}
where the notation $\tilde q^x$, $Q_x$ stands for the flavours coming from a first D7 branch being at $x$, and the same for a second D7 branch at $y$.

To understand the fate of the two phase ambiguities in the couplings $h_{1}$ and
$h_{2}$, we appeal to symmetries. The $U(1)$ action which gives $(q,\tilde q,Q,\tilde Q)$ charges $(1,-1,-1,1)$ is a symmetry explicitly broken by the flavour superpotential. The freedom of redefining the flavour fields acting with this $U(1)$ can be exploited to reduce to the case in which the phase of the two holomorphic couplings is the same. The $U(1)$ action with charges $(1,1,1,1)$ is anomalous with equal anomalies for both the gauge groups, and it can be used to absorb the phase ambiguity into a shift of the sum of Yang-Mills theta angles $\Theta_1+\Theta_2$ (while the difference holds steady). 
This is what happens for D7-branes on flat spacetime. The ambiguity that we mentioned amounts to rotations of the transverse $\mathbb{R}^2$ space, whose only effect is a shift of $C_0$. As we will show in the next subsection, the value of $C_0$ is our way of measuring the sum of theta angles through D(-1)-brane probes. Notice that if we put in our setup many separate stacks of D7-branes, all their superpotential $U(1)$ ambiguities can be reabsorbed in a single shift of $C_0$.

From a physical point of view, the smearing corresponds to put the D7-branes at different points on the two spheres, distributing each branch on one of the two-spheres.
This is done homogeneously so that there is one D7-brane at every point of $S^2$. The non-anomalous flavour symmetry is broken from $U(1)_B\times SU(N_f)_R\times SU(N_f)_L$ (localised configuration) to $U(1)_B\times U(1)^{N_f-1}_V\times U(1)^{N_f-1}_A$ (smeared configuration).%
\footnote{The axial $U(1)$ which gives charges $(1,1,-1,-1)$ to one set of fields
$(q_x,\tilde q^x,Q_x, \tilde Q^x)$ coming from a single D7-brane, is an anomalous symmetry.
For every D7-brane that we consider, the anomaly amounts to a shift of the
same two theta angles of the gauge theory. So we can combine this $U(1)$
with an axial rotation of all the flavour fields and get an anomaly
free symmetry. In total, from $N_f$ D7-branes we can find $N_f-1$ such anomaly
free axial $U(1)$ symmetries.}

Let us introduce a pair of flavour indices $(x,y)$ that naturally 
live on $S^2\times S^2$ and specify the D7-brane. 
The superpotential for the whole system of smeared D7-branes is just the sum (actually an integral) over the indices $(x,y)$ of the previous contributions:
\begin{equation} \label{wflavours}
W = \lambda \, \Tr(A_i B_k A_j B_l) \, \epsilon^{ij} \epsilon^{kl} + h_1 \: \int_{S^2} d^2x \, \tilde q^x \Big[ U_x \binom{A_1}{A_2} \Big]_1 Q_x + h_2 \int_{S^2} d^2y \, \tilde Q^y \Big[ U_y \binom{B_1}{B_2} \Big]_1 q_y \;.
\end{equation} 
Again, all the $U(1)$ ambiguities have been reabsorbed in field redefinitions and a global shift of $\Theta_1+\Theta_2$.

In this expression the $SU(2)_A\times SU(2)_B$ symmetry is manifest: rotations of the bulk fields $A_j$, $B_j$ leave the superpotential invariant because they can be reabsorbed in rotations of the dummy indices $(x,y)$.
In fact, the action of $SU(2)_A\times SU(2)_B$ on the flavours is a subgroup of the broken $U(N_f)\times U(N_f)$ flavour symmetry. In the smeared configuration, there is a D7-brane at each point of the spheres and the group $SU(2)^2$ rotates all the D7-branes in a rigid way, moving each D7 where another was. So it is a flavour transformation contained in $U(N_f)^2$. By combining this action with a rotation of $A_j$ and $B_j$, we get precisely the claimed symmetry.

Even if it is written in an involved fashion, the superpotential \eqref{wflavours} does not spoil the features of the gauge theory. In particular, the addition of a flavour mass term still would give rise to the symmetry breaking pattern
\begin{equation}
U(1)_B\times U(1)^{N_f-1}_V\times U(1)^{N_f-1}_A \quad \to \quad U(1)^{N_f}_V \;. \nn
\end{equation}

\subsubsection{Holomorphic Gauge Couplings and $\beta$-functions}

In order to extract information on the gauge theory from the supergravity
solution, we need to know the holographic relations between the gauge
couplings, the theta angles and the supergravity fields. These formulae can be
properly derived only in the
orbifold $\mathbb{R}^{1,3} \times \mathbb{C}\times
\mathbb{C}^2/\mathbb{Z}_2$, where string theory can be quantised by
considering fractional branes placed at the singularity.
The near-horizon geometry describing the IR dynamics on a stack of $N_c$
regular branes at the singularity is $AdS_5\times S^5/\mathbb{Z}_2$. The
dual gauge theory is an \Nugual{2} $SU(N_c)\times SU(N_c)$ SCFT with
bifundamental hypermultiplets. In this \Nugual{2} orbifold theory, the holographic relations (\ref{RGholography}) and (\ref{holographic_theta}) can be
derived exactly.

Usually in the literature the aforementioned holographic relations were
assumed to hold also in the conifold case.
Strassler remarked in \cite{Strassler:2005qs} that for the conifold
theory the formulae for the sum of the gauge couplings and the sum of
theta angles need to be corrected.
We expect that the formula for the sum of theta angles is
correct as far as anomalies are concerned, since anomalies
do not change in RG flows.
Instead the first formula in eq. (\ref{RGholography}) may need to be corrected in the KW theory: in general
the dilaton could be identified with some combination of the gauge and
superpotential couplings.

Let us now make contact with our supergravity solution.
In the smeared solution, since $dF_1\neq0$ at every point, it is not
possible to define a scalar potential $C_0$ such that $F_1=dC_0$. We
bypass this problem by restricting our attention to the non-compact
four-cycle defined by $(\rho, \psi, \theta_1=\theta_2,\varphi_1=2\pi-\varphi_2)$
\cite{Bertolini:2002yr}(note that it wraps the R-symmetry direction
$\psi$), so that we can take the pullback on it and write
\begin{equation}
F_1^{eff}=\frac{N_f}{4\pi} \, d\psi \,\, ,
\end{equation}
and therefore
\begin{equation} \label{C0}
C_0^{eff}=\frac{N_f}{4\pi} (\psi-\psi_0)\;.
\end{equation}
By using eqs. (\ref{RGholography}) and (\ref{holographic_theta}) we can identify now:
\begin{align}
\frac{8\pi^2}{g^2} = \pi \, e^{-\phi} &= -\frac{3N_f}{4} \rho \label{g_YM 2} \,\, , \\
\Theta_1 + \Theta_2 &= -\frac{N_f}{2} (\psi-\psi_0) \;, \label{theta_YM 2}
\end{align}
where we suppose for simplicity the two gauge couplings to be equal ($g_1=g_2\equiv g$). The
generalisation to an arbitrary constant $B_2$ is straightforward since the
difference of the inverse squared gauge couplings does not run.
Although, as discussed above,  one cannot be sure of the validity of \eqref{g_YM 2}, we can try to extract some information.

Let us first compute the \hyph{\beta}function of the gauge couplings.
The identification \eqref{RGholography} allows us to define a ``radial''
\hyph{\beta}function that we can directly compute from supergravity
\cite{Olesen:2002nh}:
\begin{equation}\label{beta sugra}
\beta_{\frac{8\pi^2}{g^2}}^{(\rho)}\equiv \frac{\partial}{\partial\rho}
\frac{8\pi^2}{g^2}= \pi \frac{\partial e^{-\phi}}{\partial\rho} =
-\frac{3N_f}{4}\;.
\end{equation}
(Compare this result with eq. \eqref{betas}).
The physical \hyph{\beta}function defined in the field theory is of
course:
\begin{equation}\label{beta FT}
\beta_{\frac{8\pi^2}{g^2}}\equiv \frac{\partial}{\partial
\log\frac{\mu}{\Lambda}} \frac{8\pi^2}{g^2}\;,
\end{equation}
where $\mu$ is the subtraction scale and $\Lambda$ is a renormalization
group invariant scale.
In order to get the precise field theory \hyph{\beta}function from the
supergravity computation one needs the \emph{energy-radius} relation
$\rho=\rho\big( \frac{\mu}{\Lambda}\big)$, from which
$ \beta = \beta^{(\rho)} \: \partial \rho / \partial \log\frac{\mu}{\Lambda}$.
In general, for non-conformal duals, the radius-energy relation depends on
the phenomenon one is interested in and accounts for the scheme-dependence
in the field theory.

Even without knowing the radius-energy relation, there is some physical
information that we can extract from the radial \hyph{\beta}function
\eqref{beta sugra}. In particular, being the energy-radius relation
$\rho=\rho\big( \frac{\mu}{\Lambda}\big)$ monotonically increasing, the
signs of the two $\beta$-functions coincide.

In our case, using $r=\frac{\mu}{\Lambda}$ and eq. \eqref{r(rho)}, one gets matching between \eqref{betas} and \eqref{beta sugra}.

\subsubsection{R-symmetry Anomaly and Vacua}

Now we move to the computation of  the $U(1)_R$ anomaly. We can perform it in field theory by using the equation (\ref{QFTanomaly}). Thus the anomaly relation in our field theory is the following:
\begin{equation}\label{anomaly}
\partial_\mu J^\mu_R  = -\frac{N_f}{2} \, \frac{1}{32\pi^2} \big(F_{\mu\nu}^a \tilde{F}_a^{\mu\nu} + G_{\mu\nu}^a \tilde{G}_a^{\mu\nu}\big) \;,
\end{equation}
or in other words, under a $U(1)_R$ transformation of parameter $\varepsilon$, for both gauge groups the theta angles transform as
\begin{equation}\label{theta shift}
\begin{split}
\Theta_i \rightarrow \Theta_i - \frac{N_f}{2} \varepsilon \;.
\end{split}
\end{equation}

On the string/gravity side a $U(1)_R$ transformation of parameter $\varepsilon$ is realised (in our conventions) by the shift $\psi \rightarrow \psi +2 \varepsilon$.
This can be derived from the transformation of the complex variables \eqref{zcoor}, which under a $U(1)_R$ rotation get $z_i\to e^{i\varepsilon}z_i$, or directly by the decomposition of the ten-dimensional spinor $\epsilon$ into four-dimensional and six-dimensional factors and the identification of the four-dimensional supercharge with the four-dimensional spinor.
By means of the dictionary \eqref{theta_YM 2} we obtain:
\begin{equation}\label{theta shift: gravity}
\Theta_1 + \Theta_2 \rightarrow \Theta_1 + \Theta_2 - 2 \,  \frac{N_f}{2} \varepsilon\;,
\end{equation}
in perfect agreement with \eqref{theta shift}.

The $U(1)_R$ anomaly is responsible for the breaking of the symmetry group but usually a discrete subgroup survives. Disjoint physically equivalent vacua, not connected by other continuous symmetries, can be distinguished thanks to the formation of domain walls among them, whose tension could also be  measured. We want to read the discrete symmetry subgroup of $U(1)_R$ and the number of vacua both from field theory and supergravity.
In field theory the $U(1)_R $ action has an extended periodicity (range of inequivalent parameters) $\varepsilon\in[0,8\pi)$ instead of the usual $2\pi$ periodicity, because the minimal charge is $1/4$. Let us remark however that when $\varepsilon$ is a multiple of $2\pi$ the transformation is not an R-symmetry, since it commutes with supersymmetry.
The global symmetry group contains the baryonic symmetry $U(1)_B$ as well, whose parameter we call $\alpha\in[0,2\pi)$. The two actions $U(1)_R$ and $U(1)_B$ satisfy the following relation: $\mathcal{U}_R (4\pi)=\mathcal{U}_B (\pi)$.
Therefore the group manifold $U(1)_R \times U(1)_B $ is parameterised by 
$\varepsilon\in[0,4\pi)$, $\alpha\in[0,2\pi)$ (this parameterisation realises a nontrivial torus) and $U(1)_B$ is a true symmetry of the theory.
The theta angle shift \eqref{theta shift} allows us to conclude that the $U(1)_R$ anomaly breaks the symmetry according to $U(1)_R\times U(1)_B \to \mathbb{Z}_{N_f}\times U(1)_B $, where the latter is given by $\varepsilon = 4n\pi/N_f \; (n=0,1,\dots,N_f-1)$, $\alpha\in[0,2\pi)$.

Coming to the string side, the solution for the metric, the dilaton and the field strengths is invariant under arbitrary shifts of $\psi$. But the nontrivial profile of $C_0$ breaks this symmetry.
The presence of DBI actions in the functional integral tells us that the RR potentials are quantised, in particular $C_0$ is defined modulo integers. Taking the formula \eqref{C0} and using the periodicity $4\pi$ of $\psi$, we conclude that the true invariance of the solution is indeed $\mathbb{Z}_{N_f}$.

One can be interested in computing the domain wall tension in the field theory by means of its dual description in terms of a D5-brane with 3 directions wrapped on a 3-sphere (see \cite{Herzog:2002ih} for a review in the conifold geometry). It is easy to see that, as in Klebanov-Witten theory, this object is stable only at $r=0$ ($\rho\to -\infty$), where the domain wall is tensionless.

\subsubsection{The UV and IR Behaviors} \label{irtheory} \label{uvtheory}

The supergravity solution allows us to extract the IR dynamics of the KW field theory with massless flavours. Really what we obtained is a class of solutions, parameterised by two integration constants $c_1$ and $c_2$.  Momentarily we will say something about their meaning but some properties are independent of them.

The fact that the \hyph{\beta}function is always positive, with the only critical point at vanishing gauge coupling, tells that the theory is irreparably driven to that point, unless the supergravity approximation breaks down before ($c_1<0$), for instance because of the presence of curvature singularities. Using the $\rho$ coordinate this is clear-cut. In the cases where the string coupling falls to zero in the IR, the gravitational coupling of the D7-branes to the bulk fields also goes to zero and the branes tend to decouple. The signature of this is in the equation \eqref{eqnf} of the BPS system: the quantity $e^\phi N_f$ can be thought of as the effective size of the flavour backreaction which indeed vanishes in the far IR. The upshot is that flavours can be considered as an ``irrelevant deformation'' of the $AdS_5\times T^{1,1}$ geometry.

The usual technique for studying deformations of an $AdS_5$ geometry was given in \cite{Gubser:1998bc,Witten:1998qj}. Looking at the asymptotic behaviour of fields in the $AdS_5$ effective theory (\ref{AdS/CFTdimension}):%
\footnote{Notice that usually the prescription (\ref{AdS/CFTdimension}) or the holographic renormalization methods are used when 
we may have flows starting from a conformal point in the UV. In this case, our conformal point is in the IR 
and one may doubt about the validity in this unconventional case. See \cite{Skenderis:2006di}
for an indication that applying the prescription in an IR point makes sense, even when the UV 
geometry is very far away from $AdS_5 \times M_5$.}
\begin{equation}
\delta \Phi = a\, r^{\Delta-4} + c\, r^{-\Delta} \;,
\label{asymptotic}
\end{equation}
we read, on the CFT side, that the deformation is $H=H_{CFT} + a\,\mathcal{O}$ with $c=\langle \mathcal{O} \rangle$ the VEV of the operator corresponding to the field $\Phi$ and $\Delta$ the quantum dimension of the operator $\cO$. Alternatively, one can compute the effective five-dimensional action and look for the masses of the fields, from which the dimension is extracted with the formula (\ref{AdS/CFTmass}).
We computed the five-dimensional effective action for the particular deformations $e^{f(r)}$, $e^{g(r)}$ and $\phi(r)$ and we included the D7-brane action terms (the details are in subsection 
\ref{SuperpotentialBPS}). After diagonalization of the effective K\"ahler potential, we got a scalar potential $V$ containing a lot of information. First of all, minima of $V$ correspond to the $AdS_5$ geometries, that is conformal points in field theory. The only minimum is formally at $e^\phi=0$ and has the $AdS_5\times T^{1,1}$ geometry. Then, expanding the potential at quadratic order the masses of the fields can be read; 
from here we deduce that 
we have operators of dimension six and eight taking VEV and a marginally irrelevant operator inserted.%
\footnote{To distinguish between a VEV and an insertion we have to appeal to the first criterium described in eq. (\ref{asymptotic}) and below. }

The operators taking VEV where already identified in \cite{Klebanov:2000nc,Benvenuti:2005qb}. The dimension eight operator is $\Tr F^4$ and represents the deformation from the conformal KW solution to the non-conformal 3-brane solution. The dimension six operator is a combination of the operators $\Tr (\cW_\alpha \bar \cW^\alpha)^2$ and represents a relative metric deformation between the $S^2\times S^2$ base and the $U(1)$ fiber of $T^{1,1}$.
The marginally irrelevant insertion is the flavour superpotential, which would be marginal at the hypothetic $AdS_5$ (conformal) point with $e^\phi=0$, but is in fact irrelevant driving the gauge coupling to zero in the IR and to very large values in the UV. 
Let us add that the scalar potential $V$ can be derived from a superpotential $W$, from which in turn the BPS system (\ref{BPSsystem}) can be obtained.

Since in the IR the flavour branes undergo a sort of decoupling, the relevant deformations dominate and their treatment is much the same as for the unflavoured Klebanov-Witten solution \cite{Klebanov:2000nc,Benvenuti:2005qb,Strassler:2005qs}. We are not going to repeat it and we will concentrate on the case $c_1=c_2=0$. The supergravity solution flows in the IR to the $AdS_5\times T^{1,1}$ solution (with corrections of relative order $1/|\log(r)|$). On one hand the R-charges and the anomalous dimensions tend to the almost conformal values:
\begin{equation} \begin{split}
R_{A,B} &= \frac{1}{2} \,\, , \\
R_{q,Q} &= \frac{3}{4} \,\, ,
\end{split} \qquad \qquad \begin{split}
\gamma_{A,B} &= -\frac{1}{2} \,\, , \\
\gamma_{q,Q} &= \frac{1}{4} \;.
\end{split} \end{equation}
Using the formula for the \hyph{\beta}function of a superpotential dimensionless coupling:
\begin{equation}
\beta_{\tilde h} = \tilde h \Big[ -3 + \sum\nolimits_\Phi \big( 1+ \frac{\gamma_\Phi}{2} \big) \Big] \;,
\end{equation} 
where $\Phi$ are the fields appearing in the superpotential term, we obtain that the total superpotential \eqref{wflavours} is indeed marginal. On the other hand the gauge coupling flows to zero. Being at an almost conformal point, we can derive the radius-energy relation through rescalings of the radial and Minkowski direction, getting $r=\mu/\Lambda$. Then, the supergravity \hyph{\beta}function coincides with the exact (perturbative) holomorphic \hyph{\beta}function (in the Wilsonian scheme):%
\footnote{Here it is manifest why the SUGRA \hyph{\beta}function computed in this context with brane probes matches the field theory one, even if this requires the absence of order $N_f/N_c$ corrections to the anomalous dimensions $\gamma_{A,B}$, which one does not know how to derive (the stress-energy tensor is linear in $N_f/N_c$). It is because those corrections are really of order $e^\phi N_f/N_c$ and in the IR $e^\phi\to 0$.}
\begin{equation}
\beta_g = - \frac{g^3}{16\pi^2} \Big[ 3N_c - 2N_c (1-\gamma_A) - N_f(1-\gamma_f) \Big] \;.
\end{equation} 

If we are allowed to trust the first orbifold relation in eq. \eqref{RGholography} relating gauge coupling constants and dilaton, we conclude that the gauge coupling flows to zero in the IR. This fact could perhaps explain the divergence of the curvature invariants in string frame \cite{Itzhaki:1998dd}, as revealed in subsection \ref{asymp analysis}. The field theory would enter the perturbative regime at this point. However, it is hard to understand why the anomalous dimensions of the fields are large while the theory seems to become perturbative.
For this reason, we question the validity in the conifold case of the first holographic relation in eq. \eqref{RGholography}, which can be derived only for the orbifold. In \cite{Benini:2006hh} we proposed an alternative interpretation of the IR regime of our field theory, based on some nice observations made in \cite{Strassler:2005qs} about the KW field theory. We argued that the theory may flow to a strongly coupled fixed point, although the string frame curvature invariant is large, as in the Klebanov-Witten solution for small values of $g_s N_c$.

Contrary to the IR limit, the UV regime of the theory is dominated by flavours and we find the same kind of behaviour for all values of the relevant deformations $c_1$ and $c_2$. The gauge couplings increase with the energy, irrespective of the number of flavours. At a finite energy scale that we conventionally fixed to $\rho=0$, the gauge theory develops a Landau pole since the string coupling diverges at that particular radius. This energy scale is finite because $\rho=0$ is at finite proper distance from the bulk points $\rho<0$.

At the Landau pole radius the supergravity description breaks down for many reasons: the string coupling diverges as well as the curvature invariants and the $\psi$ circle shrinks. An UV completion must exist and finding it is an interesting problem. One could think about obtaining a new description in terms of supergravity plus branes through various dualities. In particular T-duality will map our solution to a system of NS5, D4 and D6-branes, which could then be uplifted to M-theory. Anyway, T-duality has to be applied with care because of the presence of D-branes on a nontrivial background and we actually do not know how to T-dualize the Dirac-Born-Infeld action.

\section{Generalisations}
\label{generalisations}
\medskip
\setcounter{equation}{0}

\subsection{The BPS Equations for Any Sasaki-Einstein Space}
\label{SEBPS}

Let us now explain in detail the origin of the system of first-order differential equations (\ref{BPSsystem}). As we already said in section \ref{sect2}, the system  (\ref{BPSsystem}) is a consequence of supersymmetry. Actually, it turns out that it can be derived in the more general situation that corresponds to having  smeared D7-branes in a space of the type 
$AdS_5\times M_5$,  where $M_5$ is a five-dimensional Sasaki-Einstein (SE) manifold. Notice that the
$T^{1,1}$ space considered up to now is a SE manifold. In general,  a SE manifold can be represented as a one-dimensional bundle over a four-dimensional K\"ahler-Einstein (KE) space. Accordingly, 
we will write the $M_5$ metric  as follows
\beq
ds^2_{SE}=ds^2_{KE}+(d\tau+A)^2\,\,,
\eeq
where $\partial / \partial \tau$ is a Killing vector and $ds^2_{KE}$ stands for  the metric of the  KE space with K\"ahler form $J=dA\,/\,2$. In the case of the $T^{1,1}$ manifold the KE base is just $S^2\times S^2$, where the $S^2$'s are  parameterised by the angles $(\theta_i,\varphi_i)$ and the fiber $\tau$ is parameterised by the angle $\psi$.

Our ansatz  for ten-dimensional metric in Einstein frame will correspond to a deformation of the standard $AdS_5\times M_5$. Apart from the ordinary warp factor $h(r)$,  we will introduce some squashing between the one form dual to the Killing vector and the KE base, namely:
\beq  \label{metric1}
ds^2\,=\,\Big[\,h(r)\,\Big]^{-{1\over 2}}\,dx^2_{1,3}\,+\,
\Big[\,h(r)\,\Big]^{{1\over 2}}\,\Big[\,
dr^2\,+\,e^{2g(r)}\,ds^2_{KE}\,+\,e^{2f(r)}\,
\big(\,d\tau+A)^2\,\Big] .
\eeq
Notice that, indeed, the ansatz (\ref{metric1}) is of the same type  as the one considered in eq. (\ref{ansatzKW}) for the deformation of $AdS_5\times T^{1,1}$.  In addition our background  must have a RR five-form:
\beq  \label{F51}
F_5\,=\,K(r)\,dx^0\wedge\cdots dx^4\wedge dr\,+\,{\rm Hodge\,\,dual} ,
\eeq
and a RR one-form $F_1$ which violates Bianchi identity. Recall that this violation, which we want to be compatible with supersymmetry,  is a consequence of having a smeared D7-brane source in our system. Our proposal for $F_1$ is the following:
\beq  \label{F11}
F_1\,=\,C\,(d\tau+A)\,\, ,
\eeq
where $C$ is a constant which should be related to the number of 
flavours. Moreover, the violation of 
the Bianchi
identity is the following\footnote{We are considering that $J={1 \over 2}J_{ab}dx^a \wedge dx^b$ and that the Ricci tensor of the KE space satisfies $R_{ab}=\,6\,g_{ab}$.}:
\beq
dF_1\,=\,2 \, C\,\, J .
\label{dF_1=J}
\eeq
Notice that eq. (\ref{dF_1=J}) corresponds to taking $\Omega=-2CJ$ in our general expression 
(\ref{Bianchi-general1}). 
To proceed with 
this proposal
we should try to solve the Killing spinor equations by imposing the appropriate
projections.  Notice that the
ansatz is compatible with the K\"ahler structure of the KE  base and
this is usually related to supersymmetry.

Before going ahead, it may be useful to make contact with the explicit case studied in the previous section, namely the Klebanov-Witten model. In that case the KE base is
\beq
ds^2_{KE}\,=\,{1 \over 6} \sum_{i=1,2} ( d\theta^2_{i}\,+\,\sin^2{\theta_{i}}\,d\varphi^2_{i} ) \,\, ,
\eeq
whereas the one form dual to the Killing vector $\partial / \partial \tau$ is $d\tau=d\psi / 3$ and the form $A$ reads
\beq
A\,=\,{1 \over 3}\Big( \cos{\theta_{1}}\,d\varphi_{1} \,+\, \cos{\theta_{2}}\,d\varphi_{2}\Big)\,\,.
\eeq
Moreover, the constant $C$ was set to ${{3\,N_{f}} \over {4\pi}}$ in that case. 

Let us choose the following frame for the ten-dimensional metric:
\begin{equation} \begin{aligned}
\hat{e}^{x^{\mu}} &=\,\big[\,h(r)\,\big]^{-{1\over 4}}\,dx^{\mu} \,, &
\hat{e}^{r} &=\,\big[\,h(r)\,\big]^{{1\over 4}}\,dr\,\,, \\
\hat{e}^0 &=\,\big[\,h(r)\,\big]^{{1\over 4}}\,e^{f(r)}\,
(d\tau+A)\,\, ,  \qquad\qquad &
\hat{e}^a &=\,\big[\,h(r)\,\big]^{{1\over 4}}\,e^{g(r)}\,e^a 
\,\,,
\label{10vielbein}
\end{aligned} \end{equation}
where $e^a  \quad a=1,\ldots,4$ is the one-form basis for the KE space such that 
$ds^2_{KE}\,=\,e^a\, e^a $. In the Klebanov-Witten model the basis taken in (\ref{T11frame}) corresponds
to:
\begin{equation} \begin{aligned}
e^{1} &=\sin{\theta_{1}}\,d\varphi_{1} \,\,, \qquad\qquad &
e^{2} &=\,\,d\theta_{1} \,\,, \\
e^{3} &=\sin{\theta_{2}}\,d\varphi_{2} \,\,, &
e^{4} &=\,\,d\theta_{2} \,\,.
\end{aligned} \end{equation}
Let us write the five-form $F_5={\cal F}_5+{}^*{\cal F}_5$ of eq. (\ref{F51})
in frame components:
\bear
&&{\cal F}_5\,=\,K(r)\,\big[\,h(r)\,\big]^{{3\over 4}}\,
\hat{e}^{x^{0}}\wedge\cdots \wedge \hat{e}^{x^{3}}\wedge \hat{e}^{r} \,\,,\rc
&&{}^*{\cal F}_5\,=\,-K(r)\,\big[\,h(r)\,\big]^{{3\over 4}}\,
\hat{e}^{0}\wedge\cdots \wedge \hat{e}^{4}\,=\,-
Kh^2\,e^{4g+f}\,(d\tau+A)\wedge e^1 \wedge \dots \wedge e^4 .\rc
\eear
The equation $dF_5=0$ immediately implies:
\beq \label{Bian}
Kh^2e^{4g+f}\,=\,{\rm constant}\,=\,{(2\pi)^4 N_c\over {Vol(M_5)}}\,\,,
\eeq
where the constant has been obtained by imposing the quantisation condition (\ref{quantizationF5})
for a generic $M_5$.  It will also be useful in what follows to write the one-form $F_1$ in frame components:
\beq
F_1\,=\,C\,h^{-{1 \over 4}}\,e^{-f}\,\hat{e}^{0} .
\label{F1frame}
\eeq
Let us list the non-zero components of the spin connection:
\begin{equation} \begin{aligned}
\hat{\omega}^{x^{\mu}r} &=\,-{1\over 4}\,h'\,h^{-{5\over 4}}\,\,
\hat{e}^{x^{\mu}}\,\,,\qquad\qquad (\mu=0,\cdots, 3)\,\,, \\
\hat{\omega}^{ar} &=\,\Big[\,{1\over 4}\,{h'\over h}\,+\,g'\,\,\Big]\,
h^{-{1\over 4}}\,\,\hat{e}^{a}\,\,,\qquad\qquad (a=1,\cdots, 4)\,\,, \\
\hat{\omega}^{0r} &=\,\Big[\,{1\over 4}\,{h'\over h}\,+\,f'\,\,\Big]\,
h^{-{1\over 4}}\,\,\hat{e}^{0}\,\,\,\,, \\
\hat{\omega}^{0}_{\,\,\,\, a} &=\, e^{f-2g}  h^{-{1\over 4}}\, J_{ab}\,\hat{e}^b\,\,
\,\,, \\
\hat{\omega}^{ab} &=\,\omega^{ab}\,-\,
e^{f-2g}  h^{-{1\over 4}}\, J^{ab}\hat{e}^0 \,\, ,
\label{spinconection}
\end{aligned} \end{equation}
where $\omega^{ab}$ are components of the spin connection of the KE base.

Let us now study under which conditions our ansatz preserves some amount of supersymmetry. To address this point we must look at the supersymmetric variations of the dilatino and gravitino (\ref{Eframe1}). We will take them but using the following complex spinor notation\footnote{Notice that it is different from that taken in eq. (\ref{rule}).}:
\beq
\epsilon^*\,\leftrightarrow\,\sigma_3\,\epsilon\,\,,
\,\,\,\,\,\,\,\,\,\,\,\,\,\,\,\,\,\,\,
-i\epsilon^*\,\leftrightarrow\,\sigma_1\,\epsilon\,\,,
\,\,\,\,\,\,\,\,\,\,\,\,\,\,\,\,\,\,\,
i\epsilon\,\leftrightarrow\, i\sigma_2\,\epsilon\,\,.
\eeq

It is quite obvious from the form of our ansatz for $F_1$ in (\ref{F1frame})  that the equation resulting from the dilatino variation is:
\beq
\big ( \phi'\, - \,i\,e^{\phi}\,C\,e^{-f}\,\Gamma_{r0} \big )\, \epsilon\,=\,0\,\,.
\label{BPSdilatino}
\eeq
In eq. (\ref{BPSdilatino}), and in what follows, the indices of the $\Gamma$-matrices refer to the vielbein components (\ref{10vielbein}). 

Let us move on to the more interesting case of the gravitino transformation. The space-time and the radial components of the equation do not depend on the structure of the internal space and always yield the following two equations:
\bear
&&h'\,+\,K\,h^2\,=\,0 \,\, , \rc
&&\partial_r \epsilon \,-\, {1 \over 8}\,K\,h\, \epsilon\,=\,0\,\, .
\label{hBPS}
\eear
To get eq. (\ref{hBPS}) we have imposed  the D3-brane projection
\beq \label{proj3}
\Gamma_{x^0x^1x^2x^3}\, \epsilon\,=\, -i \,\epsilon\,,
\eeq
and we have used the fact that the ten-dimensional spinor is chiral with chirality
\beq \label{chiral}
\Gamma_{x^0\ldots x^3r01234}\,\epsilon\,=\,\,\epsilon \,\, .
\eeq
It is a simple task to integrate the second  differential equation  in (\ref{hBPS}):
\beq \label{spinor}
\epsilon\,=\,h^{-{1 \over 8}} \hat{\epsilon}\,\, ,
\eeq
where $\hat{\epsilon}$ is a spinor which can only 
depend on the coordinates of the Sasaki-Einstein space.

In order to study the variation of the SE components of the gravitino it is useful to write the covariant derivative along the SE directions
in terms of the covariant derivative in the KE space. The covariant derivative, written as a one-form for those components, $\hat{D}\equiv d \, + \, {1 \over 4} \, \hat{\omega}_{IJ}\, \Gamma^{IJ}$, is given by
\bear
&&\hat{D}\,=\, D\, -\,{1 \over 4}\,J_{ab}\,h^{-{1 \over 4}}\,e^{f-2g}\,\Gamma^{ab}\,\hat{e}^0\,-\,{1 \over 2}\,J_{ab}\,h^{-{1 \over 4}}\,e^{f-2g}\,\Gamma^{0b}\,\hat{e}^a\,+\,  \rc
&&+\, {1 \over 2}\,h^{-{1 \over 4}}\,\big ( {1 \over 4}\,{h' \over h}\,+\,g' \big )\,\Gamma^{ar}\,\hat{e}^a\,+\,{1 \over 2}\,h^{-{1 \over 4}}\,\big ( {1 \over 4}\,{h' \over h}\,+\,f' \big )\,\Gamma^{0r}\,\hat{e}^0 \,\, ,
\eear
where $D$ is the covariant derivative in the internal KE space.

The equation for the SE components of the gravitino transformation is
\beq
\hat{D}_I \, \epsilon\, - \, {1 \over 8}\,K\,h^{{3 \over 4}}\, \Gamma_{rI}\,\epsilon\,+\,{i \over 4}\,e^{\phi}\,F^{(1)}_I\,\epsilon\,=\,0.
\eeq
This equation can be split  into a part coming from the coordinates in the KE space and a part coming from the coordinate which parameterises the Killing vector. For this purpose, it is convenient to represent the frame one-forms $e^a$ and the fiber one-form $A$ in a coordinate basis of the KE space
\begin{equation} \begin{aligned}
e^a &=\,E^a_m \, dy^m\, , \\
A &=\,A_m\,dy^m\, ,
\end{aligned} \end{equation}
with $y^m \quad m=1,\ldots,4$ a set of space coordinates in the KE space.

After a bit of algebra one can see that the equation obtained for the space coordinates $y^m$ is simply
\begin{equation} \begin{aligned}  \label{eq1}
& D_m \, \epsilon \, -\,{1 \over 4}\,J_{ab}\,e^{2(f-g)}\,A_m \, \Gamma^{ab}\,\epsilon \,-\,{1 \over 2}\,J_{ab}\,h^{-{1 \over 4}}\,e^{f-2g}\,E^a_m\, \Gamma^{0b}\,\epsilon \,+\,  \\
& +\, {1 \over 2}\, h^{-{1 \over 4}}\,\big ( {1 \over 4}\,{h' \over h}\,+\,g' \big )\,E^a_m\,\Gamma^{ar}\,\epsilon\,+\,{1 \over 2}\,\big ( {1 \over 4}\,{h' \over h}\,+\,f' \big )\, e^f \, A_m \, \Gamma^{0r}\,\epsilon \,-\, \\
& -\, {1 \over 8}\, K \, h^{{3 \over 4}}\, \big ( E^a_m\, \Gamma^{ra}\,+\,h^{{1 \over 4}}\,e^f\,A_m\,\Gamma^{r0} \big )\,\epsilon \,+\, {i \over 4}\,e^{\phi}\,C\,A_m\,\epsilon\,=\,0\,\, ,
\end{aligned} \end{equation}
whereas the equation obtained for the fiber coordinate $\tau$ is given by
\begin{equation} \begin{aligned}  \label{eq2}
& {\partial \epsilon \over \partial \tau} \, -\,{1 \over 4}\,J_{ab}\,e^{2(f-g)} \, \Gamma^{ab}\,\epsilon \,+\,{1 \over 2}\,\big ( {1 \over 4}\,{h' \over h}\,+\,f' \big )\, e^f  \, \Gamma^{0r}\,\epsilon \, -\,\\
& -\, {1 \over 8}\, K \, h \, e^f \,\Gamma^{r0} \,\epsilon \,+\, {i \over 4}\,e^{\phi}\,C\,\epsilon\,=\,0\,\, .
\end{aligned} \end{equation}

Let us now solve these equations for the spinor $\epsilon$. First of all, let us consider the dilatino equation (\ref{BPSdilatino}).  Clearly, this equation implies that the spinor must be an eigenvector of the matrix $\Gamma_{r0}$. Accordingly, let us require that $\epsilon$ satisfies 
\beq
\Gamma_{r0}\,\epsilon=\,-\,i \,\epsilon \,\, .
\label{rprojection}
\eeq
Moreover, a glance at eqs. (\ref{eq1}) and (\ref{eq2}) reveals that $\epsilon$ must also be an eigenvector of the matrix $J_{ab}\Gamma^{ab}$. Actually, by combining eqs. (\ref{proj3}) , (\ref{chiral})  and (\ref{rprojection}) one easily obtains  that
\beq
\Gamma_{12}\epsilon\,=\,\Gamma_{34}\epsilon\,\,.
\label{12=34}
\eeq
To simplify matters, let us assume that we have chosen the one-form basis $e^a$ of the KE in such a way that the K\"ahler two-form $J$ takes the canonical form:
\beq
J\,=\, e^1\wedge e^2\,+\,e^3\wedge e^4\,\,.
\label{canonicalJ}
\eeq
In this basis, after using the condition (\ref{12=34}),  one trivially gets:
\beq
J_{ab}\Gamma^{ab}\,\epsilon\,=\,4\Gamma_{12}\,\epsilon\,\,.
\eeq
Thus, in order to diagonalize $J_{ab}\Gamma^{ab}$, let us impose the projection
\beq
\Gamma_{12}\,\epsilon\,=\, -i\epsilon\,\,,
\label{Gamma12}
\eeq
which implies 
\beq
\Gamma_{34}\,\epsilon\,=\,- i\epsilon\,\,,
\qquad\qquad
J_{ab}\Gamma^{ab}\,\epsilon\,=\,-4i\epsilon\,\,.
\eeq

Let us now use the well-known fact that any KE space admits a covariantly constant spinor $\eta$ satisfying:
\beq
D_m\,\eta\,=\,-{3 \over 2}\,i\,A_m\,\eta\,\, ,
\label{cov-eta}
\eeq
from which one can get a Killing spinor of the five-dimensional SE space as:
\beq
\hat{\epsilon}\,=\,e^{-i\,\,{3 \over 2}\tau}\,\eta\,\, .
\label{hatepsilon}
\eeq
Actually, in the KE frame basis we are using, $\eta$ turns out to be a constant spinor which satisfies the conditions $\Gamma_{12}\,\eta\,=\, \Gamma_{34}\,\eta\,=\, -i\eta$. Let us now insert the SE Killing spinor $\hat{\epsilon}$ of eq.  (\ref{hatepsilon}) in our ansatz (\ref{spinor}), \ie\ we take the solution of  our SUSY equations to be:
\beq
\epsilon\,=\,h^{-{1 \over 8}}\,e^{-{3 \over 2}\,i\tau}\,\eta\,\, .
\label{spinorsol}
\eeq
By plugging (\ref{spinorsol}) into eqs. (\ref{eq1}) and (\ref{eq2}), and using the projections imposed to $\epsilon$ and (\ref{cov-eta}), one can easily see that eqs. (\ref{eq1}) and (\ref{eq2}) reduce to the following  two differential equations:
\bear
&&{1 \over 4} {h' \over h}\,+\,g'\,+\,{1 \over 4}\,K\,h\,-\,e^{f-2g}\,=\,0\,\, , \rc
&&{1 \over 4} {h' \over h}\,+\,f'\,+\,{1 \over 4}\,K\,h\,+2\,e^{f-2g}\,-\,3\,e^{-f}\,+ \,{C \over 2}e^{\phi-f}\,=\,0\,\, .
\eear
By combining all equations obtained so far in this subsection we arrive at a system of first-order BPS equations for the deformation of any space of the form $AdS_5\times M_5$:
\bear  \label{BPS}
&& \phi'\, -  \, C \,e^{\phi-f}\,=\,0\,\, , \rc
&&h'\,+\,{(2\pi)^4 N_c\over {Vol(M_5)}}\,\, e^{-f-4g}\,=\,0 \,\, , \rc
&&\,g'\,-\,e^{f-2g}\,=\,0\,\, , \rc
&&\,f'\,+2\,e^{f-2g}\,-\,3\,e^{-f}\,+\,{C \over 2}e^{\phi-f}\,=\,0\,\, .
\eear
Notice that, indeed, this system reduces to the one written in eq. (\ref{BPSsystem}) for the conifold, if we take into account that for this later case the constant C is $3N_f/(4\pi)$ and $Vol(T^{1,1})=16 \pi^3 /27$. 

It is now a simple task to count the supersymmetries of the type (\ref{spinorsol})
preserved by our background: it is just thirty-two divided by the number of
independent algebraic projection imposed to the constant spinor $\eta$. As a
set of independent projections one can take the ones written in eqs.
(\ref{proj3}), (\ref{rprojection}) and (\ref{Gamma12}). It follows that our deformed background
preserves four supersymmetries generated by  Killing spinors of the type
displayed in eq. (\ref{spinorsol}).

\subsection{The BPS and Einstein Equations} \label{BPS-Einstein}

In this subsection we will prove that the BPS system implies the fulfilment of
the second-order Euler-Lagrange equations of motion for the combined
gravity plus brane system (see eq. (\ref{eom+flavour})). To begin with, let us consider the equation of
motion of the dilaton, which can be written as:
\beq
{1 \over {\sqrt{-G}}}\partial_M \Big ( G^{MN}\,
\sqrt{-G}\,\,\partial_N\phi \Big
)\,=\,e^{2\phi}\,F^2_{1}\,-\,
{2\kappa^2_{10}\over \sqrt{-G}}\,\,
{\delta\over \delta \phi}\,\,S_{DBI}\,\,,
\label{dilatoneq-general}
\eeq
where $G_{MN}$ is the ten-dimensional metric. Using the DBI action (\ref{DBIaction-general}) for the smeared D7-branes configuration, we find:
\begin{equation}
-\frac{2\kappa_{10}^2}{\sqrt{-G}} \frac{\delta}{\delta \phi} S_{DBI} = e^\phi \, \sum\nolimits_i
\big|\Omega^{(i)}\big|\;.
\end{equation}
The charge density distribution is 
$\Omega=-2CJ$ (see eq. (\ref{dF_1=J})).  
Recall that the K\"ahler form $J$ of the KE base manifold has the canonical
expression (\ref{canonicalJ}). It follows that $\Omega$ has two decomposable components
given by:
\begin{equation} \begin{aligned}
\Omega^{(1)} &=\,-2C\,e^1\wedge e^2\,=\,-2C\,h^{-{1\over 2}}\,e^{-2g}\,
\hat e^1\wedge \hat e^2\,\,, \\
\Omega^{(2)} &=\,-2C\,e^3\wedge e^4\,=\,-2C\,h^{-{1\over 2}}\,e^{-2g}\,
\hat e^3\wedge \hat e^4\,\,,
\label{dFi}
\end{aligned} \end{equation}
where the $\hat e^a$ one-forms have been defined in (\ref{10vielbein}).
Therefore, the moduli of the $\Omega^{(i)}$'s can be straightforwardly
computed:
\beq
\big|\Omega^{(1)}\big|\,=\,\big|\Omega^{(2)}\big|\,=\,
2|\,C\,|\,h^{-{1\over 2}}\,e^{-2g}\,\,.
\label{|dFi|}
\eeq
By using the explicit form of the metric, our ansatz for $F_1$ and the previous formula (\ref{|dFi|}) one can
convert eq. (\ref{dilatoneq-general}) into the following:
\beq
\phi''\,+\,(4g'+f')\,\phi'\,=\,C^2\,e^{2\phi-2f}\,+\,4\,|C|\,\,e^{\phi-2g}\,\,.
\label{dil-eq2}
\eeq
It is now a simple exercise to verify that eq. (\ref{dil-eq2}) holds if the
functions $\phi$, $g$ and $f$ solve the first-order BPS system (\ref{BPS})
and the constant $C$ is non-negative. In what follows we shall assume that
$C\ge 0$.

To check the Einstein equations we need to calculate the Ricci tensor. 
In flat coordinates the components of the Ricci tensor can be computed by
using the spin connection. The expression of the curvature two-form in terms
of the spin connection is
\beq
 R_{\hat M\hat N}\,=\,d\hat\omega_{\hat M\hat N}\,+\,
\hat\omega_{\hat M\hat P}\wedge
\hat\omega^{\hat P}_{\,\, \hat N}\,\,  ,
\eeq
with the curvature two-form defined as follows:
\beq
R^{\hat M}_{\,\, \hat N}\,=\,{1 \over 2}\,
R^{\hat M}_{\,\, \hat N \hat P\hat Q}\,e^{\hat P}
\wedge e^{\hat Q}\,\,  .
\eeq 
By using the values of the different components of the ten-dimensional spin
connection written in (\ref{spinconection}) we can easily obtain  
the Riemann tensor and, by  simple contraction of indices, we arrive at the
following flat components of the Ricci tensor:
\begin{equation} \begin{aligned}
R_{x^ix^j} &=\,h^{-{1 \over 2}}\,\eta_{x^ix^j}
\Bigg(\,\, {1 \over 4}{h'' \over h}\,-\,{1 \over 4}\Bigg({h' \over
h}\Bigg)^2\,+\,{1 \over 4}{h' \over h}f'\,+\,{h' \over h}g' \Bigg)\,\,  ,  \\
R_{rr} &=\,h^{-{1 \over 2}}\,\Bigg( -{1 \over 4}{h'' \over h}\,-\,{1
\over 4}\Bigg({h' \over h}\Bigg)^2\,-\,{1 \over 4}{h' \over h}f'
\,-\,{h' \over h}g'
\,-\,f''\,-\,(f')^2\,-\,4\,g'' \,-\,4(g')^2\,
\Bigg)\,\,  ,  \\
R_{00} &=\,h^{-{1
\over 2}}\,\Bigg( -{1 \over 4}{h'' \over h}\,+\,{1 \over 4}\Bigg({h' \over
h}\Bigg)^2\,-\,{1 \over 4}{h' \over h}f'\,-\,{h' \over h}g'
\,-\,f''\,-\,(f')^2\,-\,4\,g'\,f'\,+\,4\,e^{2f-4g} \,\Bigg)\,\,  ,  \\
R_{aa} &=\,h^{-{1 \over 2}}\,\Bigg( -{1 \over 4}{h'' \over h}\,+\,{1 \over
4}\Bigg({h' \over h}\Bigg)^2\,-\,{1 \over 4}{h' \over h}f'\,-\,{h' \over h}g'
\,-\,g''\,-\,\\
&\qquad\qquad\qquad\qquad\qquad\qquad\qquad\qquad
-\,4\,(g')^2\,-\, g'\,f'\,-\,2\,e^{2f-4g}\,+\,6\,e^{-2g}\, \Bigg)\,\,  , \\
R_{\hat M \hat N} &=\,0\,\,,
\qquad \qquad M \neq N \,\, .
\label{Ricci-components}
\end{aligned} \end{equation}
From these values it is straightforward to find 
the expression of the scalar curvature (in Einstein frame), which  is simply
\begin{multline} \label{scalar}
R\,=\,-h^{-{1 \over 2}}\,\Bigg(\, {1 \over 2}{h'' \over h}+
{1 \over 2}{h' \over h}f'+2\,{h' \over h}g'\,
+\,8\,g''+20\,(g')^2\,+\, \\
+8\,g'\,f'+2\,f''+2\,(f')^2+4\,e^{2f-4g}-24\,e^{-2g} \Bigg)\,\,  .
\end{multline}
Let us  evaluate  the different
contributions to the right-hand side of the Einstein equations (see eq. (\ref{eom+flavour})).  The
contributions from the five- and one-forms is immediately computable from our ansatz 
of eqs. (\ref{F51}) and (\ref{F11}). On the other hand, the contribution of
the DBI part of the action is just
\beq
T_{MN}\,=\,-\,{2\kappa_{10}^2 \over \sqrt{-G}}\,\,
{\delta S_{DBI} \over \delta G^{MN}}\,\,.
\eeq
By using our expression (\ref{DBIaction-general}) of $S_{DBI}$, with
$\Omega=-dF_1$, together with the definition (\ref{modulus}), one easily
arrives at the following expression of the stress-energy tensor of the
D7-brane:
\beq
T_{\hat M\hat N} = -\frac{e^\phi}{2} \,
\Big[ \eta_{\hat M\hat N} \, \sum_i \,
\big|\,\Omega^{(i)}\,\big| - 
\sum_i \, {1\over \big|\,\Omega^{(i)}\,\big|}\,\,
(\Omega^{(i)})_{\hat M\hat P}\,\,
(\Omega^{(i)})_{\hat N\hat Q} \, \eta^{\hat P\hat Q} \, \Big] \;,
\label{TMN}
\eeq
where we have used that $2\kappa_{10}^2 T_7\,=\,1$ and we have written the
result in flat components. By using in (\ref{TMN}) the values given in eqs. 
(\ref{dFi}) and (\ref{|dFi|}) of $\Omega^{(i)}$ and its modulus, we arrive
at the simple result:
\bear
&&T_{x^ix^j}\,=\,-2C\,h^{-{1\over 2}}\,e^{\phi-2g}\,\,\eta_{x^ix^j}
\,\,,\rc
&&T_{rr}\,=\,T_{00}\,=\,-2C\,h^{-{1\over 2}}\,e^{\phi-2g}\,\,,\rc
&&T_{ab}\,=\,-C\,h^{-{1\over 2}}\,e^{\phi-2g}\,\delta_{ab}\,\,,
\qquad\qquad (a,b=1,\cdots, 4)\,\,,
\label{TMNcomponents}
\eear
where the indices refer to our vielbein basis (\ref{10vielbein}). 

With all this information we can write, component by component, the set of
second-order differential equations for $h$, $g$, $f$ and $\phi$ that are
equivalent to the Einstein equations. One can then verify, after some
calculation, that these equations are satisfied if $\phi$ and the functions
of our ansatz solve the first-order system (\ref{BPS}). Therefore, we have
succeeded in proving that the background obtained from the supersymmetry
analysis is a solution of the equations of motion of the supergravity plus
Born-Infeld system. Notice that the SUSY analysis determines $F_1$, \ie\ the
RR charge distribution of the smeared D7-branes. What we have just proved is
that eq. (\ref{TMN}) gives the correct stress-energy distribution
associated to the charge distribution $\Omega=-dF_1$ of smeared
flavour branes.

To finish this subsection let us write the  DBI action in a different, and very
suggestive, fashion. It turns out that, for our ansatz, the on-shell  DBI action 
 can be written as the
integral of a ten-form and the corresponding expression is very similar to
the one for the WZ term given in equation (\ref{WZaction-general}). Actually, we show
below that
\beq
S_{DBI} =  T_7 \, \int_{{\cal M}_{10}} \, e^{\phi} \,
\Omega\wedge \Omega_8\,\,,
\label{SDBI-forms}
\eeq
where $\Omega_8$ is an eight-form which, after performing the wedge product
with the smearing two-form $\Omega$, gives rise to a volume form of the
ten-dimensional space. Let us factorise in $\Omega_8$ the factors coming
from the Minkowski directions:
\beq
\Omega_8\,=\,h^{-1}\,d^{4}x \wedge \Omega_4\, ,
\label{Omega_8}
\eeq 
where $\Omega_4$ is a four-form in the internal space. Actually, one can
check that $\Omega_4$ can be written as:
\beq \label{cal}
\Omega_4\,=\,{1 \over 2}\,\,{\cal J} \wedge {\cal J},
\eeq
where ${\cal J}$ is the following two-form:
\beq
{\cal J}\,=\,h^{{1 \over 2}}\,e^{2g}\,J\,+\,h^{{1 \over 2}} 
e^{f}\,dr\wedge(d\tau\,+\,A)\, .
\label{calJ}
\eeq
To verify this fact, let us recall that $\Omega = -2CJ$ and thus
\beq
\Omega \wedge \Omega_8 = -\,C\,h^{-1}\,
d^{4}x\,\wedge J\wedge {\cal J} \wedge {\cal J} \;.
\eeq
Taking into account that ${1\over 2}\,\,J\wedge
J$ is the volume form of the KE base of $M_5$, we readily get:
\beq
d^{4}x\,\wedge\,J\wedge {\cal J} \wedge {\cal J}\,=\,4e^{-2g}\,
h^{{1\over 2}}\,\,\sqrt{-G}\,\,d^{10}x\,\,,
\eeq
from where one can easily prove that eq. (\ref{SDBI-forms})  gives
the same result as in equation (\ref{DBIaction-general}) with
$\Omega=-dF_1$.

\subsection{A Superpotential and the BPS Equations }
\label{SuperpotentialBPS}

It is interesting to obtain the  system of first-order BPS 
equations (\ref{BPS}) by  using
an alternative approach, namely by deriving them from a superpotential. 
Generically, let us consider a one-dimensional classical mechanics system in
which  $\eta$ is the ``time" variable and 
${\cal A}(\eta)$, $\Phi^m(\eta)$ ($ m=1,2\ldots)$ are the generalised
coordinates.  Let us assume that the Lagrangian of this system takes the
form:
\beq \label{lagran}
 L\,=\,e^{\cal A} \Big [ \kappa \,\,(\partial_{\eta}{\cal A})^2\,-\,{1 \over
2}G_{mn}(\Phi)\,\partial_{\eta}\Phi^m
\,\partial_{\eta}\Phi^n\,-\,V(\Phi) \Big ]\,\, ,
\eeq
where $\kappa$ is a constant and $V(\Phi)$ is some potential, which we 
assume that is independent of the coordinate ${\cal A}$. 
If one can find a superpotential $W$ such that:
\beq \label{potential}
V(\Phi)\,=\,{1 \over 2}G^{mn}\,{{\partial W} \over 
{\partial\Phi^m}}\,{{\partial W} \over {\partial\Phi^n}}\,-\,{1
\over {4\kappa}}\,\,W^2\,\,  ,
\eeq
then the equations of motion are automatically satisfied 
by the solutions of the first-order system:
\beq \label{eqs}
{{d\,{\cal A}} \over {d\eta}}\,=\,-{1 \over {2\kappa}}W \,\,  ,  \qquad
{{d\,\Phi^m} \over {d\eta}}\,=\, G^{mn}\,{{\partial W} \over {\partial
\Phi^n}} \,\,  .
\eeq
 Let us now show how we can recover our system (\ref{BPS})
from this formalism.
 The first step is to look for an effective Lagrangian for the dilaton and
the functions of our ansatz
whose equations of  motion are the same as those obtained from the Einstein
and dilaton equations of type IIB supergravity. One can see that this
lagrangian is:
 \beq
 L_{eff}\,=\,h^{{1\over 2}}\,e^{4g+f}\, \Big 
[ R\,-\,{1\, \over 2}\,h^{-{1\over 2}}\,
(\phi')^2\,-\,{Q^2\over 2}\,\,h^{-{5\over 2}}\,e^{-8g-2f}\,-\,
{C^2\over 2}\,\,h^{-{1\over 2}}\,e^{2\phi-2f}
\,-\,4\,C\,h^{-{1\over 2}}\,e^{\phi-2g}\, \Big ]\,\, ,
 \eeq
where $R$ is the scalar curvature written in (\ref{scalar}) and
$Q$ is the constant
\beq
Q\,\equiv\,{(2\pi)^4 N_c\over {Vol(M_5)}}\,\,.
\label{Q}
\eeq
The Ricci scalar (\ref{scalar})  contains second derivatives. Up to total derivatives $L_{eff}$ takes the form:
 \bear
 L_{eff}&=&e^{4g+f} \Bigg[  -{1 \over 2}\Bigg({h'
\over h}\Bigg)^2\,+\,12\,(g')^2\,+\,8\,
g'\,f'\,-\,4\,e^{2f-4g}\,+\,24\,e^{-2g}\,-\,{1 \over 2}(\phi')^2\,\,- \rc
&&\qquad\qquad
-{Q^2\over 2}\,\,h^{-2}\,e^{-8g-2f}\,
-\,{C^2 \over 2}\,e^{2(\phi-f)}\,-\,4\,C\,e^{\phi-2g}
\Bigg]\,\, .
\label{Leff-in-r}
 \eear
We want to pass from  the lagrangian (\ref{Leff-in-r}) to that
in eq. (\ref{lagran}). With that purpose in mind let us perform 
the following redefinition of fields:
\beq \label{fields}
e^{{3\over 4}\,{\cal A}}\,=\,
h^{{1 \over 2}}e^{4g+f}\,\,,
\qquad  \qquad 
e^{2\tilde{g}}\,=\,h^{{1 \over 2}}e^{2g} , \qquad  \qquad  
e^{2\tilde{f}}\,=\,h^{{1 \over 2}}e^{2f} .
 \eeq
In addition,  we need to do the following  change of the radial variable
\footnote{The change of the Lagrangian under that change of the radial
variable is $\hat{ L}_{eff}={{dr} \over {d\eta}}\,\,
 L_{eff}$.} 
\beq \label{radial}
{{dr} \over {d\eta}}\,=\,
e^{{{\cal A}\over 4}-{8\over 3}\tilde g\,
-\,{2\over 3}\tilde f}\,\,.
\eeq
Once we have done the previous redefinitions, the Lagrangian we obtain
is:
\beq
 \hat{  L}_{eff}\,=\,e^{{\cal A}} 
\Bigg[\,\,{3\over 4}\,\, (\dot{{\cal A}})^2\,-\,{28 \over
3}\,(\dot{\tilde{g}})^2\,-\,{4
\over 3}(\dot{\tilde{f}})^2\,-\,{8 \over 3}\,
{\dot{\tilde{g}}}\,{\dot{\tilde{f}}}\,-\,{1 \over
2}(\dot{\phi})^2\,-\,V(\tilde{g},\tilde{f},\phi) \Bigg]\,\,,
\label{Leff-in-eta}
\eeq
where the dot means derivative with respect to $\eta$ and 
$V(\tilde{g},\tilde{f},\phi)$ is the following potential:
\beq
V(\tilde{g},\tilde{f},\phi)\,=\,e^{-{2 \over
3}(4\tilde{g}+\tilde{f})} \Bigg(
\,4\,e^{2\tilde{f}-4\tilde{g}}-24\,e^{-2\tilde{g}}
\,+\,{Q^2 \over 2}\,e^{-2(4\tilde{g}+\tilde{f})}
\,+\,{C^2 \over 2}\, e^{2(\phi-\tilde{f})}
\,+\,4\,C\,e^{\phi-2\tilde{g}}\,\,
\Bigg)\,\,.
\label{tilde-pot}
\eeq
The above lagrangian has the desired form (see eq. (\ref{lagran}))
 and we can identify the constant $\kappa$ and the elements of the
kinetic matrix $G_{mn}$ as:
\beq
 \kappa\,=\,{3 \over 4}\,\, , 
\quad G_{\tilde{g}\tilde{g}}\,=\,{56 \over 3}\,\, , \quad
G_{\tilde{f}\tilde{f}}\,=\,{8 \over 3}\,\, , \quad
G_{\tilde{g}\tilde{f}}\,=\,{8 \over 3}\,\, , 
\quad G_{\phi \phi}\,=\,1\,\, .
\label{constants}
\eeq
One can now check that, given the above expression of  the
potential,  the following superpotential
\beq
W\,=\,e^{-{1 \over 3}(4\tilde g+\tilde f)} \Big [ Q\,
e^{-4\tilde g-\tilde f}\,-\,4\,e^{\tilde f-2\tilde g}\,-\,6e^{-\tilde f}\,+\,Ce^{\phi-\tilde f}  \Big ]\,\, 
\eeq
satisfies  eq. (\ref{potential}) for the values of $\kappa$ and 
$G_{mn}$ written in eq. (\ref{constants}). It is now immediate to write
the first-order differential equations that stem from this superpotential. 
Explicitly we obtain:
\bear
&&\dot{{\cal A}}\,=\,-\,{2 \over 3}\,\, W \,\, , \rc
&&\dot{\tilde{g}}\,=\,{1 \over 4}e^{-{1 \over 3}(4\tilde{g}+\tilde{f})}\, 
\Big [ - Q e^{-4\tilde{g}-\tilde{f}}\,+\,4\,e^{\tilde{f}-2\tilde{g}}\,
\Big ]\,\, , \rc &&\dot{\tilde{f}}\,=\,{1 \over 4}\,e^{-{1 \over
3}(4\tilde{g}+\tilde{f})}\, \Big [ - Q
e^{-4\tilde{g}-\tilde{f}}\,-\,8\,e^{\tilde{f}-2\tilde{g}}\,+\,12\,e^{-\tilde{f}}\,-\,2\,C\,e^{\phi-\tilde{f}}
\Big ] \,\, , \rc
&&\dot{{\phi}}\,=\,C\,e^{\phi-{4 \over
3}(\tilde{g}+\tilde{f})}\,\, .
\eear
In order to verify that this system is equivalent to the one obtained from
supersymmetry, let us write down explicitly these equations in terms
of  the old radial variable (see eq. (\ref{radial})) and fields (see eqs.
(\ref{fields})). One gets:
\bear
&&{h' \over h}\,+\,8\,g'\,+\,2\,f'\,=\,-  Q\,h^{-1}\,
e^{-4g-f}\,+\,4\,e^{f-2g}\,+\,6e^{-f}\,-\,Ce^{\phi-f}  \,\, , \rc
&&{1 \over 4}{h' \over h}\,+\,g'\,=\, e^{f-2g}\,-\,
{1 \over 4}Q\,h^{-1}\,e^{-4g-f} \,\, , \rc
&&{1 \over 4}{h' \over h}\,+\,f'\,=\, 3\,e^{-f}\,-\,2\,e^{f-2g}
\,-\,{1 \over 4}Q\,h^{-1}\,e^{-4g-f}\,-\,{1 \over 2}C\,e^{\phi-f} 
\,\, , \rc &&\phi'\,=\,C\,e^{\phi-f}\,\, ,
\eear
which  are nothing else than a combination of the system of BPS
equations written in (\ref{BPS}).

Let us now use the previous results to study the five-dimensional effective action
resulting from the compactification along $M_5$ of our solution. The fields
in this effective action are the functions $\tilde f$ and $\tilde g$, which
parameterise the deformations along the fiber and the KE base of $M_5$ 
respectively, and the dilaton. Actually, in terms of the new radial variable
$\eta$ introduced in (\ref{radial}), the ten-dimensional metric can be written as:
\beq
ds^2\,=\,e^{-{2\over 3}\,\,(\,\tilde{f}\,+\,4\,\tilde g\,)}\,\,
\Big[\,e^{{{\cal A}\over 2}}\,dx^{\mu}dx_{\mu}\,+\,d\eta^2\,\Big]\,+\,
e^{2\tilde g}\,ds^2_{KE}\,+\,e^{2\tilde f}\,(d\tau+A)^2\;.
\eeq
The corresponding analysis for the unflavoured theory was
performed in \cite{Klebanov:2000nc,Benvenuti:2005qb}. 
For simplicity, let us work in units in which the $AdS_5$ radius
$L$ is one. Notice that the quantity $Q$ defined in (\ref{Q}) is just
$Q=4L^4$. Thus, in these units $Q=4$. To make contact with the analysis of
refs. \cite{Klebanov:2000nc,Benvenuti:2005qb}, let us introduce new fields
$q$ and $p$ which, in terms of $\tilde{f}$ and $\tilde{g}$ are defined as
follows%
\footnote{The function $p$ is called $f$ in refs.
\cite{Klebanov:2000nc,Benvenuti:2005qb}.}:
\beq
q\,=\,{2 \over {15}}\,(\,\tilde{f}\,+\,4\,\tilde g\,)\,\, , 
\qquad\qquad
p\,=\,-\,{1 \over {5}}\,(\,\tilde{f}\,-\,\tilde g\,)\,\, .
\label{pq-scalars}
\eeq
In terms of these new fields,  the potential (\ref{tilde-pot}) turns out to
be
\beq
V(p,q,\phi) = 4\,e^{-8q-12p}\,-\,24\,
e^{-8q-2p}\,+\,{C^2 \over
2}\,e^{2\phi-8q+8p}\,+\,8
\,e^{-20q}\,+\,4\,C\,e^{\phi-8q-2p} \;,
\label{pq-pot}
\eeq
and the effective lagrangian (\ref{Leff-in-eta}) can be written as:
\beq
\hat{L}_{eff}\,=\,\sqrt{-g_5}\,\Big[\,R_5\,-\,{1\over 2}\,\dot \phi^2\,-\,20\,\dot
p^2\,-\,30\,\dot q^2\,-V\,\Big]\,\,,
\eeq
where $g_5\,=\,-e^{2{\cal A}}$ is the
determinant of the five-dimensional metric 
$ds^2_5\,=\,e^{{\cal A}\over 2}\, \,dx^{\mu}dx_{\mu}\,+\,d\eta^2$ 
and 
$R_5\,=\,-\Big [2\,\ddot {\cal A}\,+{5\over 4}\,\dot{\cal A}^2\,\Big]$ is its
Ricci scalar. One can check that the minimum of the potential (\ref{pq-pot})
occurs only at $p=q=e^{\phi}=0$, which corresponds to the
conformal $AdS_5\times M_5$ geometry.  Moreover, by expanding $V$ around
this minimum at second order we find out that the fields $p$ and $q$ defined
in (\ref{pq-scalars}) diagonalize the quadratic potential. The corresponding
masses are $m_p^2=12$ and $m_q^2=32$. By using these values in the
mass-dimension relation (\ref{AdS/CFTmass}), we get:
\begin{equation} \begin{aligned}
m_p^2 &= 12 & \qquad\Longrightarrow\qquad \Delta_p &= 6\,\, , \\
m_q^2 &= 32 & \qquad\Longrightarrow\qquad \Delta_q &= 8 \;.
\end{aligned} \end{equation}
These scalar modes $p$ and $q$ are dual to the dimension six and eight operators
discussed in section \ref{sect2}.

\subsection{General Deformation of the Klebanov-Witten Background}
\label{generalKWdef}

In this subsection we will explore the possibility of having a more general
flavour deformation of the $AdS_5\times T^{1,1}$ background. Notice that, as 
$T^{1,1}$ is a $U(1)$ bundle over $S^2\times S^2$, there exists the
possibility of squashing with different functions each of the two $S^2$'s of
the KE base. In the unflavoured case this is precisely the type of
deformation that occurs when the singular conifold is substituted by its
small resolution. For this reason, it is worth considering this type of
metric also in our flavoured background. To be precise, 
let us adopt the following ansatz for the metric, five-form and one-form:
\begin{equation} \begin{aligned}
ds^2 &= h^{-1/2} dx_{1,3}^2 + 
h^{1/2} \Bigg ( dr^2 +
\frac{1}{6} \sum_{i=1,2} e^{2g_i}( d\theta_i^2 + \sin^2 \theta_i \,
 d\varphi_i^2)  + \frac{e^{2f}}{9} (d\psi + \sum_{i=1,2} \cos\theta_i \,
d\varphi_i)^2 \Bigg ) \,\, , \\
F_5 &= (1+\ast) \, d^4x \wedge K\, dr \,\, , \\
F_1 &= \frac{C}{3} ( d\psi + \cos\theta_2\, d\varphi_2 + 
\cos\theta_1\, d\varphi_1 ) \;,
\end{aligned} \end{equation}
where $C=3N_f/4\pi$, all functions depend on $r$ and $g_1(r)$ and $g_2(r)$
are, in general,  different (if $g_1=g_2=g$ we recover our ansatz
(\ref{ansatzKW})).  The equation $dF_5=0$ immediately implies:
\beq \label{Bian1}
Kh^2e^{2g_1+2g_2+f}\,=\,27\pi N_c\,\equiv\,Q\,\,,
\eeq
which allows to eliminate the function $K$ in favour of the other functions
of the ansatz. By following the same steps as in the $g_1=g_2$ case and requiring that the background preserve four supersymmetries, 
we  get a system of first-order BPS equations  for this
kind of deformation, namely:
\begin{equation} \begin{aligned}
\phi' &=  \, C \,e^{\phi-f}\,\, , \\
h' &=\,-Q\,e^{-f-2g_1-2g_2} \,\, , \\
g_i' &=\,e^{f-2g_i}\,\,,\qquad\qquad (i=1,2)\,\, ,\\
f' &=\,3\,e^{-f}\,-\,e^{f-2g_1}\,-\,e^{f-2g_2}\,-\,
{C \over 2}\,\,e^{\phi-f}\,\,.
\label{BPSg1g2}
\end{aligned} \end{equation}
Notice that, as it should, the system (\ref{BPSg1g2})
reduces to eq. (\ref{BPS}) when $g_1=g_2$.

It is not difficult to integrate this system of differential equations
by following the same method that was employed for the $g_1=g_2$ case. First
of all, we change the radial coordinate:
\beq
dr\,=\, e^f \,d\rho\,\,,
\label{rho-generalKW}
\eeq
what allows us to get a new system:
\begin{equation} \begin{aligned} \label{sol1}
\dot{\phi} &=  \, C \,e^{\phi}\, \,\, , \\
\dot{h} &=\,-Q\,e^{-2g_1-2g_2}\,\, , \\
\dot{g}_i &=\,e^{2f-2g_i}\,\, , 
\qquad\qquad (i=1,2)\,\,, \\
\dot f &=\,3\,-\,e^{2f-2g_1}\,-\,e^{2f-2g_2}\,-\,
{C \over 2}\,\,e^{\phi}\,\,,
\end{aligned} \end{equation}
where now the derivatives are taken with respect to the new variable $\rho$.

The equation for the dilaton in (\ref{sol1}) can be integrated immediately,
with the result:
\beq
e^{\phi}\,=\,-{1\over C}\,\,{1\over \rho}\,\,,
\qquad\qquad (\rho<0)\,\,,
\label{dil-g1g2}
\eeq
where we have absorbed an integration constant in a shift of the radial
coordinate.  Moreover, by combining the equations for $g_1$ and $g_2$ one
easily realises that the combination $e^{2g_1}\,-\,e^{2g_2}$ is constant.
Let us write:
\beq
e^{2g_1}\,=\,e^{2g_2}\,+\,a^2\,\,.
\label{g1g2}
\eeq
On the other hand, by using the solution for $\phi(r)$ just found and the
equations for the $g_i$'s in (\ref{sol1}),  the first-order equation for $f$
can be rewritten as:
\beq
\dot{f}\,=\, 3\,-\, \dot{g}_1 \,-\, \dot{g}_2\, +\,
 \frac{1}{2\rho}\,\,,
\eeq
which can be integrated immediately, to give:
\beq
e^{2f+2g_1+2g_2}\,=\,-c\rho e^{6\rho}\,\,,
\label{f-and-gs}
\eeq
with $c$ being an integration constant. This constant can be absorbed by
performing a suitable redefinition. In order to make contact with the case
in which $g_1=g_2$ let us take $c=6$. Then, by combining  (\ref{f-and-gs})
with  the equation of $g_2$, we get
\beq
e^{4g_2+2g_1}\,\dot g_2\,=\,e^{2g_1+2g_2+2f}\,=\,-6\rho e^{6\rho}\,\,,
\eeq
which, after using the relation (\ref{g1g2}), can be integrated with the
result
\beq
e^{6g_2}\,+\,{3\over 2}\,a^2\,e^{4g_2}\,=\,(1-6\rho)\,e^{6\rho}\,+\,c_1\,\,.
\label{cubic}
\eeq
Notice that, indeed, for $a=0$ this equation reduces to the $g_1=g_2$
solution (see eq. (\ref{egKW})). Moreover, by combining eqs. (\ref{g1g2}) and 
(\ref{f-and-gs}) the expression of $f$ can be
straightforwardly written in terms of $g_2$, as follows:
\beq
e^{2f}\,=\,-{6\rho e^{6\rho}\over e^{4g_2}+a^2\,e^{2g_2}}\,\,.
\eeq
It is also easy to get the expression of the warp factor $h$:
\beq
h(\rho)\,=\, - Q \,\int\,
\frac{d\rho}{e^{4g_2} \,+\, a^2\, e^{2g_2}}
\,+\,c_2\,\,.
\eeq
Thus, the full solution is determined in terms of $e^{2g_2}$ which, in turn,
can be obtained from (\ref{cubic}) by solving a cubic algebraic equation. In
order to write the  explicit value of $e^{2g_2}$, let us define the function:
\beq
\xi(\rho)\,\equiv\,(1-6\rho)\,e^{6\rho}\,+\,c_1\,\,.
\eeq
Then, one has:
\beq
 e^{2g_2}\,=\, {1\over 2}\,\,\Bigg[\,-a^2\,+\,
{a^4\over \big[\,\zeta(\rho)\,\big]^{{1\over 3}}}\,+\,
\big[\,\zeta(\rho)\,\big]^{{1\over 3}}\,\,\Bigg]\,\,,
\eeq
where the function $\zeta(\rho)$ is defined in terms of $\xi(\rho)$ as:
\beq
\zeta(\rho)\,\equiv\,4\,\xi(\rho)\,-\,a^6\,+\,
4\,\sqrt{\xi(\rho)^2\,\,-\,{a^6\over 2}\,\xi(\rho)}\,\,.
\eeq
In expanding these functions in series near the UV ($\rho \to 0$) one gets a similar behaviour to the one discussed in subsection \ref{solKW}. Very interestingly, in the IR of the field theory, that is when $\rho \to - \infty$, we get a behaviour that is ``softened'' respect to what we found in subsection \ref{solKW}. This is not unexpected, given the deformation parameter $a$. Nevertheless, the solutions are still singular. Indeed, the dilaton was not affected by the deformation $a$.

\subsection{Massive Flavors}

In the ansatz we have been using up to now, we have assumed that the density
of RR charge of the D7-branes is independent of the holographic coordinate.
This is, of course, what is expected for a flavour brane configuration which
corresponds to massless quarks. On the contrary, in the massive quark case,
a supersymmetric D7-brane has a nontrivial profile in the radial direction
\cite{Arean:2004mm} and, in particular ends at some non-zero value of the
radial coordinate. These massive embeddings have free parameters which could be
used to smear the D7-branes. It is natural to think that the corresponding
charge and mass distribution of the smeared flavour branes will depend on the
radial coordinate in a nontrivial way.

It turns out that there is a simple modification of our ansatz for $F_1$
which gives rise to a charge and mass distribution with the characteristics
required to represent smeared flavour branes with massive quarks. Indeed, let
us simply substitute in (\ref{BPS}) the constant $C$ by a function $C(r)$. In
this case:
\begin{equation} \begin{aligned}
F_1 &=\,C(r)\,(d\tau\,+\,A)\,\,, \\
dF_1 &=\,2\,C(r)\,J\,+\,C'(r) dr \wedge (d\tau\,+\,A)\, .
\end{aligned} \end{equation}
Notice that the SUSY analysis of subsection \ref{SEBPS} remains unchanged since
only $F_1$, and not its derivative, appears in the supersymmetric
variations of the dilatino and gravitino. The final result is just the same
system (\ref{BPS}) of first-order BPS equations, where now one has to
understand that 
$C$ is a prescribed function of $r$, which encodes the nontrivial profile
of the D7-brane. Notice that $C(r)$ determines the running of the dilaton
which, in turn, affects the other functions of the ansatz.

A natural question to address here is whether or not the solutions of the
modified BPS system solve the equations of motion of the supergravity plus
branes system. In order to check this fact, let us write the DBI term of
the action, following our prescription (\ref{DBIaction-general}). Notice
that, in the present case, $\Omega = -dF_1$ is the sum of three decomposable
pieces:
\beq
\Omega\,=\,\Omega^{(1)}\,+\,\Omega^{(2)}\,+\,\Omega^{(3)}\,\,,
\eeq
where $\Omega^{(1)}$ and $\Omega^{(2)}$ are just the same as in eq. \eqref{dFi},
while $\Omega^{(3)}$ is given by:
\beq
\Omega^{(3)}\,=\, -C'(r)\,dr\wedge (d\tau+A)\,=\,
-h^{-{{1 \over 2}}}\,e^{-f}\,C'(r)\,\,\, \hat{e}^{r}\wedge \hat{e}^{0}\,\,.
\eeq
The modulus of this new piece of $\Omega$ can be straightforwardly computed,
namely:
\beq
|\,\Omega^{(3)}\,|\,=\,h^{-{{1 \over 2}}}\,e^{-f}\,
|\,C'(r)\,|\,\,.
\eeq
By using this result, together with the one in (\ref{|dFi|}), one readily
gets the expression of the DBI terms of the action of the smeared D7-branes:
\beq \label{DBIM}
S_{DBI}\,=\,-\,T_7 \int_{{\cal M}_{10}}\,h^{-{{1 \over 2}}}\,
e^{\phi}\,
\Big(4\,|\,C(r)\,|\,e^{-2g}\,+\,|\,C'(r)\,|\,e^{-f}\,\Big)
\,\sqrt{-G}\,\,d^{10}x\,.
\eeq
From this action it is immediate to find the equation of motion of the
dilaton, \ie:
\beq
\phi''\,+\,(4g'+f')\,\phi'\,=\,
C^2\,e^{2\phi-2f}\,+\,4\,|C|\,\,e^{\phi-2g}\,+\,
e^{\phi-f}\,|\,C'\,|\,\,.
\label{dil-eq-massive}
\eeq
It can be verified that the first-oder BPS equations (\ref{BPS}) imply the
fulfilment of eq. (\ref{dil-eq-massive}), provided the functions
$C(r)$ and $C'(r)$ are non-negative. Notice that now, when computing
the second derivative of $\phi$ from the BPS system (\ref{BPS}) with 
$C=C(r)$, a new term containing $C'(r)$ is generated. It is easy to verify
that this new term matches precisely the last term on the right-hand side of
(\ref{dil-eq-massive}). 

It remains to verify the fulfilment of the Einstein equations. The
stress-energy tensor of the brane can be computed from eq. (\ref{TMN}), where
now the extra decomposable piece of $dF_1$  must be taken into account. The
result that one arrives at, in the  vielbein basis (\ref{10vielbein}), is a direct
generalisation of (\ref{TMNcomponents}):
\bear
&&T_{x^ix^j}\,=\,-\,e^{\phi}\,\,h^{-{1 \over 2}}\,
\Big [ 2\,|\,C(r)\,|\,e^{-2g}  \,+\,{1 \over 2}\, |\,C'(r)\,|\,e^{-f}\,\Big ]\,
\,\eta_{x^ix^j} \,\,,
\qquad (i,j=0,\ldots,3)\,\, , \rc
&&T_{ab}\,=\,-\,e^{\phi}\,
h^{-{1 \over 2}} \Big [ |\,C(r)\,|\,e^{-2g}  \,+\,{1 \over 2}\,
|\,C'(r)\,|\,e^{-f}\,\Big ]\,\,\delta_{ab}\,\,,
\, \qquad\qquad (a,b=1,\ldots,4)\,\, , \rc
&&T_{rr}\,=\,T_{00}\,=\,-\,2\,|\,C(r)\,|\,\,h^{-{1 \over 2}} \,
e^{\phi-2g}  \,\, .
\label{TMNcomponents-massive}
\eear
As it happened for the equation of motion of the dilaton, one can verify that
the extra pieces on the right-hand side of (\ref{TMNcomponents-massive})
match precisely those generated by the second derivatives appearing in the
expression (\ref{Ricci-components}) of the Ricci tensor if 
$C(r)$ and $C'(r)$ are non-negative. As a consequence,
the first-order equations (\ref{BPS}) with a function $C(r)$ also imply  the
equations of motion for the ten-dimensional metric $g_{MN}$. 
It is also interesting to point out that, if $C(r)$ and $C'(r)$ are non-negative, 
$S_{DBI}$ can also be written in the form (\ref{SDBI-forms}), where
$\Omega_8$ is exactly the same eight-form as in eqs. (\ref{Omega_8}) and
(\ref{cal}). 

Notice that, if the function $C(r)= 3N_f(r)/4\pi$ has a Heaviside-like shape  ``starting'' at some finite value of the radial coordinate, then our BPS equations and solutions will be the ones given in subsection \ref{solKW}  for values of the radial coordinate bigger than  the ``mass of the flavour''. However,  below that radial value the solution will be the one of Klebanov-Witten (or deformations of it studied in \cite{Benini:2006hh}), with a non-running dilaton. Aside from decoupling in the field theory, this is clearly indicating
that the addition of massive flavours ``resolves'' the singularity.
Physically this behaviour is expected and makes these massive flavour more interesting.

\section{Summary and Discussion}
\medskip
\setcounter{equation}{0}

In this chapter we followed the method of \cite{Casero:2006pt} 
to construct a dual to the field theory defined by Klebanov and Witten 
after  $N_f$ flavours of quarks and antiquarks have been added to both gauge groups. In 
section \ref{sect2} of this chapter, we 
wrote the BPS equations describing the dynamics of this system and found 
solutions to this first-order system that also solve all the 
second-order equations of motion. We analyzed the solutions to the BPS system and learnt 
that, even when singular, the character of the singularity permits to get 
field theory conclusions from the supergravity perspective.

We proposed a formulation for the dual field theory to these 
solutions, constructing a precise four-dimensional superpotential.
We studied these solutions making many matchings with field theory 
expectations that included the R-symmetry breaking and Wilsonian \hyph{\beta}function. Also, using the well-known (supergravity) 
superpotential approach, 
we learnt that our field theory, aside from being deformed by a marginal 
(then turned irrelevant) operator, 
modifies its dynamics by giving VEV to operators of dimension six and 
eight. We explained how to change relations between couplings and $\Theta$-angles 
in the theory, from the perspective of our solutions. 

In section \ref{generalisations} of this chapter, we presented a careful account of the many 
technical details regarding the derivation of the results in section \ref{sect2}. Using the logic and intuitions developed there, we generalised the 
approach described for {\it any} five-dimensional manifold that can 
be written as a Sasaki-Einstein space (a one-dimensional fibration over a  
K\"ahler-Einstein space). It is surprising that the same structure of BPS 
equations and  ten-dimensional superpotential repeats for all the manifolds 
described above. This clearly points to some ``universality'' of the 
behaviour of four-dimensional \Nugual{1} SCFT's with flavours.

We shortly commented on the possibility of adding to the dynamics of the 
four-dimensional field theory fundamentals with mass, presenting a 
general context to do this. It seems interesting to exploit this procedure to get a better understanding of 
our singular backgrounds, make contact with field theory results and study 
many other interesting problems. 

It would be of great interest to study the dynamics of 
moving strings in this backgrounds, details related to dibaryons, flavour 
symmetry breaking, etc. Even when technically involved, it should 
be nice to understand the backreaction of probes where the worldvolume 
fields have been turned on, since some interesting problems may be addressed.

Finding black hole solutions in our geometries is another topic that deserves to be pursued. The interest of this problem resides in 
the fact that this would produce a black hole background 
where to study, among other things, plasmas that include the dynamics of 
colour and flavour at strong coupling. This is a very well defined problem 
that we believe of much interest.

In the next chapter we are going to extend the study of backreacted flavour branes to the 
Klebanov-Tseytlin and Klebanov-Strassler solutions. The result is more interesting, since the 
fundamentals and the Klebanov-Tseytlin cascade ``push in different directions'' in the RG 
flow. We will find fine-tuned situations in which the IR dynamics is 
either that of the Klebanov-Strassler model with fundamentals or the dynamics of the case studied in this chapter.


\chapter{Unquenched Flavours in the KS Model}
\label{KS}

\medskip
\setcounter{equation}{0}
\medskip

In this chapter, we will consider the addition of flavour degrees of freedom to 
the Klebanov-Tseytlin (KT) and Klebanov-Strassler (KS) solutions introduced in subsections \ref{fractionalD3} and \ref{deformedconifold} respectively. 
These new degrees of freedom will be incorporated again in the form of flavour 
D7-branes, corresponding to fundamental matter in the dual field theory. As in the previous chapter, we will follow ideas introduced in \cite{Karch:2002sh} but we will consider the case in which the number of 
fundamental fields is of the same order as the number of adjoint or bifundamental fields, that is $N_f\sim N_c$. This means that the new (strongly coupled) dynamics of the field theory is captured by a background that includes the backreaction of the flavour branes. In order to find the new solutions, we 
follow the ideas and techniques outlined in subsection \ref{smearingprocedure}.

Let us describe the main achievements of this chapter.
We will present {\it analytic} solutions for the equations of motion of type IIB 
supergravity coupled to the DBI+WZ action of the flavour D7-branes that preserve 
four SUSY's in four dimensions; we show how to reduce these solutions to those found by Klebanov-Tseytlin/Strassler when the number of flavours is taken to zero. Using them, we make a precise matching between the field theory cascade (which, enriched by the presence of the fundamentals, is still self-similar) and the string predictions. We will also match anomalies and $\beta$-functions by using our 
new supergravity background. The behaviour of the background in the UV of the gauge theory suggests that the field theory generates a `duality wall'. We also give a nice picture of Seiberg duality as a large gauge transformation in supergravity.

\section{The setup and the ansatz}\label{section-ansatz}
\medskip
\setcounter{equation}{0}

We are interested in adding to the KT/KS cascading gauge theory a number of flavours (fundamental fields) comparable with the number of colours (adjoint and bifundamental fields). Those supergravity backgrounds were obtained by considering a stack of regular and fractional D3-branes at the tip of the conifold. After the geometric transition, the colour branes disappear from the geometry, but the closed string fluxes that they sourced remain nontrivial (see subsections \ref{fractionalD3} and \ref{deformedconifold} for a thorough explanation). The addition of flavours in the field theory, in the large $N_c$ limit considered by Veneziano \cite{Veneziano:1976wm}, amounts to introduce mesonic currents and internal quark loops in the planar diagrams that survive 't Hooft's double scaling limit \cite{'tHooft:1973jz}.

Let us then consider a system of type IIB supergravity plus $N_f$ D7-branes.
The dynamics of the latter will be governed by the corresponding
Dirac-Born-Infeld and Wess-Zumino actions (\ref{actionintrodEF}). Our solution will have a nontrivial metric and
dilaton $\phi$ and, as in any cascading background,  non-vanishing
RR three- and five-forms $F_3$ and $F_5$, as well as a nontrivial NSNS
three-form $H_3$. In addition, the D7-branes act as a source for (the Hodge
dual of) the RR one-form $F_1$ through the WZ coupling:
\begin{equation}
S_{WZ}^{D7}\,=\,T_7\,\sum_{N_f}\,\int_{{\cal
M}_8}\,\hat C_{8}\,+\,\cdots\,\,,
\end{equation}
which induces a violation of the Bianchi identity $dF_1=0$, as we showed in subsection \ref{smearingprocedure}.
Therefore our configuration will also  necessarily have a
non-vanishing value of $F_1$. The ansatz that we shall adopt for the
Einstein frame metric is the following:
\bear
\label{metric}
&&ds^2\,=\,\Big[\,h(r)\,\Big]^{-\frac{1}{2}}\,dx^2_{1,3}\,+
\,\Big[\,h(r)\,\Big]^{\frac{1}{2}}\,\Bigg[\,dr^2\,+\,
e^{2G_1(r)}\,(\sigma_1^2\,+\,\sigma_2^2)\,+\,\,
\rc
&&\,+\,e^{2G_2(r)} \bigg(
(\omega_1\,+\,g(r)\,\sigma_1)^2\,+\,(\omega_2\,+\,g(r)\,\sigma_2)^2
\bigg)\,+\,{{e^{2G_3(r)}}\over 9}\,
\big(\omega_3\,+\,\sigma_3)^2\,\Bigg ] \,\, ,
\eear
where $dx^2_{1,3}$ denotes the four-dimensional Minkowski metric and
$\sigma_i$ and $\omega_i$ ($i=1,2,3$) are the one-forms displayed in equation (\ref{basisforms}). 

Notice that our metric ansatz (\ref{metric}) depends on five unknown
radial functions $G_i(r)$ ($i=1,2,3$), $g(r)$ and $h(r)$. The ansatz for
$F_5$ has the standard form, namely:
\begin{equation}
\label{F5}
F_5\,=\,d h^{-1}(r)\wedge dx^0\wedge\cdots\wedge dx^3\,+\,{\rm Hodge\,\,dual} \;.
\end{equation}
As expected for flavour branes, we will take D7-branes extended along the
four Minkowski coordinates as well as other four internal coordinates. The kappa symmetric embedding of the D7-branes that we start from will be discussed in section \ref{sect: Field Th}. 
In order to simplify the computations, following the approach of subsection \ref{smearingprocedure}, we will smear the D7-branes in their two transverse directions in such a way that the symmetries of the unflavoured background are recovered. As we explained in subsection \ref{smearingprocedure}, this smearing amounts to the following
generalisation of the WZ term of the D7-brane action:
\begin{equation}
\label{WZ-smearing}
S_{WZ}^{D7}\,=\,T_7\,\,\sum_{N_f}\,\,\int_{{\cal M}_8}\,\,\hat C_8\,\,+\,
\cdots
\qquad \rightarrow \qquad
\,\,T_7\,\,\int_{{\cal M}_{10}}\,
\Omega\wedge C_8\,\,+\,
\cdots,
\end{equation}
where $\Omega$ is a two-form which determines the distribution of the RR
charge of the D7-brane and ${\cal M}_{10}$ is the full ten-dimensional
manifold. Notice that $\Omega$ acts as a magnetic charge source for $F_1$
which generates the violation of its Bianchi identity (see eq. (\ref{Bianchi-general1})). In what follows we will assume that the flavours introduced by the
D7-brane are massless, which is equivalent to require that the flavour
brane worldvolume reaches the origin in the holographic direction. Under
this condition one expects a radial coordinate independent D7-brane charge density. 
Moreover, the D7-brane  embeddings that we will smear imply that $\Omega$ is symmetric
under the exchange of the two $S^2$'s parameterised  by
$(\theta_1, \varphi_1)$ and $(\theta_2, \varphi_2)$, and independent of
$\psi$ (see section \ref{sect: Field Th}). The smeared charge density distribution is the one already
adopted in chapter \ref{KW}, namely:
\begin{equation} \label{smearing}
dF_1\,= - {N_f\over 4\pi}\,(\sin\theta_1\,d\theta_1\wedge d\varphi_1\,+\,
\sin\theta_2\,d\theta_2\wedge d\varphi_2\,)\,=\,
{N_f\over 4\pi}\,\, (\omega_1 \wedge \omega_2 \,-\,
\sigma_1 \wedge \sigma_2)\,\,,
\end{equation}
where the coefficient $N_f/4\pi$ is determined by normalization. With
this ansatz for $\Omega$, the modified Bianchi identity
(\ref{Bianchi-general1}) determines the value of $F_1$, namely:
\begin{equation}  \label{F1}
F_1\,=\,{N_f\over 4\pi}(\omega_3\,+\,\sigma_3)\,\,.
\end{equation}

The ansatz for the RR and NSNS three-forms that we propose is an extension of the one
given by Klebanov and Strassler (see eq. (\ref{KTKSforms})) and it is simply (in this chapter we set for convenience $\alpha'=1$):
\begin{equation}\begin{aligned}  \label{ansatz}
B_2 &= \frac{M}{2} \Bigl[ f\, g^1 \wedge g^2\,+\,k\, g^3 \wedge 
g^4 \Bigr]  \,\,, \\
H_3 &= dB_2\,=\, \frac{M}{2} \, \Bigl[ dr \wedge (f' \,g^1 \wedge g^2\,+\,
k'\,g^3 \wedge g^4)\,+\,{1 \over 2}(k-f)\, g^5 \wedge (g^1
\wedge g^3\,+\,g^2 \wedge g^4) \Bigr]\,\,,   \\
F_3 &= \frac{M}{2} \Big\{ g^5\wedge \Big[ \big( F+\frac{N_f}{4\pi}f\big)g^1\wedge g^2 + \big(1- F+\frac{N_f}{4\pi}k\big)g^3\wedge g^4 \Big] +F' dr \wedge \big(g^1\wedge g^3 + g^2\wedge g^4   \big)\Big\} , 
\end{aligned} \end{equation}
where $M$ is a constant, $f(r)$, $k(r)$ and $F(r)$ are functions of the radial coordinate, and the  $g^i$'s are the set of one-forms given in (\ref{gbasis}).
The forms $F_3$, $H_3$ and $F_5$ must satisfy the set of Bianchi written in equation (\ref{BI}).
Notice that the equations for $F_3$ and $H_3$ are automatically satisfied
by our ansatz (\ref{ansatz}). However, the Bianchi identity for $F_5$
gives rise to the following differential equation:
\begin{equation}
{d \over {dr}} \Big [  h'\, e^{2G_1+2G_2+G_3} \Big ] = -
{3 \over 4}M^2
\Big[ (1-F\,+\,{N_f\over 4\pi}\,k)f'+(F+{N_f\over 4\pi}f)k'
+(k-f) F'\Big ]\,\, ,
\end{equation}
which can be integrated, with the result:
\begin{equation}
 h'\, e^{2G_1+2G_2+G_3}\,=\,-
{3 \over 4}M^2\Big[f-(f-k)F+{N_f\over 4\pi}fk\Big]\,+\,
{\rm constant}\,\,.
\label{Kh2-fkF}
\end{equation}
Let us now parameterise $F_5$ (see eq. (\ref{five-form})) as
\begin{equation}
F_5\,=\,{\pi\over 4}\,N_{eff} (r)\,
g^1\wedge g^2\wedge g^3\wedge g^4\wedge g^5\,+\,
{\rm Hodge\,\,dual}\,\,,
\label{F5-Neff}
\end{equation}
and let us define the five-manifold ${\cal M}_5$ as the one that is obtained
by taking the Minkowski coordinates and $r$ fixed to a constant value.
As $\int_{{\cal M}_5}F_5\,=\,(4\pi^2)^2\,N_{eff} (r)$, it follows that
$N_{eff} (r)$ can be interpreted as the effective D3-brane charge at the
value $r$ of the holographic coordinate. From our ansatz (\ref{F5}), it
follows that:
\begin{equation}
N_{eff} (r)\,=   \,-{4\over 3\pi}\, h'\, e^{2G_1+2G_2+G_3}\,\,,
\label{Neff-Kh2}
\end{equation}
and taking into account (\ref{Kh2-fkF}), we can write
\begin{equation} \label{Neff}
N_{eff} (r) \equiv \frac{1}{(4\pi^2)^2} \int_{{\cal M}_5} F_5 \,=  \,N_0\,+\,{M^2\over \pi}\,\Big[\,
f-(f-k)F+{N_f\over 4\pi}fk\Big]\,\,,
\end{equation}
where $N_0$ is a constant. It follows from \eqref{Neff} that the RR
five-form $F_5$ is determined once the functions $F$, $f$ and $k$ that
parameterise the three-forms are known. Moreover, eq. \eqref{Kh2-fkF} allows to compute the warp factor once the functions $G_i$ and the three-forms are determined. Notice also that the effective D5-brane charge is obtained by integrating the gauge-invariant field strength $F_3$ over 
the 3-cycle $S^3$ (see eq. (\ref{KTquantization})): $\theta_2=\text{const.}$, $\varphi_2=\text{const.}$. The result is:
\begin{equation} \label{Meff}
M_{eff}(r) \equiv \frac{1}{4\pi^2} \int_{S^3} F_3 = M \Bigl[ 1 + \frac{N_f}{4\pi} (f+k) \Bigr]\;.
\end{equation}

The strategy to proceed further
is to look at  the conditions imposed by supersymmetry. We will smear,
as in chapter \ref{KW}, D7-brane embeddings that are kappa symmetric and,
therefore, the supersymmetry requirement (\ref{Eframe1}) gives
 rise to a large number of BPS first-order ordinary differential equations for
the dilaton and the different functions that parameterise the metric and
the forms. 
In the end, one can check that the first-order differential equations imposed by supersymmetry imply the second-order differential equations of motion.
In particular, from the variation of the dilatino we get the
following differential equation for the dilaton:
\begin{equation}
\phi'\,=\,{3N_f\over 4\pi}\,e^{\phi-G_3}\;.
\end{equation}
A detailed analysis of the conditions imposed by supersymmetry shows that
the fibering function $g$ in eq. \eqref{metric} is subjected to the following algebraic
constraint:
\begin{equation}
g \Big [ g^2\,-\,1\,+\,e^{2(G_1-G_2)} \Big ] \,=\,0 \,\,,
\label{g-constraint}
\end{equation}
which has obviously two solutions. The first of these solutions is $g=0$
and, as it is clear from our metric ansatz (\ref{metric}), it corresponds
to the cases of the flavoured singular and resolved conifolds. 
 In the second solution $g$ is such
that the term in brackets on the right-hand side of (\ref{g-constraint})
vanishes. This solution gives rise to the flavoured version of the warped
deformed conifold. 
The flavoured KT solution will be presented in section \ref{KTflavoured}, whereas the flavoured KS solution will be analyzed in section \ref{KSflavoured}.

\subsection{Maxwell and Page charges}

Before presenting the explicit solutions for the metric and the forms of the supergravity equations, let us discuss the different charges carried out by our solutions. In theories, like type IIB supergravity, that have  Chern-Simons terms in the action (which give rise to modified Bianchi identities), it is possible  
to define more than one notion of  charge associated with a given gauge
field. Let us discuss here, following the presentation of ref. \cite{MarolfCB}, two particular definitions of this quantity, namely the so-called Maxwell and Page charges \cite{Page}.
Given a gauge invariant field strength $F_{8-p}$, the
(magnetic) Maxwell current associated to it is defined through  the following
relation:
\beq
d\,F_{8-p}\,=\,\star j^{Maxwell}_{D_{p}}\,\, ,
\label{Maxwell-j}
\eeq
or equivalently, the Maxwell charge in a volume $V_{9-p}$ is
 given by:
\beq
Q^{Maxwell}_{D_p}\,\sim \, \int_{V_{9-p}} \star j^{Maxwell}_{D_{p}}\,\, ,
\eeq
with a suitable normalization. Taking $\partial V_{9-p}=M_{8-p}$ and using 
(\ref{Maxwell-j}) and Stokes theorem,  we can rewrite the previous expression as:
\beq
Q^{Maxwell}_{D_p}\,\sim  \int_{M_{8-p}}\,\,F_{8-p}\,\,.
\eeq

This notion of current is gauge invariant and conserved and it has 
other properties that are discussed in \cite{MarolfCB}. In particular, it
is not ``localised" in the sense that for a solution of pure supergravity
(for which $d\,F_{8-p}=H_3 \wedge F_{6-p}$) this current does not vanish.
These are the kind of charges that we have calculated so far \eqref{Neff}-\eqref{Meff}, namely:
\bear
&&Q^{Maxwell}_{D5}\,=\,M_{eff}\,=\, 
{1 \over {4 \pi^2}} \int F_3 \, \, , \rc
&&Q^{Maxwell}_{D3}\,=\, N_{eff}\,=\, 
{1 \over {(4 \pi^2)^2}} \int F_5 \,\, .
\eear
An important issue regarding these charges is that, in general, they
are not quantised. Indeed, we have  checked explicitly  that
$Q^{Maxwell}_{D5}=M_{eff}$ and 
$Q^{Maxwell}_{D3}=N_{eff}$ vary continuously with the holographic
variable $r$ (see eqs. (\ref{Meff}) and (\ref{Neff})).

Let us move on to the notion of Page charge. The idea is first to write 
the Bianchi identities for $F_3$  and $F_5$ 
as the exterior derivatives of some differential
form, which in general will not be gauge invariant. Page currents can then be introduced as magnetic sources on the right-hand side, thus violating the Bianchi identities.  In  our
case, we can define the following (magnetic) Page currents:
\begin{equation}
\begin{split}
&d(F_3\,-\,B_2 \wedge F_1)\,=\, \star j^{Page}_{D5}\,\, , \\
&d(F_5\,-\,B_2 \wedge F_3 \,+\,{1 \over 2} B_2 \wedge B_2 \wedge F_1)\,=\, 
\star j^{Page}_{D3}\,\, .
\end{split} \label{Jpage}
\end{equation}
Notice that the
currents defined by the previous expression are ``localised" as a
consequence of the Bianchi identities satisfied by $F_3$ and $F_5$, namely
$dF_3\,=\,H_3\wedge F_1$ and $dF_5\,=\,H_3\wedge F_3$.
The Page charges $Q^{Page}_{D5}$ and $Q^{Page}_{D3}$ are just  defined
as the integrals of $\star j^{Page}_{D5}$ and $\star j^{Page}_{D3}$
with the appropriate normalization,  \ie:
\begin{equation}
\begin{split}
Q^{Page}_{D5}\,&=\,  {1 \over {4 \pi^2}} \int_{V_4} \star j^{Page}_{D5} 
\,\, , \\
Q^{Page}_{D3}\,&=\, {1 \over {(4 \pi^2)^2}} \int_{V_6} \star j^{Page}_{D3}\,\,,
\end{split}
\end{equation}
where $V_4$ and $V_6$ are submanifolds in the transverse space to the D5- and D3-branes respectively, which enclose the branes. By using the expressions of the currents $\star j^{Page}_{D5} $ and $\star j^{Page}_{D3}$ given in (\ref{Jpage}),  and by applying Stokes
theorem, we get:
\begin{equation}\label{D5D3Page}
\begin{split}
Q^{Page}_{D5}\,&=\,  {1 \over {4 \pi^2}} \int_{S^3} 
\Big(F_3\,-\,B_2 \wedge F_1\Big)\,\, , \\
Q^{Page}_{D3}\,&=\, {1 \over {(4 \pi^2)^2}} \int_{\mathcal{M}_5} \Big(\,
 F_5\,-\,B_2 \wedge F_3 \,+\,{1 \over 2} B_2 \wedge B_2 \wedge F_1
 \,\Big)\,\,,
\end{split}
\end{equation}
where $S^3$ and $\mathcal{M}_5$ are the same manifolds used to compute the
Maxwell charges in eqs. (\ref{Meff}) and (\ref{Neff}).  It is not difficult to establish the topological nature of these Page charges. Indeed, let us consider, for concreteness, the expression of $Q^{Page}_{D5}$ in (\ref{D5D3Page}). Notice that the three-form under the integral can be {\it locally} represented as the exterior derivative of a two-form, since
$F_3-B_2\wedge F_1=dC_2$, with $C_2$ being the RR two-form potential.  If $C_2$ were well defined globally on the $S^3$, the Page charge $Q^{Page}_{D5}$ would vanish identically as a consequence of Stokes theorem. Thus, $Q^{Page}_{D5}$  can be naturally interpreted as a monopole number and it can be non-vanishing only in the case in which the gauge field is topologically nontrivial. For the D3-brane Page charge 
$Q^{Page}_{D3}$ a similar conclusion can be reached. 

Due to the topological nature of the Page charges defined above, one naturally expects that they are quantised and, as we shall shortly verify, they are independent of the holographic coordinate. This shows that they are the natural objects to compare with the numbers of branes that create the geometry in these backgrounds with varying flux. 
However, as it is manifest from the fact that $Q^{Page}_{D5}$ and $Q^{Page}_{D3}$ are given in (\ref{D5D3Page}) in terms of the $B_2$ field and not in terms of its  field strength $H_3$, the Page charges are not gauge invariant. In subsection \ref{Large-gauge} we will relate this non-invariance to the Seiberg duality of the field theory dual.

Let us now calculate the associated Page charges  for our ansatz (\ref{ansatz}) . We shall start by computing the D5-brane Page charge for the three-sphere $S^3$
defined by $\theta_2, \varphi_2={\rm constant}$. We already know  the value of the
integral of $F_3$, which gives precisely $M_{eff}$ (see eq. (\ref{Meff})). Taking into account that
\beq
\int_{S^3}\,g^5\wedge g^1\wedge g^2\,=\,\int_{S^3}
g^5\wedge g^3\wedge g^4\,=\,8\pi^2\,\,, \label{intS3}
\eeq
we readily get:
\beq
{1\over 4\pi^2}\,\int_{S^3}\,B_2\wedge F_1\,=\,{MN_f\over 4\pi}\,\,
(f+k)\,\,,
\eeq
and therefore:
\beq
Q^{Page}_{D5}\,=\,M_{eff}\,-\,{MN_f\over 4\pi}\,\,
(f+k)\,\,.
\label{QD5-Meff}
\eeq
Using the expression of $M_{eff}$ given in (\ref{Meff}), we obtain:
\beq
Q^{Page}_{D5}\,=\, M\,\,,
\label{QD5=M}
\eeq
which is certainly quantised and independent of the radial coordinate.

Let us now look at the D3-brane Page charge, which can be computed as an integral  over the angular manifold $M_5$. Taking into account that
\beq
\int_{\mathcal{M}_5}\,g^1\wedge g^2\wedge g^3\wedge g^4\wedge g^5\,=\,(4\pi)^3\,\,,
\label{int5g}
\eeq
we get that, for our ansatz (\ref{ansatz}):
\begin{equation}
\begin{split}
&{1 \over {(4 \pi^2)^2}} \int_{\mathcal{M}_5} B_2\wedge F_3\,=\,
{M^2\over \pi}\,\,\Big[\,f\,-\,(f-k)\,F\,+\,{N_f\over 2\pi}\,fk\,\Big]
\,\,,\\
&{1 \over {(4 \pi^2)^2}} \int_{\mathcal{M}_5} B_2\wedge B_2\wedge F_1\,=\,
{M^2\over \pi}\,{N_f\over 4\pi}\,fk\,\,,
\end{split}
\end{equation}
and, thus
\beq
Q^{Page}_{D3}\,=\,N_{eff}\,-\,
{M^2\over \pi}\,\,\Big[\,f\,-\,(f-k)\,F\,+\,{N_f\over 4\pi}\,fk\,\Big]\,\,.
\label{QD3-Neff}
\eeq
Using the expression of $N_{eff}$, we obtain
\beq
Q^{Page}_{D3}\,=\,N_0\,\,,
\label{QD3=N0}
\eeq
which is again independent of the holographic coordinate. 
Recall that these Page charges are not gauge invariant and we will study in 
subsection \ref{Large-gauge} how they change under a large gauge transformation.

We now proceed to present the solutions to the BPS equations of motion.

\section{Flavored warped deformed conifold} \label{KSflavoured}
\medskip
\setcounter{equation}{0}

Let us now consider the following solution of the algebraic constraint (\ref{g-constraint}):
\begin{equation}
\label{g}
g^2\,=\,1\,-\,e^{2(G_1-G_2)} \,\,.
\end{equation}
In order to write the equations for the metric and dilaton in this case, let us 
perform the following change of variable:
\begin{equation}
3\, e^{-G_3}\,dr\,=\,d\tau\,\, . 
\end{equation}
In terms of this new variable, the differential equation for the 
dilaton is simply:
\begin{equation}
\dot{\phi}\,=\,{N_f\over 4\pi}\,e^{\phi}\,\,,
\label{dotphi}
\end{equation}
where the dot means derivative with respect to $\tau$. This equation can be
straightforwardly integrated, namely:
\begin{equation}
{N_f\over 4\pi}\,e^{\phi}\,=\,{1\over \tau_0-\tau}\,\,,
\qquad\qquad
0\le \tau\le \tau_0\,\,,
\label{phi(tau)}
\end{equation}
where $\tau_0$ is an integration constant. 
Let us now write the equations imposed by supersymmetry  to the metric functions $G_1$, $G_2$
and $G_3$, which are:
\bear
&&\dot{G}_1\,-\,{1 \over 18}e^{2G_3-G_1-G_2}\,-\,{1 \over 2}
e^{G_2-G_1}\,+\,{1 \over 2}e^{G_1-G_2}\,=\,0\,\, , \rc
&&\dot{G_2}\,-\,{1 \over 18}e^{2G_3-G_1-G_2}\,+\,{1 \over 2}
e^{G_2-G_1}\,-\,{1 \over 2}e^{G_1-G_2}\,=\,0\,\, , \rc
&&\dot{G_3}\,+\,{1 \over 9}e^{2G_3-G_1-G_2}\,-\, e^{G_2-G_1}\,
+\,{N_f \over 8\pi}e^{\phi}\,=\,0\,\,.
\eear
In order to write the solution of this system of equations, let us 
define the following function
\begin{equation}
\Lambda(\tau)\,\equiv\,{
\Big[\,2(\tau-\tau_0)(\tau-\sinh 2\tau)\,+\,\cosh
(2\tau)\,-\,2\tau\tau_0\, -\,1\,\Big]^{{1\over 3}}\over
\sinh\tau}\,\,.
\end{equation}
Then, the metric functions $G_i$ are given by:
\bear
&&e^{2G_1}\,=\,{1\over  4}\,\,\mu^{{4\over 3}}\,
{\sinh^2\tau\over \cosh\tau}\,\Lambda(\tau)\,\,,
\qquad\qquad
e^{2G_2}\,=\,{1\over 4}\,\,\mu^{{4\over 3}}\,
\cosh\tau\,\Lambda(\tau)\,\,,\rc\rc
&&e^{2G_3}\,=\,6\,\mu^{{4\over 3}}\,\,{\tau_0-\tau\over
\big[\,\Lambda(\tau)\,\big]^2}\,\,,
\label{G-sol}
\eear
where $\mu$ is an integration constant. Notice that the range of 
$\tau$ variable chosen in (\ref{phi(tau)}) is the one that makes the dilaton and the metric
functions real. Moreover, for the solution we have found,
the  fibering function $g$ is given by:
\begin{equation}
g\,=\,{1\over \cosh\tau}\,\,.
\end{equation}
By using this result, we can write the metric as:
\bear  \label{metric2}
&&ds^2\,=\,\Big[\,h(\tau)\,\Big]^{-{1\over2}}\,dx^2_{1,3}\,+
\,\Big[\,h(\tau)\,\Big]^{{1\over2}}\,ds^2_{6}\,\,,
\eear
where $ds^2_{6}$ is the metric of the `flavoured' deformed conifold, namely
\bear
&&ds^2_{6}\,=\,{1\over 2}\,\,\mu^{{4\over 3}}\,\,\Lambda(\tau)\,\,
\Bigg[\,{4(\tau_0-\tau)\over 
3\Lambda^3(\tau)}\,\,\big(\,d\tau^2\,+\,(g^5)^2\,\big)\,+\,
  \cosh^2\Big({\tau\over 2}\big)\,\Big(\,(g^3)^2\,+\, 
(g^4)^2\,\Big)\,+\,\,\rc
  &&\qquad\qquad\qquad\qquad\qquad\qquad\qquad
  +\,\sinh^2\Big({\tau\over 2}\Big)\,
\Big(\,(g^1)^2\,+\, (g^2)^2\,\Big)\,
\,\Bigg]\,\,.
\label{flav-def-metric}
\eear
Notice the similarity between the metric (\ref{flav-def-metric}) and the one corresponding to the `unflavoured' deformed conifold (\ref{unflavourdeform}). To further analyze this similarity, let us study the $N_f\to 0$ limit of our solution.  By looking at the expression of the dilaton in (\ref{phi(tau)}), one realises that this limit is only sensible if one also sends $\tau_0\to+\infty$ with 
$N_f\tau_0$ fixed. Indeed, by performing this scaling and neglecting $\tau$ versus 
$\tau_0$, one gets a constant value for the dilaton. Moreover, the function $\Lambda(\tau)$ reduces in this limit to $\Lambda(\tau)\approx (4\tau_0)^{{1\over 3}}\,\,K(\tau)$, where $K(\tau)$ is the function (\ref{KSfunction}) which appears in the metric of the deformed conifold. By using this result one easily verifies that, after redefining $\mu\to\mu/(4\tau_0)^{{1\over 4}}$, the metric (\ref{flav-def-metric}) reduces to the one in equation (\ref{unflavourdeform}) for the unflavoured system.

The requirement of supersymmetry imposes the following differential equations for the functions $k$,
$f$ and $F$ appearing in the fluxes of our ansatz:
\bear
&&\dot{k}\,=\,e^{\phi}\,\Big(\,F\,+\,{N_f\over 4\pi}\,f\,\Big)\,
\coth^2{{\tau\over 2}}\,\,,\rc
&&\dot{f}\,=\,e^{\phi}\,\,\Big(\,1\,-\,F\,+\,{N_f\over 4\pi}\,k\,\Big)\,
\tanh^2{\tau\over 2}\,\,,  \rc
&&\dot{F}\,=\,{1 \over 2}e^{-\phi} (k-f)\,\,.
\label{kfF-with-tau}
\eear
Notice, again, that for $N_f=0$ the system (\ref{kfF-with-tau}) reduces to (\ref{firstorder}). Moreover, for $N_f\not=0$ this system can be solved as:
\bear
\label{sol}
&&e^{-\phi}\,f\,=\,{{\tau \coth{\tau}-1} \over {2 \,
\sinh{\tau}}}(\cosh{\tau}-1)\,\,,
\qquad\qquad
e^{-\phi}\,k\,=\,{{\tau \coth{\tau}-1} \over {2 \, \sinh{\tau}}}
(\cosh{\tau}+1)\,\, , \rc
&&\qquad\qquad\qquad\qquad\qquad\qquad
F\,=\,{{\sinh{\tau}-\tau} \over {2\, \sinh{\tau}}}\,\,,
\eear
where $e^{\phi}$ is given in eq. (\ref{phi(tau)}).
By using the solution displayed in (\ref{G-sol}) and (\ref{sol}) in the
general eq.  (\ref{Kh2-fkF}) we can immediately obtain the expression
of the warp factor $h(\tau)$.  Actually, if we require that $h$ is
regular at $\tau=0$, the integration constant $N_0$ in \eqref{Neff} must
be chosen to be zero. In this case, we get:
\begin{equation}
h(\tau)=-{{\pi \, M^2} \over {4\,\mu^{8/3} N_f}}
\int^{\tau} dx {{x \coth{x}-1} \over {(x\,-\,\tau_0)^2 \sinh^2{x}}}
{{-\cosh{2x}\,+\,4x^2\,-\,4x\tau_0\,+\,1\,-\,(x\,-\,2\tau_0)\sinh{2x}} \over
{(\cosh{2x}\,+\,2x^2\,-\,4x\tau_0\,-\,1\,-\,2(x\,-\,\tau_0)\sinh{2x})^{2/3}}} 
\,\, .
\end{equation}
The integration constant can be fixed by requiring that the analytic continuation of $h(\tau)$ goes to zero as $\tau \to\ +\infty$, to connect with the Klebanov-Strassler solution in the unflavoured (scaling) limit. Then, close to the tip of the geometry, $h(\tau) \sim h_0 - \mathcal{O}(\tau^2)$.

We should emphasize now an important point: even though at first sight this solution may look smooth in the IR ($\tau\sim 0$), where all the components of our metric approach the same limit as those of the KS solution (up to a suitable redefinition of parameters, see eqs. (\ref{KSIRlimit}) and (\ref{KSIRlimit1})),  there is actually a curvature singularity.
Indeed, in Einstein frame the curvature scalar behaves as $R_E \sim 1/\tau$.%
\footnote{The simplest example of this kind of singularity appears at $r=0$ in a 2-dimensional  manifold whose metric is $ds^2= dr^2 + r^2 (1+r)d\varphi^2$.}
This singularity of course disappears when taking the unflavoured limit, using the scaling described above. A more detailed analysis of the singularity was done in \cite{Benini:2007gx}.

The solution presented above is naturally interpreted as the addition of fundamentals to the KS background. In the next section, we will present a solution that can be understood as the addition of flavours to the KT background.

\section{Fractional branes in the singular conifold with flavour} \label{KTflavoured}
\medskip
\setcounter{equation}{0}

Let us now consider the solutions with $g=0$. First of all, let us change
the radial variable from $r$ to $\rho$, where the later is defined by the
relation  $dr=e^{G_3}\,d\rho$. The equation for the dilaton can be
integrated trivially:
\begin{equation}
e^{\phi}\,=\,-{4\pi\over 3N_f}\,\,{1\over \rho}\,\,,
\qquad\qquad \rho<0\,\,.
\end{equation}
The supersymmetry requirement imposes now that
the metric functions $G_i$ satisfy  in this case the following system of
differential equations:
\bear
&&\dot G_i\,=\,{1\over 6}\,e^{2G_3-2G_i}\,\,,
\qquad\qquad (i=1,2)\,\,,\rc\rc
&&\dot G_3\,=\,3\,-\,{1\over 6}\,e^{2G_3-2G_1}
\,-\,{1\over 6}\,e^{2G_3-2G_2}\,-\,{3N_f\over 8\pi}\,e^{\phi}\,\,,
\eear
where now the dot refers to the derivative with respect to $\rho$.
This system is equivalent to the one analyzed in chapter \ref{KW} for the
Klebanov-Witten model with flavours,  concretely in equations (\ref{eqng}) and (\ref{eqnf}). In what follows we will restrict
ourselves to the particular  solution with $G_1=G_2$ given by (see eqs. (\ref{egKW}) and (\ref{efKW})):
\begin{equation}
e^{2G_1}\,=\,e^{2G_2}\,=\,{1\over 6}\,(1-6\rho)^{{1\over 3}}\,
e^{2\rho}\,\,,
\qquad\qquad
e^{2G_3}\,=\,-6\rho\,(1-6\rho)^{-{2\over 3}}\,e^{2\rho}\,\,.
\end{equation}
Notice that, as in chapter \ref{KW}, the range of values of $\rho$ for
which the metric is well defined is $-\infty<\rho<0$. The equations for
the flux functions  $f$, $k$  and $F$ are now:
\bear
&&\dot f\,-\,\dot k\,=\,2e^{\phi}\dot F\,\,,\rc
&&\dot f\,+\,\dot k\,=\,3e^{\phi}
\Big[\,1\,+\,{N_f\over 4\pi}(f+k)\,\Big]\,\,,\rc
&&F\,=\,{1\over 2}\,\Big[\,1\,+\,
\Big(\,e^{-\phi}\,-\,{N_f\over 4\pi}\,\Big)\,
(f-k)\,\Big]\,\,.
\eear
We will focus on the particular solution of this system
such that $f=k$ and $F$ is constant, namely:
\begin{equation}
F={1\over 2}\,\,,
\qquad\qquad
f\,=\,k\,=\,-{2\pi\over N_f}\,\Bigg(\,1\,-\,
{\Gamma\over \rho}\,\Bigg)\,\,,
\label{KT-Ffk-sol}
\end{equation}
where $\Gamma$ is an integration constant. By substituting these values
of $F$, $f$ and $k$ in our ansatz (\ref{ansatz}) we obtain the form of
$F_3$ and $H_3$. Notice that the constants $M$ and $\Gamma$ only appear 
in the combination
$M\Gamma$. Accordingly, let us define ${\cal M}$ as
${\cal M}\,\equiv\,M\Gamma$. We will write the result in terms of the
function:
\begin{equation}
M_{eff}(\rho)\equiv {{\cal M}\over \rho}\,\,.
\end{equation}
One has:
\begin{equation}
\begin{split}
F_3 \,&=\,{M_{eff}(\rho)\over 4}\,\,g^5\wedge
\big(g^1\wedge g^2\,+\,g^3\wedge g^4\big)\,\,,\\
H_3\,&=\,-{\pi\over N_f}\,{M_{eff}(\rho)\over \rho}\,\,d\rho\wedge
\big(g^1\wedge g^2\,+\,g^3\wedge g^4\big)\,\,.
\end{split}\label{F3 H3 KT}
\end{equation}
Moreover, the RR five-form $F_5$ can be written as in (\ref{F5-Neff}) in
terms of the effective D3-brane charge defined in \eqref{Neff}. For the
solution  (\ref{KT-Ffk-sol}) one gets:
\begin{equation}
N_{eff}(\rho)\,=\,N_c\,+\,{{\cal M}^2\over N_f}\,{1\over \rho^2}\,\,,
\end{equation}
where  $N_c\,\equiv\,N_0\,-\,{ M^2\over N_f}$. By using eq.
(\ref{Kh2-fkF}), one can obtain the expression of the warp factor, namely:
\begin{equation}
h(\rho)\,=\,-27\pi \int\,d\rho\,
\Bigg[\,N_c\,+\,{{\cal M}^2\over N_f}\,{1\over \rho^2}\,\Bigg]\,
{e^{-4\rho}\over (1-6\rho)^{{2\over 3}}}\,\,.
\end{equation}
To interpret the solution just presented, it is interesting to study it in the deep IR region 
$\rho\to-\infty$. Notice that in this limit the three-forms $F_3$ and $H_3$ vanish. Actually, it is easy to verify that for $\rho\to-\infty$ the solution obtained here reduces to the one studied in chapter \ref{KW}, corresponding to the Klebanov-Witten model with flavours. Indeed, in this IR region it is convenient to go back to our original radial variable $r$. The relation between $r$ and $\rho$ for $\rho\to-\infty$ is $r\approx (-6\rho)^{{1\over 6}}\,e^{\rho}$ (see eq. (\ref{r(rho)})). Moreover, one can prove that for 
$\rho\to-\infty$  (or equivalently $r\to 0$), the warp factor $h$ and the metric functions $G_i$ become:
\begin{equation}
h(r)\approx {27\pi N_c\over 4}\,{1\over r^4}\,\,,
\qquad\qquad
e^{2G_1}\,=\,e^{2G_2}\approx {r^2\over 6}\,\,,
\qquad\qquad
e^{2G_3}\,\approx\,r^2\,\,,
\end{equation}
which implies that the IR Einstein frame metric is $AdS_5\times T^{1,1}$ plus logarithmic corrections, exactly as the solution that we found in equation (\ref{asympKW}).
The interpretation of the RG flow of the field theory dual to this solution will be explained in sections \ref{sect: Field Th} and \ref{cascade-SUGRA}.

Finally, let us stress that the UV behaviour of this solution (coincident with that of the solution presented in section \ref{KSflavoured}) presents a divergent dilaton at the point $\rho=0$ (or $\tau=\tau_0$ for the flavoured warped deformed conifold). Hence the supergravity approximation fails at some value of the radial coordinate that we will associate in section \ref{cascade-SUGRA} with the presence of a duality wall \cite{Strassler:1996ua} in the cascading field theory.

\section{The field theory dual: a cascade of Seiberg dualities} \label{sect: Field Th}
\medskip
\setcounter{equation}{0}

The field theory dual to our supergravity solutions can be engineered by putting stacks of two kinds of fractional D3-branes (colour branes) and two kinds of fractional D7-branes (flavour branes) on the singular conifold.
The smeared charge distribution introduced in the previous sections can be realised by homogeneously distributing D7-branes among a class of localised kappa symmetric embeddings. The complex structure of the deformed conifold is described by the equation (\ref{zcoorKS}). Recall that the $U(1)_R$ action is broken to $\bbZ_2$ by the deformation parameter. Consider the embedding \cite{Kuperstein:2004hy}:
\begin{equation} \label{embedding eq}
z_3 + z_4 = 0 \;.
\end{equation}
This is invariant under $U(1)_R$ and a diagonal $SU(2)_D$ of the isometry group of the deformed conifold (and a $\bbZ_2$ which exchanges $z_3 \leftrightarrow z_4$). Moreover it is free of $C_8$ tadpoles and it was shown to be kappa symmetric in \cite{Kuperstein:2004hy}. It could be useful to write it in the angular coordinates of the previous section (see also eq. (\ref{zcoorKSangular})): $\theta_1=\theta_2$, $\varphi_1 = \varphi_2$, $\forall \psi,\forall\tau$. We can obtain other embeddings with the same properties by acting on it with the broken generators. One can show that the charge distribution obtained by homogeneously spreading the D7-branes in this class is \eqref{smearing}:
\begin{equation}
\Omega = \frac{N_f}{4\pi} \bigl( \sin\theta_1 \, d\theta_1\wedge d\varphi_1 + \sin\theta_2 \, d\theta_2 \wedge d\varphi_2 \bigr) \;,
\end{equation}
where $N_f$ is the total number of D7-branes.

Notice that one could have considered the more general embedding: $z_3 + z_4 = m$, where $m$ corresponds in field theory to a mass term for quarks. However we will not consider these embeddings and their corresponding supergravity solutions. 

\begin{figure}[ht]
\begin{center}
\includegraphics[width=0.8\textwidth]{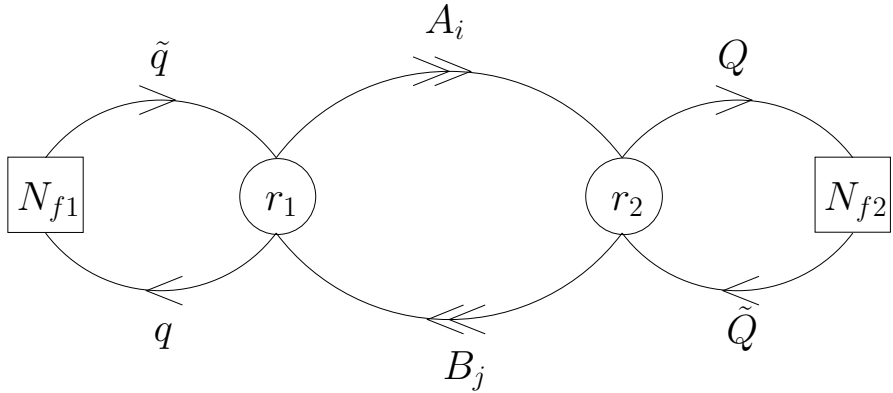}
\end{center}
\caption[quiverN]{The quiver diagram of the gauge theory. Circles are gauge groups, squares are flavour groups and arrows are bifundamental chiral superfields. $N_{f1}$ and $N_{f2}$ sum up to $N_f$. \label{quiverN}} 
\end{figure}

Different techniques have been developed to identify the field theory dual to our type IIB plus D7-branes background, which can be engineered by putting $r_l$ fractional D3-branes of the first kind, $r_s$ fractional D3-branes of the second kind, $N_{fl}$ fractional D7-branes of the first kind, and $N_{fs}$ fractional D7-branes of the second kind ($l,s=1,2$) on the singular conifold, before the deformation has dynamically taken place. The properties of the different kinds of fractional branes will be explained at the end of this section and in section \ref{cascade-SUGRA}; what matters for the time being is that this brane configuration will give rise to a field theory with gauge groups $SU(r_l)\times SU(r_s)$ and flavour groups $SU(N_{fl})$ and $SU(N_{fs})$ for the two gauge groups respectively, with the  matter content displayed in Fig. \ref{quiverN}.
The most convenient technique for our purpose has been that of performing a T-duality along the isometry $(z_1,z_2)\to(e^{i\alpha}z_1,e^{-i\alpha}z_2)$. Once the system is mapped into type IIA, the spectrum is directly read off and the superpotential comes from the analysis of its moduli space \cite{Radu}.

The field content of the gauge theory can be read from the quiver diagram of Fig. \ref{quiverN}. It is an extension of the Klebanov-Strassler field theory with nonchiral flavours for each gauge group.
The superpotential is%
\footnote{Sums over gauge and flavour indices are understood.}
\begin{equation}
\begin{split}
W & = \lambda (A_1 B_1 A_2 B_2 -A_1 B_2 A_2 B_1) + h_1 \, \tilde{q} (A_1 B_1 + A_2 B_2)q + h_2 \, \tilde{Q} (B_1 A_1 + B_2 A_2)Q + \\
&\quad +\alpha \, \tilde{q} q \tilde{q} q + \beta \, \tilde{Q} Q \tilde{Q} Q  \;.
\end{split}
\end{equation}
The factors $A_1B_1 + A_2B_2$ directly descend from the embedding equation \eqref{embedding eq}, while the quartic term in the fundamental fields is derived from type IIA.
This superpotential explicitly breaks the $SU(2) \times SU(2)$ global symmetry of the unflavoured theory to $SU(2)_D$, but this global symmetry is recovered after the smearing.

The $N_f$ flavours are split into $N_{fl}$ and $N_{fs}$ groups, according to which gauge group they are charged under. Both sets come from D7-branes along the embedding \eqref{embedding eq}.%
\footnote{The embedding is in fact invariant under the $\bbZ_2$ ($z_3 \leftrightarrow z_4$) that exchanges the two gauge groups.} 
The only feature that discriminates between these two kinds of (fractional) D7-branes is their coupling to the RR $C_2$ and $C_4$ gauge potentials. On the singular conifold, before the dynamical deformation, there is a vanishing two-cycle, living at the singularity, which the D7-branes are wrapping. According to the worldvolume flux on it, the D7-branes couple either to one or the other gauge group. Since this flux is stuck at the origin, far from the branes we can only measure the D3, D5 and D7-charges produced. Unfortunately three charges are not enough to fix four ranks. This curious ambiguity will show up again in section \ref{cascade-SUGRA}.

\subsection{The cascade} \label{cascade subsection}

One can assume that, as in the unflavoured case discussed in subsection \ref{fractionalD3}, the $\beta$-functions of the two gauge couplings have opposite sign. When the gauge coupling of the gauge group with larger rank is very large, one can go to a Seiberg-dual description \cite{Seiberg:1994pq}: remarkably, it is straightforward to see that the quartic superpotential is such that the field theory is self-similar, namely the field theory in the dual description is a quiver gauge theory with the same field content and superpotential, except for changes in the ranks of the groups.

Let us define the theory at some energy scale to be an $SU(r_l)\times SU(r_s)$ gauge theory (where $l$ stands for the larger gauge group and $s$ for the smaller: $r_l > r_s$), with flavour group $SU(N_{fl})$ ($SU(N_{fs})$) for $SU(r_l)$ ($SU(r_s)$). In the beginning we can set, conventionally, $r_l=r_1$, $r_s=r_2$, $N_{fl}=N_{f1}$, $N_{fs}=N_{f2}$; after a Seiberg duality on the gauge group with the larger rank, the field theory is $SU(2r_2-r_1+N_{f1})\times SU(r_2)$, with again $N_{f1}$ and $N_{f2}$ flavours respectively. In identifying which gauge group is now the larger and which is the smaller, we have to exchange the labelling of the groups, so that we get  $r'_l=r_2$, $r'_s=2r_2-r_1+N_{f1}$, $N'_{fl}=N_{f2}$ and $N'_{fs}=N_{f1}$.
The assumption leads to an RG flow which is described by a cascade of Seiberg dualities, analogous to that described in subsection \ref{fractionalD3}. In the UV the ranks of the gauge groups are much larger than their disbalance, which is much larger than the number of flavours. Hence the assumption of having $\beta$-functions with opposite sign is justified in the UV flow of the field theory.

The supergravity background of section \ref{KSflavoured} is dual to a quiver gauge theory where the cascade goes on until the IR, with nonperturbative dynamics at the end, as in the Klebanov-Strassler solution discussed in subsection \ref{deformedconifold}. 

In the background of section \ref{KTflavoured}, the cascade does not take place anymore below some value of the radial coordinate, and it asymptotes to the flavoured Klebanov-Witten solution studied in chapter \ref{KW}. In the field theory, this reflects the fact that, because of a suitable choice of the ranks, the last step of the cascade leads to a theory where the $\beta$-functions of both  gauge couplings are positive. The IR dynamics is the one discussed in chapter \ref{KW}, but with a quartic superpotential for the flavours.

The description of the duality cascade in our solutions and its interesting UV behaviour will be the content of the next section.

\section{The cascade: supergravity side}\label{cascade-SUGRA}
\medskip
\setcounter{equation}{0}

We claim that our supergravity solutions are dual to the class of quiver gauge theories with backreacting fundamental flavours introduced in the previous section. Indeed we will show that the effective brane charges, the R-anomalies and the $\beta$-functions of the gauge couplings that we can read from the supergravity solutions precisely match the picture of a cascade of Seiberg dualities that describes the RG flow of the field theories. Thus we are generalising the results of \cite{Klebanov:2000nc,Klebanov:2000hb} to gauge theories which include dynamical flavours.

\subsection{Effective brane charges and ranks}
\label{charges and ranks}

By integrating fluxes over suitable compact cycles, we can compute three effective D-brane charges in our solutions, which are useful to pinpoint the changes in the ranks of gauge  groups when the field theory undergoes a Seiberg duality: one of them (D7) is dual to a quantity which is constant along the RG flow, whereas two of them (D3, D5) are not independent of the holographic coordinate and are  dual to the nontrivial part of the RG flow. Recall that the (Maxwell) charges of D3- and D5-brane ($N_{eff}$ and $M_{eff}$)  for our ansatz were  already calculated in section \ref{section-ansatz} (see eqs. (\ref{Neff}) and (\ref{Meff})). 
Let us now compute the D7-brane charge, integrating \eqref{smearing} on a two-manifold with boundary   which is intersected once by all the smeared D7-branes (e.g. $\mathcal{D}_2$: $\theta_2=\text{const.}$, $\varphi_2=\text{const.}$, $\psi=\text{const.}$). This charge is conserved along the RG flow because no fluxes appear on the right hand side of \eqref{smearing}. The D7-brane charge, which we interpret as the total number of flavours added to the Klebanov-Strassler gauge theory, is indeed:
\begin{equation} \label{Nflav}
N_{flav}\equiv \int_{\mathcal{D}_2} dF_1 = N_f\;.
\end{equation}

Another important quantity was already introduced in (\ref{SDfield}) and it is the integral of $B_2$ over the nontrivial two-cycle $S^2$: $\theta_1 = \theta_2 \equiv \theta$, $\varphi_1 = 2\pi -\varphi_2 \equiv \varphi$, $\psi=\text{const.}$:
\begin{equation} \label{b_0}
b_0(\tau)\equiv \frac{1}{4\pi^2} \int_{S^2} B_2 =  \frac{M}{\pi} \Bigl( f \sin^2\frac{\psi}{2} + k\cos^2\frac{\psi}{2} \Bigr)\;.
\end{equation}
This quantity is important because string theory is invariant as it undergoes a shift of 1. Recall that in the KW background it amounts to a Seiberg duality, and the same happens here.
So we will shift this last quantity by one unit, identify a shift in the radial variable $\tau$ that realises the same effect and see what happens to $M_{eff}$ and $N_{eff}$. This was the process that we already followed in subsection \ref{fractionalD3} when we studied the cascade of Seiberg dualities of the theory without flavours. Actually, the cascade will not work along the whole flow down to the IR but only in the UV asymptotic (below the UV cut-off $\tau_0$ obviously). Notice that the same happens for the unflavoured solutions (see subsection \ref{deformedconifold}). This is expected since the last step of the cascade is not a Seiberg duality. Thus we will not be worried and compute the cascade only in the UV asymptotic for large $\tau$ which also requires $\tau_0 \gg 1$ (we neglect $\mathcal{O}(e^{-\tau})$): in that regime the functions $f$ and $k$ become equal and $b_0$ is $\psi$-independent.\\
Actually, we will not compute the explicit shift in $\tau$ but rather the shift in the functions $f$ and $k$.
We have:
\begin{equation}
\begin{split}
b_0 (\tau) \to b_0(\tau') = b_0 (\tau) - 1 
\end{split}
\quad 
\Longrightarrow \quad 
\begin{split}
f(\tau) &\to f(\tau') = f(\tau)  - \frac{\pi}{M} \,\, , \\
k(\tau) &\to k(\tau') = k(\tau)  - \frac{\pi}{M} \,\, .
\end{split} \label{shift f k}
\end{equation} 

Correspondingly, after a Seiberg duality step from $\tau$ to $\tau'<\tau$, that is going towards the IR, we have:
\begin{align} 
N_f &\to N_f \label{scaling sugra 1} \,\, ,\\
M_{eff} (\tau) &\to M_{eff} (\tau') = M_{eff} (\tau) - \frac{N_f}{2} \label{scaling sugra Meff}\,\, ,\\
N_{eff} (\tau) &\to N_{eff} (\tau') = N_{eff} (\tau) - M_{eff} (\tau) + \frac{N_f}{4} \,\, . \label{scaling sugra 2}
\end{align}
This result is valid for all of our solutions.

We would like to compare this result with the action of Seiberg duality in field theory, as computed in section \ref{sect: Field Th}. We need an identification between the brane charges computed in supergravity and the ranks of the gauge and flavour groups in the field theory.

The field theory of interest for us has gauge groups $SU(r_l) \times SU(r_s)$ ($r_l>r_s$) and flavour groups $SU(N_{fl})$ and $SU(N_{fs})$ for the gauge groups $SU(r_l)$ and $SU(r_s)$ respectively. It is engineered, at least effectively at some radial distance, by the following objects: $r_l$ fractional D3-branes of one kind (D5-branes wrapped on the shrinking two-cycle), $r_s$ fractional D3-branes of the other kind ($\overline{\text{D}5}$-branes wrapped on the shrinking cycle, supplied with $-1$ quanta of gauge field flux on the two-cycle), $N_{fs}$ fractional D7-branes without gauge field strength on the two-cycle, and $N_{fl}$ fractional D7-branes with $-1$ units of gauge field flux on the shrinking two-cycle. This description is good for $b_0\in [0,1]$.

This construction can be checked explicitly in the case of the $\mathbb{C} \times \mathbb{C}^2/\mathbb{Z}_2  $ orbifold \cite{Bertolini:2001qa, Bertolini:2000dk}, where one is able to quantise the open and closed string system for the case $b_0=\frac{1}{2}$ \cite{Aspinwall:1995zi}. That is the $\mathcal{N}=2$ CFT, which flows to the field theory that we are considering when equal and opposite masses are given to the adjoint chiral superfields (the geometric description of this relevant deformation is a blowup of the orbifold singularity)  \cite{Klebanov:1998hh,Morrison:1998cs}. Fractional branes are those branes which couple to the twisted closed string sector.%
\footnote{Notice that one can build a regular D3-brane (\emph{i.e.} not coupled to the twisted sector) by means of a fractional D3-brane of one kind and a fractional D3-brane of the other kind. This regular brane can move outside the orbifold singularity. On the contrary, there is no regular D7-brane: the two kinds of fractional D7-branes, extending entirely along the orbifold, cannot bind into a regular D7-brane that does not touch the orbifold fixed locus and is not coupled to the twisted sector \cite{Bertolini:2001qa}.}

Here we will consider a general background value for $B_2$. In order to compute the charges, we will follow quite closely the computations in \cite{Grana:2001xn}.

We will compute the charges of D7-branes and wrapped D5-branes on the singular conifold (\ref{zcoor}). The D5 Wess-Zumino action (see eq. (\ref{WZDpbraneaction})) is
\begin{equation}
S_{D5} = T_5 \int_{M^4\times S^2} \Bigl\{ C_6 + (2\pi\, F_2 + B_2)\wedge C_4 \Bigr\}\;,
\end{equation} 
where $S^2$ is the only two-cycle in the conifold, vanishing at the tip, that the D5-brane is wrapping. We write also a worldvolume gauge field $F_2$ on $S^2$. Then we expand:
\begin{equation}
B_2 = 2\pi\, \theta_B\,  W_2 \, ,\qquad\qquad \theta_B = 2\pi \, b_0 \, ,\qquad\qquad F_2 = \Phi\, W_2\;,
\end{equation}
where $W_2$ is the two-form\footnote{This two-form is a rescaling of the two-form introduced in eq. (\ref{Upsilon}).} on the two-cycle, which satisfies $\int_{S^2} W_2 = 1$. In this conventions, $b_0$ has period 1, and $\Phi$ is quantised in $2\pi \, \mathbb{Z}$. We obtain (using $T_p (4\pi^2)=T_{p-2}$, see eq. (\ref{Dptension})):
\begin{equation}
S_{D5} = T_5 \int_{M^4\times S^2} C_6 + \frac{T_3}{2\pi} \int_{M^4} (\Phi + \theta_B) \, C_4 \;.
\end{equation}
The first fractional D3-brane \cite{Polchinski:2000mx} is obtained with $\Phi=0$ and has D3-charge $b_0$ and D5-charge 1. The second fractional D3-brane is obtained either as the difference with a D3-brane or as an anti-D5-brane (global $-$ sign in front) with $\Phi=-2\pi$. It has D3-charge $1-b_0$ and D5-charge -1. These charges are summarised in Table \ref{Table:general charges}.

Now consider a D7-brane along the surface $z_3 + z_4 = 0$. It describes a $z_1 z_2 + z_3^2=0$ inside the conifold (\ref{zcoor}), which is a copy of $\mathbb{C}^2/\mathbb{Z}_2$.  The D7 Wess-Zumino action (\ref{WZDpbraneaction}) is (up to a curvature term considered below)
\begin{equation}
S_{D7} = T_7 \int_{M^4\times \Sigma} \Bigl\{ C_8 + (2\pi\, F_2 + B_2) \wedge C_6 + \frac{1}{2} (2\pi\, F_2 + B_2) \wedge (2\pi\, F_2 + B_2) \wedge C_4 \Bigr\}\;.
\end{equation}
The surface $\Sigma = \mathbb{C}^2/\mathbb{Z}_2$ has a vanishing two-cycle at the origin. Since the conifold has only one two-cycle, these two must be one and the same and we can expand on $\Sigma$ using $W_2$ again. Moreover, being $2\,W_2$ the Poincar\'e dual to the two-cycle on $\Sigma$, 
\begin{equation}
\int_\Sigma W_2 \wedge \alpha_2 = \frac{1}{2} \int_{S^2} \alpha_2
\end{equation} 
holds for any closed two-form $\alpha_2$. The fact that the Poincar\'e dual to the two-cycle $S^2$ on $\Sigma = \mathbb{C}^2/\mathbb{Z}_2$ is $2 \, W_2$ follows from our normalisation $\int_{S^2} W_2=1$ and from the self-intersection number of the $S^2$ living at the singularity, namely:\footnote{The minus sign on the right-hand side of (\ref{self-intersectionnumber}) comes from the sign of the pullback volume form on $S^2$ and $\Sigma$.} 
\beq  \label{self-intersectionnumber}
-2\,=\, \# (S^2,S^2)\, \equiv - \, \int_{\Sigma} (2 \, W_2) \wedge (2 \, W_2) \,\, . 
\eeq

There is another contribution of induced D3-charge coming from the curvature coupling \cite{Bershadsky:1995qy}:
\begin{equation}
\frac{T_7}{96} (2\pi)^2 \int_{M^4\times \Sigma} C_4 \wedge \Tr \, \mathcal{R}_2\wedge \mathcal{R}_2 = -T_3 \int_{M^4\times \Sigma} C_4 \wedge \frac{p_1(\mathcal{R})}{48}\;,
\end{equation}
where $\mathcal{R}_2$ is the curvature two-form and $p_1(\mathcal{R})=$ is the first Pontryagin class of the manifold $M^4\times \Sigma$. This can be computed in the following way. On K3, $p_1(\mathcal{R})= 48$ and the induced D3-charge is $-1$. In the orbifold limit K3 becomes $T^4/\mathbb{Z}_2$ which has sixteen orbifold singularities. Thus on $\mathbb{C}^2/\mathbb{Z}_2$ the induced D3-charge is $-1/16$. 
Putting all together we get:
\begin{equation}
S_{D7} = T_7 \int_{M^4\times \Sigma} C_8 + \frac{T_5}{4\pi} \int_{M^4 \times S^2} (\Phi + \theta_B) \, C_6 + \frac{T_3}{16\pi^2} \int_{M^4} \Bigl[ (\Phi + \theta_B)^2 - \pi^2 \Bigr] \, C_4\;.
\end{equation} 
The second fractional D7-brane (the one that couples to the second gauge group) is obtained with $\Phi=0$ and has D7-charge 1, D5-charge $\frac{b_0}{2}$ and D3-charge $(4b_0^2-1)/16$. With $\Phi=2\pi$ we get a non-SUSY or non-minimal object (see \cite{Polchinski:2000mx} for some discussion of this). The first fractional D7-brane (coupled to the first gauge group) has $\Phi=-2\pi$ and has D7-charge 1, D5-charge $\frac{b_0-1}{2}$ and D3-charge $(4(b_0-1)^2-1)/16$. This is summarised in Table \ref{Table:general charges}. 
Which fractional D7-brane provides flavours for the gauge group of which fractional D3-brane can be determined from the orbifold case with $b_0=\frac{1}{2}$ (compare with \cite{Bertolini:2001qa}).

\begin{table}[ht]
\begin{center}
\begin{tabular}{c|cccc}
 Object & frac D3 (1) & frac D3 (2) & frac D7 (1) & frac D7 (2) \\
\hline

D3-charge                    & $b_0$       & $1-b_0$     & $\dfrac{4(b_0-1)^2-1}{16}$ & $\dfrac{4b_0^2-1}{16}$ \\
\\
D5-charge                    & 1           & $-1$        & $\dfrac{b_0-1}{2}$ & $\dfrac{b_0}{2}$ \\
\\
D7-charge                    & 0           & 0           & 1           & 1\\
 \hline
 Number of objects & $r_l$ & $r_s$ & $N_{fl}$ & $N_{fs}$\\
\end{tabular}
\caption{Charges of fractional branes on the conifold. \label{Table:general charges}}
\end{center}
\end{table}

Given these charges, we can compare with the field theory cascade. First of all we construct the dictionary:
\begin{align}
N_f &= N_{fl} + N_{fs} \,\, ,\\
M_{eff} &= r_l - r_s + \frac{b_0-1}{2} N_{fl} + \frac{b_0}{2} N_{fs}\label{dictionaryM} \,\, ,\\
N_{eff} &= b_0 \, r_l + (1-b_0) \, r_s + \frac{4(1-b_0)^2-1}{16} N_{fl} + \frac{4b_0^2-1}{16} N_{fs} \,\, . \label{dictionary}
\end{align}
To derive this, we have only used that the brane configuration that engineers the field theory that we consider consists of $r_l$ fractional D3 of the first kind, $r_s$ fractional D3 of the second kind, $N_{fl}$ fractional D7 of the first kind and $N_{fs}$ fractional D7 of the second kind. Recall that, by convention, $r_l>r_s$ and $N_{fl}$ ($N_{fs}$) are the flavours for $SU(r_l)$ ($SU(r_s)$).

It is important to remember that $b_0$ is defined modulo 1, and shifting $b_0$ by one unit amounts to go to a Seiberg dual description in the field theory. At any given energy scale in the cascading gauge theory, there are infinitely many Seiberg dual descriptions of the field theory because Seiberg duality is exact along the RG flow \cite{Strassler:2005qs}. Among these different pictures, there is one which gives the best effective description of the field theory degrees of freedom around that energy scale (this is also the  description with positive squared gauge couplings): it is the one where $b_0$ has been redefined, by means of a large gauge transformation, so that $b_0\in[0,1]$ (see subsection \ref{Large-gauge}). This is the description that we will use when we effectively engineer the field theory in terms of branes in some range of the RG flow that lies between two adjacent Seiberg dualities. 

In field theory, as before, we start with gauge group $SU(r_1) \times SU(r_2)$ and $N_{f1}$ flavours for $SU(r_1)$, $N_{f2}$ flavours for $SU(r_2)$, with $r_1>r_2$. The gauge group $SU(r_1)$ flows towards strong coupling. When its gauge coupling diverges we turn to a Seiberg dual description.
After the Seiberg duality on the larger gauge group,  we get $SU(2r_2 - r_1 + N_{f1}) \times SU(r_2)$ and the flavour groups are left untouched. 

The effective D5- and D3-brane charges of a brane configuration that engineers this field theory {\it before} the duality are: 
\begin{equation}
\begin{split}
M_{eff} &= r_1-r_2+ \frac{b_0-1}{2}N_{f1} + \frac{b_0}{2}N_{f2}\;,\\
N_{eff} &= b_0 r_1 + (1-b_0)r_2 + \frac{4(1-b_0)^2-1}{16}N_{f1} + \frac{4b_0^2-1}{16}N_{f2}\;.
\end{split}
\end{equation} 
{\it After} the duality they become: 
\begin{equation}
\begin{split}
M'_{eff} &= -r_2 + r_1 - N_{f1} + \frac{b_0-1}{2}N_{f2} + \frac{b_0}{2}N_{f1}= M_{eff}-\frac{N_f}{2}\;,\\
N'_{eff} &= b_0r_2 + (1-b_0)(2r_2-r_1+N_{f1}) + \frac{4(1-b_0)^2-1}{16}N_{f2} + \frac{4b_0^2-1}{16}N_{f1}=\\ &=N_{eff}-M_{eff}+\frac{N_f}{4}\;.
\end{split}
\end{equation}
They \emph{exactly} reproduce the SUGRA behaviour \eqref{scaling sugra 1}-\eqref{scaling sugra 2}.
Notice that the matching of the cascade between supergravity and field theory is there, irrespective of how we distribute the flavours between the two gauge groups. From the three charges and the cascade, we are not able to determine how the flavours are distributed but only their total number.

We conclude with some remarks. Even though the effective brane charges computed in supergravity are running and take integer values only at some values of the holographic coordinate, the ranks of gauge and flavour groups computed from them are constant and integer (for suitable choice of the integration constants) in the whole range of radial coordinate dual to the energy range where we use a specific field theory description. This range of scales is $b_0 \in [0,1]$ mod 1. At the boundaries of this region, we perform a Seiberg duality and go into a new more effective description. In particular, if ranks are integer before the duality, they still are after it; meanwhile we shift $b_0$ by one unit.
Hence the field theory description of the cascade is perfectly matched by the ranks that we can compute from our supergravity solution.

Notice also that the fact that $M_{eff}$ shifts by $N_f/2$ instead of $N_f$ confirms that the flavoured version of the Klebanov-Strassler theory that we are describing has nonchiral flavours (with a quartic superpotential) rather than chiral flavours (with a cubic superpotential) like in \cite{Ouyang:2003df, Benini:2007kg}, where the shift goes with units of $N_f$. 

Finally, we want to stress again that we are engineering a field theory with four objects but we have only three charges to recognize them. The comparison of the cascade between SUGRA and field theory, surprisingly enough, does not help.


\subsection{Seiberg duality as a large gauge transformation}
\label{Large-gauge}

We have argued that a shift by a unit of the normalised flux $b_0$ 
as we move towards the IR along the holographic direction 
is equivalent to performing a Seiberg duality step on the field theory side (see equations \eqref{shift f k}-\eqref{scaling sugra 2}). Moreover, we have checked that, under this shift of $b_0$,  the change of the effective (Maxwell) charges $M_{eff}$ and $N_{eff}$ of supergravity is exactly the same as the one computed in the field theory engineered with fractional branes on the singular conifold. 

In this subsection we will present an alternative way of understanding, in supergravity, Seiberg duality at a {\it fixed} energy scale. As we know, for a given value of the holographic coordinate $\tau$, the value of $b_0$ lies generically outside the interval $[0,1]$, where a good field theory description exists. However, the flux of the $B_2$ field is not a gauge invariant quantity in supergravity and can be changed under a large gauge transformation. Indeed, let us take the two-form $\Upsilon_2$ defined in equation (\ref{Upsilon}) and let us change $B_2$ as follows: 
\beq
B_2\to B_2+\Delta B_2\,\,,
\qquad\qquad
\Delta B_2\,=\,-\pi n \Upsilon_2\,\,,
\qquad\qquad
n\,\in \mathbb{Z}\,\,.
\label{large}
\eeq
As $d\Upsilon_2=0$, the field strength $H_3$ does not change and our
transformation is a gauge  transformation of the NSNS field. However the
flux of $B_2$ does change as:
\beq
\int_{S^2}B_2 \to \int_{S^2}B_2 \,-\, 4 \pi^2 n \,\, ,
\label{flux-change}
\eeq
or, equivalently $b_0 \to b_0 - n$. This non-invariance of the flux shows
that this transformation of $B_2$ is a large gauge transformation which
cannot be globally written as $\Delta  B_2=d\Lambda$. Moreover, as always
happens with large gauge transformations, it is
quantised. 
If we want that our transformation
(\ref{large})  be a gauge transformation of supergravity, it should leave the RR
field strength $F_3$ invariant. Defining the potential $C_2$ as
$dC_2=F_3-B_2 \wedge F_1 $, we see that $dC_2$ must change as:
\beq
dC_2 \to dC_2 \,+\, {{n N_f} \over 4} g^5 \wedge \Upsilon_2 \,\, .
\eeq
One can verify that this change of $dC_2$ can be obtained if the
variation of $C_2$ is (see equations \eqref{ansatz} and \eqref{F3 H3 KT}):
\beq
\Delta C_2\,=\,{nN_f\over 8}\,\,\Big[\,
(\psi-\psi^*)\,(\,\sin\theta_1 d\theta_1\wedge d\varphi_1\,-\,
\sin\theta_2\,d\theta_2\wedge d\varphi_2\,)\,-\,\cos\theta_1\cos\theta_2\,
d\varphi_1\wedge d\varphi_2\,\Big]\,\,,
\label{changeC2}
\eeq
where $\psi^*$ is a constant.  In the study of the R-symmetry anomaly of
the next subsection it will be convenient
to know the change of $C_2$ on the submanifold $S^2$:
$\theta_1=\theta_2=\theta$,  $\varphi_1=2\pi-\varphi_2=\varphi$. Denoting
by  $C_2^{eff}$  the RR potential $C_2$ restricted to this
cycle, we get from (\ref{changeC2}) that: 
\beq
\Delta C_2^{eff}\,=\,{{n N_f} \over 4} (\psi - \psi^*) \, 
\sin\theta d\theta\wedge d\varphi\,\,. \label{DeltaC2}
\eeq

Let us now study how the Page charges change under these large
gauge transformations. From the expressions written in (\ref{D5D3Page}), we
obtain:
\begin{equation}
\begin{split}
\Delta Q^{Page}_{D5}&=  -{1 \over {4 \pi^2}} \int_{S^3} 
\Delta B_2 \wedge F_1\,\, ,\\
\Delta Q^{Page}_{D3}&={1 \over {(4 \pi^2)^2}} \int_{M_5} 
\Big(-\Delta B_2 \wedge F_3 + \Delta B_2 \wedge B_2 \wedge F_1
+{1\over 2}\,\Delta B_2\wedge\Delta B_2\wedge F_1\,\Big)\,.
\end{split}   \label{DeltaQs}
\end{equation}
By using  in (\ref{DeltaQs}) our ansatz for $F_3$ and $B_2$ \eqref{ansatz}, together
with the expression of $ \Delta B_2 $ given in (\ref{large}) as well as
the relations \eqref{intS3} and \eqref{int5g}, one readily gets:
\begin{equation}
\begin{split}
\Delta Q^{Page}_{D5}  \,&=\, n {N_f \over 2} \,\, ,\\
\Delta Q^{Page}_{D3} \,&=\, n \, M \,+\,n^2 {N_f \over 4}
\, \,. 
\end{split} \label{Delta-Page}
\end{equation}
Thus, under a  large gauge transformation (\ref{large}) with $n=1$, the Page charges
transform as:
\begin{equation}
\begin{split}
Q^{Page}_{D5} &\to Q^{Page}_{D5} \,+\,  {N_f \over 2} \,\, ,\\
Q^{Page}_{D3} &\to Q^{Page}_{D3} \,+\, \, M \,+ {N_f \over 4}
\,\,.
\end{split}\label{Page-change}
\end{equation}
Recall that for our ansatz $Q^{Page}_{D5}=M$ and $Q^{Page}_{D3}=N_0$ (see eqs.
(\ref{QD5=M}) and (\ref{QD3=N0})). Thus, eq. (\ref{Page-change}) gives how these constants change under a large gauge transformation. At a given holographic scale $\tau$ we should perform
as many large transformations as needed to have $b_0\in[0,1]$.  Given that $b_0$ grows when 
the holographic coordinate increases, the transformation (\ref{Page-change}) should correspond to the change of ranks under a Seiberg duality when we flow towards the UV. By comparing (\ref{Page-change}) with our previous expressions one can show that this is the case. Actually, one can get an explicit expression of $Q^{Page}_{D5} $ and $Q^{Page}_{D3}$ in terms of the ranks $r_l$ and $r_s$ and the number of flavours $N_{fl}$ and $N_{fs}$. In order to verify this fact, let us suppose that we are in a region of the holographic coordinate such that the two functions $f$ and $k$ of our ansatz are equal. Notice that for the flavoured KS solution this happens in the UV, while for the flavoured KT this condition 
holds for all values of the radial coordinate. If $f=k$ the normalised flux $b_0$ in (\ref{b_0}) can be written as:
\beq
b_0(\tau)\,=\,{M\over \pi}\,f(\tau)\,\,.
\eeq
Using this expression we can  write the D5-brane Page charge (\ref{QD5-Meff}) as:
\beq
Q^{Page}_{D5}\,=\,M_{eff}\,-\,{N_f\over 2}\,\,b_0\;.
\label{QD5UV}
\eeq
Notice also that the supergravity expression (\ref{Meff}) of $M_{eff}$ can be written  when $f=k$ as:
\beq
M_{eff}\,=\,M\,+\,{N_f\over 2}\,\,b_0\,\,.
\label{Meff-M-UV}
\eeq
Let us next assume that we have chosen our gauge such that, at the given holographic scale, 
$b_0\in [0,1]$.  In that case we can use the value of $M_{eff}$ obtained by the field theory calculation of subsection \ref{charges and ranks} to evaluate the Page charge $Q^{Page}_{D5}$. Actually, by plugging the value of 
 $M_{eff}$ given in (\ref{dictionaryM}) on the right-hand side of (\ref{QD5UV}) we readily get the following
 relation between $Q^{Page}_{D5}$ and the field theory data:
 \beq
 Q^{Page}_{D5}\,=\,r_l\,-\,r_s\,-\,{N_{fl}\over 2}\,\,.
  \label{QD5-ranks}
 \eeq
Similarly, for $f=k$, one can express the D3-brane Page charge (\ref{QD3-Neff}) as:
\beq
 Q^{Page}_{D3}\,=\,N_{eff}\,-\,b_0M\,-\,{b_0^2\over 4}\,N_f\,\,,
 \eeq
which, after using the relation (\ref{Meff-M-UV}), can be written in terms of $M_{eff}$ as:
\beq
 Q^{Page}_{D3}\,=\,N_{eff}\,-\,b_0M_{eff}\,+\,{N_f\over 4}\,b_0^2\,\,.
 \eeq
Again, if we assume that $b_0\in [0,1]$ and use the field theory expressions (\ref{dictionary}) and
 (\ref{dictionaryM})  of $N_{eff}$ and $M_{eff}$, we get:
 \beq
 Q^{Page}_{D3}\,=\,r_s\,+\,{3N_{fl}-N_{fs}\over 16}\,\,.
 \label{QD3-ranks}
 \eeq
Notice that, as it should,  the expressions (\ref{QD5-ranks}) and  (\ref{QD3-ranks})  of 
$Q^{Page}_{D5}$ and $Q^{Page}_{D3}$ that 
we have just found are independent of $b_0$, as far as $b_0\in [0,1]$. Moreover, one can verify that under a field theory Seiberg duality the right-hand sides of 
(\ref{QD5-ranks}) and  (\ref{QD3-ranks}) transform  as the left-hand sides do 
under a large gauge transformation of supergravity.

Finally, let us point out that in this approach Seiberg duality is performed at a fixed 
energy scale and $M_{eff}$ and $N_{eff}$ are left 
invariant (recall that Maxwell charges are gauge invariant). Indeed, by
looking at our ansatz for $B_2$ one easily concludes that the change of
$B_2$ written in (\ref{large}) is equivalent to the following change in
the functions $f$ and $k$
\beq
f\to f-{\pi\over M}\,n\,\,,
\qquad\qquad
k\to k-{\pi\over M}\,n\,\,,
\label{fk-changes}
\eeq
and one can verify that the changes (\ref{Delta-Page}) and
(\ref{fk-changes}) leave the expressions of $M_{eff}$ and $N_{eff}$,
as written in eqs. (\ref{QD5-Meff}) and (\ref{QD3-Neff}),  invariant. 
From eqs.  (\ref{QD5-ranks}) and (\ref{QD3-ranks}) it is clear that the Page charges provide a clean
way to extract the ranks and number of flavours of the corresponding (good)
field theory dual at a given energy scale. Actually, the ranks of this good field
theory description change as  step-like functions along the RG flow, due to the fact that
$b_0$ varies continuously and needs to suffer a large gauge transformation every time that,
flowing towards the IR,  it reaches the value $b_0=0$ in the good gauge. This large gauge transformation changes $Q^{Page}_{D5}$ and $Q^{Page}_{D3}$ in the way described above, which realises in supergravity the change of the ranks under a Seiberg duality in field theory.

Let us now focus on a different way of matching the behaviour of the field theory and our solutions.

\subsection{R-symmetry anomalies and $\beta$-functions}

We can compute the $\beta$-functions (up to the energy-radius relation) and the R-symmetry anomalies for the two gauge groups both in supergravity and in field theory in the spirit of subsection \ref{fractionalD3}. In the UV, where the cascade takes place, they nicely match. For the comparison we make use again of the holographic formulae ((\ref{RGholography}) and (\ref{holographic_theta})) derived in the $\mathcal{N}=2$ orbifold case. It will be useful to what follows to write them down again:
\begin{equation} \begin{split} \label{holographic relations}
\frac{4\pi^2}{g_l^2} + \frac{4\pi^2}{g_s^2} &= \pi\, e^{-\phi} \,\, ,\\
\frac{4\pi^2}{g_l^2} - \frac{4\pi^2}{g_s^2} &= \frac{e^{-\phi}}{2\pi} \Bigl[ \int_{S^2} B_2 - 2\pi^2 \; (\text{mod } 4\pi^2) \Bigr] \,\, ,
\end{split}
\qquad\qquad
\begin{split}
\Theta_l + \Theta_s &= - 2\pi \, C_0 \,\, ,\\
\Theta_l - \Theta_s &= \frac{1}{\pi} \int_{S^2} C_2 \;.
\end{split}
\end{equation}
 
Recall that strictly speaking, these formulae need to be corrected for small values of the gauge couplings and are only valid in the large 't Hooft coupling regime (see \cite{Dymarsky:2005xt, Benvenuti:2005wi, Strassler:2005qs}), which is the case under consideration. Moreover, they give positive squared couplings only if $b_0=\frac{1}{4\pi^2}\int_{S^2} B_2$ is in the range $[0,1]$. This is the physical content of the cascade: at a given energy scale we must perform a large gauge transformation on $B_2$ in supergravity to shift $\int B_2$ by a multiple of $4\pi^2$ to get a field theory description with positive squared couplings.

We have adapted the indices in \eqref{holographic relations} to the previous convention for the gauge group with the larger (the smaller) rank. Let us restrict our attention to an energy range, between two subsequent Seiberg dualities, where a field theory description in terms of specific ranks holds. In this energy range the gauge coupling $g_l$ of the gauge group with larger rank flows towards strong coupling, while the gauge coupling $g_s$ of the gauge group with smaller rank flows towards weak coupling.
Indeed, as formulae \eqref{holographic relations} confirm, the coupling $g_l$ was not touched by the previous Seiberg duality and starts different from zero. It flows to $\infty$ at the end of this range where a Seiberg duality on its gauge group is needed. The coupling $g_s$ of the gauge group with smaller rank is the one which starts very large (actually 
divergent) after the previous Seiberg duality on its gauge group and then flows toward weak coupling.

In supergravity, due to the presence of magnetic sources for $F_1$, we cannot define a potential $C_0$. Therefore we project our fluxes on the submanifold $\theta_1 = \theta_2\equiv\theta$, $\varphi_1 = 2\pi-\varphi_2\equiv\varphi$, $\forall\, \psi,\tau$ before integrating them. Recalling that $F_3 = dC_2 + B_2\wedge F_1$, what we get from \eqref{F1}-\eqref{ansatz} (in the UV limit) are the effective potentials
\begin{equation} \label{C0 C2 eff}
C_0^{eff} = \frac{N_f}{4\pi} \, (\psi - \psi_0^*) \,\, , \qquad\qquad \tilde{C}_2^{eff} = \Bigl[\frac{M}{2} + \frac{n N_f}{4}\Bigr] \, (\psi - \psi_2^*) \, \sin{\theta}\,d\theta\wedge d\varphi \;.
\end{equation}
The integer $n$ in $\tilde{C}_2^{eff}$ comes from a large gauge transformation on $B_2$ (Seiberg duality in field theory, see eq. (\ref{DeltaC2})) which shifts $b_0 (\tau) \in [n,n+1]$ by $n$ units - so that the gauge transformed $\tilde{b}_0 (\tau)= b_0 (\tau) - n$ is between 0 and 1 - and at the same time shifts $d C_2^{eff}\to d \tilde{C}_2^{eff} = d C_2^{eff} + \pi n \frac{N_f}{4\pi} \sin\theta \,d\theta\wedge d\varphi\wedge d\psi$, since $F_3$ is gauge-invariant, but leaves $C_0$ invariant.

The field theory possesses an anomalous R-symmetry which assigns charge $\frac{1}{2}$ to all chiral superfields.
The field theory R-anomalies are easily computed using equation (\ref{QFTanomaly}). Continuing to use $r_l$ ($r_s$) for the larger (smaller) group rank and $N_{fl}$ ($N_{fs}$) for the corresponding flavours (see Fig. \ref{quiverN}), the anomalies under a $U(1)_R$ rotation of parameter $\varepsilon$ are:
\begin{equation}
\text{Field theory:} \qquad\qquad
\begin{aligned}
\delta_\varepsilon \Theta_l &= [2(r_l - r_s) - N_{fl}] \, \varepsilon \,\, ,\\
\delta_\varepsilon \Theta_s &= [-2(r_l - r_s) - N_{fs}] \, \varepsilon \;.
\end{aligned} \end{equation}
Along the cascade of Seiberg dualities, the coefficients of the anomalies for the two gauge groups change when we change the effective description; what does not change is the unbroken subgroup of the R-symmetry group.  
Because we want to match them with the supergravity computations, it will be convenient to rewrite the field theory anomalies in the following form:
\begin{equation}
\text{Field theory:} \qquad\qquad
\begin{aligned}
\delta_\varepsilon (\Theta_l+\Theta_s) &= - N_{f}\, \varepsilon \,\, ,\\
\delta_\varepsilon (\Theta_l-\Theta_s) &= [4(r_l - r_s) + N_{fs}-N_{fl}] \, \varepsilon \;. \label{anomalies FT}
\end{aligned} \end{equation}

An infinitesimal $U(1)_R$ rotation parameterised by $\varepsilon$ in field theory corresponds to a shift $\psi \to \psi + 2\varepsilon$ in the geometry. Therefore, making use of \eqref{C0 C2 eff}, we find on the supergravity side:
\begin{equation}
\text{SUGRA:} \qquad\qquad
\begin{aligned}
\delta_\varepsilon (\Theta_l + \Theta_s) &=  - N_f \, \varepsilon \,\, , \\
\delta_\varepsilon (\Theta_l - \Theta_s) &= [4M + 2n \, N_f] \, \varepsilon \;.
\end{aligned} \label{anomalies sugra} \end{equation}
These formulae agree with those computed in the field theory.
For the difference of the anomalies, what we can compute and compare is its change after a step in the duality cascade. 
Notice indeed that the difference of the anomalies, as computed in \eqref{anomalies sugra}, gives a step function: as we flow towards the IR, after some energy scale (the scale of a Seiberg duality along the cascade) we need to perform a large gauge transformation in supergravity to turn to the correct Seiberg dual description of the field theory (the only one with positive squared gauge couplings). This corresponds to changing $n \to n-1$ in \eqref{anomalies sugra}, therefore the coefficient of the difference of the R-anomalies decreases by $2N_f$ units. 
This result is reproduced exactly by the field theory computation \eqref{anomalies FT}. In field theory the difference of the anomalies depends on the quantity $4(r_l-r_s) + N_{fs}-N_{fl} $. Keeping the same conventions adopted in  subsection \ref{cascade subsection} and repeating the same reasoning, it is easy to see that after a step of the cascade towards the IR, this quantity decreases exactly by $2N_f$ units.

The dictionary \eqref{holographic relations} allows us also to compute the $\beta$-functions of the two gauge couplings and check further the picture of the duality cascade.

Since we will be concerned in the cascade, we will make use of the flavoured Klebanov-Tseytlin solution of section \ref{KTflavoured}, to which the flavoured Klebanov-Strassler solution of section \ref{KSflavoured} reduces in the UV limit.

We shall keep in mind that, at a fixed value of the radial coordinate, we want to shift $b_0=\frac{1}{4\pi^2}\int_{S^2} B_2 $ by means of a large gauge transformation in supergravity in such a way that its gauge transformed $\tilde{b}_0=b_0-[b_0]\equiv b_0-n$ 
\footnote{$n$ is a step-like function of the radial coordinate.}
belongs to $[0,1]$: in doing so, we are guaranteed to be using the good description in terms of a field theory with positive squared gauge couplings.

Recall that  
\begin{align}
e^{-\phi}&=\frac{3N_f}{4\pi}(-\rho) \,\, , \\
b_0&=\frac{2M}{N_f}\bigg(\frac{\Gamma}{\rho}-1\bigg)\,\,,
 \label{b_0 seiberg dualities}
\end{align}
and the dictionary \eqref{holographic relations}, that we rewrite as:
\begin{align}
\frac{8\pi^2}{g_+^2}\equiv \frac{8\pi^2}{g_l^2}+\frac{8\pi^2}{g_s^2}&=2\pi e^{-\phi} \,\, , \label{g+}\\
\frac{8\pi^2}{g_-^2}\equiv \frac{8\pi^2}{g_l^2}-\frac{8\pi^2}{g_s^2}&=2\pi e^{-\phi}(2\tilde{b}_0-1)\;,\label{g-}
\end{align}
where $\tilde{b}_0\equiv b_0- [b_0] \in [0,1]$ comes from integrating on the two-cycle the suitably gauge transformed Kalb-Ramond potential.

Then we can compute the following `radial' $\beta$-functions from the gravity dual: 
\begin{align}
\beta^{(\rho)}_+ \equiv \beta^{(\rho)}_{\frac{8\pi^2}{g_+^2}} &\equiv\frac{d}{d\rho}\frac{8\pi^2}{g_+^2}\,\, ,\\
\beta^{(\rho)}_- \equiv \beta^{(\rho)}_{\frac{8\pi^2}{g_-^2}} &\equiv\frac{d}{d\rho}\frac{8\pi^2}{g_-^2}\;,
\end{align}
and we would like to match these with the field theory computations.\\
Using the expressions \eqref{g+}-\eqref{g-}, we can conclude that 
\begin{align}  \label{beta grav1} 
\beta^{(\rho)}_+ &=-3 \frac{N_f}{2} \,\, ,\\   
\beta^{(\rho)}_- &= 3\bigg(\frac{N_f}{2}+Q\bigg) \;,   \label{beta grav2}
\end{align} 
where $Q=N_f [b_0(\rho)]+2M = N_f n(\rho) +2M $ is a quantity which undergoes a change $Q \rightarrow Q-N_f$ as $b_0(\rho)\rightarrow b_0(\rho')=b_0(\rho)-1$ (one Seiberg duality step along the cascade towards the IR), or equivalently $n(\rho)\rightarrow n(\rho')=n(\rho)-1$. 
Up to an overall factor of 2, $Q$ is the same quantity appearing in the difference of the R-anomalies in \eqref{anomalies sugra}.

The field theory computations of the $\beta$-functions (in the Wilsonian scheme) give:
\begin{align}
\beta_l &\equiv \beta_{\frac{8\pi^2}{g_l^2}}= 3 r_l-2r_s (1-\gamma_A) - N_{fl} (1-\gamma_q)\,\, , \\
\beta_s &\equiv \beta_{\frac{8\pi^2}{g_s^2}}= 3 r_s-2r_l (1-\gamma_A) - N_{fs} (1-\gamma_q)\;,
\end{align}
with the usual conventions. Hence
\begin{align}
\beta_+ &\equiv \beta_{l}+\beta_s= (r_l+r_s)(1+2\gamma_A)-N_f(1-\gamma_q)\,\, ,\\
\beta_- &\equiv \beta_{l}-\beta_s= (5-2\gamma_A)(r_l-r_s)+(N_{fs}-N_{fl})(1-\gamma_q)\;.
\end{align}
In order to match the above quantities with the gravity computations \eqref{beta grav1}-\eqref{beta grav2}, an energy-radius relation is required. This is something we miss here.
Although it is not really needed to extract from our supergravity solutions the qualitative information on the running of the gauge couplings, we are going initially to make two assumptions, which can be viewed as an instructive simplification. Let us then assume that the radius-energy relation is $\rho=\ln\frac{\mu}{E_{UV}}$, where $E_{UV}$ is the scale of the UV cutoff dual to the maximal value of the radial coordinate $\rho=0$, and that the anomalous dimensions do not acquire subleading corrections. Matching $\beta_+$ implies $\gamma_A=\gamma_q=-\frac{1}{2}$. Matching $\beta_-$, once we insert these anomalous dimensions, implies that $Q=2(r_l-r_s)-N_{fl}$. This quantity correctly shifts as $Q \rightarrow Q-N_f$ when $b_0\rightarrow b_0-1$. 
This last observation allows us to check the consistency of the cascade of Seiberg dualities also against the running of the gauge couplings.

Actually, the qualitative picture of the RG flow in the UV can be extracted from our supergravity solution  even without knowing the precise radius-energy relation, but simply recalling that the radius must be a monotonic function of the energy scale.

It is interesting to notice the following phenomenon: as we flow up in energy and  approach the far UV
$\rho\to 0^-$ in \eqref{b_0 seiberg dualities}, a large number of Seiberg dualities is needed to keep $b_0$ varying in the interval $[0,1]$. The Seiberg dualities pile up the more we approach the UV cut-off $E_{UV}$.
Meanwhile, formula \eqref{beta grav2} reveals that, when going towards the UV cutoff $E_{UV}$, the `slope' in the plots of $\frac{1}{g_i^2}$ versus the energy scale becomes larger and larger, and \eqref{beta grav1} reveals that the sum of the inverse squared gauge coupling goes to zero at this UV cutoff. At the energy scale $E_{UV}$ the effective number of degrees of freedom needed for a weakly coupled description of the gauge theory becomes infinite.
Since $\rho=0$ is at finite proper radial distance from any point placed in the interior $\rho<0$, $E_{UV}$ is a finite energy scale. 

\begin{figure}[ht]
\begin{center}
\includegraphics[width=0.7\textwidth]{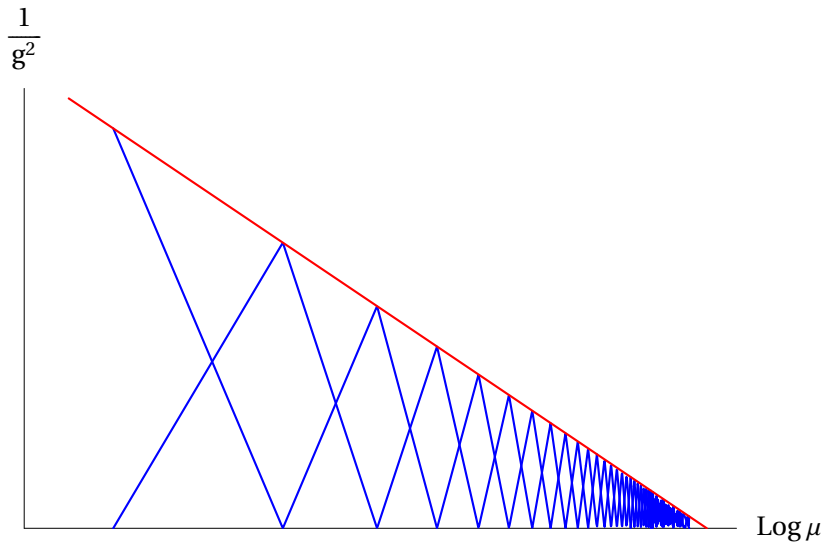}
\end{center}
\caption[wall]{Qualitative plot of the running gauge couplings as functions of the logarithm of the energy scale in our cascading gauge theory. The blue lines are the inverse squared gauge couplings, while the red line is their sum. \label{wall}}
\end{figure}

The picture which stems from our flavoured Klebanov-Tseytlin/Strassler solution is that $E_{UV}$ is a so-called ``Duality Wall", namely an accumulation point of energy scales at which a Seiberg duality is required in order to have a weakly coupled description of the gauge theory \cite{Strassler:1996ua}. Above the duality wall, Seiberg duality does not proceed and a weakly coupled dual description of the field theory is not known. See Fig. \ref{wall}.
 
Duality walls were studied in \cite{Fiol:2002ah} in the framework of quiver gauge theories with only bifundamental chiral superfields and the study was restricted to the field theory.

To our knowledge, our solutions are the first explicit realisations of this exotic UV phenomenon on the supergravity side of the gauge/gravity correspondence.

\section{Summary and Discussion} \label{Sect: final remarks}
\medskip
\setcounter{equation}{0}

In this chapter we have presented a very precise example of the duality between field theories with flavours and string solutions that include the dynamics of (flavour) branes. We focused on the Klebanov-Tseytlin/Strassler case, providing a well defined dual field theory, together with different matchings that include the cascade of Seiberg dualities, $\beta$-functions and anomalies.
Indeed, we have shown in detail how the ranks of the gauge groups change from a string theory viewpoint (in perfect agreement with the usual field theory prescription), providing also a rigorous definition of the gauge groups ranks in terms of Page charges.
We have also shown how the runnings of the gauge couplings are matched by the string background and how global anomalies are also captured by our solution.

It would be interesting to provide more general solutions to our system of BPS equations and analyze the details of their dual dynamics, focusing mainly on the IR (the last steps of the cascade leading to a baryonic branch of the field theory, behavior of the Wilson loop, etc).


\chapter{SUSY defects in the Maldacena-N\'u\~nez 
background}
\label{MN}
\medskip
\setcounter{equation}{0}
\medskip

In this chapter we will make a rather systematic search for
possible supersymmetric embeddings for D5-brane probes in a concrete
model, the Maldacena-Nu\~nez  background (MN) introduced in subsection \ref{MNbackground}. As shown in ref. \cite{Nunez:2003cf}, the MN  background  has a rich structure of
submanifolds along which one can wrap a D5-brane probe without breaking supersymmetry
completely. Hence we continue the analysis of \cite{Nunez:2003cf} by studying the configurations of
D5-brane probes which are a codimension one or two defect in the gauge theory directions. The main tool used will be again kappa symmetry (see subsection \ref{kappasymmetry}). By
imposing the equation $\Gamma_{\kappa}\,\epsilon=\epsilon$ one can systematically determine the
supersymmetric embeddings of the  probe and it is possible to identify the fraction of
supersymmetry  preserved by the configuration. It will become clear that only the D5-brane probes can have supersymmetric embeddings of the type we are interested in and that they preserve two of the four supersymmetries of the background. We will demonstrate that the solutions found in sections \ref{Walldefects} and \ref{twodimdefect} saturate certain energy bound \cite{GGT}.

\section{Supersymmetric Probes in the Maldacena-N\'u\~nez background}  \label{SPMN}
\medskip
\setcounter{equation}{0}

The idea, as we have already explained before in subsection \ref{kappasymmetry}, is to consider a D5-brane probe embedded in the MN background. We will use the notation given in subsection \ref{MNbackground} where we set $g_s=1, \alpha'=1$ and $N_c=1$ for simplicity and irrelevance for the analysis of this chapter. 

We will assume that there are not worldvolume gauge fields on the D5-brane, which is consistent
with the equations of motion of the probe if there are not source terms which could induce them.
These source terms must be linear in the gauge field and can only be originated in the
Wess-Zumino part of the probe action (\ref{actionintrod}). For the cases considered below we will verify that the RR potentials of the MN background do not act as source of the worldvolume gauge fields and,
therefore, the latter can be consistently put to zero. If this is the case, the kappa symmetry
matrix of a Dp-brane in the type IIB theory is the one written in (\ref{gammakappa}). 

The kappa symmetry equation $\Gamma_{\kappa}\,\epsilon\,=\,\epsilon$ imposes a condition on the
Killing spinors which should be compatible with the ones required by the supersymmetry of the
background. These latter conditions are precisely the ones written in eq.
(\ref{fullprojection}). In particular the spinor $\epsilon$ must
be such that $\epsilon=i\epsilon^*$, which in the real notation (\ref{rule}) is equivalent to 
$\sigma_1\epsilon=\epsilon$. Notice that the Pauli matrix appearing in the
expression of $\Gamma_{\kappa}$ in  (\ref{gammakappa}) is $\sigma_1$ or $\sigma_2$, depending on the dimensionality of the probe. Clearly, the conditions $\Gamma_{\kappa}\,\epsilon\,=\,\epsilon$ and
$\sigma_1\epsilon=\epsilon$ can only  be compatible if $\Gamma_{\kappa}$ contains the Pauli matrix
$\sigma_1$. By inspecting eq. (\ref{gammakappa}) one readily realises that this happens for
$p=1,5$.  Moreover, we want our probes to be extended both along the spatial Minkowski and
internal directions, which is not possible for Lorentzian D1-branes and leaves us with the
D5-branes as the only case to be studied.  Notice that for the MN background the only couplings
of the Wess-Zumino term of the action (\ref{WZDpbraneaction}) linear in the worldvolume gauge field $F$ are of the form $C^{(2)}\wedge F$ and $C^{(6)}\wedge F$, where $C^{(2)}$ and $C^{(6)}$ are the RR potentials. By simple counting of the degree of these forms one immediately concludes that these terms are not present in the action of a D5-brane and, thus, the gauge fields can be consistently taken to be
zero, as claimed above. 

Coming back to the complex notation for the spinors, and taking into account the fact that the
Killing spinors of the MN background satisfy the condition $\epsilon=i\epsilon^*$, one can write
the matrix $\Gamma_{\kappa}$ for a D5-brane probe as:
\beq
\Gamma_{\kappa}\,=\,{1\over 6!}\,\,{1\over \sqrt{-g}}\,\,
\epsilon^{\mu_1\cdots \mu_6}\,\,\gamma_{\mu_1\cdots \mu_6}\,\,.
\label{GammaD5}
\eeq

As we explained in detail in section \ref{ypq}, for a general embedding, the kappa symmetry condition
$\Gamma_{\kappa}\,\epsilon=\epsilon$ imposes a new projection to the Killing spinor $\epsilon$.
This new projection is not, in general, consistent with the conditions (\ref{fullprojection}),
since it involves matrices which do not commute with those appearing in (\ref{fullprojection}).
The only way of making the equation $\Gamma_{\kappa}\,\epsilon=\epsilon$ and
(\ref{fullprojection}) consistent with each other is by requiring the vanishing of the
coefficients of those non-commuting matrices. On the contrary, the terms in $\Gamma_{\kappa}$
which commute with the projections (\ref{fullprojection}) should act on the Killing spinors as
the unit matrix. These conditions will give rise to a set of first-order
BPS differential equations. By solving these BPS equations we will determine the
embeddings of the D5-brane we are interested in, namely those which preserve some
fraction of the background supersymmetry. The configurations found by solving these equations also solve the equations of motion derived from the Dirac-Born-Infeld action of the
probe and, actually, we will verify that they saturate a bound for the energy,
as it usually happens in the case of worldvolume solitons. We have already seen this bound in the configurations studied in chapters \ref{Ypq} and \ref{Labc} but it is worth recalling the procedure for the MN model. The lagrangian density for a D5-brane probe (see eq. (\ref{actionintrod})) in the MN background is given by:
\beq
{\cal L}\,=\,-e^{-\phi}\,\sqrt{-g}\,-\,P[\,C^{(6)}\,]\,\,,
\label{lagrangian}
\eeq
where we have taken the string tension equal to one and $P[\,C^{(6)}\,]$ denotes the pullback of
the RR potential written in eqs. (\ref{C6}) and (\ref{calC}). In eq. (\ref{lagrangian}) we have
already taken into account that we are considering configurations of the probe with vanishing
worldvolume gauge field. For  static embeddings, such as the ones we will consider in this chapter, the
hamiltonian density ${\cal H}$ is just ${\cal H}= -{\cal L}$. We will verify that, for the systems studied in sections \ref{Walldefects} and \ref{twodimdefect}, ${\cal H}$ satisfies a lower  bound,
which is saturated just when the corresponding BPS equations are satisfied. Actually, we will
show that, for a generic embedding, ${\cal H}$ can be written as:
\beq
{\cal H}\,=\,{\cal Z}\,+\,{\cal S}\,\,,
\label{H=Z+S}
\eeq
where ${\cal Z}$ is a total derivative and ${\cal S}$ is non-negative:
\beq
{\cal S}\,\ge\,0\,\,.
\label{calS}
\eeq
From eqs. (\ref{H=Z+S}) and (\ref{calS}) it follows immediately that ${\cal H}\ge {\cal Z}$,
which is the energy bound that we have stated above. Moreover, we will check that 
${\cal S}=0$ precisely when the BPS equations obtained from kappa symmetry are satisfied, which
means that the energy bound is saturated for these configurations.

\setcounter{equation}{0}
\section{Wall defects}   \label{Walldefects}
\medskip
In this section we are going to find supersymmetric configurations of a D5-brane probe which,
from the point of view of the four-dimensional gauge theory, are codimension one objects.
Accordingly, we extend the D5-brane along three of the Minkowski coordinates $x^{\mu}$ (say 
$x^0$, $x^1$, $x^2$) and along a three dimensional submanifold of the internal part of the
metric. To describe these configurations it is convenient to choose 
the following set of worldvolume coordinates:
\beq
\xi^{m}\,=\,(x^0,x^1,x^2,r,\theta_1,\phi_1)\,\,.
\label{DWvwcoordinates}
\eeq
Moreover, we will adopt the following ansatz for the dependence of the remaining
ten-dimensional coordinates on the $\xi^{\mu}$'s:
\bear
&&x^3=x^3(r),\,\,\rc
&&\theta_2=\theta_2(\theta_1,\phi_1)\,\,,
\,\,\,\,\,\,\,\,\,\,\,\,\,
\phi_2=\phi_2(\theta_1,\phi_1)\,\,,\rc
&&\psi=\psi_0={\rm constant}\,\,. 
\label{DWansatz}
\eear
In subsection \ref{morewalldefects} we will explore other possibilities and, in particular, we will study
configurations for which $\psi$ is not constant. For the set of worldvolume coordinates
(\ref{DWvwcoordinates}) the kappa symmetry matrix acts on the Killing spinors $\epsilon$ as:
\beq
\Gamma_{\kappa}\,\epsilon\,=\,{1\over \sqrt{-g}}\,
\gamma_{x^0 x^1x^2 r\theta_1\phi_1}\,\epsilon\,\,.
\label{DWGammak}
\eeq
The induced gamma matrices appearing on the right-hand side of eq. (\ref{DWGammak}) can be
straightforwardly computed from the general expression (\ref{wvgamma}). One gets:
\bear
&&e^{-{\phi\over 2}}\,\gamma_{x^{\mu}}=\,\Gamma_{x^{\mu}}\,\,,
\,\,\,\,\,\,\,\,\,\,\,\,\,\,\,\,\,\,
(\mu=0,1,2),\,\,\rc
&&e^{-{\phi\over 2}}\,\gamma_r\,=\,\Gamma_r\,+\,\partial_rx^3\Gamma_{x^3}\,\,,\rc
&&e^{-{\phi\over 2}}\,\gamma_{\theta_1}\,=\,e^{h}\Gamma_1\,+\,
\big(\,V_{1\theta}+{a\over 2}\,\big)\,\hat\Gamma_1\,+\,V_{2\theta}\,\hat\Gamma_2\,+\,
V_{3\theta}\,\hat\Gamma_3\,\,,\rc
&&{e^{-{\phi\over 2}}\over \sin\theta_1}\,\gamma_{\phi_1}\,=\,e^{h}\Gamma_2\,+\,
V_{1\phi}\,\hat\Gamma_1\,+\,
\big(\,V_{2\phi}-{a\over 2}\,\big)\,\hat\Gamma_2\,+\
V_{3\phi}\,\hat\Gamma_3\,\,,
\label{DWDiracma}
\eear
where the $V$'s are the quantities:
\bear
&&V_{1\theta}\equiv {1\over 2}\,\Big[\cos\psi_0\,\partial_{\theta_1}\theta_2\,+\,
\sin\psi_0\,\sin\theta_2\,\partial_{\theta_1}\phi_2\,\Big]\,\,,\rc
&&V_{2\theta}\equiv {1\over 2}\,\Big[-\sin\psi_0\,\partial_{\theta_1}\theta_2\,+\,
\cos\psi_0\,\sin\theta_2\,\partial_{\theta_1}\phi_2\,\Big]\,\,,\rc
&&V_{3\theta}\equiv {1\over 2}\,\cos\theta_2\,\partial_{\theta_1}\phi_2\,\,,\rc
&&\sin\theta_1V_{1\phi}\equiv {1\over 2}\,\Big[\cos\psi_0\,
\partial_{\phi_1}\theta_2\,+\,
\sin\psi_0\,\sin\theta_2\,\partial_{\phi_1}\phi_2\,\Big]\,\,,\rc
&&\sin\theta_1\,V_{2\phi}\equiv {1\over 2}\,\Big[-\sin\psi_0\,
\partial_{\phi_1}\theta_2\,+\,
\cos\psi_0\,\sin\theta_2\,\partial_{\phi_1}\phi_2\,\Big]\,\,,\rc
&&\sin\theta_1\,V_{3\phi}\equiv {1\over 2}\,\Big[\cos\theta_1\,+\,
\cos\theta_2\,\partial_{\phi_1}\phi_2\,\Big]\,\,.
\label{Vs}
\eear
Notice that the $V$'s depend on the angular part of the embedding (\ref{DWansatz}), \ie\ on the
functional dependence of $\theta_2$, $\phi_2$ on $(\theta_1, \phi_1)$. Using the expressions of
the $\gamma$'s given in eq. (\ref{DWDiracma}), one can write the action of $\Gamma_{\kappa}$ on 
$\epsilon$ as:
\beq
\Gamma_{\kappa}\,\epsilon\,=\,{e^{2\phi}\over \sqrt{-g}}\,
\Gamma_{x^0x^1x^2}\,\big[\,\Gamma_r\,+\,\partial_rx^3\Gamma_{x^3}\,\big]\,
\gamma_{\theta_1\phi_1}\,\epsilon\,\,.
\eeq
Moreover, by using the projection $\Gamma_{12}\,\epsilon\,=\,\hat\Gamma_{12}\,\epsilon$ (see eq.
(\ref{fullprojection})), $\gamma_{\theta_1\phi_1}\,\epsilon$ can be written as:
\bear
&&{e^{-\phi}\over \sin\theta_1}\,\gamma_{\theta_1\phi_1}\,\epsilon=\,
\big[\,
c_{12}\,\Gamma_{12}\,+\,c_{1\hat 2}\,\Gamma_{1}\hat\Gamma_{2}\,+\,
c_{1\hat 1}\,\Gamma_{1}\hat\Gamma_{1}\,+\,
\rc\rc
&&\,\,\,\,\,\,\,\,\,\,\,\,\,\,\,\,\,\,\,\,\,\,\,\,\,\,\,\,\,\,\,\,\,\,\,\,\,
\,+\,
c_{1\hat 3}\,\Gamma_{1}\hat\Gamma_{3}\,+\,c_{\hat 1\hat 3}\,\hat\Gamma_{13}\,
+\,c_{\hat 2\hat 3}\,\hat\Gamma_{23}\,
+\, c_{2\hat 3}\,\Gamma_{2}\hat\Gamma_{3}\,\big]\,\epsilon\,\,,
\label{DWces}
\eear
with the $c$'s  given by:
\bear
&&c_{12}\,=\,e^{2h}\,+\,\Big(\,V_{1\theta}+{a\over 2}\,\Big)
\Big(\,V_{2\phi}-{a\over 2}\,\Big)\,-\,V_{2\theta}\,V_{1\phi}\,\,,\rc
&&c_{1\hat 2}\,=\,e^{h}\,\Big(\,V_{2\phi}\,-\,V_{1\theta}\,-\,a\Big)\,\,,\rc
&&c_{1\hat 1}\,=\,e^{h}\,\Big(\,V_{1\phi}\,+\,V_{2\theta}\,\Big)\,\,,\rc
&&c_{1\hat 3}\,=\,e^{h}\,V_{3\phi}\,\,,\rc
&&c_{\hat 1\hat 3}\,=\,\Big(\,V_{1\theta}+{a\over 2}\,\Big)\,V_{3\phi}\,-\,
V_{1\phi}\,V_{3\theta}\,\,,\rc
&&c_{ \hat 2\hat 3}\,=\,V_{2\theta}\,V_{3\phi}\,-\,
\Big(\,V_{2\phi}-{a\over 2}\,\Big)\,V_{3\theta}\,\,,\rc
&&c_{2\hat 3}\,=\,-e^{h}\,V_{3\theta}\,\,.
\label{DWcexpression}
\eear
As mentioned at the end of section \ref{SPMN}, we have to ensure that the kappa symmetry projection 
$\Gamma_{\kappa}\,\epsilon=\epsilon$ is compatible with the conditions (\ref{fullprojection}).
In particular,  it should be consistent with the second projection
written in  (\ref{fullprojection}), namely
$\Gamma_{12}\,\epsilon\,=\,\hat\Gamma_{12}\,\epsilon$. It is rather obvious that the terms in
(\ref{DWces}) containing the matrix $\hat\Gamma_3$ do not fulfil this requirement. Therefore
we must impose the vanishing of their coefficients, \ie:
\beq
c_{1\hat 3}\,=\,c_{\hat 1\hat 3}\,=\,c_{\hat 2\hat 3}\,=\,c_{ 2\hat 3}\,=\,0\,\,.
\label{DWhat3conditions}
\eeq
By inspecting the last four equations in (\ref{DWcexpression}) one readily realises that the
conditions (\ref{DWhat3conditions}) are equivalent to:
\beq
V_{3\theta}\,=\,V_{3\phi}\,=0\,\,.
\eeq
Moreover, from the expression of $V_{3\theta}$ in (\ref{Vs}) we conclude that the condition 
$V_{3\theta}=0$ implies that
\beq
\phi_2=\phi_2(\phi_1)\,\,.
\label{phi2dep}
\eeq
Furthermore (see eq. (\ref{Vs}) ), $V_{3\phi}=0$ is equivalent to the following differential
equation:
\beq
{\partial\phi_2\over\partial\phi_1}\,=\,-{\cos\theta_1\over \cos\theta_2}\,\,.
\eeq
Let us now write
\beq
{\partial\phi_2\over\partial\phi_1}\,=\,m(\phi_1)\,\,,
\label{partialphi}
\eeq
where we have already taken into account the functional dependence written in eq.
(\ref{phi2dep}). By combining the last two equations we arrive at:
\beq
\cos\theta_2\,=\,-{\cos\theta_1\over m(\phi_1)}\,\,.
\label{costheta2}
\eeq
By differentiating eq. (\ref{costheta2})  we get
\beq
{\partial \theta_2\over \partial\theta_1}\,=\,-
{\sin\theta_1\over m(\phi_1)\sin\theta_2}\,\,.
\label{partialtheta2}
\eeq

Then, if we define
\bear
&&\Delta(\theta_1,\phi_1)\equiv{1\over 2}\,\Bigg[\,{\sin\theta_2\over \sin\theta_1}\,
\partial_{\phi_1}\phi_2\,-\,\partial_{\theta_1}\theta_2\,\Bigg]\,\,,\rc
&&\tilde \Delta(\theta_1,\phi_1)\equiv{1\over 2}\,\,
{\partial_{\phi_1}\theta_2\over \sin\theta_1}\,\,,
\label{DWdeltas}
\eear
the $c$ coefficients can be written in terms of $\Delta$ and $\tilde\Delta$, namely:
\bear
&&c_{12}\,=\,e^{2h}\,-{a^2\over 4}\,-\,{1\over 4}\,+\,
{a\Delta\over 2}\,\cos\psi_0
\,-\,{a\tilde \Delta \over 2}\,\,\sin\psi_0
\,\,,\rc
&&c_{1\hat 2}\,=\,e^{h}\,\big[\,\Delta\cos\psi_0
\,-\,\tilde \Delta\,\sin\psi_0\,
\,-\,a\,\big]\,\,,\rc
&&c_{1\hat 1}\,=\,e^{h}\,\big[\,\Delta\,\sin\psi_0\,+\,
\tilde \Delta\,\cos\psi_0\,\,\big]\,\,,
\label{csdelta}
\eear
where we have used eqs. (\ref{phi2dep})-(\ref{partialtheta2}) and  the fact that
\beq
V_{1\theta}V_{2\phi}\,-\,V_{2\theta}V_{1\phi}\,=\,-{1\over 4}\,\,.
\eeq
Moreover,
by using the values of the derivatives $\partial_{\phi_1}\phi_2$ and 
$\partial_{\theta_1}\theta_2$ written in eqs. (\ref{partialphi}) and (\ref{partialtheta2}),
together with  eq. (\ref{costheta2}), it is easy to find 
 $\Delta(\theta_1,\phi_1)$ in terms of the function 
$m(\phi_1)$:
\beq
\Delta(\theta_1,\phi_1)\,=\,{{\rm sign}(m)\over 2}\,\,\Bigg[\,
\Bigg[1\,+\,{m(\phi_1)^2-1\over \sin^2\theta_1}\Bigg]^{{1\over 2}}
\,+\,
\Bigg[1\,+\,{m(\phi_1)^2-1\over \sin^2\theta_1}\Bigg]^{-{1\over 2}}\,\,\Bigg]\,\,,
\label{Delta}
\eeq
an expression which will be very useful in what follows.

\subsection{Abelian worldvolume solitons}

The expression of $\Gamma_{\kappa}\,\epsilon$ that we have found above is rather complicated. In
order to tackle the general problem of finding the supersymmetric embeddings for the ansatz
(\ref{DWansatz}), let us consider the simpler problem of solving the condition
$\Gamma_{\kappa}\,\epsilon=\epsilon$ for  the abelian background\footnote{See the discussion about the abelian limit of the MN background in subsection \ref{MNbackground}.}, for which  $a=\alpha=0$. First
of all  let us define the following matrix:
\beq
\tilde{\Gamma}_{*}\equiv \Gamma_{x^0x^1x^2}\,\Gamma_r\,\Gamma_1\,\hat\Gamma_2\,\,.
\eeq
Using the fact that for the abelian background
$\Gamma_{x^0x^1x^2x^3}\,\Gamma_{12}\epsilon=\epsilon$ (see eq. (\ref{fullprojection})), one can
show that
\bear
&&\Gamma_{\kappa}\,\epsilon\,=\,{e^{3\phi}\over \sqrt{-g}}\,\sin\theta_1\,\,
\Bigg[\,\partial_rx^3\,c_{12}\,+\,c_{1\hat 2}\,\tilde{\Gamma}_{*}\,+\,
c_{1\hat 1}\,\hat\Gamma_{12}\,\tilde{\Gamma}_{*}\,+\,\rc
&&\,\,\,\,\,\,\,\,\,\,\,\,\,\,\,\,\,\,\,\,\,\,\,
\,\,\,\,\,\,\,\,\,\,\,\,\,\,\,\,\,\,\,\,\,\,\,\,\,\,\,\,\,\,
\,+\,\big(\,c_{12}\,\tilde{\Gamma}_{*}\,+\,\partial_rx^3\,c_{1\hat 2}\,\big)\,
\Gamma_1\hat\Gamma_1\,-\,\partial_rx^3\,c_{1\hat 1}\,\Gamma_1\hat\Gamma_2\,\Bigg]\,
\epsilon\,\,.
\label{abeliankappa}
\eear
The first three terms on the right-hand side commute with the projection 
$\Gamma_{r}\hat \Gamma_{123}\,\epsilon\,=\,\epsilon$. Let us write them in detail:
\beq
\big[\partial_rx^3\,c_{12}\,+\,c_{1\hat 2}\,\tilde{\Gamma}_{*}\,+\,
c_{1\hat 1}\,\hat\Gamma_{12}\,\tilde{\Gamma}_{*}\,\big]\,\epsilon\,=
\big[\,\partial_rx^3\,c_{12}\,+\,e^h\Delta e^{\psi_0\hat\Gamma_{12}}\,\tilde{\Gamma}_{*}
\,+\,e^h\tilde\Delta \hat \Gamma_{12}\,e^{\psi_0\hat\Gamma_{12}}\,\tilde{\Gamma}_{*}\,\big]\,\epsilon\,\,.
\label{DWcomm}
\eeq
The matrix inside the brackets must act diagonally on $\epsilon$. In order to fulfil this
requirement we have to impose an extra projection to the spinor $\epsilon$. Let us define the
corresponding projector as:
\beq
{\cal P}_*\,\equiv\,\beta_1\,\tilde{\Gamma}_{*}\,+\,\beta_2\,\hat\Gamma_{12}\,\tilde{\Gamma}_{*}\,\,,
\eeq
where $\beta_1$ and $\beta_2$ are constants. We will require that $\epsilon$ satisfies the
condition:
\beq
{\cal P}_*\,\epsilon\,=\,\sigma\,\epsilon\,\,,
\label{extraDW}
\eeq
where $\sigma=\pm 1$.  For consistency ${\cal P}_*^2=1$, which, as the matrices $\tilde{\Gamma}_{*}$ and
$\hat\Gamma_{12}\,\tilde{\Gamma}_{*}$ anticommute, implies that $\beta_1^2\,+\,\beta_2^2\,=\,1$.
Accordingly, let us parameterise $\beta_1$ and $\beta_2$ in terms of a constant angle $\beta$ as
$\beta_1=\cos\beta$ and $\beta_2=\sin\beta$. The extra projection (\ref{extraDW}) takes the form:
\beq
e^{\beta\hat\Gamma_{12}}\,\tilde{\Gamma}_{*}\,\epsilon\,=\,\sigma\epsilon\,\,.
\label{extraDWbeta}
\eeq
Making use of the condition (\ref{extraDWbeta}), we can write the right-hand side of eq. 
(\ref{DWcomm}) as:
\beq
\big[\partial_rx^3\,c_{12}\,+\,e^{h}\,e^{(\psi_0\,-\,\beta)\hat\Gamma_{12}}\,\,
(\Delta\,+\,\tilde\Delta\,\hat\Gamma_{12}\,)\,\big]\,\epsilon\,\,.
\label{DWcomm2}
\eeq
We want that the matrix inside the brackets in (\ref{DWcomm2}) acts diagonally. Accordingly, we
must require that the coefficient of $\hat\Gamma_{12}$ in (\ref{DWcomm2}) vanishes which, in
turn, leads to the relation:
\beq
\tan (\beta-\psi_0)\,=\,{\tilde\Delta\over \Delta}\,\,.
\label{tanbeta}
\eeq
In particular eq. (\ref{tanbeta}) implies that $\tilde\Delta/\Delta$ must be constant. Let us
write:
\beq
{\tilde\Delta\over \Delta}\,=\,p\,=\,{\rm constant}\,\,.
\label{Deltarel}
\eeq
Let us now consider the
last three terms in (\ref{abeliankappa}), which contain matrices that do not commute with the
projection  $\Gamma_{r}\hat \Gamma_{123}\,\epsilon\,=\,\epsilon$. By using the projection 
(\ref{extraDWbeta})
these terms can be written as:
\bear
&&\bigg[\,\big(\,c_{12}\tilde{\Gamma}_{*}\,+\,\partial_rx^3\,c_{1\hat 2}\,\big)\,
\Gamma_1\hat\Gamma_1\,-\,\partial_rx^3\,c_{1\hat 1}\,\Gamma_1\hat\Gamma_2\,\,\bigg]\,
\epsilon\,=\,
\rc
&&\,\,\,\,\,\,\,\,\,\,\,=\,
\bigg[\,(\partial_rx^3\,c_{1\hat 2}\,-\,\sigma c_{12}\cos\beta\,)\,
\Gamma_1\hat\Gamma_1\,+\,(\sigma c_{12}\sin\beta\,-\,\partial_rx^3\,c_{1\hat 1})\,
\Gamma_1\hat\Gamma_2\,
\bigg]\,\epsilon\,\,.
\label{noncommuting}
\eear
This contribution should vanish.  By inspecting the right-hand side of eq. 
(\ref{noncommuting}) one immediately concludes that this vanishing condition determines the
value of $\partial_rx^3$, namely:
\beq
\partial_rx^3\,=\,\sigma\,c_{12}\,{\cos\beta\over c_{1\hat 2}}\,=\,
\sigma\,c_{12}\,{\sin\beta\over c_{1\hat 1}}\,\,.
\label{partialx3first}
\eeq
The compatibility between the two expressions of $\partial_rx^3$ in eq. (\ref{partialx3first})
requires that $\tan\beta=c_{1\hat 1}/c_{1\hat 2}$. By using the values of  $c_{1\hat 1}$ and
$c_{1\hat 2}$ written in eq. (\ref{csdelta})  it is easy to verify that this compatibility
condition is equivalent to (\ref{tanbeta}). Moreover, one can write eq. (\ref{partialx3first})
as:
\beq
\partial_rx^3\,=\,{\sigma\over \Delta}\,\,e^{-h}\,\,
\big[\,e^{2h}\,-\,{1\over 4}\,\big]\,\,
{\cos\beta\over \cos\psi_0\,-\,p\sin\psi_0}\,\,.
\label{partialx3second}
\eeq
Notice that $\Delta$ only depends on the angular variables $(\theta_1,\phi_1)$.
However, since in our ansatz $x^3=x^3(r)$,  eq. (\ref{partialx3second}) is only consistent if
$\Delta$ is independent of
$(\theta_1,\phi_1)$,
\ie\ when $\Delta$ is constant. By looking at eq. (\ref{Delta}) one readily realises that this
can only happen if $m^2=1$, \ie:
\beq
m\,=\,\pm 1\,\,.
\eeq
In this case (see eq. (\ref{Delta})) $\Delta$ is given by
\beq
\Delta=m\,\,.
\eeq
Moreover, as $\tilde\Delta\,=\,p\Delta$ (see eq. (\ref{Deltarel})), it follows that
$\tilde\Delta$ must be constant. A glance at the definition of $\tilde\Delta$ in (\ref{DWdeltas})
reveals that $\tilde\Delta$ can only be constant if it vanishes. Thus, we must have:
\beq
\tilde\Delta\,=\,0\,\,.
\eeq 
Notice that this implies that $\theta_2$ is independent of $\phi_1$ and, therefore:
\beq
\theta_2=\theta_2(\theta_1)\,\,.
\eeq
When $\tilde\Delta=0$, eq. (\ref{tanbeta}) can be solved by putting $\beta=\psi_0+n\pi$ with 
$n\in\zet$. Without loss of generality we can take $n=0$ or, equivalently, $\beta=\psi_0$. Then,
it follows from (\ref{extraDWbeta}) that  we must require that $\epsilon$ be an eigenvector of
$e^{\psi_0\hat\Gamma_{12}}\,\tilde{\Gamma}_{*}$, namely
\beq
e^{\psi_0\hat\Gamma_{12}}\,\tilde{\Gamma}_{*}\,\epsilon\,=\,\sigma\epsilon\,\,.
\label{DWprojection}
\eeq
Moreover, by putting $\Delta=m$, $\beta=\psi_0$ and $p=0$, eq. (\ref{partialx3second}) becomes: 
\beq
\partial_rx^3\,=\,\sigma m e^{-h}\,\Big[\,e^{2h}-{1\over 4}\,\Big]\,\,.
\label{BPSx3}
\eeq
Let us now check that the BPS equations for the embedding that we have found (eqs.
{(\ref{partialphi}) and  ({\ref{costheta2}) with $m=\pm 1$ and eq. (\ref{BPSx3})), together with
some election for the signs
$\sigma$ and $m$, are enough to guarantee the fulfilment of the kappa symmetry condition $\Gamma_{\kappa}\,\epsilon=\epsilon$. First of all, for a general configuration with arbitrary functions
$\theta_2=\theta_2(\theta_1)$, 
$\phi_2=\phi_2(\phi_1)$ and $x^3=x^3(r)$, the determinant of the induced metric is:
\bear
&&\sqrt{-g}\,=\,e^{3\phi}\,\big[\,1\,+\,(\partial_rx^3)^2\,\big]^{{1\over 2}}\,
[\,e^{2h}\,+\,{1\over 4}\,(\partial_{\theta_1}\theta_2)^2\,\big]^{{1\over 2}}
\,\times\rc
&&\,\,\,\,\,\,\,\,\,\,\,\,\,\,\,\,\,\,\,\,\,\,\,\,
\times\,\big[\,e^{2h}\sin^2\theta_1\,+\,{\cos^2\theta_1\over 4}\,+\,
{cos\theta_1\cos\theta_2\over 2}\,\partial_{\phi_1}\phi_2\,+\,
{1\over 4}\,(\partial_{\phi_1}\phi_2)^2\,\,\big]^{{1\over 2}}\,\,.\qquad
\eear
Moreover, when $x^3$ satisfies (\ref{BPSx3}), it is straightforward to prove that:
\beq
1\,+\,(\partial_rx^3)^2_{\,\,|BPS}\,=\,e^{-2h}\,\big[\,e^{2h}\,+\,{1\over 4}\,\big]^2\,\,.
\eeq
If, in addition, the angular embedding is such that  $\cos\theta_2=-m\cos\theta_1$,
$\sin\theta_2=\sin\theta_1$, 
$\partial_{\theta_1}\theta_2=-m$ with $m=\pm 1$ (see eqs. (\ref{costheta2}) and
(\ref{partialtheta2})),  one can demonstrate that:
\beq
\sqrt{-g}_{\,\,|BPS}\,=\,e^{3\phi-h}\,\sin\theta_1\,
\big[\,e^{2h}\,+\,{1\over 4}\,\big]^2\,\,.
\eeq
Moreover, in this abelian background, one can verify that:
\beq
\big[\partial_rx^3\,c_{12}\,+\,c_{1\hat 2}\,\tilde{\Gamma}_{*}\,+\,
c_{1\hat 1}\,\hat\Gamma_{12}\,\tilde{\Gamma}_{*}\,\big]
\,\epsilon_{\,\,|BPS}\,=\,\sigma m e^{-h}\,
\big[\,e^{2h}\,+\,{1\over 4}\,\big]^2\,\epsilon\,\,.
\eeq
By using these results, we see that $\Gamma_{\kappa}\epsilon=\epsilon$ if the sign $\sigma$ is
such  that
\beq
\sigma=m\,\,.
\eeq
The corresponding configurations preserve two supersymmetries, characterized by the extra
projection
\beq
e^{\psi_0\hat\Gamma_{12}}\,\tilde{\Gamma}_{*}\,\epsilon\,=\,m\epsilon\,\,,
\label{DWlastprojection}
\eeq
while $x^3(r)$ is determined by the first-order BPS differential equation
\beq
{dx^3\over dr}\,=\,e^{-h}\,\big[\,e^{2h}\,-\,{1\over 4}\,\big]\,\,.
\label{abelianx3bps}
\eeq
\subsubsection{Integration of the first-order equations}

When $m=\pm 1$, the equations (\ref{partialphi}) and (\ref{costheta2})  that determine the
angular part of the embedding are trivial to solve. The result is:
\bear
&&\theta_2=\pi-\theta_1\,\,,
\,\,\,\,\,\,\,\,\,\,\,\,\,\,\,\,\,\,\,\,
\phi_2=\phi_1\,\,,
\,\,\,\,\,\,\,\,\,\,\,\,\,\,\,\,\,\,\,\,
(m=+1)\,\,,\rc
&&\theta_2=\theta_1\,\,,
\,\,\,\,\,\,\,\,\,\,\,\,\,\,\,\,\,\,\,\,
\phi_2=2\pi-\phi_1\,\,,
\,\,\,\,\,\,\,\,\,\,\,\,\,\,\,\,\,\,\,\,
(m=-1)\,\,.
\label{DWangular}
\eear
Moreover,  by using the value of $e^{2h}$ for the abelian metric given in eq. (\ref{abelianh}),
it is also immediate to get the form of
$x^3(r)$ by direct integration of eq. (\ref{abelianx3bps}):
\beq
x^3(r)\,=\,{2\over 3}\,\,\Big(\,r-{1\over 4}\,\Big)^{{3\over 2}}\,-\,
{1\over 2}\,\,\Big(\,r-{1\over 4}\,\Big)^{{1\over 2}}\,+\,{\rm constant}\,\,.
\label{DWx3ab}
\eeq

\subsection{Non-Abelian worldvolume solitons}

Let us now deal with the full non-abelian background. We will require that the non-abelian
solutions coincide with the abelian one in the asymptotic UV. As displayed in eq.
(\ref{epsiloneta}), the non-abelian Killing spinor $\epsilon$ is related to the asymptotic one
$\epsilon_0=f(r)\eta$ by means of a rotation 
\beq
\epsilon\,=\,e^{{\alpha \over 2}\,\Gamma_1\hat\Gamma_1}\,\,\epsilon_0\,\,,
\label{epsilon0}
\eeq
where $\alpha$ is the angle of (\ref{alpha}) and
$\epsilon_0$ satisfies the same projections as in the abelian case, namely
\beq
\Gamma_{r}\hat \Gamma_{123}\,\epsilon_0\,=\,
\Gamma_{x^0x^1x^2x^3}\,\Gamma_{12}\,\epsilon_0=\epsilon_0\,\,.
\label{proj-epsilon0}
\eeq
By using the relation between the spinors $\epsilon$ and $\epsilon_0$,
the kappa symmetry condition $\Gamma_{\kappa}\,\epsilon\,=\,\epsilon$ can be recast as a
condition on $\epsilon_0$:
\beq
e^{-{\alpha \over 2}\,\Gamma_1\hat\Gamma_1}\,\Gamma_{\kappa}\,\epsilon\,=\,\epsilon_0\,\,,
\label{kappaepsilon0}
\eeq
where the left-hand side is given by:
\bear
&&e^{-{\alpha \over 2}\,\Gamma_1\hat\Gamma_1}\,\Gamma_{\kappa}\,\epsilon\,=\,
{e^{3\phi}\over \sqrt{-g}}\,
\Gamma_{x^0x^1x^2}\,\sin\theta_1\,
\big[\,\Gamma_r\,+\,\partial_rx^3\Gamma_{x^3}\,\big]\,\times\rc
&&\,\,\,\,\,\,\,\,\,\,\,\,\,\,\,\,\,\,\,\,\,\,\,\,\,\,\,\,\,\,\,
\times\big[\,
c_{12}\,e^{-\alpha\,\Gamma_1\hat\Gamma_1}\,\Gamma_{12}\,+
\,c_{1\hat 2}\,e^{-\alpha\,\Gamma_1\hat\Gamma_1}\,\Gamma_{1}\hat\Gamma_{2}\,+\, 
c_{1\hat 1}\,\Gamma_{1}\hat\Gamma_{1}\,\big]\,\epsilon_0\,\,.
\eear
Proceeding as in the abelian case, and using the projections (\ref{proj-epsilon0}), one
arrives at:
\bear
&&e^{-{\alpha \over 2}\,\Gamma_1\hat\Gamma_1}\,\Gamma_{\kappa}\,\epsilon\,=\,
{e^{3\phi}\over \sqrt{-g}}\,\sin\theta_1\,\,\Bigg[\,c_{12}\,
e^{-\alpha\,\Gamma_1\hat\Gamma_1}\,\tilde{\Gamma}_{*}\,\Gamma_1\hat\Gamma_1\,+\,
c_{1\hat 2}\,e^{-\alpha\,\Gamma_1\hat\Gamma_1}\,\tilde{\Gamma}_{*}\,+\,
c_{1\hat 1}\,\hat \Gamma_{12}\,\tilde{\Gamma}_{*}\,+\rc
&&\,\,\,\,\,\,\,\,\,\,\,\,\,\,\,\,\,\,\,\,\,\,\,\,\,\,\,\,\,\,\,\,\,\,\,
+\,\,\,\partial_r x^3 c_{12}\, e^{-\alpha\,\Gamma_1\hat\Gamma_1}\,+\,
\partial_r x^3 c_{1\hat 2}\, e^{-\alpha\,\Gamma_1\hat\Gamma_1}\,\Gamma_1\hat\Gamma_1\,-\,
\partial_r x^3 c_{1\hat 1}\,\,\Gamma_1\hat\Gamma_2\,\Bigg]\,\epsilon_0\,\,.
\label{con-nocon}
\eear
In order to verify eq. (\ref{kappaepsilon0})
we shall impose to $\epsilon_0$ the same projection as in the abelian solution, namely:
\beq
e^{\psi_0\hat\Gamma_{12}}\,\tilde{\Gamma}_{*}\,\epsilon_0\,=\,\sigma\epsilon_0\,\,.
\label{asymDWprojection}
\eeq
Moreover, by expanding the exponential $e^{-\alpha\,\Gamma_1\hat\Gamma_1}$ on the
right-hand side of eq. (\ref{con-nocon}) as 
$e^{-\alpha\,\Gamma_1\hat\Gamma_1}=\cos\alpha-\sin\alpha\Gamma_1\hat\Gamma_1$ we find two
types of terms. The terms involving a matrix that commutes with the projections 
(\ref{proj-epsilon0}) are given by:
\bear
&&\bigg[\,\partial_rx^3\,(c_{12}\,\cos\alpha\,+\,c_{1\hat 2}\,\sin\alpha\,)\,+\,
(\,c_{1\hat 2}\,\cos\alpha\,-\,c_{12}\,\sin\alpha\,)\tilde{\Gamma}_{*}\,+\,
c_{1\hat 1}\,\hat\Gamma_{12}\,\tilde{\Gamma}_{*}\,\bigg]\,\epsilon_0\,\equiv\rc\rc
&&\,\,\,\,\,\,\,\,\,\,\,\,\,\,\,\,\,\,\,\,\,\,\,\,\,\,\,\,\,\,\,\,\,\,\,
\,\,\,\,\,\,\,\,\,\,\,\,\,\,\,\,\,\,\,\,\,\,\,\,\,\,\,\,\,\,\,\,\,\,\,
\equiv\,\bigg(\,{\cal A}_I\,+\,{\cal A}_{\hat 1\hat 2}\,
\hat\Gamma_{12}\,\bigg)\,\epsilon_0\,\,,
\label{calA}
\eear
while those with a matrix which does not commute with the projections are:
\bear
&&-\Gamma_1\hat\Gamma_1\,\Bigg[\,\bigg(\,
c_{12}\,\cos\alpha\,+\,c_{1\hat 2}\,\sin\alpha\,\bigg)\,\tilde{\Gamma}_{*}\,-\,
(\,c_{1\hat 2}\,\cos\alpha\,-\,c_{12}\,\sin\alpha\,)\,\partial_r x^3\,+\,
c_{1\hat 1}\,\partial_r x^3\,\hat\Gamma_{12}\,\Bigg]\epsilon_0\,=\,\rc
&&\,\,\,\,\,\,\,\,\,\,\,\,\,\,\,\,\,\,\,\,\,\,\,\,\,\,\,\,\,\,\,\,\,\,\,
\,\,\,\,\,\,\,\,\,\,\,\,\,\,\,\,\,\,\,\,\,\,\,\,\,\,\,\,\,\,\,\,\,\,\,
\equiv\,-\Gamma_1\hat\Gamma_1\,
\bigg(\,{\cal B}_I\,+\,{\cal B}_{\hat 1\hat 2}\,
\hat\Gamma_{12}\,\bigg)\,\epsilon_0\,\,.
\label{calB}
\eear
The coefficients ${\cal A}$ and ${\cal B}$ defined in eqs. (\ref{calA}) and 
(\ref{calB}) can be read from the left-hand side of these equations after substituting the
value of $\tilde{\Gamma}_{*}$ from eq. (\ref{asymDWprojection}). They are given by:
\bear
&&{\cal A}_I=\partial_rx^3\,(c_{12}\,\cos\alpha\,+\,c_{1\hat 2}\,\sin\alpha\,)\,+\,
\sigma(\,c_{1\hat 2}\,\cos\alpha\,-\,c_{12}\,\sin\alpha\,)\cos\psi_0\,+\,
\sigma c_{1\hat 1}\sin\psi_0\,\,,\rc
&&{\cal A}_{\hat 1\hat 2}=\sigma c_{1\hat 1}\cos\psi_0\,-\,
\sigma(\,c_{1\hat 2}\,\cos\alpha\,-\,c_{12}\,\sin\alpha\,)\sin\psi_0\,\,,\rc
&&{\cal B}_I=\sigma(c_{12}\,\cos\alpha\,+\,c_{1\hat 2}\,\sin\alpha\,)\cos\psi_0\,-\,
(\,c_{1\hat 2}\,\cos\alpha\,-\,c_{12}\,\sin\alpha\,)\,\partial_rx^3\,\,,\rc
&&{\cal B}_{\hat 1\hat 2}=c_{1\hat 1}\,\partial_rx^3\,-\,\sigma
(c_{12}\,\cos\alpha\,+\,c_{1\hat 2}\,\sin\alpha\,)\sin\psi_0\,\,.
\eear
 Since we are looking for
solutions which must coincide with the abelian ones in the UV, we can restrict ourselves to
the case in which $\theta_2=\theta_2(\theta_1)$, \ie\ with $\tilde\Delta=0$. It is easy to
prove that in this case  the combinations of $c_{12}$ and $c_{1\hat 2}$ appearing
above reduce to:
\bear
&&c_{12}\,\cos\alpha\,+\,c_{1\hat 2}\,\sin\alpha\,=\,\bigg[\,
r\coth 2r\,-\,{1\over 2}\,\bigg]\,\bigg[\,\coth 2r\,-\,
{\Delta\cos\psi_0\over \sinh 2r}\,\bigg]\,\,,\rc
&&c_{1\hat 2}\,\cos\alpha\,-\,c_{12}\,\sin\alpha\,=\,e^h\,\bigg[\,
\Delta\cos\psi_0\coth 2r\,-\,{1\over \sinh 2r}\,\Bigg]\,\,.
\eear
To derive this result  we have used the following useful relations:
\bear
&&e^h\sin\alpha\,+\,{a\over 2}\,\cos\alpha\,=\,{1\over \sinh 2r}\,
\bigg[\,{1\over 2}\,-\,r\coth 2r\,\bigg]\,\,,\rc
&&e^h\cos\alpha\,-\,{a\over 2}\,\sin\alpha\,=\,e^h\,\coth 2r\,\,,\rc
&&\bigg(\, e^{2h}\,-\,{a^2\over 4}\,-\,{1\over 4}\,\bigg)\,\sin\alpha\,+\,
ae^h\cos\alpha\,=\,{e^h\over \sinh2r}\,\,,\rc
&&\bigg(\, e^{2h}\,-\,{a^2\over 4}\,-\,{1\over 4}\,\bigg)\,\cos\alpha\,-\,
ae^h\sin\alpha\,=\,\coth 2r\,\bigg[r\coth 2r\,-\,{1\over 2}\,\bigg]\,\,,
\label{MNidentities}
\eear
which can be easily demonstrated by using eqs. (\ref{MNsol}) and (\ref{alpha}). Clearly, in order
to satisfy (\ref{kappaepsilon0}) we must require that
\beq
{\cal A}_{\hat 1\hat 2}={\cal B}_I={\cal B}_{\hat 1\hat 2}=0\,\,.
\eeq
Let us now consider the  ${\cal A}_{\hat 1\hat 2}=0$   equation first. It is easy to
conclude  that this equation reduces to:
\beq
\sin\psi_0\,\bigg[\,(1-\coth 2r)
\Delta\cos\psi_0\,+\,{1\over \sinh 2r}
\,\Bigg]\,=\,0\,\,.
\eeq
If $\sin\psi_0\not=0$ the above equation can be used to obtain an expression of $\Delta$
with a nontrivial dependence on the radial variable $r$, which is in contradiction with
eq. (\ref{Delta}). Thus we conclude that $\sin\psi_0$ must vanish, \ie\ only four values
of $\psi_0$ are possible, namely:
\beq
\psi_0\,=\,0,\pi,2\pi, 3\pi\,\,.
\label{DWpsi_0}
\eeq
Let us denote
\beq
\lambda\equiv\cos\psi_0\,=\,\pm 1\,\,.
\label{DWlambda}
\eeq
Then, the condition ${\cal B}_{\hat 1\hat 2}=0$ is automatically satisfied when 
$\sin\psi_0=0$, while ${\cal B}_I=0$ leads to the following equation for $\partial_rx^3$:
\beq
\partial_rx^3\,=\,\lambda\sigma\,\,e^{-h}\,\,\,
{\cosh 2r\,-\,\Delta\lambda\over \Delta\lambda \cosh 2r\,-\,1}
\,\,\bigg[r\coth 2r\,-\,{1\over 2}\,\bigg]\,\,.
\label{DWnonabeBPSx3}
\eeq
As in the abelian case, the consistency of the above equation with our ansatz for $x^3$
requires that $\Delta$ be constant which, in turn, only can  be achieved if $m=\pm 1$ and
$\Delta=m$. Notice that this implies that the angular equations for the embedding are
exactly those written in eq. (\ref{DWangular}) for the abelian case. Moreover, when
$\theta_2=\theta_2(\theta_1)$ and $\phi_2=\phi_2(\phi_1)$ are given as in eq. 
(\ref{DWangular}), the determinant of the induced metric is
\beq
\sqrt{-g}\,=\,e^{3\phi}\,\sin\theta_1\,{r\over \sinh 2r}\,\bigg[\,
\cosh 2r\,-\,\lambda m\,\bigg]\,\sqrt{1\,+\,(\partial_r x^3)^2}\,\,.
\label{DWnonabeliandet}
\eeq
When $x^3$ satisfies the differential equation (\ref{DWnonabeBPSx3}), one can easily demonstrate
that:
\beq
\sqrt{1\,+\,(\partial_r x^3)^2}_{\,\,|BPS}\,=\,re^{-h}\
\label{x3bpsidentity}\,\,,
\eeq
and, using this result to evaluate the right-hand side of (\ref{DWnonabeliandet}), one arrives
at: 
\beq
{e^{3\phi}\,\sin\theta_1\,{\cal A}_I}_{\,\,|BPS}\,=\,\sigma m \sqrt{-g}_{\,\,|BPS}\,\,.
\eeq
Therefore,  one must take $\sigma=m$ in order to satisfy eq. (\ref{kappaepsilon0}). 
When $\sin\psi_0=0$, the
extra projection (\ref{asymDWprojection}) on the asymptotic spinor $\epsilon_0$ is
\beq
\tilde{\Gamma}_{*}\,\epsilon_0\,=\,\lambda\,m\,\epsilon_0\,\,,
\eeq
which is equivalent to the following projection on the complete spinor $\epsilon$:
\beq
e^{\alpha\Gamma_1\hat\Gamma_1}\,\tilde{\Gamma}_{*}\epsilon\,=\,\lambda\,m\,\epsilon\,\,.
\eeq
Moreover, the differential equation which determines $x^3(r)$ is:
\beq
{dx^3\over dr}\,=\,e^{-h}\,\bigg[\,r\coth 2r \,-\,{1\over 2}\,\bigg]\,\,.
\label{DWnonabeBPS}
\eeq
It is straightforward to demonstrate that this equation coincides with the abelian one in the UV.
Actually, in Fig. \ref{fig1} we represent the result of integrating eq. (\ref{DWnonabeBPS}) and
we compare this result with that given by the function $x^3(r)$ for the abelian background (eq.
(\ref{DWx3ab})). Moreover, if we fix
the embedding $\theta_2=\theta_2(\theta_1)$, $\phi_2=\phi_2(\phi_1)$ and 
$x^3=x^3(r)$ we have two possible projections, corresponding to the two possible values
of $\lambda$. Each of these values of $\lambda$ corresponds to  two values of the angle
$\psi_0$, which again shows that the $U(1)$ symmetry of the abelian theory is broken to
$\zet_2$. One can check that the embeddings characterized by eqs. (\ref{DWangular}),
(\ref{DWpsi_0}) and (\ref{DWnonabeBPS}) satisfy the equations of motion derived from the
Dirac-Born-Infeld action of the probe (\ref{actionintrod}). 

\begin{figure}
\centerline{\epsffile{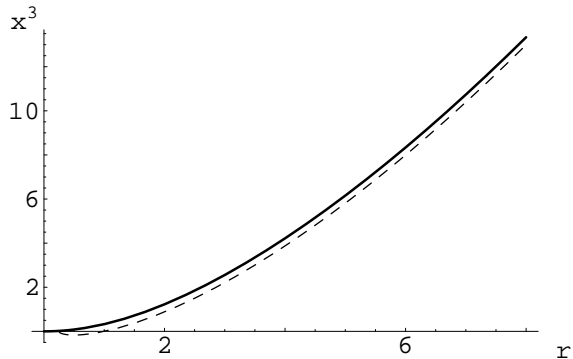}}
\caption{In this figure we represent the function $x^3(r)$ for the  wall defect in the
non-abelian background (solid line). The dashed line represents $x^3(r)$ for the abelian
background  as given by eq. (\ref{DWx3ab}). In both cases the constant of integration has been
fixed by requiring that the minimal value of $x^3$ is 0. 
}
\label{fig1}
\end{figure}

\subsection{Energy bound for the  wall solutions}

 The embeddings that we have just found
saturate an energy bound, as expected for BPS worldvolume solitons. 
Let us consider a D5-brane probe in the non-abelian MN background and let us
choose the same worldvolume coordinates as in eq. (\ref{DWvwcoordinates}) and the ansatz
(\ref{DWansatz}) for the embedding.  For simplicity we will consider the angular embeddings
$\theta_2(\theta_1,\phi_1)$ and 
$\phi_2(\theta_1,\phi_1)$ written in  eq. (\ref{DWangular}) and we will consider a completely
arbitrary function $x^{3}(r)$. Using
the value of $\sqrt{-g}$ given in (\ref{DWnonabeliandet}), one gets:
\beq
{\cal H}\,=\,-
{\cal L}\,=\,e^{2\phi}\,\sin\theta_1\,\Bigg[\,
{r\over \sinh 2r}\,
\big(\,
\cosh 2r\,-\,\lambda m\,\big)\,\sqrt{1\,+\,(\partial_r x^3)^2}\,-\,
{\lambda m\over 4}\,a'\,\partial_r x^3\,\Bigg]\,\,,\qquad
\eeq
where $m=\pm 1$ is the same as in eq. (\ref{DWangular}) and $\lambda=\cos\psi_0\,=\,\pm 1$ (see
eq. (\ref{DWlambda})). In order to write ${\cal H}$ as in eq. (\ref{H=Z+S}) , let us
define the function 
\beq
\Lambda_r\equiv e^{-h}\,\bigg[\,r\coth 2r \,-\,{1\over 2}\,\bigg]\,\,.
\eeq
Notice that the BPS equation for $x^{3}(r)$ (eq. (\ref{DWnonabeBPS})) is just 
$\partial_r x^3=\Lambda_r$.  Furthermore, $a'$ can be written in terms
of $\Lambda_r$ as:
\beq
{a'\over 4}\,=\,-{e^h\Lambda_r\over \sinh2r}\,\,.
\eeq
Using this last result, we can write ${\cal H}$ as :
\beq
{\cal H}\,=\,\sin\theta_1\,{e^{2\phi}\over \sinh 2r}\,\Bigg[\,r\,
\big(\,\cosh 2r\,-\,\lambda m\,\big)\,\sqrt{1\,+\,(\partial_r x^3)^2}\,+\,
\lambda m e^{h}\,\Lambda_r\,\partial_r x^3\,\Bigg]\,\,.
\eeq
Let us now write ${\cal H}$ as in eq. (\ref{H=Z+S}), with:
\beq
{\cal Z}\,=\,\sin\theta_1\,{e^{2\phi+h}\over \sinh 2r}\,\Bigg[\,\cosh 2r\,
\Lambda_r\,\partial_r x^3\,+\,\cosh 2r\,-\,\lambda m\,\Bigg]\,\,.
\eeq
By using eq. (\ref{x3bpsidentity}), one can prove that
\beq
{\cal H}_{\,\,|BPS}\,=\,
{\cal Z}_{\,\,|BPS}\,\,.
\eeq
Moreover, ${\cal Z}$ can be written as a total derivative, \ie\ 
${\cal Z}\,=\,\partial_r{\cal Z}^r\,+\,\partial_{\theta_1}{\cal Z}^{\theta_1}$, with
\bear
&&{\cal Z}^r\,=\,\sin\theta_1\,{e^{2\phi+h}\over \sinh 2r}\,\Bigg[\,\cosh 2r\,
\Lambda_r\,x^3\,+\,{\sinh 2r\over 2}\,-\,\lambda m r\,\Bigg]\,\,,\rc
&&{\cal Z}^{\theta_1}\,=\,\cos\theta_1\,e^{2\phi}\,\Bigg[\,2e^{2h}\,+\,
{1-a^4\over 8}\,e^{-2h}\,\Bigg]\,x^3\,\,.
\eear
To derive this result it is useful to remember that $e^{2\phi+h}/\sinh 2r$ is constant and
use the relation
\beq
\partial_r\,\bigg[\,\cosh 2r \Lambda_r\,\bigg]\,=\,e^{-h}\,\sinh 2r\,
\bigg[\,2e^{2h}\,+\,{1-a^4\over 8}\,e^{-2h}\,\bigg]\,\,,
\eeq
which can be proved by direct calculation. Moreover, taking into account that 
$r=e^{h}\,\sqrt{1+\Lambda_r^2}$ (see eq. (\ref{x3bpsidentity})), one can write ${\cal S}$
as:
\beq
{\cal S}\,=\,\sin\theta_1\,{e^{2\phi+h}\over \sinh 2r}\,
\big(\,\cosh 2r\,-\,\lambda m\,\big)\,\Bigg[\,
\sqrt{1+\Lambda_r^2}\,\sqrt{1\,+\,(\partial_r x^3)^2}\,-\,
(\,1\,+\,\Lambda_r\partial_r x^3\,)\,\Bigg]\,\,,
\eeq
and it is straightforward to verify that ${\cal S}\ge 0$ is equivalent to
\beq
(\,\partial_r x^3\,-\,\Lambda_r\,)^2\ge 0\,\,,
\eeq
which is obviously always satisfied for any function $x^3(r)$ and reduces to an equality when the
BPS equation (\ref{DWnonabeBPS}) holds.

\setcounter{equation}{0}
\section{Two-dimensional defects}   \label{twodimdefect}
\medskip
In this section we will determine BPS configurations of a D5-brane which extends along two
Minkowski coordinates (say $x^0$ and $x^1$) and along a four-dimensional submanifold embedded in
the internal part of the metric (\ref{metricMN}).  Such branes would be a two-dimensional
object from the gauge theory perspective and, actually, we will find that they preserve the same
supersymmetries as a D1-string stretched along $x^1$. In order to find these configurations from
the kappa symmetry condition $\Gamma_{\kappa}\,\epsilon\,=\,\epsilon$ let us choose the
following set of worldvolume coordinates for the D5-brane:
\beq
\xi^\mu\,=\,(x^0,x^1,\theta_1,\phi_1, \theta_2,\phi_2)\,\,,
\label{Stringcoordinates}
\eeq
and let us consider an embedding of the type
\beq
r\,=\,r(\theta_1,\theta_2)\,\,,
\,\,\,\,\,\,\,\,\,\,\,\,\,\,\,\,\,
\psi=\psi(\phi_1,\phi_2)\,\,,
\label{Stringansatz}
\eeq
with $x^2$ and $x^3$ being constant\footnote{For two-dimensional defects obtained with a
different election of worldvolume coordinates and ansatz, see subsection \ref{morestringdefects}.}. From our general
expression (\ref{GammaD5}) it is straightforward to prove that in this case
$\Gamma_{\kappa}\,\epsilon$ is given by:
\beq
\Gamma_{\kappa}\,\epsilon\,=\,{e^{\phi}\over \sqrt{-g}}\,\,\Gamma_{x^0x^1}\,
\gamma_{\theta_1\phi_1 \theta_2\phi_2}\,\epsilon\,\,.
\eeq
The induced Dirac matrices $\gamma_{\theta_i}$ and $\gamma_{\phi_i}$ are easily obtained by
using our ansatz in eq. (\ref{wvgamma}).
With the purpose of writing these matrices in a convenient form, let us define the quantities:
\beq
\Delta_i\equiv {1\over 2}\,{\cos\theta_i\,+\,\partial_{\phi_i}\psi\over \sin\theta_i}\,\,,
\label{Deltas}
\eeq
in terms of which the  $\gamma$-matrices are:
\bear
e^{-{\phi\over 2}}\,\gamma_{\theta_1}&=&e^h\Gamma_1\,+\,{a\over 2}\,\hat\Gamma_1\,+\,
\partial_{\theta_1}r\,\Gamma_r\,\,,\rc
{e^{-{\phi\over 2}}\over \sin\theta_1}
\,\gamma_{\phi_1}&=&e^h\,\Gamma_2\,-\,{a\over 2}\,
\hat\Gamma_2\,+\,\Delta_1\,
\hat\Gamma_3\,\,,\rc
e^{-{\phi\over 2}}\,\gamma_{\theta_2}&=&{1\over 2}\cos\psi\,\hat\Gamma_1\,-\,{1\over 2}\,
\sin\psi\,\hat\Gamma_2\,+\,\partial_{\theta_2}r\,\Gamma_r\,\,,\rc
{e^{-{\phi\over 2}}\over\sin\theta_2} \,
\gamma_{\phi_2}&=&{1\over 2}\sin\psi\,\hat\Gamma_1\,+\,
{1\over 2}\cos\psi\,\hat\Gamma_2\,+\,
\Delta_2\,
\hat\Gamma_3\,\,.
\label{indgammastr}
\eear
By using eqs. (\ref{indgammastr}) and (\ref{fullprojection}) the action of the antisymmetrised
product of the $\gamma$'s on the Killing spinors $\epsilon$ can be readily obtained. It is of
the form:
\bear
{e^{-2\phi}\over \sin\theta_1\sin\theta_2}\,\,
\gamma_{\theta_1\phi_1\theta_2\phi_2}\,\epsilon\,&=&\,\big[\,b_I\,+\,
b_{2\hat 2}\,\Gamma_{2}\hat\Gamma_{2}\,+\,
b_{12}\,\Gamma_{12}\,+\,b_{1\hat 2}\,\Gamma_{1}\hat\Gamma_{2}\,+\,
b_{1\hat 3}\,\Gamma_{1}\hat\Gamma_{3}\,+\,
\rc
&&+\,b_{\hat 1\hat 3}\,\hat\Gamma_{13}\,
+\,b_{\hat 2\hat 3}\,\hat\Gamma_{23}\,
+\, b_{2\hat 3}\,\Gamma_{2}\hat\Gamma_{3}\,\big]\,\epsilon\,\,,
\label{Stringces}
\eear
where the $b$'s are functions whose expression depends on the embedding of the probe. In order
to write them more compactly let us define $\Lambda_1$ and $\Lambda_2$ as follows:
\bear
&&\Lambda_1\equiv {1\over 4}\,
\Big[\,\partial_{\theta_1}r-a\cos\psi\,\partial_{\theta_2}r\,\Big]\Delta_1\,+\,
\,\Big[(\,e^{2h}-{a^2\over 4}\,)\partial_{\theta_2}r\,+\,
{a\over 4}\cos\psi\,\partial_{\theta_1}r\,\Big]\,\Delta_2\,,\rc
&&\Lambda_2\equiv -{e^h\over 2}\,
\Big[\,\cos\psi\,\partial_{\theta_1}r\,-\,2a\,\partial_{\theta_2}r\,\Big]\,\Delta_2\,+\,
{e^h\over 2}\,\cos\psi\,\partial_{\theta_2}r\,\Delta_1\,\,,
\eear
where $\Delta_1$ and $\Delta_2$ have been defined in eq. (\ref{Deltas}). Then, the coefficients
of the different matrix structures appearing on the right-hand side of eq. (\ref{Stringces}) are:
\bear
&&b_I\,=\,\Lambda_1\cos\alpha\,-\,\Lambda_2\sin\alpha\,-\,
{e^{2h}\over 4}\,\,\,,\rc
&&b_{2\hat 2}\,=\,\Lambda_1\sin\alpha\,+\,\Lambda_2\cos\alpha\,\,,\rc
&&b_{1 2}\,=\,-\,{e^h\sin\psi\over 2}\,
\Big[\,\partial_{\theta_2}r\,\Delta_1\,-\,
\,\partial_{\theta_1}r\,\Delta_2\,\Big]\,\sin\alpha\,\,,\rc
&&b_{1\hat 2}\,=\,{e^h\sin\psi\over 2}\,
\Big[\,\,\partial_{\theta_2}r\,\Delta_1\,-\,
\partial_{\theta_1}r\,\Delta_2\,\Big]\,\cos\alpha\,\,,\rc
&&b_{1\hat 3}\,=\,{e^h\over 2}\,\sin\psi\Big[\,
\partial_{\theta_2}r\,(\,e^h\sin\alpha\,+\,{a\over 2}\cos\alpha\,)
\,-\,{a\over 2}\,\Delta_2\,\Big]\,\,,\rc
&&b_{\hat 1\hat 3}\,=\,{e^h\over 2}\,\sin\psi\Big[\,
\,\partial_{\theta_2}r\,(\,e^h\cos\alpha\,-\,{a\over 2}\sin\alpha\,)
\,+\,e^h\,\Delta_2\,\Big]\,\,,\rc
&&b_{\hat 2\hat 3}\,=\,{e^h\over 2}\,\cos\psi\Big[\,
\partial_{\theta_2}r\,(\,e^h\cos\alpha\,-\,{a\over 2}\sin\alpha\,)
\,+\,e^h\,\Delta_2\,\Big]\,+\,
{e^{h}\over 4}\,\partial_{\theta_1}r\,\sin\alpha\,\,,\rc
&&b_{2\hat 3}\,=\,{e^h\over 2}\,\cos\psi
\Big[\,{a\over 2}\,\Delta_2\,-\,\partial_{\theta_2}r\,
(\,e^h\sin\alpha\,+\,{a\over 2}\cos\alpha\,)\,\Big]\,+\,
{e^h\over 4}\,\Big[\,\Delta_1\,+\,
\partial_{\theta_1}r\,\cos\alpha\,\Big]\,\,.\qquad\qquad
\eear
By inspecting the right-hand side of eq. (\ref{Stringces}) one immediately realises that the
terms containing the matrix $\hat\Gamma_3$ give rise to contributions to $\Gamma_{\kappa}$ which
do not commute with the projection $\Gamma_{12}\,\epsilon\,=\,\hat\Gamma_{12}\,\epsilon$
satisfied by the Killing spinors (see eq. (\ref{fullprojection})). Then, if we want that the
supersymmetry preserved by the probe be compatible with that of the background, the coefficients
of these terms must vanish. Moreover, 
we would like to obtain embeddings of the D5-brane probe which preserve the same supersymmetry
as a D1-string extended along the $x^1$ direction. Accordingly\footnote{From a detailed analysis 
of the form of the $b$'s one can show that  the requirement of the vanishing of the
coefficients of the terms containing the matrix $\hat \Gamma_3$ implies the vanishing of 
$b_{2\hat 2}$, $b_{1 2}$ and $b_{1\hat 2}$. Therefore, we are not loosing generality by imposing
(\ref{cesnull}).}, we shall require the
vanishing of all terms on the right-hand side of eq. (\ref{Stringces}) except for the one
proportional to the unit matrix, \ie:
\beq
b_{2\hat 2}\,=\,b_{1 2}\,=\
b_{1\hat 2}\,=\,b_{1\hat 3}\,=\,b_{\hat 1\hat 3}\,=\,b_{\hat 2\hat 3}\,=\,
b_{2\hat 3}\,=\,0\,\,\,.
\label{cesnull}
\eeq
By plugging the explicit form of the $b$'s in (\ref{cesnull}), one gets a system of differential
equations for the embedding which will be analyzed in the rest of this section.

\subsection{Abelian worldvolume solitons}

The above equations  (\ref{cesnull})
are quite complicated. In order to simplify the problem, let us consider first the equations
for the embedding in the abelian background, which can be obtained from the general ones by  
putting $a=\alpha=0$. In this case from  $b_{\hat 2\hat 3}=b_{2\hat 3}=0$ we get 
$\,\partial_{\theta_i} r=-\Delta_i$, where the $\Delta_i$'s have been defined in eq.
(\ref{Deltas}). More explicitly:
\beq
\partial_{\theta_i} r\,=\,-{1\over 2}\,
{\cos\theta_i+\partial_{\phi_i}\psi\over \sin\theta_i}\,\,.
\label{abelianBPS}
\eeq
One can verify that the other $b's$ in (\ref{cesnull}) vanish if these differential
equations are satisfied. Let us see the form of the kappa symmetry condition when the BPS
equations   (\ref{abelianBPS}) are satisfied. For the abelian background, the
determinant of the induced metric is given by:
\beq
\sqrt{-g}\,=\,{e^{3\phi}\over 4}\,\,\sin\theta_1\,\sin\theta_2\,
\Big[\,(\partial_{\theta_1}r)^2\,+\,4e^{2h}\,(\partial_{\theta_2}r)^2\,+\,e^{2h}\,
\Big]^{{1\over 2}}\,\,
\Big[\,\Delta_1^2\,+\,4e^{2h}\,\Delta_2^2\,+\,
e^{2h}\,\Big]^{{1\over 2}}\,\,,
\eeq
and the coefficient $b_I$ is:
\beq
b_I\,=\,{1\over 4}\,\partial_{\theta_1}r\,\Delta_1\,+\,
e^{2h}\,\partial_{\theta_2}r\,\Delta_2\,-\,
{e^{2h}\over 4}\,\,\,.
\eeq
If the BPS equations $\,\partial_{\theta_i} r=-\Delta_i$ hold, one can verify by inspection
that:
\beq
e^{-3\phi}\,\sqrt{-g}_{\,|_{BPS}}\,=\,-\sin\theta_1\sin\theta_2\,b_{I\,|_{BPS}}\,\,,
\label{sqrt-cI}
\eeq
and, thus,  the kappa symmetry condition $\Gamma_{\kappa}\,\epsilon=\epsilon$ becomes
\beq
\Gamma_{x^0x^1}\epsilon\,=-\,\epsilon\,\,,
\label{D1projection}
\eeq
which indeed corresponds to a D1 string extended along $x^1$. In this abelian case the
spinors
$\epsilon$ and $\eta$ in eq. (\ref{epsiloneta}) differ in a function which commutes with
everything. Therefore, the condition (\ref{D1projection}) translates into the same condition for
the constant spinor $\eta$, namely:
\beq
\Gamma_{x^0x^1}\eta\,=-\,\eta\,\,.
\label{D1constprojection}
\eeq
It follows that this configuration is 1/16 supersymmetric: it preserves the two supersymmetries
determined by eqs. (\ref{constantfullpro}) and (\ref{D1constprojection}).

\subsubsection{Integration of the first-order equations}

The BPS equations (\ref{abelianBPS}) relate the partial derivatives of $r$ with those of $\psi$.
According to our ansatz (\ref{Stringansatz}) the only dependence on $\phi_1$ and $\phi_2$ in 
(\ref{abelianBPS})   comes from the derivatives of $\psi$. Therefore, 
for consistency of eq. (\ref{abelianBPS}) with our ansatz we must have:
\beq
\partial_{\phi_1}\psi\,=\,n_1\,\,,
\,\,\,\,\,\,\,\,\,\,\,\,\,\,\,\,\,\,\,\,\,\,\,\,
\partial_{\phi_2}\psi\,=\,n_2\,\,,
\eeq
where $n_1$ and $n_2$  are two constant numbers. Thus, $\psi$ must be given by:
\beq
\psi\,=\,n_1\phi_1\,+\,n_2\phi_2\,+\,{\rm constant}\,\,.
\label{winding}
\eeq
Using this form of $\psi(\phi_1,\phi_2)$ in eq. (\ref{abelianBPS}), one can easily integrate
$r(\theta_1,\theta_2)$, namely:
\beq
e^{2r}\,=\,{C\over
\Big(\,\sin{\theta_1\over 2}\,\Big)^{n_1+1}\,\,
\Big(\,\cos{\theta_1\over 2}\,\Big)^{1-n_1}\,\,
\Big(\,\sin{\theta_2\over 2}\,\Big)^{n_2+1}\,\,
\Big(\,\cos{\theta_2\over 2}\,\Big)^{1-n_2}}\,\,,
\label{rtheta}
\eeq
where $C$ is a constant.  From the analysis of eq. (\ref{rtheta}) one easily concludes that not
all the values of the constants $n_1$ and $n_2$ are possible. Indeed, the left-hand side of eq.
(\ref{rtheta}) is always greater than one, whereas the right-hand side always vanishes for some
value of $\theta_i$ if $|n_i|>1$. Actually, we will verify in the next subsection that only when
$n_1=n_2=0$ (\ie\ when $\psi={\rm constant}$) we will be able to generalise the embedding to the
non-abelian geometry. Therefore, from now on we will concentrate only in this case, which we
rewrite as:
\beq
e^{2r}\,=\,{e^{2r_*}\over \sin\theta_1\sin\theta_2}\,\,,
\qquad\qquad
(n_1=n_2=0)\,\,,
\label{rthetazerons}
\eeq
where $r_*=r(\theta_1=\pi/2, \theta_2=\pi/2)$ is the minimal value of $r$. It is clear from
(\ref{rthetazerons}) that $r$ diverges at $\theta_i=0,\pi$. Therefore our effective strings
extend infinitely in the holographic coordinate $r$.

\subsection{Non-Abelian worldvolume solitons}

Let us consider now the more complicated case of the non-abelian background. We are going
to argue that the kappa symmetry condition can only be solved if $\psi$ is constant and
$\sin\psi=0$. Indeed, let us assume that $\sin\psi$ does not vanish. If this is the case,
by combining the conditions $b_{\hat 1\hat 3}=0$ and $b_{\hat 2\hat 3}=0$ one gets 
$\partial_{\theta_1}r=0$. Using this result in the equation $b_{1 2}=0$, one concludes that
$\partial_{\theta_2}r=0$ (notice that the functions $\Delta_i$ can never vanish). However, 
if $r$ is independent of the $\theta_i$'s the equation $b_{ 1\hat 3}=0$ can never be
fulfiled. Thus,  we arrive at a contradiction that can only be resolved if $\sin\psi=0$.
Then, one must have:
\beq
\psi=0,\pi,2\pi,3\pi=0 \,\,(\mod\, \pi)\,\,.
\label{Strnonabpsi}
\eeq
Let us now  define
\beq
\lambda\equiv \cos\psi=\pm 1\,\,.
\label{Strlambda}
\eeq
Thus, in this non-abelian case we are only going to have zero-winding embeddings, \ie, as
anticipated above,  only the solutions with $n_1=n_2=0$ in eq. (\ref{rtheta})  generalise to the
non-abelian case. Since $\psi$ is constant, we now have
\beq
\Delta_i={1\over 2}\,\cot\theta_i\,\,.
\eeq
When $\sin\psi=0$ the equations 
$b_{1 2}=b_{1\hat 2}=b_{\hat 1\hat 3}=b_{1\hat 3}=0$ are automatically satisfied.
Moreover, the conditions $b_{\hat 2\hat 3}=b_{ 2\hat 3}=0$ reduce to:
\bear
&&\sin\alpha\,\partial_{\theta_1}r\,+\,2\lambda\,\big(\,e^h\cos\alpha\,-\,
{a\over 2}\,\sin\alpha\,\big)\,\partial_{\theta_2} r\,+\,\lambda e^h\,\cot\theta_2
\,=\,0\,\,,\rc
&&\cos\alpha\,\partial_{\theta_1}r\,-\,2\lambda\,\big(\,e^h\sin\alpha\,+\,
{a\over 2}\,\cos\alpha\,\big)\,\partial_{\theta_2} r\,+\,
 {\lambda a\over 2}\,\cot\theta_2\,+\,{1\over 2}\,\cot\theta_1\,=\,0\,\,.\qquad
\label{partialsr}
\eear
From  eq. (\ref{partialsr}) one can obtain the values of the partial derivatives of $r$. Indeed,
let us define
\bear
\Delta_{\theta_1}&\equiv&{1\over 2}\,\cot\theta_1\coth(2r)\,+\,{\lambda\over 2}\,\,
{\cot\theta_2\over \sinh (2r)}\,\,,\rc
\Delta_{\theta_2}&\equiv&{1\over 2}\,\cot\theta_2\coth(2r)\,+\,{\lambda\over 2}\,\,
{\cot\theta_1\over \sinh (2r)}\,\,.
\eear
Then, one has
\beq
\partial_{\theta_i}r\,=\,-\Delta_{\theta_i}\,\,.
\label{nonabelianBPS}
\eeq
To derive this result we have used  some of the identities written in eq. (\ref{MNidentities}).
Notice that $\Delta_{\theta_i}\to \Delta_i$ when $r\to\infty$ and the non-abelian
BPS equations (\ref{nonabelianBPS}) coincide with the abelian ones in eq.
(\ref{abelianBPS}) for $n_1=n_2=0$ in this limit. 
After some calculation one can check that $b_{2\hat 2}$ also vanishes as a consequence of
(\ref{nonabelianBPS}). Indeed, one can prove that $b_{2\hat 2}$ can be written:
\beq
b_{2\hat 2}\,=\,{\lambda e^h\over 2}\,\Big[\,\Delta_{\theta_1}\,\partial_{\theta_2}r\,-\,
\Delta_{\theta_2}\,\partial_{\theta_1}r\,\Big]\,\,,
\eeq
which clearly vanishes if eq. (\ref{nonabelianBPS}) is satisfied. 

For a general function $r(\theta_1,\theta_2)$, when the angle $\psi$ takes the values
written in eq. (\ref{Strnonabpsi}), the determinant of the induced metric takes the form:
\bear
&&\sqrt{-g}\,=\,{e^{3\phi}\over 4}\,\sin\theta_1\,\sin\theta_2\Big[
(\partial_{\theta_1}r)^2\,+\,4(e^{2h}\,+\,{a^2\over 4}\,)
(\partial_{\theta_2}r)^2\,-\,2a\lambda\partial_{\theta_1}r\partial_{\theta_2}r
\,+\,e^{2h}\,\Big]^{{1\over 2}}\times\rc
&&\,\,\,\,\,\,\,\,\,\,\,\,\,\,\,\,\,\,\,\,\,\,\,\,\,\,\,\,\,\,\,\,\,\,\,\,\,\,
\times
\Big[\,\Delta_{\theta_1}^2\,+\,4(e^{2h}\,+\,{a^2\over 4}\,)\,
\Delta_{\theta_2}^2\,-\,2 a\lambda\Delta_{\theta_1}\Delta_{\theta_2}
\,+\,e^{2h}\,\Big]^{{1\over 2}}\,\,.
\label{sqrtg}
\eear
If the BPS equations (\ref{nonabelianBPS}) are satisfied, the two factors under the square
root on the right-hand side of eq. (\ref{sqrtg})  become equal. Moreover,  one can prove
that:
\beq
b_I\,=\,{1\over 4}\,\partial_{\theta_1}r\,(\Delta_{\theta_1}-\lambda a \Delta_{\theta_2})
\,+\,{1\over 4}\,\partial_{\theta_2}r\,
\Big(4\,(e^{2h}\,+\,{a^2\over 4}\,)\,\Delta_{\theta_2}
-\lambda a \Delta_{\theta_1}\Big)\,-\,{e^{2h}\over 4}\,\,.
\eeq
Using this result one can demonstrate, after some calculation, that eq. 
(\ref{sqrt-cI}) is also satisfied in this non-abelian case. As a consequence, the
kappa symmetry projection reduces to the one written in eq. (\ref{D1projection}), \ie\ to
that corresponding to a D1-brane.

\subsubsection{Integration of the first-order equations}

In order to integrate the first-order equations (\ref{nonabelianBPS})  for
$r(\theta_1,\theta_2)$, let us define the new variable $y(r)$ as:
\beq
y(r)\equiv \cosh(2r)\,\,.
\eeq
In terms of $y$, the BPS system (\ref{nonabelianBPS}) can be greatly simplified, namely:
\bear
\partial_{\theta_1}y\,+\,\cot\theta_1\,y&=&-\lambda\cot\theta_2\,\,,\rc
\partial_{\theta_2}y\,+\,\cot\theta_2\,y&=&-\lambda\cot\theta_1\,\,,
\eear
which can be easily integrated by the method of variation of constants. In terms of the
original variable $r$ one has:
\beq
\cosh(2r)\,=\,{\cosh(2r_*)\,+\,\lambda\cos\theta_1\cos\theta_2\over
\sin\theta_1\sin\theta_2}\,\,,
\label{nonabeliansolutions}
\eeq
where $r_*\equiv r(\theta_1=\pi/2, \theta_2=\pi/2)$ is the minimal value of $r$. This is a
remarkably simple solution for the very complicated system of kappa symmetry equations.
Notice that there are two solutions for $r(\theta_1,\theta_2)$, which correspond to the two
possible values of $\lambda$ on the right-hand side of (\ref{nonabeliansolutions}). If
$\lambda=+1$ ($\lambda=-1$) the angle $\psi$ is fixed to $\psi=0,2\pi$ ($\psi=\pi,3\pi$).
Thus, the $U(1)$ symmetry $\psi\to\psi+\epsilon$ of the abelian case is broken to $\zet_2$,
reflecting the same breaking that occurs in the geometry. 
Moreover, it follows from (\ref{nonabeliansolutions}) that $r$ diverges at 
$\theta_{1,2}=0,\pi$. It is easily proved that the embedding written in eqs. (\ref{Strnonabpsi})
and (\ref{nonabeliansolutions}) satisfies the equations of motion of the probe. 

\subsection{Energy bound for the effective string solutions}

We will now consider the configurations that we have just studied and we will show that they saturate an energy bound, as expected for BPS worldvolume solitons. Accordingly, let us choose
worldvolume coordinates as in (\ref{Stringcoordinates}) and an embedding of the type 
displayed in eq. (\ref{Stringansatz}) in the non-abelian MN background, where, for simplicity, we
will take the angle
$\psi$ to be a constant such that $\sin\psi=0$ (see eq. (\ref{Strnonabpsi})). In this case it is easy to prove that the hamiltonian density can be written as in eq. (\ref{H=Z+S}), where
for an arbitrary function $r(\theta_1,\theta_2)$, ${\cal Z}$ is a total derivative and 
${\cal S}\ge 0$. In order to verify these facts,
let us take ${\cal Z}$  to be:
\bear
&&{\cal Z}\,=\,{e^{2\phi}\over 4}\,\sin\theta_1\,\sin\theta_2\,\Big[\,
e^{2h}-(\Delta_{\theta_1}-\lambda a \Delta_{\theta_2})\partial_{\theta_1} r
\,-\,\Big(4\,(e^{2h}\,+\,{a^2\over 4}\,)\,\Delta_{\theta_2}
-\lambda a \Delta_{\theta_1}\Big)\partial_{\theta_2} r\,\Big]\,\,.\qquad\rc
\label{StrcalZ}
\eear

One can prove that ${\cal Z}$ is  a total derivative. Indeed, let us
introduce  the functions $z_1(r)$ and $z_2(r)$ as the solutions of the equations:
\bear
&&{dz_1\over dr}\,=\,\cos\alpha\,{e^{2\phi}\over 8}\,\,,\rc\
&&{dz_2\over dr}\,=\,-\,\Big[\,a\cos\alpha\,+\,2e^{h}\sin\alpha\,\Big]\,{e^{2\phi}\over 8}\,\,,
\eear
where $h$, $\phi$  and  $\alpha$, are the functions of the radial coordinate displayed in eqs. 
(\ref{MNsol}) and (\ref{alphaexplicit}). Then, one can verify that 
${\cal Z}\,=\,\partial_{\theta_1}\,{\cal Z}^{\theta_1}\,+\,
\partial_{\theta_2}\,{\cal Z}^{\theta_2}$, where 
\bear
{\cal Z}^{\theta_1}&=&-\cos\theta_1\sin\theta_2 \,z_1\,+\,
\lambda\sin\theta_1\cos\theta_2\,z_2\,\,,\rc
{\cal Z}^{\theta_2}&=&-\sin\theta_1\cos\theta_2 \Big[\,
{e^{2\phi+2h}\over 4}\,-\,z_1\,\Big]\,-\,\lambda\cos\theta_1\sin\theta_2\,z_2\,\,.
\eear
In order to prove this result the following relation:
\beq
{d\over dr}\,\Big[\,e^{2\phi+2h}\,\Big]\,=\,2re^{2\phi}\,\,,
\eeq
is quite useful. 

It is straightforward to prove that for these configurations the pullback of $C^{(6)}$ vanishes.
Therefore (see eq. (\ref{lagrangian})), the hamiltonian density in this case is just 
${\cal H}=e^{-\phi}\,\sqrt{-g}$, with $\sqrt{-g}$ given in eq. (\ref{sqrtg}). 
Once  ${\cal Z}$ is known and given by the expression written in eq. (\ref{StrcalZ}), 
${\cal S}$ is defined as  ${\cal H}-{\cal Z}$. 
One can verify that ${\cal S}\ge 0$ is equivalent to the condition
\bear
&&\Big[\,\partial_{\theta_1} r+\Delta_{\theta_1}\,-\,\lambda a
(\partial_{\theta_2} r+\Delta_{\theta_2})\,\Big]^2\,+\,4e^{2h}\,
\Big[\,\partial_{\theta_2} r+\Delta_{\theta_2} \,\Big]^2\,+\,
4\Big[\,\Delta_{\theta_2}\partial_{\theta_1}r-\Delta_{\theta_1}\partial_{\theta_2}r
\,\Big]^2\,\ge 0\,\,,\rc
\eear
which is obviously satisfied and reduces to an identity when the BPS equations
(\ref{nonabelianBPS}) hold.  It is easy to compute the central charge ${\cal Z}$ for the BPS
configurations. The result is:
\beq
{\cal Z}_{\,|_{BPS}}\,=\,{e^{2\phi}\over 4}\,\sin\theta_1\,\sin\theta_2\,\Big[\,
(\Delta_{\theta_1}-\lambda a \Delta_{\theta_2})^2\,+\,4e^{2h}\,\Delta_{\theta_2}^2
\,+\,e^{2h}\,\Big]\,\,.
\eeq
It follows from the above expression that ${\cal Z}_{\,|_{BPS}}$ is always non-negative.

\setcounter{equation}{0}
\section{More defects}   \label{moredefects}
\medskip

\subsection{Wall defects}  \label{morewalldefects}

Let us  find more supersymmetric configurations of the D5-brane probe which
behave as a codimension one defect from the gauge theory point of view. In particular, we are
interested in trying to obtain embeddings for which the angle $\psi$ is not constant. To insure
this fact we will include $\psi$ in our set of worldvolume coordinates. Actually, we will choose
the $\xi$'s as:
\beq
\xi^{\mu}\,=\,(x^0,x^1,x^2,\theta_2,\phi_2,\psi)\,\,,
\eeq
and we will adopt the following ansatz for the embedding:
\bear
&&\theta_1=\theta_1(\theta_2),\,\,
\,\,\,\,\,\,\,\,\,\,\,\,\,
\phi_1=\phi_1(\phi_2),\,\,\rc
&&x^3=x^3(\psi)\,\,,
\,\,\,\,\,\,\,\,\,\,\,\,\,
r\,=\,r(\psi)\,\,.
\label{DW2ansatz}
\eear
For these configurations the kappa symmetry matrix (\ref{GammaD5})
acts on the Killing spinors $\epsilon$ as:
\beq
\Gamma_{\kappa}\,\epsilon\,=\,{1\over \sqrt{-g}}\,
\gamma_{x^0 x^1x^2 \theta_2\phi_2\psi}\,\epsilon\,\,.
\eeq
Now the induced gamma matrices are:
\bear
&&e^{-{\phi\over 2}}\,\gamma_{x^{\mu}}\,=\,\Gamma_{x^{\mu}}\,\,,
\,\,\,\,\,\,\,\,\,\,\,\,\,\,\,\,\,\,
(\mu=0,1,2),\,\,\rc
&&e^{-{\phi\over 2}}\,\gamma_{\theta_2}\,=\,e^h\,\partial_{\theta_2}\theta_1\,
\Gamma_{1}\,+\,W_{1\theta}\,\hat\Gamma_{1}\,+\,W_{2\theta}\,\hat\Gamma_{2}\,\,,\rc
&&e^{-{\phi\over 2}}\,\gamma_{\phi_2}\,=\,e^h\,\sin\theta_1
\partial_{\phi_2}\phi_1\,
\Gamma_{2}\,+\,W_{1\phi}\,\hat\Gamma_{1}\,+\,W_{2\phi}\,\hat\Gamma_{2}\,
+\,W_{3\phi}\,\hat\Gamma_{3}\,\,,\rc
&&e^{-{\phi\over 2}}\,\gamma_{\psi}\,=\,{1\over 2}\,\hat\Gamma_{3}\,+\,
\partial_{\psi}\,r\,\Gamma_r\,+\,\partial_{\psi}\,x^3\,\Gamma_{x^3}\,\,,
\label{exDWindgamma}
\eear
where the $W$'s are the following quantities:
\bear
&&W_{1\theta}\,=\,{1\over 2}\,[\cos\psi\,+\,a\partial_{\theta_2}\theta_1\,]\,\,,\rc
&&W_{2\theta}\,=\,-{1\over 2}\,\sin\psi\,\,,\rc
&&W_{1\phi}\,=\,{1\over 2}\,\sin\theta_2\sin\psi\,\,,\rc
&&W_{2\phi}\,=\,{1\over 2}\,[\,\sin\theta_2\cos\psi\,-\,a\sin\theta_1
\partial_{\phi_2}\phi_1\,]\,\,,\rc
&&W_{3\phi}\,=\,{1\over 2}\,[\cos\theta_2\,+\,\partial_{\phi_2}\phi_1
\cos\theta_1\,]\,\,.
\eear

\subsubsection{Embeddings at $r=0$}

Let us analyze first the possibility of taking in our previous equations $r=0$ and an arbitrary 
constant value of $x^3$. Since $e^h\to 0$, $a\to 1$ and $\phi\to\phi_0$ when $r\to 0$, one has in
this case $\gamma_{\theta_2}=e^{{\phi_0\over 2}}
[W_{1\theta}\,\hat\Gamma_{1}\,+\,W_{2\theta}\,\hat\Gamma_{2}]$, 
$\gamma_{\phi_2}=e^{{\phi_0\over 2}}[
W_{1\phi}\,\hat\Gamma_{1}\,+\,W_{2\phi}\,\hat\Gamma_{2}\,
+\,W_{3\phi}\,\hat\Gamma_{3}]$ and $\gamma_{\psi}=e^{{\phi_0\over 2}}\hat\Gamma_{3}/2$ and one
immediately gets:
\beq
\gamma_{\theta_2\phi_2\psi}\,\epsilon={e^{{3\over 2}\phi_0}\over 8}\,\Big[
\sin\theta_2+(\sin\theta_2\,\partial_{\theta_2}\theta_1-\sin\theta_1
\partial_{\phi_2}\phi_1)\cos\psi-\sin\theta_1\,
\partial_{\theta_2}\theta_1\,\partial_{\phi_2}\phi_1)\,\Big]
\hat\Gamma_{123}\,\epsilon\,\,.\qquad
\eeq
On the other hand, it is easy to compute the value of the determinant of the induced metric for
an embedding of the type (\ref{DW2ansatz}) at $r=0$ and constant  $x^3$. By using this result
one readily gets the action of $\Gamma_{\kappa}$ on $\epsilon$. Indeed, 
let us define $s(\theta_2,\phi_2,\psi)$ to be the following sign:
\beq
s(\theta_2,\phi_2,\psi)\equiv {\rm sign}\,
\Big[
\sin\theta_2+(\sin\theta_2\,\partial_{\theta_2}\theta_1-\sin\theta_1
\partial_{\phi_2}\phi_1)\cos\psi-\sin\theta_1\,
\partial_{\theta_2}\theta_1\,\partial_{\phi_2}\phi_1)\,\Big]\,\,.
\eeq
Then, one has:
\beq
\Gamma_{\kappa}\epsilon_{\,\,|_{r=0}}\,=\,s\,
\Gamma_{x^0x^1x^2}\hat \Gamma_{123}\epsilon_{\,\,|_{r=0}}\,\,.
\eeq
It follows that the condition $\Gamma_{\kappa}\epsilon=\epsilon$ is equivalent to the
projection:
\beq
\Gamma_{x^0x^1x^2}\hat \Gamma_{123}\epsilon_{\,\,|_{r=0}}\,=\,s\,\epsilon_{\,\,|_{r=0}}\,\,.
\label{r=0proj}
\eeq
Notice that  the right-hand side of (\ref{r=0proj}) only depends on the angular part of the
embedding through the sign $s$.  Let us rewrite eq. (\ref{r=0proj}) in  terms of the spinor
$\epsilon_0$ defined in  eq. (\ref{epsilon0}). First of all, let us introduce the matrix
$\hat\Gamma_{*}$ as:
\beq
\hat\Gamma_{*}\,=\,\Gamma_{x^0x^1x^2}\Gamma_1\hat \Gamma_{23}\,\,.
\eeq
Recall from (\ref{epsilon0}) that
$\epsilon\,=\,e^{{\alpha \over 2}\,\Gamma_1\hat\Gamma_1}\,\,\epsilon_0$. As 
$\alpha(r=0)=-\pi/2$, see eqs. (\ref{alpha}) and (\ref{alphaexplicit}),  the above condition
reduces to:
\beq
\hat\Gamma_{*}\,\epsilon_0\,=\,s\,\epsilon_0\,\,.
\label{originproj}
\eeq
It is easy to verify that this condition commutes with the projections satisfied by  
$\epsilon_0$, which are the same as those satisfied by the constant spinor $\eta$ (see eq.
(\ref{constantfullpro})). Moreover, it is readily checked that these configurations satisfy the
equations of motion of the probe. Notice that the angular embedding is undetermined. However, 
the above projection only makes sense if
$s(\theta_2,\phi_2,\psi)$ does not depend on the angles. Although the angular embedding is not
uniquely determined, there are some embeddings that can be discarded. For example if we take 
$\theta_1=\theta_2$, $\phi_1=\phi_2$ the corresponding three-cycle has vanishing volume and
$s$ is not well defined. For $\theta_1=$ constant, $\phi_1=$ constant one has $s=1$. The same
value of $s$ is obtained if $\theta_1=\pi-\theta_2$, $\phi_1=\phi_2$ or when
$\theta_1=\theta_2$, $\phi_1=2\pi -\phi_2$.
Notice that this configuration consists of a D5-brane, which is finite in the 
internal directions, wrapping the finite $S^3$ inside the geometry, which 
has minimal volume at $r=0$. This object is thought to correspond to a 
domain wall of the field theory \cite{Maldacena:2000yy, Loewy}.
However,  the physics of domain walls is yet not fully understood in this 
model.

\subsubsection{General case}
Let us now come back to the general case. By using the relation between the spinors $\epsilon$
and $\epsilon_0$, the kappa symmetry equation  
$\Gamma_{\kappa}\,\epsilon\,=\,\epsilon$
can be rephrased as the following condition  on the spinor $\epsilon_0$:
\beq
e^{-{\alpha \over 2}\,\Gamma_1\hat\Gamma_1}\Gamma_{\kappa}
e^{{\alpha \over 2}\,\Gamma_1\hat\Gamma_1}\,\,\epsilon_0\,=\,\epsilon_0\,\,.
\label{DW2kappa-epsilon0}
\eeq
Let us evaluate the left-hand side of this equation by imposing the projection 
(\ref{originproj}), \ie\ the same projection as the one satisfied by the supersymmetric
configurations at $r=0$. After some calculation one gets an expression of the type:
\bear
&&e^{-{\alpha \over 2}\,\Gamma_1\hat\Gamma_1}\Gamma_{\kappa}
e^{{\alpha \over 2}\,\Gamma_1\hat\Gamma_1}\,\,\epsilon_0\,=\,
{e^{3\phi}\over \sqrt{-g}}\,\,\Big[\,
d_I\,+\,d_{1 \hat 1}\,\Gamma_1\hat\Gamma_1\,+\,
d_{\hat 1 \hat 2}\,\hat\Gamma_1\hat\Gamma_2\,+\,
d_{ 1 \hat 2}\,\Gamma_1\hat\Gamma_2\,+\rc
&&\qquad\qquad \qquad\qquad+
d_{\hat 2 \hat 3}\,\hat\Gamma_2\hat\Gamma_3\,+\,
d_{ 2 \hat 3}\,\Gamma_2\hat\Gamma_3\,+\,
d_{\hat 1 \hat 3}\,\hat\Gamma_1\hat\Gamma_3\,+\,
d_{ 1 \hat 3}\,\Gamma_1\hat\Gamma_3\,\Big]\,\epsilon_0\,\,,
\label{Gamma-dIs}
\eear
where the $d$'s depend on the embedding (see below). Clearly, in order to satisfy
eq. (\ref{DW2kappa-epsilon0})
we must require the conditions:
\beq
d_{1 \hat 1}\,=\,d_{\hat 1 \hat 2}\,=\,d_{ 1 \hat 2}\,=\,
d_{\hat 2 \hat 3}\,=\,d_{ 2 \hat 3}\,=\,d_{\hat 1 \hat 3}\,=\,
d_{ 1 \hat 3}\,=\,0\,\,.
\eeq
The expressions of the $d$'s are quite involved. In order to write them in a compact form
let us define the quantities ${\cal P}_1$, ${\cal P}_2$ and ${\cal P}_3$ as:
\bear
&&{\cal P}_1\,\equiv\,W_{1\theta}\,W_{2\phi}\,-\,W_{1\phi}\,W_{2\theta}\,+\,
e^{2h}\,\sin\theta_1\,\partial_{\theta_2}\theta_1\,\partial_{\phi_2}\phi_1\,\,,\rc
&&{\cal P}_2\,\equiv\,e^h\,\Big(\,W_{2\phi}\,\partial_{\theta_2}\theta_1\,-\,
W_{1\theta}\,\sin\theta_1\,\partial_{\phi_2}\phi_1\,\Big)\,\,,\rc
&&{\cal P}_3\,\equiv\,e^h\,\Big(\,W_{1\phi}\,\partial_{\theta_2}\theta_1\,+\,
W_{2\theta}\,\sin\theta_1\,\partial_{\phi_2}\phi_1\,\Big)\,\,.
\label{Ps}
\eear
Then the coefficients of the terms that do not contain the matrix $\hat\Gamma_3$ are:
\bear
&&d_I\,=\,{s\over 2}\,\,\Big[\,{\cal P}_2\cos\alpha\,-\,{\cal P}_1\sin\alpha\,\Big]\,+\,
\Big[\,{\cal P}_1\cos\alpha\,+\,{\cal P}_2\sin\alpha\,\Big]\,\partial_{\psi}x^3\,+\,
s\,{\cal P}_3\,\partial_{\psi}r\,,\rc
&&d_{1 \hat 1}\,=\,-{s\over 2}\,\,
\Big[\,{\cal P}_1\cos\alpha\,+\,{\cal P}_2\sin\alpha\,\Big]\,+\,
\Big[\,{\cal P}_2\cos\alpha\,-\,{\cal P}_1\sin\alpha\,\Big]\,\partial_{\psi}x^3\,,\rc
&&d_{\hat 1 \hat 2}\,=\,{s\over 2}\,{\cal P}_3\,-\,s
\Big[\,{\cal P}_2\cos\alpha\,-\,{\cal P}_1\sin\alpha\,\Big]\,\partial_{\psi}r\,,\rc
&&d_{ 1 \hat 2}\,=\,s\,
\Big[\,{\cal P}_1\cos\alpha\,+\,{\cal P}_2\sin\alpha\,\Big]\,\partial_{\psi}r\,-\,
{\cal P}_3\,\partial_{\psi}x^3\,\,.
\label{dIs}
\eear
From the conditions $d_{1 \hat 1}\,=\,d_{\hat 1 \hat 2}\,=\,0$ we get the BPS equations
that determine $\partial_{\psi}x^3$ and $\partial_{\psi}r$, namely:
\beq
\partial_{\psi}x^3\,=\,{s\over 2}\,{
{\cal P}_1\cos\alpha\,+\,{\cal P}_2\sin\alpha\over 
{\cal P}_2\cos\alpha\,-\,{\cal P}_1\sin\alpha}\,\,,
\qquad\qquad
\partial_{\psi}r\,=\,{1\over 2}\,{{\cal P}_3\over 
{\cal P}_2\cos\alpha\,-\,{\cal P}_1\sin\alpha}\,,
\label{BPSpsi}\,\,
\eeq
while the equation $d_{ 1 \hat 2}=0$ is satisfied if the  differential equations  (\ref{BPSpsi}) 
hold.

The expressions of the coefficients of the terms with the matrix $\hat \Gamma_{3}$ are:
\bear
d_{\hat 2 \hat 3}&=&-W_{3\phi}\,\Big[\,W_{1\theta}\cos\alpha\,+\,
e^{h}\,\partial_{\theta_2}\theta_1\sin\alpha\,\Big]\,\partial_{\psi}x^3\,\,,\rc
d_{ 2 \hat 3}&=&W_{3\phi}\,\Big[\,\Big(\,
W_{1\theta}\,\sin\alpha\,-\,e^{h}\,\partial_{\theta_2}\theta_1\cos\alpha\,\Big)
\partial_{\psi}x^3\,-\,s\,W_{2\theta}\,\partial_{\psi}r\,\Big]\,\,,\rc
d_{ \hat 1 \hat 3}&=&W_{3\phi}\,\Big[\,W_{2\theta}\,\partial_{\psi}x^3\,-\,s
\Big(\,e^{h}\,\partial_{\theta_2}\theta_1\cos\alpha\,-\,W_{1\theta}\,\sin\alpha
\,\Big)\,\partial_{\psi}r\,\Big]\,\,,\rc
d_{1 \hat 3}&=&sW_{3\phi}\,\Big[\,
W_{1\theta}\cos\alpha\,+\, e^{h}\,\partial_{\theta_2}\theta_1\sin\alpha\,\Big]\,
\partial_{\psi}r\,\,.
\label{3hat}
\eear

Let us impose now the vanishing of the coefficients (\ref{3hat}). Clearly, this condition
can be achieved by requiring that $r$ and $x^3$ be constant. It is easy to see from
the vanishing of the  right-hand side of eq. (\ref{BPSpsi})  that this only happens at 
$r=0$ and, therefore, the configuration reduces to the one studied above. Another
possibility is to impose  $W_{3\phi}=0$, which is equivalent to the following differential
equation:
\beq
-{\cos\theta_2\over \cos\theta_1}\,=\,\partial_{\phi_2}\phi_1\,\,.
\eeq
For consistency, both sides of the equation must be equal to a constant which we will denote
by $m$:
\beq
\partial_{\phi_2}\phi_1\,=\,m\,\,,
\qquad\qquad\qquad
\cos\theta_1\,=\,-{\cos\theta_2\over m}\,\,.
\label{separation}
\eeq
Moreover, 
by differentiating the above relation between $\theta_1$ and $\theta_2$, we immediately obtain:
\beq
\partial_{\theta_2}\theta_1\,=\,-{\sin\theta_2\over m
\sin\theta_1}\,=\,-{\rm sign}(m)\,
{\sin\theta_2\over \sqrt{\sin^2\theta_2\,+\,m^2-1}}\,\,.
\label{partialtheta}
\eeq
Moreover, by using eqs. (\ref{separation}) and (\ref{partialtheta}) one can easily find the
following expression of the ${\cal P}$'s:
\bear
{{\cal P}_1\over \sin\theta_2}&=&{1\over 4}\,
\Big[\,1\,+\, a\,\Big(\,\partial_{\theta_2}\theta_1\,+\,
{1\over \partial_{\theta_2}\theta_1}\,\Big)\,\cos\psi\,+\,a^2\,\Big]\,-\,e^{2h}\,\,,\rc
e^{-h}\,{{\cal P}_2\over \sin\theta_2}&=&{1\over 2}\,
\Big(\,\partial_{\theta_2}\theta_1\,+\,
{1\over \partial_{\theta_2}\theta_1}\,\Big)\,\cos\psi\,+\,a\,\,,\rc
e^{-h}\,{{\cal P}_3\over \sin\theta_2}&=&{1\over 2}\,
\Big(\,\partial_{\theta_2}\theta_1\,+\,
{1\over \partial_{\theta_2}\theta_1}\,\Big)\,\sin\psi
\label{BPSPs}\,\,.
\eear

For consistency with our ansatz, the right-hand side of the equation for 
$\partial_{\psi}r$ in (\ref{BPSpsi}) must necessarily be independent of $\theta_2$. By inspecting
the right-hand side of (\ref{BPSPs}) it is evident that this only happens if 
$\partial_{\theta_2}\theta_1$ is constant which, in view of eq. (\ref{partialtheta}) can
only occur if $m^2=1$, \ie\ when $m=\pm 1$. In this case $\partial_{\theta_2}\theta_1=-m$
and the angular embedding is:  
\bear
&&\theta_1=\pi-\theta_2\,\,,
\,\,\,\,\,\,\,\,\,\,\,\,\,\,\,\,\,\,\,\,
\phi_1=\phi_2\,\,,
\,\,\,\,\,\,\,\,\,\,\,\,\,\,\,\,\,\,\,\,
(m=+1)\,\,,\rc
&&\theta_1=\theta_2\,\,,
\,\,\,\,\,\,\,\,\,\,\,\,\,\,\,\,\,\,\,\,
\phi_1=2\pi-\phi_2\,\,,
\,\,\,\,\,\,\,\,\,\,\,\,\,\,\,\,\,\,\,\,
(m=-1)\,\,.
\label{angularemb}
\eear
Notice that the functions in  (\ref{angularemb}) are just the same as those corresponding to the
embeddings with constant $\psi$ (eq. (\ref{DWangular})). Moreover, taking
$\partial_{\theta_2}\theta_1=-m$ in the expression of the ${\cal P}_i$'s in eq. (\ref{BPSPs}),
and substituting this result on the right-hand side of eq. (\ref{BPSpsi}), one finds the
following BPS differential equations for $r(\psi)$ and $x^3(\psi)$:

\bear
&&\partial_{\psi}r\,=\,
{1\over 2}\,{\sinh (2r)\sin\psi\over \cosh(2r)\cos\psi\,-\,m}\,\,,\rc
&&\partial_{\psi}x^3\,=\,{sm\over 2}\,\,e^{-h}\,\,
\Big(\,r\coth(2r)\,-\,{1\over 2}\,\Big)\,\,
{\cosh(2r)\,-\,m\cos\psi
\over \cosh(2r)\cos\psi\,-\,m}\,\,.
\label{rx3psi}
\eear
Lets us now verify that the BPS equations written  above are enough to guarantee that 
(\ref{DW2kappa-epsilon0}) holds. With this purpose in mind, let us compute the only
non-vanishing term of the right-hand side of eq. (\ref{Gamma-dIs}), namely $d_I$.
By plugging the BPS equations (\ref{BPSpsi}) into the expression of $d_I$ in eq. (\ref{dIs}), one
gets:
\beq
{d_I}_{\,\,|BPS}\,=\,s\,\,{{\cal P}_1^2\,+\,{\cal P}_2^2\,+\,{\cal P}_3^2\over
{\cal P}_2\cos\alpha\,-\,{\cal P}_1\sin\alpha}\,\,.
\label{dIBPS}
\eeq
From  eq. (\ref{dIBPS}) one can check that:
\beq
\sqrt{-g}_{\,\,|BPS}\,=\,e^{3\phi}\,\Big|\,{d_I}_{\,\,|BPS}\,\Big|\,\,.
\eeq
In order to verify that the kappa symmetry condition (\ref{DW2kappa-epsilon0}) is satisfied we
must check that the sign of  ${d_I}_{\,\,|BPS}$ is positive. It can be verified that:
\beq
{\rm sign} \Big[\,{\cal P}_2\cos\alpha\,-\,
{\cal P}_1\sin\alpha \,\Big]_{\,\,|BPS}\,=\,
-m\,{\rm sign} (\cos\psi)\,\,,
\eeq
and  therefore (see eq. (\ref{dIBPS})), the condition ${\rm sign}\,({d_I}_{\,\,|BPS})=+1$ holds
if the sign $s$ of the projection (\ref{originproj}) is such that:
\beq
s\,=\,-m\,\, {\rm sign}(\cos\psi)\,\,.
\label{signrelation}
\eeq
Then, given an angular embedding (\ie\ for a fixed value of $m$), we must restrict $\psi$ to a
range in which  the sign of $\cos\psi$
does not change and the sign $s$ of the projection must be chosen according to 
(\ref{signrelation}). Moreover,  one can show that the equations of
motion are satisfied if the first-order equations (\ref{rx3psi}) hold.

\subsubsection{Integration of the BPS equations}
After a short calculation one can demonstrate that
the equation for $r(\psi)$ in (\ref{rx3psi}) can be rewritten as:
\beq
\partial_{\psi}\Big[\,\cos\psi \sinh(2r)\,-\,2mr\,\Big]\,=\,0\,\,.
\eeq
In this form the BPS equation for $r(\psi)$ can be immediately integrated, namely:
\beq
\cos\psi\,=\,{C+2mr\over \sinh(2r)}\,\,,
\label{anyC}
\eeq
where $C$ is a constant. Moreover, once the function $r(\psi)$ is known, one can get $x^3(\psi)$
by direct integration of the right-hand side of the second equation in 
(\ref{rx3psi}).

Let us study the above solution for different signs of $\cos\psi$. Consider first the
region in which $\cos\psi\ge 0$, which corresponds to 
$\psi\in[-\pi/2, \pi/2]\,\,\mod\,\, 2\pi$. If the constant $C> 0$, let us represent it
in terms of a new constant $r_*$ as $C=\sinh(2r_*)-2mr_*$. Then, the above solution can be
written as:
\begin{figure}
\centerline{\epsffile{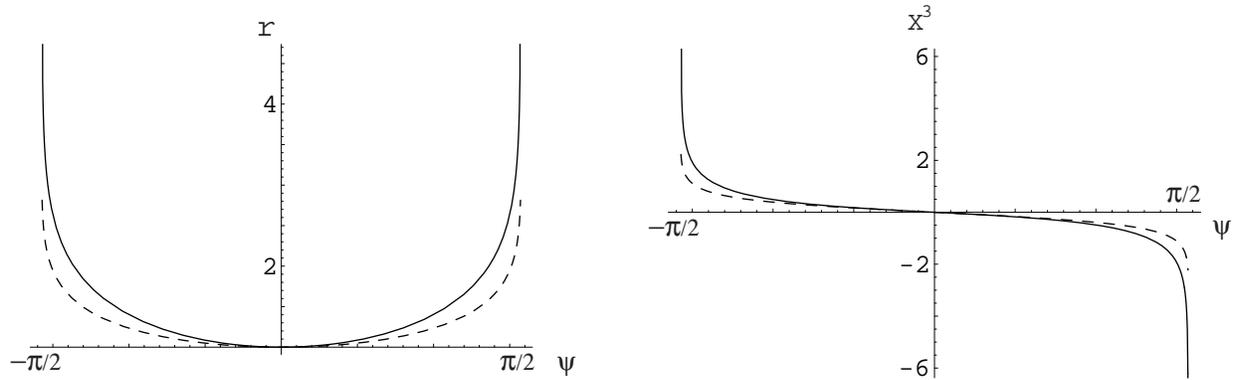}}
\caption{The functions $r(\psi)$ and $x^3(\psi)$ for the solutions (\ref{posC}) in the interval 
$\psi\in[-\pi/2, \pi/2]\,\,\mod\,\, 2\pi$. The continuous line represents the embedding with
$m=+1$, while the dashed line corresponds to $m=-1$. In this latter case $r(\psi)$ and
$x^3(\psi)$ remain finite, while for $m=+1$ they diverge at $\psi=\pm \pi/2$. }
\label{fig2}
\end{figure}
\beq
\cos\psi\,=\,{\sinh(2r_*)\,+\,2m(r-r_*)\over \sinh(2r)}\,\,,
\label{posC}
\eeq
from which it is clear that $r_*$ is the value of $r$ such that $\cos\psi=1$. The functions
$r(\psi)$ for $m=\pm 1$ written in eq. (\ref{posC}), 
and the corresponding $x^3(\psi)$, have been plotted in Fig. \ref{fig2}. 
If $m=+1$, 
the solution (\ref{posC}) is such that $r\to\infty$  and $|x^3|\to \infty$ when 
$\psi\to \pm \pi/2\,\,\mod\,\,2\pi$. However, if  $m=-1$ the radial coordinate $r$ grows
from its minimal value $r_*$ at $\psi=0\,\,\mod\,\,2\pi$ to a  maximal value 
$\hat r\,=\, r_*\,+\,{\sinh(2r_*)\over 2}$ at $\psi=\pm \pi/2\,\,\mod\,\,2\pi$, while 
$x^3(\psi)$ remains finite when $\psi\in[-\pi/2, \pi/2]\,\,\mod\,\, 2\pi$.

 If $C<0$,
it is clear from (\ref{anyC}) that, as we are considering the region  $\cos\psi\ge 0$, only
the solution with
$m=+1$ is possible. Defining $2\tilde r=-C$, the solution in this case can be written as
\beq
\cos\psi\,=\,{2(r-\tilde r)\over \sinh(2r)}\,\,,\qquad\qquad (m=+1). 
\eeq
This solution has two branches such that $r\to \tilde r, \infty$ when $\psi\to\pm\pi/2$.
Finally, if $C=0$ only the $m=+1$ solution makes sense. In this case the solution grows
from $r=0$ at $\psi=0$ to $r=\infty$ at $\psi=\pm\pi/2$.

In the region $\cos\psi\le 0$, \ie\ for  $\psi\in[\pi/2, 3\pi/2]\,\,\mod\,\, 2\pi$,
the solutions can be found from those for
$\cos\psi\ge 0$ by means of the following symmetry of the solution (\ref{anyC}):
\beq
\psi\to \pi-\psi\,\,,
\qquad\qquad
C\to-C\,\,,
\qquad\qquad
m\to-m\,\,.
\label{trasnfor}
\eeq

Then, one can get solutions in the range $\psi\in [0,2\pi]$ by joining one solution in the 
region $\cos\psi\ge 0$ to the one obtained by means of the  transformation 
(\ref{trasnfor}). Notice that the resulting solutions preserve supersymmetry at the cost
of changing the angular embedding, \ie\ by making $m\to -m$,  when the sign of $\cos\psi$
changes. In particular, in the solution obtained from the one in (\ref{posC}) when $m=-1$
the coordinate $r$ does not diverge. One can apply this construction to a single brane probe
with a singular embedding or, alternatively, one can consider two different brane probes
preserving the same supersymmetry with different angular embeddings and lying on disjoint regions
of $\psi$.

\subsection{Two-dimensional defects}  \label{morestringdefects}

In analogy with what we have just done with the wall defect solitons, let us find some
codimension two embeddings of the D5-brane probe in which the angle $\psi$ is not constant. We
shall take the following set of worldvolume coordinates:
\beq
\xi^{\mu}\,=\,(x^0,x^1,r,\theta_2,\phi_2,\psi)\,\,,
\eeq
and we will adopt an  ansatz in which $x^2$ and $x^3$ are constant and
\beq
\theta_1=\theta_1(\theta_2),\,\,
\,\,\,\,\,\,\,\,\,\,\,\,\,
\phi_1=\phi_1(\phi_2). \,\,\rc
\eeq
The induced gamma matrices $\gamma_{x^\mu}$ $(\mu=0,1)$, $\gamma_{\theta_2}$ and 
$\gamma_{\phi_2}$ are exactly those written in eq. (\ref{exDWindgamma}), while $\gamma_{r}$ and
$\gamma_{\psi}$ are given by:
\beq
e^{-{\phi\over 2}}\,\gamma_{r}\,=\,\Gamma_r\,\,,\qquad
e^{-{\phi\over 2}}\,\gamma_{\psi}\,=\,{1\over 2}\,\hat\Gamma_{3}\,\,.
\eeq
Let us try to implement the kappa symmetry condition in the form displayed in eq.
(\ref{DW2kappa-epsilon0}). For this case, the left-hand side of (\ref{DW2kappa-epsilon0}) can be
written as:
\beq
e^{-{\alpha \over 2}\,\Gamma_1\hat\Gamma_1}\Gamma_{\kappa}
e^{{\alpha \over 2}\,\Gamma_1\hat\Gamma_1}\,\,\epsilon_0\,=\,
{e^{3\phi}\over 2\sqrt{-g}}\,\,\Gamma_{x^0 x^1}\,\Big[\,
f_I\,+\,f_{1\hat 1}\,\Gamma_1\hat\Gamma_{1}\,+\,
f_{1\hat 2}\,\Gamma_1\hat\Gamma_{2}\,\Big]\,\epsilon_0\,\,,
\eeq
where the $f$'s are expressed in terms of the ${\cal P}_i$ functions of (\ref{Ps}) as:
\bear
&&f_I\,=\,\cos\alpha\,{\cal P}_1\,+\,\sin\alpha\,{\cal P}_2\,\,,\rc
&&f_{1\hat 1}\,=\,-\sin\alpha\,{\cal P}_1\,+\,\cos\alpha\,{\cal P}_2\,\,,\rc
&&f_{1\hat 2}\,=\,-{\cal P}_3\,\,.
\eear
Since the matrices $\Gamma_1\hat\Gamma_1$ and $\Gamma_1\hat\Gamma_2$ do not commute with the
projection (\ref{proj-epsilon0}), it is clear that we must impose:
\beq
f_{1\hat 1}\,=\,f_{1\hat 2}\,=\,0\,\,.
\eeq
From the condition $f_{1\hat 1}\,=\,0$, we get:
\beq
{{\cal P}_2\over {\cal P}_1}\,=\,\tan\alpha\,\,,
\label{p2p1bps}
\eeq
while $f_{1\hat 2}\,=\,0$ is equivalent to the vanishing of ${\cal P}_3$, which implies:
\beq
\sin\theta_2\,\partial_{\theta_2}\theta_1\,=\,\sin\theta_1\partial_{\phi_2}\phi_1\,\,.
\label{exstringangular}
\eeq
By using this condition for the angular part of the embedding, we can write the ratio between
the functions ${\cal P}_1$  and ${\cal P}_2$ as:
\beq
{{\cal P}_2\over {\cal P}_1}\,=\,
{r\sin\alpha\,\big(\partial_{\theta_2}\theta_1\big)^2\over
r\cos\alpha\,+\,\Big(e^{2h}\,-\,{a^2\over 4}\,\Big)
\Big(\big(\partial_{\theta_2}\theta_1\big)^2-1\Big)}\,\,.
\label{p2p1bps2}
\eeq
The consistency between the expressions (\ref{p2p1bps}) and (\ref{p2p1bps2}) requires that  
$\partial_{\theta_2}\theta_1\,=\,\pm 1$. Moreover, by separating variables in the angular
embedding equation (\ref{exstringangular}) one concludes  that 
$\partial_{\phi_2}\phi_1=m$ , with  $m$ constant. Proceeding as in the previous subsection, one
easily verifies that the only consistent solutions of (\ref{exstringangular}) with 
$\partial_{\theta_2}\theta_1$ constant are:
\bear
&&\theta_1=\theta_2\,\,,
\,\,\,\,\,\,\,\,\,\,\,\,\,\,\,\,\,\,\,\,
\,\,\,\,\,\,\,\,\,\,
\phi_1=\phi_2\,\,,
\,\,\,\,\,\,\,\,\,\,\,\,\,\,\,\,\,\,\,\,
\,\,\,\,\,\,\,\,\,\,\,\,\,
(m=+1)\,\,,\rc
&&\theta_1=\pi-\theta_2\,\,,
\,\,\,\,\,\,\,\,\,\,\,\,\,\,\,\,\,\,\,\,
\phi_2=2\pi-\phi_1\,\,,
\,\,\,\,\,\,\,\,\,\,\,\,\,\,\,\,\,\,\,\,
(m=-1)\,\,.
\label{Strangularemb}
\eear
Notice the difference between (\ref{Strangularemb}) and (\ref{angularemb}). One can verify that
this embedding is a solution of the equations of motion of the probe. Moreover, by computing
$\sqrt{-g}$ and $f_I$ for the embeddings (\ref{Strangularemb}), one readily proves that the
kappa symmetry condition is equivalent to the following projection on $\epsilon_0$:
\beq
\Gamma_{x^0 x^1}\,\epsilon_0\,=\,\epsilon_0\,\,.
\eeq

\setcounter{equation}{0}
\section{Summary and Discussion}  \label{remarksMN}
\medskip

In this chapter we have systematically studied the possibility of adding supersymmetric
configurations of D5-brane probes in the MN background in such a way that they create a
codimension one or two defect in the gauge theory directions. The technique consists of using kappa symmetry to look for a system of first-order equations which guarantee that the supersymmetry
preserved by the worldvolume of the probe is consistent with that of the background. Although the general system of equations obtained from kappa symmetry is very involved, the solutions we have found are remarkably simple. For a given election of worldvolume coordinates and a given ansatz for the embedding, chosen for their simplicity and physical significance, the result is unique. 
 
In order to extract consequences of our results in the gauge theory dual, some additional work
must be done. First of all, one can study the fluctuations of the probes around the
configurations found here and one can try to obtain the dictionary between these fluctuations
and the corresponding operators in the field theory side, along the lines of refs. \cite{DeWolfe:2001pq, CEGK}. In the analysis of these fluctuations we will presumably find the difficulties associated
with the UV blowup of the dilaton, which could be overcome by using the methods employed in 
ref. \cite{Nunez:2003cf} in the case of flavour branes. Once this
fluctuation-operator dictionary is obtained we could try to give some meaning to the functions
$x^3(r)$ and $r(\theta_1, \theta_2)$ of eqs. (\ref{DWnonabeBPS}) and (\ref{nonabeliansolutions})
respectively, which should encode some renormalization group flow of the defect theory. This analysis could shed light on the exact nature of the deformation introduced by the defect. Another possible way of getting information of this subject could be trying to go beyond the probe approximation and to study how the defect modifies the geometry. From the behaviour of this backreacted geometry one could possibly learn about the type of deformation that we have on the field theory side.

Interestingly, our non-abelian solitons of sections \ref{Walldefects} and \ref{twodimdefect} select certain values of the R-symmetry coordinate $\psi$ (see eqs. (\ref{DWpsi_0}) and (\ref{Strnonabpsi})). This seems to suggest that the mechanism of spontaneous breaking of R-symmetry (explaining in subsection \ref{MNbackground}) is also acting on our defects, perhaps by forming a condensate of the fields living on the defect. Notice that we have also found in section \ref{moredefects} other defect solitons in which $\psi$ is not constant. Although the interpretation is less clear, the defects of subsection \ref{morewalldefects} at $r=0$ might correspond to domain walls which interpolate between different vacua of the field theory dual. Notice that the tension of these domain walls remains finite for some particular choice of the embedding (see eq. (\ref{posC}) with $m=-1$ and $C>0$), as it should be for an object of that nature in field theory. 

Let us also point out that one could explore with the same techniques employed here some other
supergravity backgrounds (such as the one obtained in \cite{MaldaNas}, which are dual to ${\cal
N}=1$, $d=3$ super Yang-Mills theory) and try  to find the configurations of probes which
introduce supersymmetric defects in the field theory. It is also worth  mentioning that, although
we have focussed here on the analysis of the supersymmetric objects in the MN background, we
could have stable non-supersymmetric configurations, such as the confining strings of ref.
\cite{Herzog:2001fq}, which are constructed from D3-branes wrapping a two-sphere. Another
example of an interesting non-supersymmetric configuration is the baryon vertex, which consists
of a D3-brane wrapped on a three-cycle which captures the RR flux \cite{Hartnoll:2004yr}.


\chapter{Final Conclusions}
\label{conclusions}
\medskip
\setcounter{equation}{0}

In this last chapter we will briefly summarise the main achievements of this Ph.D. thesis. At the end of each chapter we have already discussed the partial results obtained and some possible extensions to the research performed in that chapter. However we would like to finish with an overview of the work presented in this thesis.

We have concentrated on the study of some aspects of the (extensions of the) AdS/CFT correspondence to more realistic theories \cite{Aharony:2002up, Bertolini:2003iv}. Let us recall that the AdS/CFT correspondence states that type IIB string theory on $AdS_5 \times S^5$ is dual to four-dimensional ${\cal N}=4$ superconformal Yang-Mills theory with $SU(N_c)$ gauge group living at the boundary of $AdS_5$ \cite{Maldacena:1997re}. This is a holographic duality in the sense that the boundary of the $AdS_5$ space where the gauge theory lives encodes all the bulk information \cite{hologram}.  By extensions of the above conjecture to ``more realistic theories" we mean extensions to less supersymmetric theories. In particular we have payed attention to supersymmetric solutions of type IIB supergravity which are dual to ${\cal N}=1$ supersymmetric gauge theories in four dimensions. These theories are more interesting from a phenomenological point of view when the conformal invariance is broken. They present some features analogous to QCD, such as for instance confinement.

We have searched for the possibility of adding supersymmetric D-branes in those supergravity backgrounds and we have analysed which nontrivial information of the gauge theory dual we are capturing with these additional degrees of freedom. It is worth pointing out that the addition of extra D-branes to a supergravity background can have two different goals. On one side, as it was firstly proposed by Witten \cite{ba0}, the dual of a field theory cannot be simple supergravity but it must contain extra D-branes. These D-branes (wrapped on nontrivial cycles) correspond to solitonic-like states in the large $N_c$ gauge theory dual. On the other side, the addition of D-branes extended infinitely in the holographic direction modifies the lagrangian of the field theory since we are adding degrees of freedom to the boundary of the $AdS$ space \cite{MAGOO}. This modification could be due to the addition of a new operator to the lagrangian or could be interpreted as though this new operator takes a VEV. The nature of the new operator which enters into the game depends on the kind of D-brane that we are adding and on the way that we add it.

In this Ph.D. thesis we have explored the two goals of the addition of D-branes to supergravity backgrounds dual to ${\cal N}=1$ supersymmetric gauge theories in four dimensions. The main tool that we have used to introduce the extra D-branes is a local fermionic symmetry of the worldvolume theory on the branes called kappa symmetry. By looking for configurations of D-branes which preserve this local symmetry, we have explicitly determined their embedding in the supergravity background and we have read the fraction of supersymmetry that they preserve. The system of first-order BPS equations that the kappa symmetry condition gives rise fulfils the second-order equations of motion for the worldvolume bosonic fields. Actually, this system saturates a bound for the energy, as it usually happens in the case of worldvolume solitons.

Once we know how to include in a supersymmetric way the D-branes in a given supergravity background, we should extract the information about the stringy spectrum that these additional degrees of freedom source. The final goal would be to introduce extra D-branes in a supergravity background and to take into account, not only the effects that the D-branes feel coming from the background fields, but also the backreaction undergone by the supergravity background due to the presence of these extra D-branes. However this is a very involved problem from a technical point of view and it is still not clear how to solve it for a D-brane of arbitrary dimension. Thus, the first thing that one may do is to tackle this problem in the probe approximation and to discard the backreaction undergone by the supergravity background. We have seen in chapters \ref{Ypq}, \ref{Labc} and \ref{MN} that, even in the probe approximation, we can still capture nontrivial information of the gauge theory dual. 

In chapters \ref{Ypq} and \ref{Labc} we have performed a systematic analysis of the possible supersymmetric D-brane configurations in backgrounds dual to ${\cal N}=1$ four-dimensional superconformal field theories. We have focused our attention on the recently found extensions to the Klebanov-Witten model, firstly on the $AdS_5 \times Y^{p,q}$ solution of type IIB supergravity in chapter \ref{Ypq} and then on a further generalisation of it, namely $AdS_5 \times L^{a,b,c}$ in chapter \ref{Labc}.  In both cases we have also been able to identify the configurations of D3-branes wrapping a three-cycle dual to the dibaryonic operators of the gauge theory dual. The study of the BPS fluctuations of these D3-brane configurations (dibaryons) was performed in the $AdS_5 \times Y^{p,q}$ case and it was shown to match the gauge theory results. In both chapters we have found configurations of D3-branes wrapping a two-cycle which could describe a ``fat" string in the gauge theory dual, configurations of D5-branes wrapping a two-cycle which could be suitable to couple a defect conformal field theory and (non-supersymmetric) D5-branes which look like domain walls (if they wrap a three-cycle) or a baryon vertex (if they wrap a five-cycle) in the field theory dual. We have also studied in both chapters configurations of spacetime filling D7-branes, which could be used to add flavour to the gauge theory, as well as configurations of D7-branes wrapping a five-cycle which are dual to one-dimensional defects in field theory.

A systematic analysis of the possible supersymmetric D5-brane configurations dual to defects (of codimension one and two) in four-dimensional ${\cal N}=1$ Super Yang-Mills theory was also carried out in the framework of the Maldacena-N\'u\~nez solution in chapter \ref{MN}. We also found the configuration of a D5-brane wrapping a three-cycle, which should be dual to a domain wall.

In order to extract more consequences of our results of chapters \ref{Ypq}, \ref{Labc} and \ref{MN} in the gauge theory dual, some additional work must be done. First of all, one could study the fluctuation of the brane probes around the configurations found there and one could try to obtain the dictionary between these fluctuations and the corresponding operators in the field theory side, as we did in subsection \ref{BPSfluctuationdibaryons}.

Another way of getting information would be to go beyond the probe approximation and to study how the geometry is modified by the D-branes included. This analysis was performed in chapters \ref{KW} and \ref{KS}, where the extra D-branes added to the geometry were suitable to account flavour degrees of freedom. Recall that the final goal in the study of realistic extensions of the AdS/CFT correspondence would be to attain the best possible dual description of theories similar to QCD, like its supersymmetric extension. Understanding fields on the fundamental representation is essential for this purpose. The construction and analysis of duals to gauge theories with flavours in the so-called Veneziano limit \cite{Veneziano:1976wm}, {\it i.e.} $N_c\to \infty$ with $\frac{N_f}{N_c}$ fixed, where $N_f$ is the number of flavours and $N_c$ is the number of colours, hence becomes of capital importance. The interesting fact about this limit (unlike the 't Hooft limit \cite{'tHooft:1973jz} where $N_f$ is kept fixed) is that the quantum effects associated to the existence of
fundamental quarks are not {\it quenched, i.e.} they are not 
suppressed by the large $N_c$ limit. In chapters \ref{KW} and \ref{KS}, which addressed respectively
${\cal N}=1$ superconformal field theories and their extension to cascading theories in four dimensions (in particular, to the Klebanov-Strassler model), techniques were developed in order to take into account the backreaction in the geometry of the $N_f$ fundamentals. In both cases, several gauge theory features were matched, like $\beta$-functions and anomalies. Moreover, in chapter \ref{KS} we managed to provide an analysis of the duality cascade which describes the RG flow of the field theory. The approach that we considered consists of finding solutions of supergravity coupled to D-brane sources.

In the framework of chapters \ref{KW} and \ref{KS}, it would be interesting to study the fluctuations of the flavour branes since they can be identified with the dynamical mesons of the gauge theory dual. Another stimulating problem in the same context would be to find a black hole in the backreacted geometry of chapters \ref{KW} and \ref{KS} where to study plasmas which include the dynamics of colour and flavour at strong coupling. Finally, it would be of great interest to study the possibility of softly breaking SUSY in these backgrounds with flavour degrees of freedom.

As a final conclusion, it seems that in the absence of a string theory formulation on backgrounds dual to ${\cal N}=1$ field theories in four dimensions, the addition of extra D-branes to a supergravity background (in the probe approximation or more interestingly, taking into account their backreacted effects) captures nontrivial stringy information of the gauge theory dual. This is an important theme in the context of the gauge/gravity correspondence which deserves further research.

\vspace{1cm}

\centerline{{\large \textbf{Acknowledgements}}}

\vspace{1cm}

First of all, I would like to thank my supervisor
Alfonso V\'azquez Ramallo for giving me the opportunity
to carry out this Ph. D. thesis work at the University of
Santiago de Compostela. I am also indebted to Francesco Benini, Stefano Cremonesi, Jos\'e Edelstein, Paolo Merlatti, Carlos N\'u\~nez, Leopoldo A. Pando Zayas, \'Angel Paredes and Diana Vaman. I have really enjoyed collaborating with them during this years. I would also like to express my more sincere gratitude to Dario Martelli and again to Carlos N\'u\~nez by the attention that they paid to me and by all the physics that they taught me during my short-term visits to the CERN and to the University of Swansea respectively. Finally, I wish to thank 
Daniel Are\'an, I\~naki Garc\'ia-Etxebarr\'ia, Javier Mas, Frank Meyer and Jose M. Mu\~noz for stimulating discussions.



\end{document}